%% file: main.tex
\documentclass[11pt]{report}
\usepackage{upennstyle} 
\usepackage{tgpagella}
\usepackage{aas_macros}
\usepackage{xcolor}
\usepackage{multirow}
\usepackage{ifthen}
\usepackage{mathtools}
\usepackage{amssymb}
\usepackage{amsmath}
\usepackage{amsthm}

\input{misc/macros}

\input{misc/std_macros}

\usepackage[nonamebreak,round]{natbib}
\title{TOWARDS PRECISION PHOTOMETRIC TYPE IA SUPERNOVA COSMOLOGY WITH MACHINE LEARNING}
\author{Helen Qu}
\gradgroup{Physics and Astronomy} 
\date{2024} 
\supervisor{Masao Sako}
\supervisortitle{Professor, Physics and Astronomy}
\gradchair{Ravi Sheth, Professor, Physics and Astronomy}
\committee{Bhuvnesh Jain, Professor, Physics and Astronomy}
\committee{Gary Bernstein, Professor, Physics and Astronomy}
\committee{Mariangela Bernardi, Professor, Physics and Astronomy} 
\committee{Dylan Rankin, Assistant Professor, Physics and Astronomy} 

\authorlegal{Helen Qu} 

\dedication{For my family, whose love and encouragement means the universe.} 

\acknowledgement{
It has been an incredible privilege to be supported by countless people during the course of my PhD and the completion of this work.

This thesis and my entire academic career was made possible by my advisor, Masao Sako. Masao taught my first year undergraduate physics class in 2013 and gave me the confidence to pursue a degree in engineering. When I was contemplating graduate school, he took me in with little to no advanced physics education on just a promise that I could figure things out. His unwavering trust in my potential has made an irreconcilable difference in my life, and for that I am immeasurably grateful.

My committee members deserve additional thanks: Bhuv Jain, Gary Bernstein, Dylan Rankin, and Mariangela Bernardi, who have all enriched my graduate school experience far beyond your committee roles. Thanks in particular to Bhuv Jain, who has been incredibly supportive both in my academic forays into machine learning and on-the-ground development of a machine learning community within the department. Finally, many thanks to our faculty and staff for providing the most welcoming environment a graduate student could ask for.

My path to graduate school was shaped by many important individuals. In my final year of college, Prof. Max Mintz first incepted the idea of returning to school for physics, and his tireless love for learning continues to inspire me every day. My colleagues and mentors at Academia.edu have taught me so much, from the ins-and-outs of vim to the importance of stepping out of my comfort zone. In particular, I am incredibly grateful to Yuri Niyazov, Alexei Chernikov, and Davis Vu, whose impacts on my life reach far beyond my time at Academia.edu and my software engineering career. 

I have also been privileged to learn from and collaborate with a multitude of remarkable people over the years. Many thanks to wonderful friends, colleagues, and mentors in the DES-SN WG and LSST DESC - these have been the most welcoming communities to me and I always look forward to our biannual in-person meetings. I feel particularly fortunate to have been a part of the final DES-SN analysis, the culmination of decades of hard work by many. 

I am so grateful for the extraordinary researchers and human beings of Penn Physics and Astronomy. The support of my cohort of graduate students during our first year was invaluable: Justin Bracks, Yvonne Zagzag, Charlotte Slaughter, Eric Chong, Avi Kahn, Arpit Arora, Aaron Winn. I knew I could always ask for help, and I still think back to our long homework sessions together fondly. Since COVID reopening, I have been so lucky to find a community in the past and present residents of DRL 4N. Jason Lee, Qiuyue Liang, Minsu Park, Karen Perez Sarmiento, Julia Robe, Neha Joshi, Shivam Pandey, Guram Kartvelishvili, Tanvi Karwal, Cyrille Doux, Carles Sanchez, Marco Gatti, and Marco Raveri: our times together have been some of my fondest memories of graduate school, and I can't thank you all enough for your support in all aspects of my life.

My friends and family have been a vital source of unwavering love, encouragement, and support. Thank you to the residents of 1062 Oakridge for 1.5 years of unexpected time together: in the midst of a global pandemic, your companionship meant the world. Thank you to my mother, Hong, the most selfless person I know; and my father, Rongda, who sparked my interest in science from an early age. Their unconditional support has accompanied me every step of the way.

Finally, thank you to my partner Michael, who shows me daily that it is possible to strive for great things while giving selflessly to those you love.

} 

\abstract{
The revolutionary discovery of dark energy and accelerating cosmic expansion was made with just 42 type Ia supernovae (SNe Ia) in 1999. Since then, large synoptic surveys, e.g., Dark Energy Survey (DES), have observed thousands more SNe Ia and the upcoming Rubin Legacy Survey of Space and Time (LSST) and Roman Space Telescope promise to deliver millions in the next decade. This unprecedented data volume can produce the required precision to unambiguously test concordance cosmology which could represent a monumental shift in our understanding of dark energy and its role in cosmic history. However, extracting a pure SN Ia sample with accurate redshifts for such a large dataset will be a challenge. Specifically, spectroscopic classification will not be possible for the vast majority of discovered objects, and only $\sim 25$\% will have spectroscopic redshifts. This thesis presents a series of observational and methodological studies designed to address the questions associated with this new era of photometric SN Ia cosmology.

First, we present a machine learning method for photometric classification of SNe, Supernova Classification with a COnvolutional Neural Network (SCONE). Photometric classification enables SNe with no spectroscopic information to be confidently categorized, and is a critical component of current and future analysis pipelines. SCONE achieves $>99$\% accuracy distinguishing simulated SNe Ia from non-Ia SNe, and has been integrated into DES, LSST, and Roman analysis pipelines. We also demonstrate the efficacy of SCONE on early-time photometric classification, which will be vital for optimal allocation of spectroscopic resources. We show that SCONE can distinguish between 6 SN types with 75\% accuracy on the night of initial discovery, comparable to results in the literature for full-phase SNe.

Next, we study current methods for estimating SN Ia redshifts and propose a machine learning alternative that uses SN photometry alone to extract redshift information. Most SNe Ia inherit redshift information from their host galaxy, but the process of matching SNe to the correct host galaxy can be challenging. We systematically analyze the impact of incorrect redshifts from host galaxy mismatch on 5 years of DES SN data, and conclude that improved host matching or redshift estimation methods can reduce our systematic errors by $\sim 10$\%. In response to this finding, we present a SN photometry-only method for estimating redshifts independent of host galaxy information, Photo-zSNthesis. We show that Photo-zSNthesis redshift estimates are accurate to within 2\% across the full redshift range of LSST, a first in the literature.

Finally, we focus on the robustness of machine learning (ML) algorithms for real-world and scientific applications. ML models generalize poorly beyond their training set and often experience severe performance degradation when deployed on new data. We demonstrate a general method for improving robustness that achieves new state-of-the-art results on astronomical object classification, wildlife identification, and tumor detection.

} 

\begin{document}
\maketitle 
\setcounter{page}{2}

\makecopyright 

\makededication 

\makeacknowledgement 

\makeabstract
\tableofcontents

\clearpage \phantomsection \addcontentsline{toc}{chapter}{LIST OF TABLES} \begin{singlespacing} \listoftables \end{singlespacing}

\clearpage \phantomsection \addcontentsline{toc}{chapter}{LIST OF ILLUSTRATIONS} \begin{singlespacing} \listoffigures \end{singlespacing}


\begin{mainf} 
\chapter{Introduction}
\input{chapters/intro}

\chapter{SCONE: Supernova Classification with a Convolutional Neural Network}
\input{chapters/scone}

\chapter{Photometric Classification of Early-Time Supernova Lightcurves with SCONE}
\input{chapters/scone_early}

\chapter{The Dark Energy Survey Supernova Program: Cosmological Biases from Host Galaxy Mismatch of Type Ia Supernovae}
\input{chapters/host_mismatch}

\chapter{Photo-$z$SNthesis: Converting Type Ia Supernova Lightcurves to Redshift Estimates via Deep Learning}
\input{chapters/photoz}

\chapter{Connect Later: Improving Fine-tuning for Robustness with Targeted Augmentations}
\input{chapters/connect_later}

\chapter{Conclusions and Future Directions}
\input{chapters/conclusion}

\end{mainf}

\begin{append}

\end{append}

\begin{bibliof}
\bibliography{bibliography}
\end{bibliof}
\end{document}

%% file: misc/macros.tex
\usepackage{bm}

\def\vb{{\bm{b}}}

\def\vt{{\bm{t}}}

\def\vw{{\bm{w}}}

\def\mX{{\bm{X}}}

\newcommand{\loss}{\ell}
\newcommand{\losstext}{\text{loss}}
\newcommand{\pstar}{p^*}

\newcommand{\classification}{\textsc{AstroClassification}\xspace}
\newcommand{\redshifts}{\textsc{Redshifts}\xspace}
\newcommand{\iwildcam}{\textsc{iWildCam-WILDS}\xspace}
\newcommand{\camelyon}{\textsc{Camelyon17-WILDS}\xspace}
\newcommand{\sft}{standard fine-tuning\xspace}
\newcommand{\Sft}{Standard fine-tuning\xspace}

\newcommand{\augX}{\mX'}
\newcommand{\augXerr}{\mX'_{\text{err}}}
\newcommand{\Xin}{\mX}
\newcommand{\Xinerr}{\mX_{\text{err}}}
\newcommand{\znew}{z'}
\newcommand{\zorig}{z_{\text{orig}}}

\newcommand{\sourcedist}{P_S}
\newcommand{\targetdist}{P_T}
\newcommand{\unlabeldist}{P_U}

\newcommand{\inputx}{x}

\newcommand{\labelx}{y_x}
\newcommand{\inputxp}{{x'}}

\newcommand{\domainx}{d_x}

\newcommand{\numcls}{r}

\newcommand{\embeddim}{k}

\newcommand{\posx}{\inputx^+}
\newcommand{\pospairdist}{S_+}
\newcommand{\dattract}{d_+}
\newcommand{\drepel}{d_-}

\newcommand{\inputspace}{\sX}
\newcommand{\labelspace}{\sY}

\newcommand{\aug}{\mathcal{A}_{\text{pre}}}
\newcommand{\augft}{\mathcal{A}_{\text{ft}}}

\newcommand{\augproba}{{\alpha'}}
\newcommand{\augprobb}{{\beta'}}
\newcommand{\augprobg}{{\gamma'}}
\newcommand{\augprobr}{{\rho'}}

\newcommand{\pairproba}{{\alpha}}
\newcommand{\pairprobb}{{\beta}}
\newcommand{\pairprobg}{{\gamma}}
\newcommand{\pairprobr}{{\rho}}

\newcommand{\posclass}{1}
\newcommand{\negclass}{-1}

\newcommand{\lossft}{\losstext_{\text{ft}}}

\newcommand{\encoder}{\phi}
\newcommand{\empencoder}{\widehat{\encoder}}

\newcommand{\head}{h}

\newcommand{\emphead}{\widehat{\head}}

\newcommand{\clf}{f}
\newcommand{\empclf}{\widehat{\clf}}
\newcommand{\linmat}{B}
\newcommand{\emplinmat}{\widehat{\linmat}}
\newcommand{\empclferm}{\widehat{\clf}_\text{erm}}

\newcommand{\lpretrain}{\sL_\text{pretrain}}

\newcommand{\lfinetune}{\sL}

\newcommand{\Lzeroone}{\sL_{0-1}}
\newcommand{\Lerm}{\sL_\text{ERM}}

\DeclareMathOperator*{\argmin}{arg\,min}
\DeclareMathOperator*{\argmax}{arg\,max}

\newlength{\widebarargwidth}
\newlength{\widebarargheight}
\newlength{\widebarargdepth}

%% file: misc/std_macros.tex
\newcommand\sA{\ensuremath{\mathcal{A}}}

\newcommand\sL{\ensuremath{\mathcal{L}}}

\newcommand\sR{\ensuremath{\mathcal{R}}}
\newcommand\sS{\ensuremath{\mathcal{S}}}
\newcommand\sT{\ensuremath{\mathcal{T}}}

\newcommand\sX{\ensuremath{\mathcal{X}}}
\newcommand\sY{\ensuremath{\mathcal{Y}}}
\newcommand\sZ{\ensuremath{\mathcal{Z}}}

\newcommand\R{\ensuremath{\mathbb{R}}} 
\newcommand\Z{\ensuremath{\mathbb{Z}}} 

\ifthenelse{\isundefined{\definition}}{\newtheorem{definition}{Definition}}{}
\ifthenelse{\isundefined{\assumption}}{\newtheorem{assumption}{Assumption}}{}
\ifthenelse{\isundefined{\hypothesis}}{\newtheorem{hypothesis}{Hypothesis}}{}
\ifthenelse{\isundefined{\proposition}}{\newtheorem{proposition}{Proposition}}{}
\ifthenelse{\isundefined{\theorem}}{\newtheorem{theorem}{Theorem}}{}
\ifthenelse{\isundefined{\lemma}}{\newtheorem{lemma}{Lemma}}{}
\ifthenelse{\isundefined{\corollary}}{\newtheorem{corollary}{Corollary}}{}
\ifthenelse{\isundefined{\alg}}{\newtheorem{alg}{Algorithm}}{}
\ifthenelse{\isundefined{\example}}{\newtheorem{example}{Example}}{}
\newcommand{\E}{\ensuremath{\mathbb{E}}} 

%% file: chapters/intro.tex
\section{We live in an expanding universe!}
This year marks the $100^{\text{th}}$ anniversary of Edwin Hubble's monumental discovery that the universe doesn't stop at the edge of our Milky Way -- in fact, it contains ``hundreds of thousands'' of other stellar systems just like our own \citep{nyt}. In other words, the universe is vast beyond imagination. 

This marked the first of many notable discoveries made with \textit{standard candle} distance measurements: a distance estimation technique using apparent brightness measurements of objects with known luminosity. Hubble used Cepheid variable stars, which had been established as standard candles following the discovery of the period-luminosity relation \citep{leavitt}, enabling luminosity standardization for these stars. Estimated distances to 36 Cepheids in modern-day Andromeda and Triangulum galaxies revealed that these ``nebulae'' (now called galaxies) were far more distant than the edge of the Milky Way. Hubble concluded that these ``extra-galactic nebulae [...] are now recognized as systems complete in themselves'' \citep{extragalactic}: that ``spiral nebulae [...] are in reality distant stellar systems, or `island universes'" \citep{nyt}. 

That same year, Swedish astronomer Knut Lundmark became the first to measure distances to extragalactic nebulae, albeit using an unproven method of comparing apparent diameters to that of Andromeda, which he assumed had a ``known'' distance. He concluded that ``more distant spirals have higher space-velocity'' \citep{lundmark}, a foreshadowing of Hubble's next impactful discovery.

\begin{figure}
    \centering
    \includegraphics[scale=0.5]{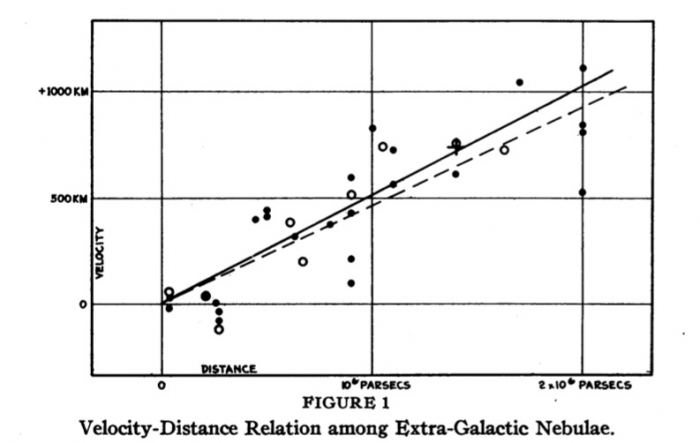}
    \caption{The near-linear velocity-distance relation discovered by Hubble through observations of ``standard candle'' Cepheid variable stars in 24 extragalactic nebulae \citep[adapted from][]{hubble_law}.}
    \label{fig:hubble-law}
\end{figure}

In the next five years, Hubble estimated distances to Cepheids in 24 extragalactic nebulae and combined these with radial velocities estimated from redshifts \citep{slipher} to discover that there was an approximately linear relationship between distance ($d$) and radial velocity ($v$) -- nebulae are moving away from us at a rate proportional to their distance \citep{hubble_law}. This simple relationship,
\begin{equation}
    v = H_0 d,
\end{equation}
was theorized two years earlier by a Belgian priest named Georges Lemaître \citep{lemaitre} and is known as the Hubble-Lemaître Law. These results provided the first observational proof for an expanding universe (Figure~\ref{fig:hubble-law}).

This conclusion was completely revolutionary for a scientific community, including Albert Einstein, that was convinced our universe was static. The theorized static universe consisted of gravitational attraction between massive bodies perfectly counterbalanced by a repulsive vacuum energy component, or ``cosmological constant'', a mathematical trick introduced by Einstein in the equations of general relativity to allow for static solutions. In light of Hubble's findings, Einstein abandoned his static universe and famously remarked in 1932 that the introduction of the cosmological constant was his ``greatest blunder''.

The notion of a cosmological constant was largely dismissed for the ensuing 60 years as the scientific consensus settled uneasily to that of a contracting universe, in which gravitational attraction above a critical matter density will eventually slow and reverse the expansion, causing the universe to collapse in on itself. However, measuring this deceleration parameter requires high-powered telescopes that can look further out and detect slight deviations to the linear velocity-distance relation found by Hubble. In the 1990s, when a combination of new technology (the 4-meter Victor Blanco telescope at Cerro Tololo) and new science (the replacement of Cepheids with their brighter cousins, type Ia supernovae, for distance measurement) finally made the first measurement of the deceleration parameter possible, scientists found the exact opposite of what they expected. Analyses by \citet{riess} and \citet{perlmutter} showed that the deceleration parameter was negative, indicating that the expansion of the universe is, in fact, \textit{accelerating}. This groundbreaking result once again revolutionized our understanding of the cosmos and was awarded the 2011 Nobel Prize in Physics. Today, this result has been bolstered by observations of thousands more type Ia supernovae as well as other cosmological probes, but the explanation for the accelerating expansion remains a mystery. A leading hypothesis is that of ``dark energy'', an unknown form of energy that drives accelerated cosmic expansion and may be explained by the infamous cosmological constant postulated by Einstein more than a century ago. 

\subsection{Cosmological Framework}
The theoretical groundwork for an expanding universe was laid in 1922, when Alexander Friedmann derived the now-famous Friedmann equations from general relativity. A universe that is homogeneous and isotropic on large scales can be described by the Friedmann-Lemaître-Robertson-Walker (FLRW) metric, in which the expansion (or contraction) of space is governed by a time-dependent scale factor, $a(t)$. Assuming the FLRW metric and a universe filled with a perfect fluid with energy density $\rho$ and pressure $p$, Friedmann reduced Einstein's field equations to two equations that model cosmic evolution:
\begin{equation}
    \frac{\ddot{a}}{a} = \frac{-4 \pi G}{3} (\rho(a) + 3p(a)) 
\end{equation}
\begin{equation}
\label{eq:friedmann}
    \Big( \frac{\dot{a}}{a} \Big)^2 \equiv H(a)^2 = \frac{8 \pi G}{3} \rho(a) - \frac{\kappa}{a^2}.
\end{equation}
Here, we set $c=1$, we define the Hubble parameter as $H(a) \equiv \frac{\dot{a}}{a}$, $\kappa$ parameterizes the curvature of the universe, and $G$ is Newton's gravitational constant.

Moreover, the perfect fluid assumption relates $\rho$ and $p$ via the equation of state $p = w \rho$, where $w$ depends on the nature of the fluid. Then, the solution to the fluid equation for arbitrary $w$ is 
\begin{equation}
    \rho(a) \propto \text{exp} \Big(3 \int_{a}^{1} \frac{1+w(a')}{a'} da' \Big).
\end{equation}
We know that the universe contains non-relativistic matter ($w_m = 0$) which dilutes as $\rho_m \propto a^{-3}$, and radiation ($w_r = 1/3$) which dilutes as $\rho_r \propto a^{-4}$. The current standard model of cosmology also postulates the presence of dark energy, which may be a cosmological constant ($w_{\Lambda}=-1$) but the possibility of a time-varying $w_{\Lambda}$ cannot be ruled out with the current data.

\begin{figure}
    \centering
    \includegraphics[scale=0.8]{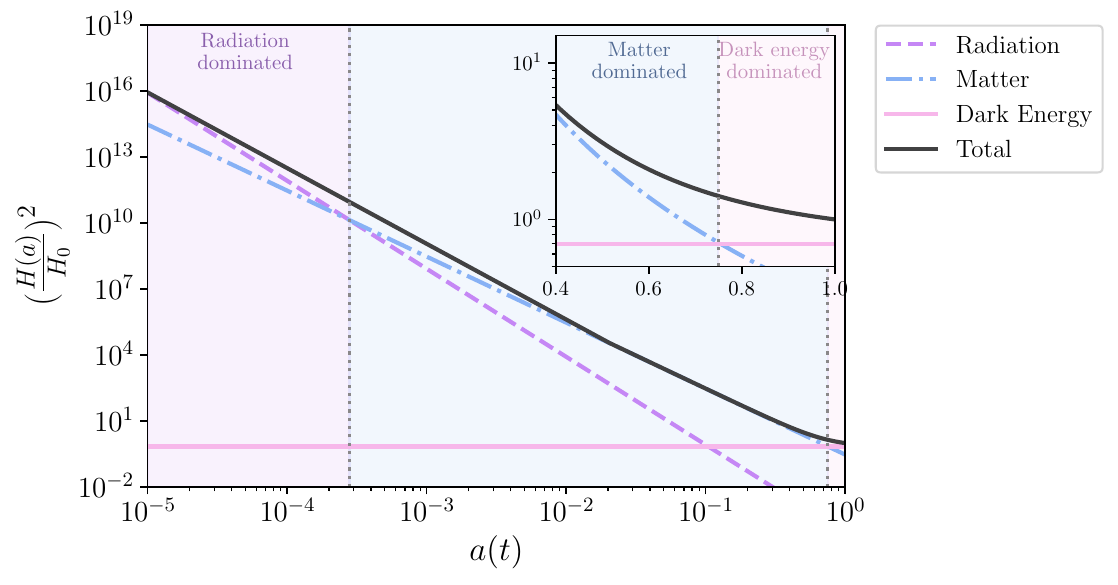}
    \caption{Evolution of radiation, matter, and dark energy (modeled as a cosmological constant) energy density as a function of the scale factor $a$.}
    \label{fig:universe-history}
\end{figure}

Finally, we introduce a convenient notation that expresses energy densities as a fraction of the critical energy density $\rho_c$, i.e. the total energy density for the universe to be flat ($\kappa=0$) given the current value of the Hubble parameter $H_0$:
\begin{equation}
     \Omega = \frac{\rho_0}{\rho_{c,0}} = \frac{8 \pi G \rho_0}{3H_0^2}
 \end{equation}
The 0 subscripts denote present-day values of each parameter.
We can now rewrite Equation~\ref{eq:friedmann} as
\begin{align}
\label{eq:hubble}
 H(a)^2 = H_0^2 (\Omega_m a^{-3} + \Omega_r a^{-4} + 
 \begin{cases}
     \Omega_{\Lambda} & \text{Cosmological Constant} \\
     \Omega_{\Lambda} a^{-3(1+w_{\Lambda})} & w_{\Lambda} \neq -1 \\
     \Omega_{\Lambda}\text{exp}(-3 \int_{a}^{1} \frac{1+w(a')}{a'} da') & \text{Time-varying}\; w_{\Lambda}
 \end{cases}
)
\end{align}

These 3 possibilities for $w_{\Lambda}$ and the nature of dark energy correspond to 3 of the most widely-accepted cosmological models today:
\begin{itemize}
    \item Flat $\Lambda$CDM: a flat universe consisting of non-relativistic matter, ultra-relativistic radiation, and a cosmological constant ($w_{\Lambda}=-1)$
    \item Flat $w$CDM: a flat universe consisting of non-relativistic matter, ultra-relativistic radiation, and dark energy described by a constant equation of state ($w_{\Lambda} \neq -1$ but constant)
    \item Flat $w_0 w_a$CDM: a flat universe consisting of non-relativistic matter, ultra-relativistic radiation, and dark energy described by a time-varying equation of state, specifically of the form $w_{\Lambda}(a) = w_0 + w_a (1-a)$
\end{itemize}

This formulation of $H(a)$ allows us to partition cosmic history into three discrete phases: a radiation-dominated phase at early times, followed by an extended matter-dominated phase, and finally, in the last $\sim 5$ billion years we've entered a dark energy-dominated phase (Figure~\ref{fig:universe-history}). Uncovering the nature and characteristics of dark energy, much of which remains a mystery today, is key to better understanding the future of our universe. A significant thrust of cosmology research today is towards increasingly precise measurements of $w_{\Lambda}$ to constrain these possibilities.

\subsection{Standard Candle Cosmology}
Leavitt and Hubble pioneered the use of standard candles, celestial objects with a known luminosity, to measure distances at cosmological scales more than a century ago. This technique enabled the discoveries of cosmic expansion as well as its acceleration, and is still used today to make state-of-the-art measurements of cosmological parameters, such as $\Omega_m, w_0,$ and $w_a$.

We know that the comoving distance $r$ traveled by light in an FLRW universe can be described by
\begin{equation}
    r = -c \int_{a_e}^{1} \frac{da}{a^2 H(a)},
\end{equation}
where $a_e$ is the scale factor at emission and $H(a)$ is defined as in Equation~\ref{eq:hubble}.
After a change of variables to the cosmological redshift, $z \equiv \frac{1}{a}-1$, we obtain
\begin{equation}
    r = \frac{c}{H_0} \int_{0}^{z_e} \frac{dz}{E(z)},
\end{equation}
where $E(z) \equiv H(z)/H_0$.

Now that we have an expression relating cosmological parameters to distances, how does the standard candle method measure distance?

We know that the luminosity distance to an object is expressed by
\begin{equation}
    d_L \equiv \sqrt{\frac{L}{4 \pi f}},
\end{equation}
where $L$ is an object's intrinsic luminosity and $f$ is its observed flux. In a flat, expanding universe, 
\begin{equation}
    r = d_L (1+z).
\end{equation}
Since intrinsic luminosity is known for standard candles, we can leverage this machinery to constrain the cosmological parameters encoded in $E(z)$.

\subsection{Complementary Cosmological Probes}
While standard candles facilitated many groundbreaking cosmological discoveries, our understanding of cosmology today rests on the foundation of multiple independent cosmological probes.

\paragraph{Cosmic Microwave Background} \mbox{}\\
The cosmic microwave background (CMB), discovered in 1965 by Arno Penzias and Robert Wilson, is the oldest light in the universe. It is a relic from the surface of last scattering, the moment when photons decoupled from the primordial plasma and the universe became transparent, around 370,000 years after the Big Bang or $z \sim 1100$. The CMB is one of the strongest pieces of observational evidence for the hot Big Bang theory and paved the way for further extensions (e.g. inflation) that explain the flatness, homogeneity, and isotropy of the universe.

\begin{figure}
    \centering
    \includegraphics[scale=0.4,trim={0 0.3cm 0 0},clip=true]{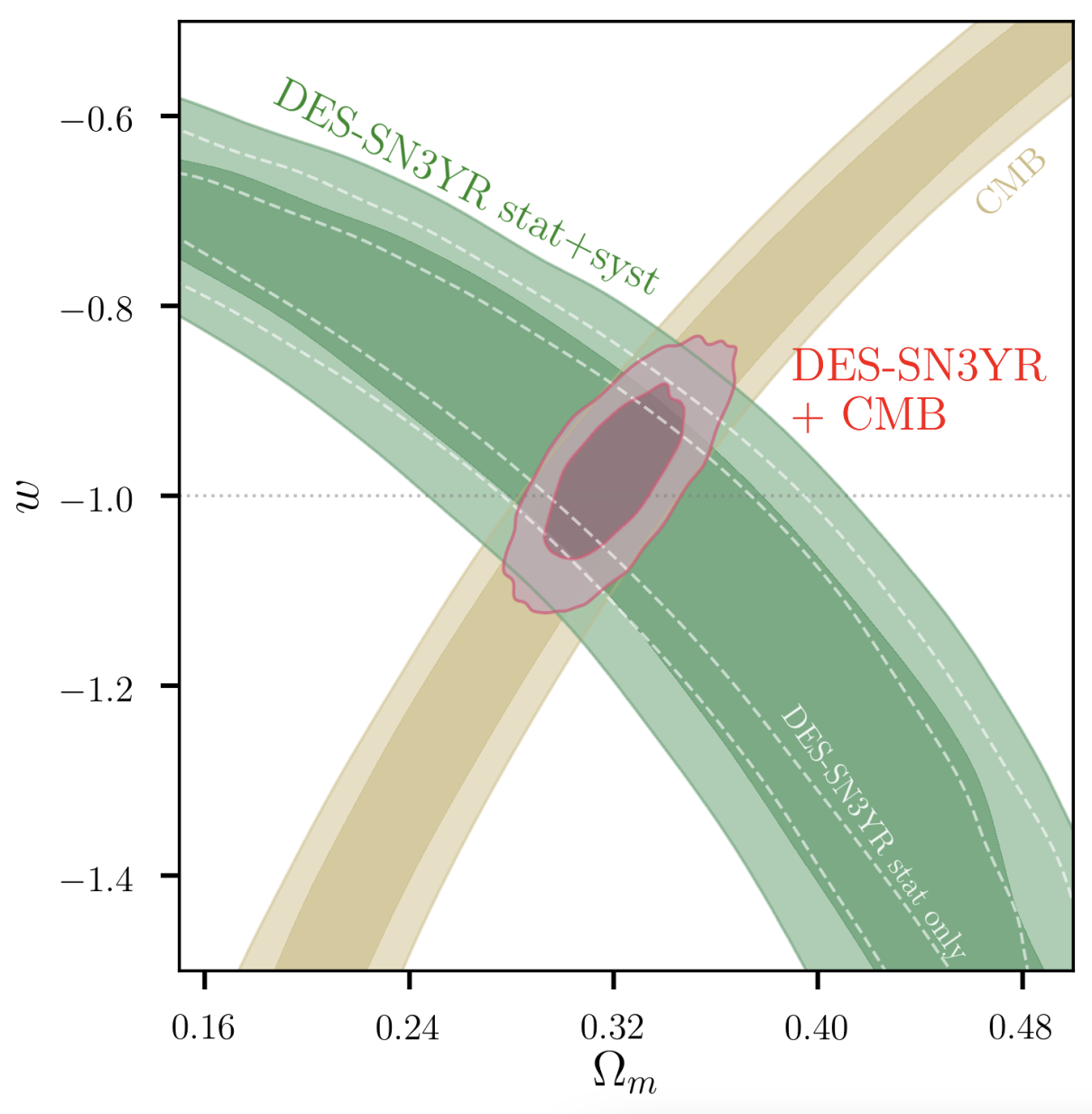}
    \caption{Constraints on $\Omega_m$-$w$ for the flat $w$CDM model (68\% and 95\% confidence intervals). SN contours are shown with only statistical uncertainty (white-dashed) and with total uncertainty (green-shaded). Constraints from CMB (brown) and 3-year DES-SN+CMB combined (red) are also shown. Adapted from \citet{des3yr}.}
    \label{fig:cmb-plus-sne}
\end{figure}

Several decades later, it was discovered that the CMB was not perfectly isotropic, and the temperature anisotropies of order 0.1\% are imprints of the primordial dark matter density distribution \citep{cmb_aniso}. The first peak of the CMB temperature power spectrum carries information about curvature as well as baryon fraction of the universe, a complementary probe to standard candles, which are more sensitive to total matter density and dark energy. Figure~\ref{fig:cmb-plus-sne} shows the complementary constraining power of CMB measurements compared with that of type Ia supernovae.

\paragraph{Baryon Acoustic Oscillations} \mbox{}\\
Analogous to the standard candle distance measurement technique where distance information is extracted from comparing intrinsic vs. apparent luminosity, ``standard rulers'' are known intrinsic length scales that can be compared with observations of that length scale. An example of this is the size of the sound horizon at the epoch of recombination. Matter oscillations under the opposing effects of gravity and outward radiation pressure prior to photon decoupling created a universal clustering scale, imprinted as the first peak in the matter density power spectrum. A single BAO measurement requires surveying orders of magnitude more galaxies than are needed for standard candle measurements, but are much less affected by systematic uncertainties and can achieve higher precision \citep[e.g.,][]{BAO}.

\section{Cosmology with Type Ia Supernovae}
Observations of type Ia supernovae (SNe Ia) led to the discovery of accelerating cosmic expansion \citep{riess, perlmutter}, and have become an established choice for cosmology with standard candles. They are standardizable ($<0.15$ mag scatter after standardization), bright and more easily visible at high redshifts ($\sim -19$ mag in rest-frame $B$ band), as well as frequent ($\sim$1 per galaxy per 2 centuries). 

Astrophysically, they are thermonuclear explosions of white dwarfs and spectroscopically identified as SNe lacking hydrogen but showing strong silicon absorption lines. Little else can be firmly established about the SN Ia progenitor population or the physical mechanism of explosion. They are traditionally thought to arise from two possible progenitor scenarios: (1) ``single degenerate'', in which a white dwarf accretes mass in a binary system and approaches the Chandrasekhar mass limit ($M_{\star} \sim 1.4 M_{\odot}$); and (2) ``double degenerate'', in which two white dwarfs in a binary system merge with combined mass greater than the Chandrasekhar limit. However, observations indicate a significant fraction of sub-Chandrasekhar mass explosions \citep[see][for a review]{progenitors}, clouding the picture.

\begin{figure}
    \centering
    \includegraphics[scale=0.5]{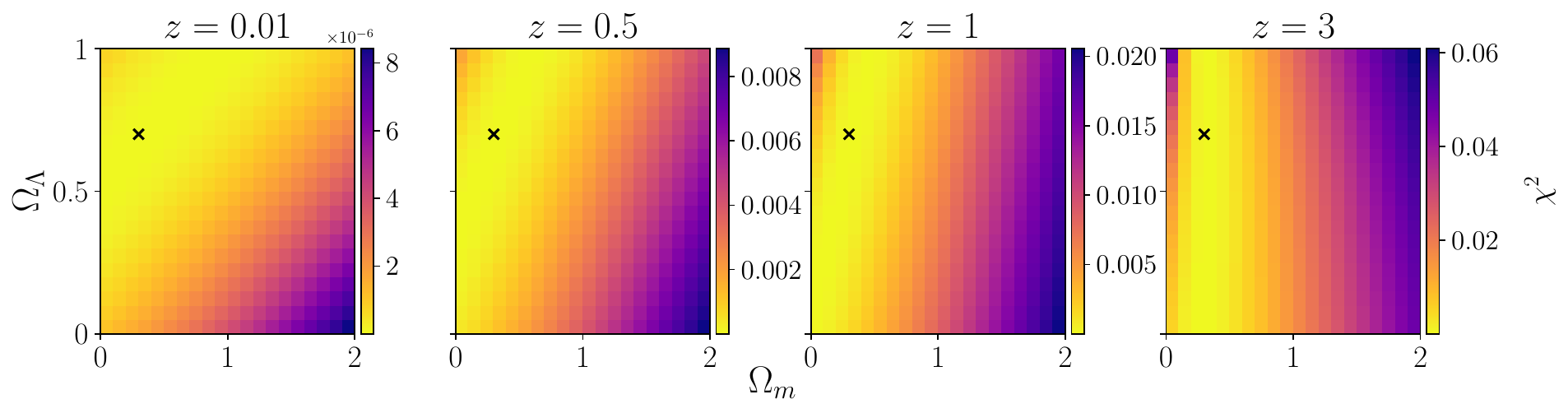}
    \caption{$\chi^2$ contours representing constraints on the $\Lambda$CDM model with a single observed SN Ia at different redshifts.}
    \label{fig:single-sn-constraints}
\end{figure}

\begin{figure}
    \centering
    \includegraphics[scale=0.38]{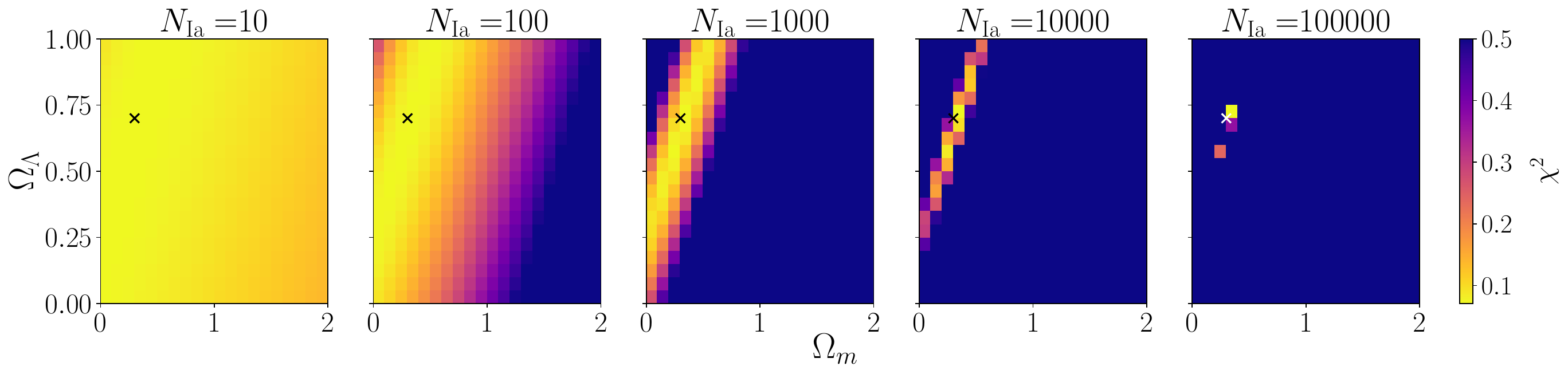}
    \caption{$\chi^2$ contours representing constraints on the $\Lambda$CDM model with $N_{\rm Ia}$ observed SNe Ia at uniformly sampled redshifts $z \in [0, 1]$.}
    \label{fig:many-sn-constraints}
\end{figure}

\subsection{Cosmological Parameter Constraints from SNe Ia}
The constraining power of SNe Ia is explored in Figures~\ref{fig:single-sn-constraints} and~\ref{fig:many-sn-constraints}. Figure~\ref{fig:single-sn-constraints} shows the cosmological constraining power of a single SN Ia observed at various redshifts. A single high redshift SN Ia is able to establish constraints on $\Omega_m$, but is not very effective at constraining $\Omega_{\Lambda}$. On the other hand, Figure~\ref{fig:many-sn-constraints} shows how the size of SNe Ia samples observed in the same redshift range affects their constraining power. 

\begin{figure}
    \centering
    \includegraphics[scale=0.35,trim={0 1cm 1cm 0},clip=true]{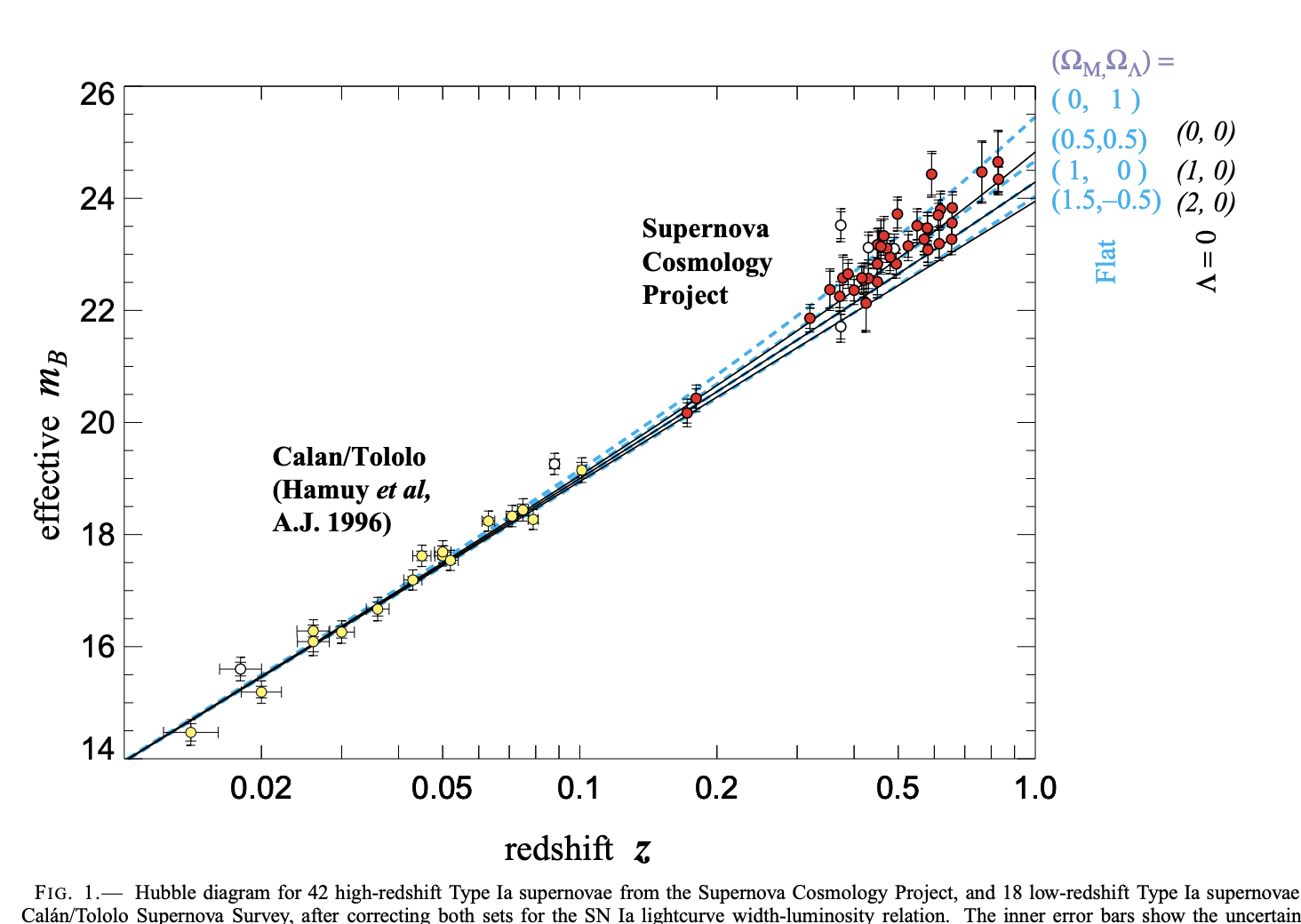}
    \includegraphics[scale=0.35, trim={0 1cm 1cm 0cm},clip=true]{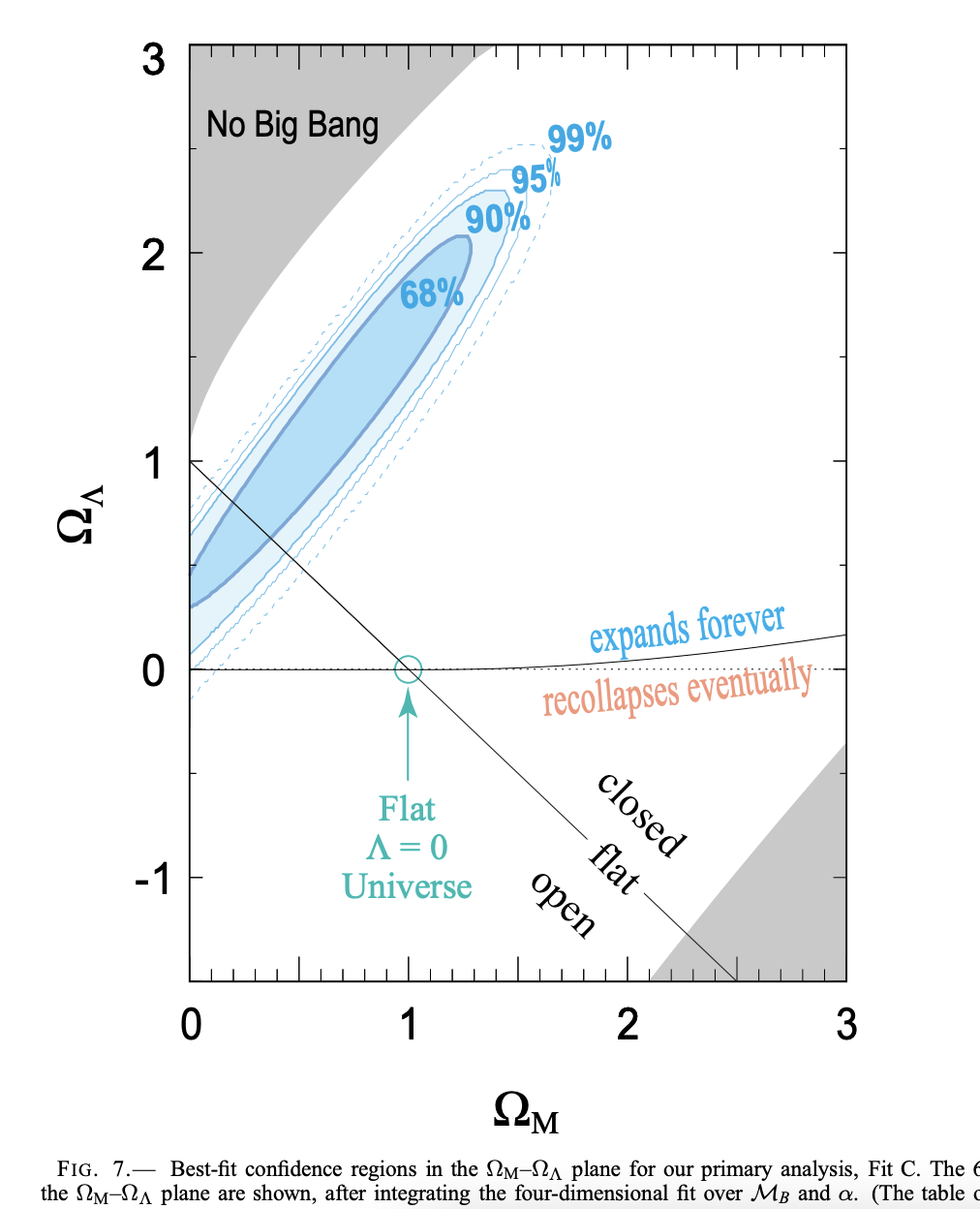}
    \includegraphics[scale=0.35, trim={0 1cm 1cm 0cm},clip=true]{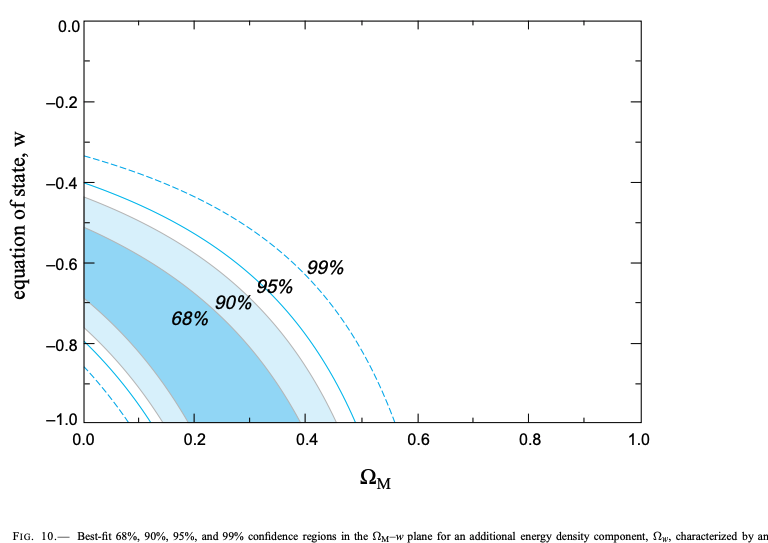}
    \caption{Results from \citet{perlmutter}: (top left) Hubble diagram of 42 high redshift SNe Ia from the Supernova Cosmology Project plotted over theoretical expectations for different choices of $(\Omega_{\Lambda}, \Omega_m)$; (top right) Constraints on $\Omega_{\Lambda}-\Omega_m$ for the $\Lambda$CDM model; (bottom) constraints on $w-\Omega_m$ for the flat $w$CDM model.}
    \label{fig:perlmutter-cosmo}
\end{figure}


The initial discovery of dark energy and accelerating cosmic expansion was made with only 42 \citep{perlmutter} and 16 \citep{riess} high-redshift ($0.18 < z < 0.82$) SNe Ia. The Hubble diagram made from the SNe Ia used in \citet{perlmutter} is reproduced in Figure~\ref{fig:perlmutter-cosmo} (top left). These results confidently rejected the dark energy-free, flat $\Lambda = 0$ universe model (Figure~\ref{fig:perlmutter-cosmo}, top right) and found $\Omega_m = 0.28^{+0.09}_{-0.08}$ assuming the flat $\Lambda$CDM model, but are not able to meaningfully constrain $(\Omega_{\Lambda}, \Omega_m)$ for the $\Lambda$CDM model or $(w, \Omega_m)$ for the flat $w$CDM model (Figure~\ref{fig:perlmutter-cosmo}, bottom).

We have observed thousands of SNe Ia in the 25 years since through surveys such as the SuperNova Legacy Survey \citep[SNLS,][]{snls}, Sloan Digital Sky Survey II \citep[SDSS-II][]{sdss-cosmo}, and Pan-STARRS \citep{ps1-cosmo}. These SNe Ia samples from individual surveys have been combined by e.g., the Joint Lightcurve Analysis \citep{jla} and Pantheon/Pantheon+ \citep{pantheon,pantheonplus} to derive strong cosmological constraints by creating the largest SNe Ia samples to date. 

However, though impressive SNe Ia discovery rates have steadily driven down statistical uncertainty, this unprecedented data volume as well as our improved understanding of SNe Ia continue to create new sources of systematic uncertainty. One example is the development and adoption of photometric classifiers for SN Ia type confirmation, replacing reliance on spectroscopic follow-up. However, classification error is significantly more likely with photometric data and has thus become an important systematic in recent analyses \citep[e.g.,][]{des5yr}. A thorough description of spectroscopic vs. photometric samples is given in Section~\ref{sec:spec-vs-phot}, and a review of photometric classification can be found in Section~\ref{sec:ml-cosmo}.

\begin{figure}[htp]
    \centering
    \includegraphics[scale=0.3,trim={0 1cm 1cm 0},clip=true]{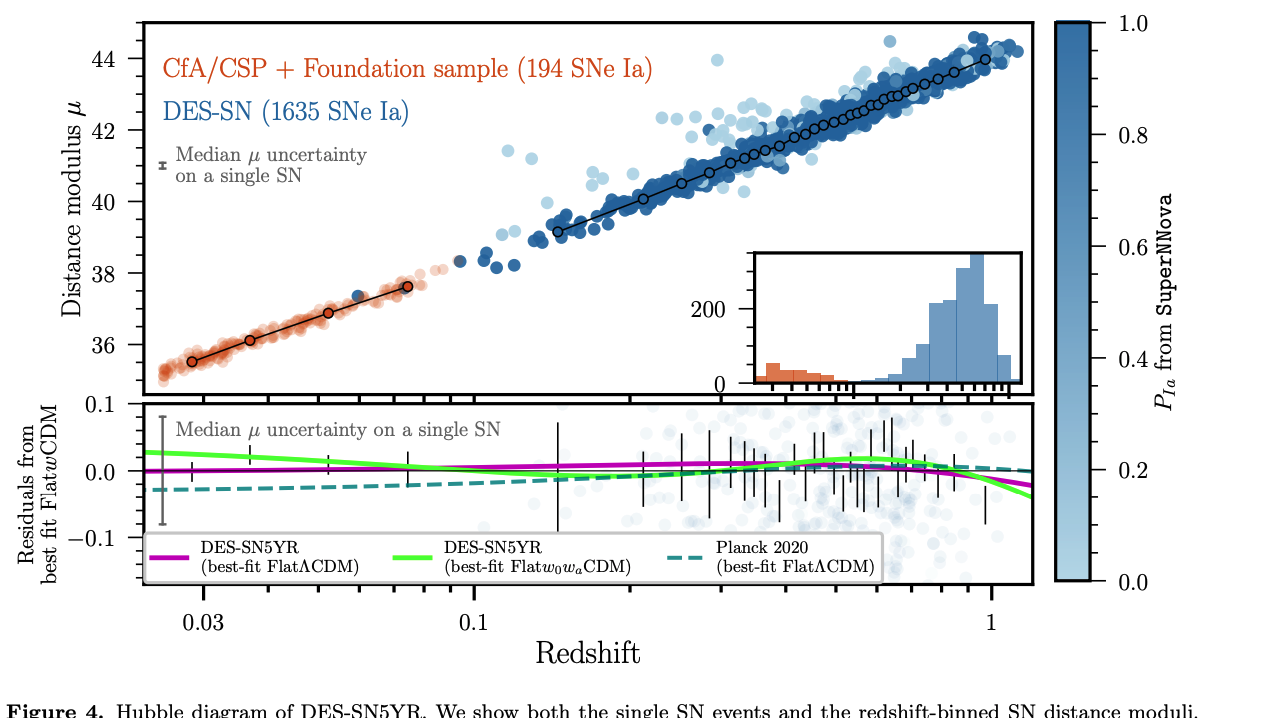}
    \includegraphics[scale=0.35, trim={0 1cm 1cm 0cm},clip=true]{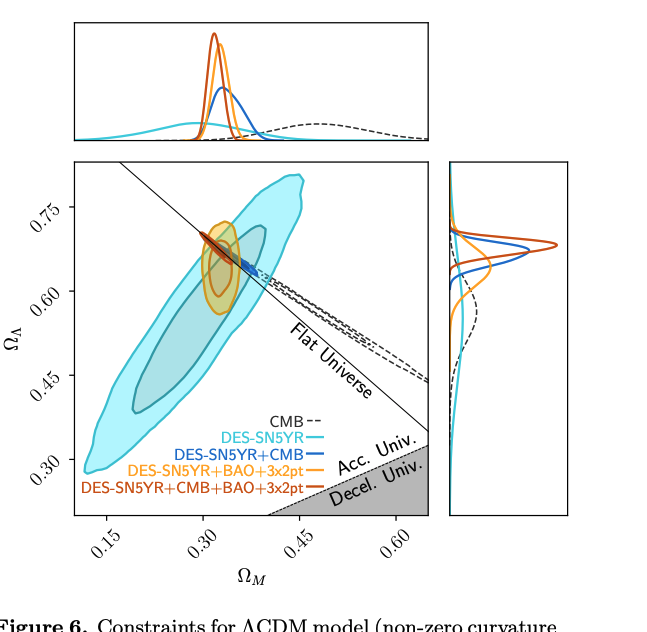}
    \includegraphics[scale=0.35, trim={0 1cm 1cm 0cm},clip=true]{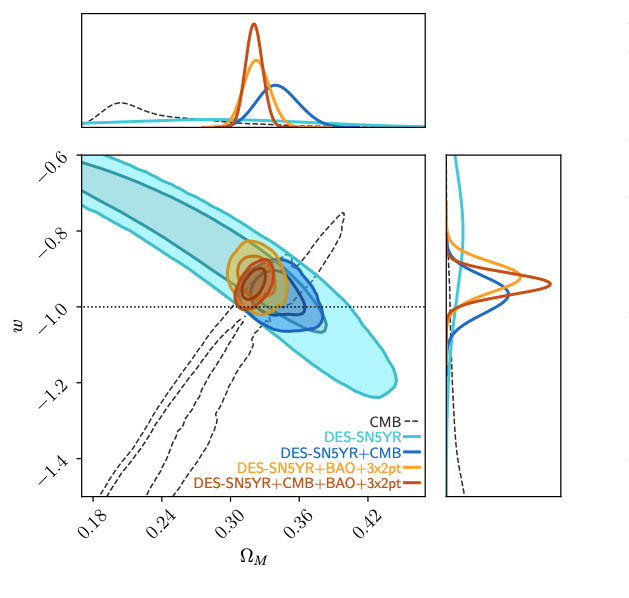}
    \caption{Results from \citet{des5yr}: (top) Hubble diagram of 1,635 high redshift SNe Ia from the DES 5-year (DES5YR) sample colored by the SN Ia probability assigned by the photometric classifier; (bottom left) Constraints on $\Omega_{\Lambda}-\Omega_m$ for the $\Lambda$CDM model; (bottom right) constraints on $w-\Omega_m$ for the flat $w$CDM model.}
    \label{fig:des-cosmo}
\end{figure}

This month, the Dark Energy Survey collaboration published its final cosmology analysis with all 5 years of observed SNe Ia, representing the largest sample of SNe Ia from a single survey to date \citep{des5yr}. This sample includes 1,635 photometrically classified SNe Ia in the range $0.10 < z < 1.13$, and reports $\Omega_m = 0.352 \pm 0.017$ for flat $\Lambda$CDM and $(w, \Omega_m) = (-0.80^{+0.14}_{-0.16},0.264^{+0.074}_{-0.096})$ for flat $w$CDM (Figure~\ref{fig:des-cosmo}), representing $>5\times$ improvement over the \citet{perlmutter,riess} results.

\begin{figure}
    \centering
    \includegraphics[scale=0.6, trim={0 0.5cm 0 0},clip=true]{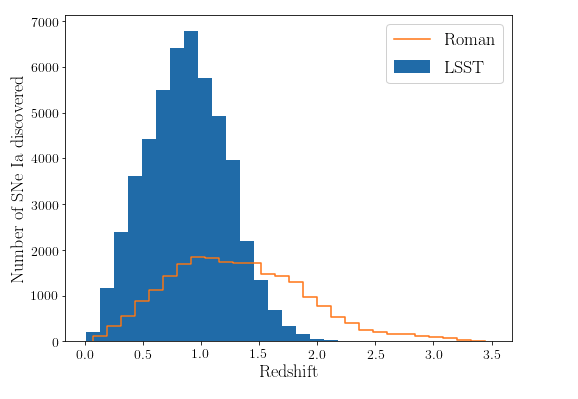}
    \caption{Projected redshift distribution of SNe Ia discovered by LSST and the Roman Space Telescope in one observing year.}
    \label{fig:lsst-roman-zdist}
\end{figure}

This is an exciting time for SN Ia cosmology, as two next-generation telescopes are scheduled to come online in the next 3 years: the Legacy Survey of Space and Time (LSST) at the Vera C. Rubin Observatory and the Nancy Grace Roman Space Telescope. Expected redshift distributions for one observing year are shown in Figure~\ref{fig:lsst-roman-zdist}. LSST is expected to observe 50,000 SNe Ia in a single year (one tenth of its planned observing lifetime), which is estimated to deliver better than 5\% constraints on $w$ (assumed constant) and constraints on $w_0$ to 0.05, $w_a$ to order unity for the $w_0 w_a$CDM model \citep{lsstsciencecollaboration2009lsst}. Roman will observe up to $10^4$ SNe Ia, significantly fewer than expected from LSST, but Roman SNe Ia will cover an unprecedented redshift range of up to $z \sim 3$. Roman data alone is expected to produce 1\% constraints on $w_0$ and 10\% constraints on $w_a$ \citep{rose2021reference}. 

\subsection{SN Ia Standardization}
Given our limited knowledge about SN Ia progenitors and explosion mechanisms, our luminosity standardization machinery is largely empirically derived. Intrinsically, SNe Ia are a relatively diverse population with luminosity scatter $\sim 0.8$ mag. However, two relations were observed that enabled standardization to within $\sim 0.15$ mag and firmly established SNe Ia as standardizable candles.

\begin{figure}
    \centering
    \includegraphics[scale=0.7,trim={0 0.25cm 0 0.25cm},clip=true]{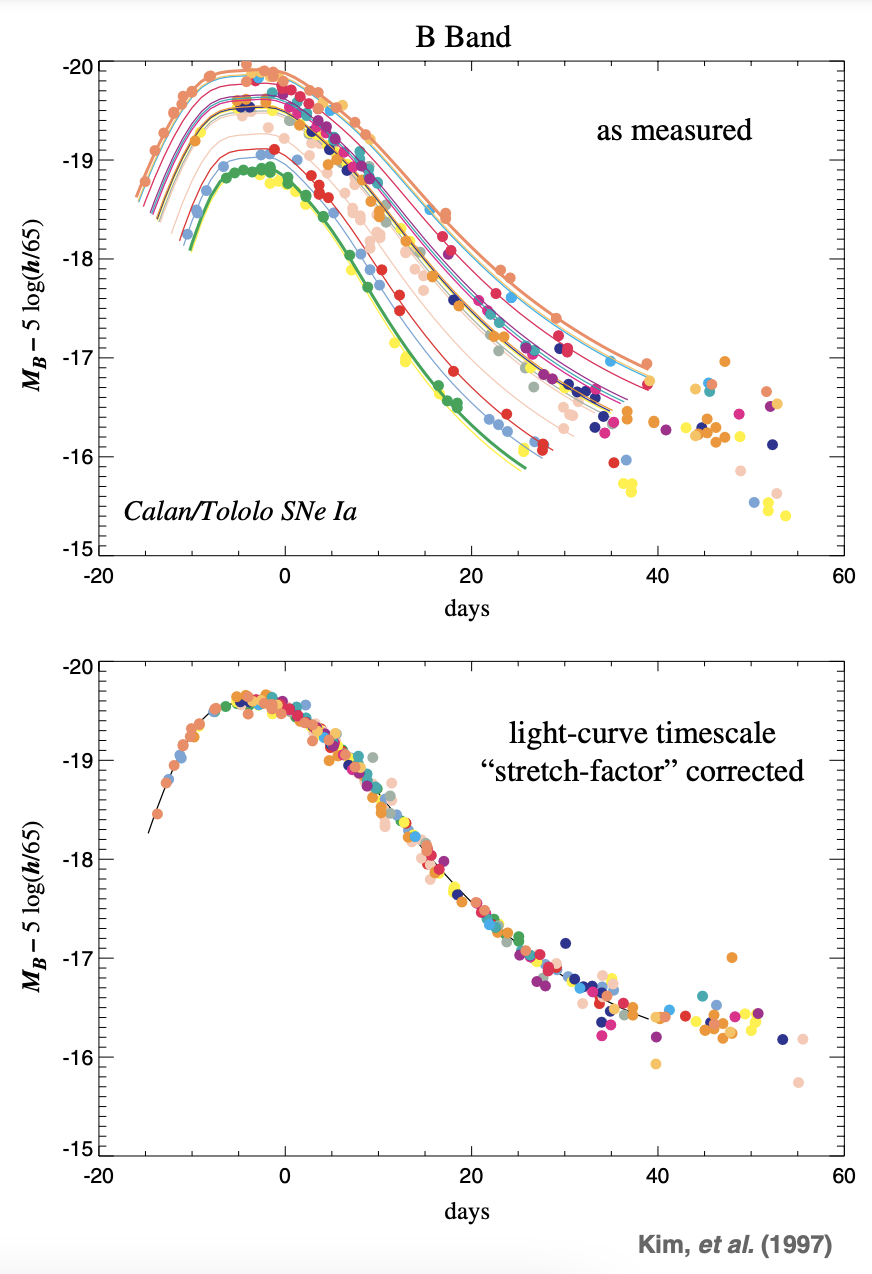}
    \caption{\textit{Upper panel:} Lightcurves of low-redshift supernovae discovered by the Calan/Tololo Supernova Survey. \textit{Lower panel:} The same lightcurves after calibrating the supernova brightness using the “stretch” of the timescale of the lightcurve as an indicator of brightness and the color at peak as an indicator of dust absorption. Adapted from \citet{perlmutter-nobel}.}
    \label{fig:standardization}
\end{figure}

\paragraph{Stretch-luminosity correction.}
\citet{stretch-l} observed that decline rate and lightcurve width (i.e. explosion duration) was well-correlated with SN Ia luminosity. \citet{stretch} later showed that equivalently, fitting a ``stretch'' parameter representing the stretching of the timescale of an event could be used to effectively normalize SNe luminosities to within $\sim 0.19$ mag (Figure~\ref{fig:standardization}).

\paragraph{Color-luminosity correction.}
$B-V$ color was also found to correlate with SNe Ia luminosity, with bluer SNe appearing brighter than redder ones. This may be due to intrinsic physical properties of SNe Ia or to effects of dust in the host galaxy, but is able to be corrected with a simple linear relationship to decrease scatter to within $\sim 0.15$ mag \citep{tripp}.

While these corrections significantly improved the standardizability of SNe Ia, intrinsic scatter is still a sizable component of our modeling error budget today and an active area of study.

\subsection{From Fluxes to Distances to Cosmology}
\label{sec:dist-to-cosmo}
We compute the standardized peak flux of a SN Ia by applying the stretch and color corrections described above, as well as a host galaxy correction:
\begin{equation}
\label{eq:standardization}
    m_B^{\text{standardized}} = m_B + \alpha x_1 - \beta c + \gamma G_{\text{host}}.
\end{equation}
Here, $m_B$ is the measured flux in magnitudes, $x_1,c$ are the stretch and color parameters described above, and $\alpha, \beta$ set the amplitude of these corrections. $\gamma G_{\text{host}}$ is a further correction originating from observations that SNe Ia occurring in high-mass galaxies ($M_{\star} > 10^{10} M_{\odot}$) are $\sim 0.07$ mag fainter than SNe in lower mass galaxies \citep{kelly2010, mass_step, jla, smith2020, kelsey2021, kelsey2023}. We model this dependence as 
\begin{equation}
\label{eq:mass-step}
    \gamma G_{\text{host}} = \begin{cases}
        +1/2 \gamma & M_{\star} > M_{\text{step}}\\
        -1/2 \gamma & \text{otherwise},
    \end{cases}
\end{equation}
where $\gamma$ sets the size of the correction and  $M_{\text{step}}$ is generally fixed to $10^{10} M_{\odot}$.

\paragraph{Lightcurve Fitting.} Since we observe SNe Ia at various redshifts in fixed rest-frame wavelength bands and often miss the epoch of peak brightness, we leverage a lightcurve fitting approach to standardize these observations. An established lightcurve fitting method is SALT2 \citep{salt2}, which attempts to model SNe Ia spectral evolution with time by finding best-fit matrices $F_0(t, \lambda), F_1(t, \lambda)$ of flux estimates on a grid of times $t$ and wavelengths $\lambda$. Once trained, these matrices are combined with coefficients $x_0, x_1$ that represent the overall scale of each component, and $c$, a coefficient to the fitted color correction relation. These fitted parameters are used in Equation~\ref{eq:standardization} for luminosity standardization.

\paragraph{Distance estimation.} To represent SN distances, we define the observed distance modulus as 
\begin{equation}
    \mu_{\text{obs}} = m^{\text{standardized}} - M + \Delta \mu_{\text{bias}},
\end{equation}
where $m^{\text{standardized}}$ is the standardized SN peak magnitude described above, $M$ is the absolute magnitude of a SN Ia with $x_1 = 0, c=0$, and $\Delta \mu_{\text{bias}}$ is a bias correction term applied to address survey selection effects \citep[see e.g.,][for a current approach]{bbc}. $\mu_{\text{obs}}$ is simply defined for convenience and related to luminosity distance as $\mu_{\text{obs}} \sim \textrm{log}_{10}\;d_L$.

\paragraph{Cosmological parameter measurement.} To compare our observed distance moduli with theoretically predicted distance moduli $\mu_{\rm theory}(\Theta)$ computed from assumed cosmological parameters $\Theta$, we calculate
\begin{equation}
    \Delta \mu_i = \mu_{\text{obs},i} - \mu_{\rm theory}(\Theta, z_i).
\end{equation}
for each $i^{\rm th}$ SN Ia.
We then minimize 
\begin{equation}
    \chi^2 = \Delta \mu_i \mathcal{C}_{ij}^{-1} \Delta \mu_j^{\rm T},
\end{equation}
where $\mathcal{C}^{-1}$ is the inverse covariance matrix, to determine the best fit cosmological parameters. The covariance matrix accounts for all of our statistical and modeling-based (systematic) uncertainties \citep{Conley_2011}.

Finally, we note that SN Ia absolute magnitude $M$ is completely degenerate with the Hubble constant $H_0$, leading SN Ia-only analyses to define $\mathcal{M} \equiv M + 5 \text{log}_{10} (c/H_0)$ and marginalize over this term. 

\subsection{Spectroscopic vs. Photometric Samples}
\label{sec:spec-vs-phot}
Historically, spectroscopic follow-up observation has been crucial for determining a host of properties for any observed object. \textit{Spectroscopy} provides a granular view of object flux as a function of wavelength, which can reveal chemical components of observed objects (e.g., emission/absorption lines), velocity (via broadening of emission lines), and redshift (via shifts in spectral features compared to a template spectrum). However, collecting enough light in each fine-grained wavelength bin requires long integration times [X min for Y mag object], which can be prohibitive for faint, distant objects. In contrast, broad-band \textit{photometry} provides an inexpensive [Z min for Y mag object] but coarse-grained view. Most survey telescopes primarily collect data photometrically through 1-6 broad-band filters that cover complementary ranges of wavelengths. In this regime, information about specific spectral features is difficult to extract, rendering redshift estimation and object typing a challenge.

Traditional SN Ia cosmology analyses relied on spectroscopic follow-up for two reasons: (1) confirmation that an object is in fact a SN Ia, and (2) a precise redshift measurement for either the SN or its host galaxy. However, as time domain surveys become wider, faster, and deeper, our ability to photometrically discover new SNe Ia as well as other time domain events is far outpacing our spectroscopic resource availability.  $<0.1$\% of SNe Ia observed by the upcoming Rubin Legacy Survey of Space and Time (LSST) will be followed up spectroscopically. This has led to two distinct types of SN samples:
\begin{itemize}
    \item Spectroscopic SN Ia samples, composed of spectroscopically confirmed SNe Ia;
    \item Photometric SN samples, composed of SNe that do not have spectroscopic type confirmation but generally have a spectroscopic redshift from its likely host galaxy.
\end{itemize}
Spectroscopic samples have high guaranteed purity but a severely limited sample size, driving up statistical uncertainty of parameter estimates. Photometric samples, on the other hand, are much larger but may contain objects that are not SNe Ia. These non-Ia ``contaminants'' are not standard candles and will bias cosmological parameter estimates, impacting our systematic uncertainty.

The past decade has seen a significant effort to develop accurate photometric SN classification methods, ranging from template matching \citep{psnid} to deep learning \citep{snn} (see Section~\ref{sec:ml-cosmo} for further details). In conjunction with new statistical methods for marginalizing over non-Ia contamination in the cosmological parameter fitting process \citep{bbc}, these classifiers have made precision cosmology with photometric samples possible.

\subsection{Host Galaxy Correlations}
The type and properties of SNe have been shown to depend heavily on its host galaxy environment. Broadly, since core-collapse SNe have massive ($>8 M_{\odot}$) progenitors, they are almost exclusively hosted by gas-rich, star-forming galaxies. In contrast, SNe Ia arise from white dwarfs and have been observed to occur in a variety of host galaxy environments. In fact, \citet{foleymandel} and \citet{ghost} found that host galaxy information alone is sufficient to perform accurate classification of supernova types.

In the context of SN Ia standardization, host galaxy correlations have played an important role. Numerous studies have observed that higher stellar mass, passive galaxies tend to host brighter/slower declining SNe Ia, whereas SNe Ia found in star-forming galaxies are dimmer/fast declining \citep[e.g.,][] {lampeitl, sullivan_dependence}. This so-called ``mass step'' has persisted over a decade of study though its physical origins remain mysterious, and has become enshrined in the canon of SN Ia standardization (Equation~\ref{eq:mass-step}).

Host galaxies are also the primary source of accurate redshifts for photometrically classified SNe Ia, which is vital for determining the distance-redshift relation that we use to estimate cosmological parameters. Procuring spectroscopic redshift measurements for large numbers of SN host galaxies has been tractable so far, since spectroscopic follow-up of galaxies is not time-sensitive and large (and growing) catalogs of galaxy redshifts are available in the literature.

Finally, identifying the correct host galaxy for each SN Ia is crucial to ensuring accurate host galaxy correlations and redshifts. Host galaxy matching is nontrivial, however, since SNe can occur in crowded fields or in faint galaxies that are below the detection limit. In certain cases, host galaxy mismatch can even lead to nontrivial biases in estimated cosmological parameters. I provide an overview of current host matching methods as well as the impact of mismatch on cosmology in Chapter 4.

\section{Cosmology in the Era of Big Data}
In the next three years, both the Rubin Legacy Survey of Space and Time (LSST) and the Roman Space Telescope will be streaming science-grade data. This is an exciting time for time domain studies and SN Ia cosmology: these surveys promise to deliver millions of time domain objects, including hundreds of thousands of SNe Ia, over their observing lifetimes. This data deluge represents $100\times$ the number of time domain objects ever observed, and presents immense promise as well as new challenges. Machine learning has a unique ability to derive insights from massive amounts of data, and may be the key to maximizing science returns from future survey data.

\subsection{The Role of Machine Learning}

Machine learning is a non-parametric, data-driven approach to function approximation. Specifically, a machine learning algorithm defines a transformation $\hat{f}(x,\theta)$ to approximate the true function $f(x)$. The objective of the learning process is to learn parameters $\theta$ that minimize a loss function $L(f(x), \hat{f}(x,\theta))$. Traditional machine learning techniques generally operate on tabular data, where an input (e.g., an SN Ia event) is summarized by a collection of expert-designed \textit{features} (e.g., peak magnitude, stretch, color, etc.). On the other hand, \textit{deep learning} provides the additional benefit of learning the best features for a particular task, allowing raw data (e.g., supernova time-series, galaxy image) to act as the input with minimal expert guidance. Deep learning models can generally be thought of as a composition of functions, e.g., $\hat{f} = \hat{f}^{(n)} \circ ... \circ \hat{f}^{(2)} \circ \hat{f}^{(1)}$, where $\hat{f}^{(1)}$ defines the first \textit{layer} of the model, etc. This deep, layered architecture allows intermediate layers to build off of previous layer outputs and learn more abstract or complicated concepts. 

The power and flexibility of deep learning approaches has led to a proliferation of architectures tailored to diverse data types. For example, convolutional neural networks \citep{lecun, alexnet, resnet} are designed to leverage the spatial hierarchy and local patterns inherent in images, while transformers \citep{vaswani2017attention} are well-suited for sequential data due to their ability to capture long-range dependencies and relationships within sequences.

\paragraph{Robust Machine Learning.} In real-world scenarios, machine learning models are often deployed on data that differ from the training data. Machine learning is well known to generalize poorly outside of its training data distribution, and performance on these out-of-distribution (OOD) datasets can drop dramatically \citep{quinonero2009dataset, koh2021wilds}. This is of utmost importance to scientific applications -- we often have large unlabeled datasets (e.g., real data from telescopes) that we use machine learning models to sort through (e.g., photometric SN classification). However, these machine learning models are traditionally trained on a different data distribution (e.g., simulated or a labeled subset of telescope data), with no means of precisely measuring model performance on the real data. In astronomy, the Photometric LSST Astronomical Time-Series Classification Challenge \citep[PLAsTiCC,][]{theplasticcteam2018photometric} dataset attempts to simulate this by creating a training set modeled after a spectroscopically confirmed dataset of time-varying objects (lower redshift, brighter), while the test set represents the full photometric dataset of these objects (higher redshift, fainter). Improving robustness to OOD inputs is an active area of study in the machine learning community and is the subject of Chapter 6.

\subsection{Machine Learning in Observational Cosmology}
\label{sec:ml-cosmo}
In recent years, the unprecedented data volume of wide-field sky surveys has highlighted the need for fast, automated, and accurate algorithms at every stage of the pipeline, from object detection in telescope images to cosmological parameter inference.

\paragraph{Transient detection.} The current technique to detect time-varying objects in astronomical survey images, difference imaging, involves subtracting pixel values of each telescope image from a ``template'' image of only constant light sources (e.g., galaxies) \citep{diffimg}. This process requires precise alignment of the image coordinates and point spread functions, which is extremely computationally expensive and prone to failure. A class of machine learning-based ``real-bogus'' detection algorithms emerged to distinguish difference imaging artifacts (bogus detections) from actual time-varying object detections \citep[e.g.,][]{autoscan}. Recent work has focused on real-bogus detection with the search and template images only \citep{Acero_Cuellar_2023}, opening the door to potential future approaches that can perform transient detection without difference imaging altogether.


\paragraph{Transient classification.} After detection, objects must be classified by type for further analysis. In Section~\ref{sec:spec-vs-phot}, I described the importance of spectroscopic follow-up for determining SN type information, and the limitations of scaling spectroscopic resources to modern photometric data volumes. In the absence of spectroscopic information for the vast majority of discovered objects, attention has turned in recent years to the role of computational methods and machine learning to extract that information from photometry. Early methods used a template fitting approach that compares an input against known SN Ia and core-collapse SN templates to find the best match \citep[e.g.,][]{psnid}. Recent years have seen significant interest in machine learning-based photometric classifiers with increasingly impressive reported performance \citep{snn,rapid,avocado,Villar_2020,parsnip,villar2023hierarchical}. Some of these models \citep[e.g.,][]{psnid, snn} have been successfully incorporated into SN Ia cosmology analysis pipelines to extract a photometrically classified SN Ia sample \citep{Campbell_2013,jones_2017,Sako_2018,des5yr}. However, practical concerns such as long training times/large training sets, ease of use, and generalizability from simulated to real data are still common issues plaguing these approaches.

\paragraph{Host Galaxy Matching.} Information about an event's host galaxy environment can be very useful for understanding its properties and physical progenitor scenarios. Host galaxy matching can be a nontrivial problem, however, and traditional methods, such as the directional light radius method \citep[DLR,][]{dlr}, are prone to occasional catastrophic failure. \citet{Gupta_2016} showed that adding a random forest classifier to detect incorrect matches on top of DLR-identified host galaxies can improve host matching accuracy. A few recent studies introduce novel machine learning-based solutions to host galaxy matching, \citep[e.g.,][]{ghost,delight}, and could be an interesting avenue for future work. 

\paragraph{Distance/Redshift Estimation for SNe Ia.} Redshifts for SNe Ia have traditionally come from either live spectra of the SNe themselves or spectroscopic redshifts from their matched host galaxies. In parallel with improving host galaxy matching, another line of work has emerged on photometric redshift estimation from SNe themselves. An extension to the SALT2 lightcurve fitting model is a popular and well-studied approach \citep{lcfit+z}, but degeneracies between redshift and other fitted properties (e.g., color) can lead to poor fits. \citet{oliveira} explored applying machine learning to this problem, but most redshift estimation techniques in the literature suffer heavily from redshift-dependent bias (i.e., predictions tend toward the mean). Estimation of a joint posterior over distance and redshift using machine learning techniques (e.g. neural density estimators) is a potential direction for future work.

\paragraph{Cosmological Parameter Estimation.} As described in Section~\ref{sec:dist-to-cosmo}, traditional parameter estimation uses $\chi^2$ minimization to determine a posterior distribution over best-fit cosmological parameters given the data. However, this makes the strong assumption that our likelihood is Gaussian. An alternative approach known as \textit{simulation-based inference} (SBI) does not require such assumptions (see \citet{Cranmer_2020} for a review). This method was explored for SN Ia cosmology by \citet{jennings2016new}, an approximate Bayesian computation approach that requires a ``distance metric'' over the simulated vs. observed raw data, which is nontrivial to define. Exploring modern machine learning-based SBI techniques for principled posterior inference is a promising direction for future work.

\section{Motivation and Overview}
In the preceding sections, I presented a brief overview of SN Ia cosmology today and my view of the opportunities and challenges ahead. In summary, I believe that the data deluge of LSST and the Roman Space Telescope will pose substantial challenges for our statistical frameworks and infrastructure, especially our heavy reliance on spectroscopy to elucidate SN properties. 
Machine learning techniques offer a way forward by learning spectroscopic properties from photometric data, and have already enabled the shift away from spectroscopic confirmation of supernova type \citep{des5yr}. 
Photometric SN cosmology in the LSST era, however, will also require an optimized spectroscopic targeting program and accurate redshift estimates for SNe Ia. Further gains in cosmological constraints could come from further emphasis on robust machine learning, e.g., for photometric classification; and improved statistical methodology for parameter inference, such as with machine learning-accelerated simulation-based inference.
The development of the photometric supernova cosmology toolkit will rely on harnessing the power of machine learning and other cutting-edge technologies. My research, outlined in this thesis, has focused on developing practical, impactful machine learning tools to accelerate this transition:
\begin{enumerate}
    \item I developed SCONE, a quick-to-train, lightweight convolutional neural network approach to photometric SN classification (Chapter 2);
    \item I demonstrated SCONE's classification performance on early epoch SN photometry for use cases in spectroscopic targeting and allocation (Chapter 3);
    \item I analyzed the impact of incorrect SN redshifts from mismatched host galaxies on inferred cosmological parameters in the DES 5-year SN analysis (Chapter 4);
    \item I developed Photo-zSNthesis, a novel photomteric redshift estimation algorithm for SNe Ia (Chapter 5);
    \item I developed Connect Later, a general framework for improving model robustness under distribution shift (Chapter 6).
\end{enumerate}

%% file: chapters/scone.tex
\newcommand{\scone}{{\small SCONE}}

\section*{Abstract}
    We present a novel method of classifying Type Ia supernovae using convolutional neural networks, a neural network framework typically used for image recognition. Our model is trained on photometric information only, eliminating the need for accurate redshift data. Photometric data is pre-processed via 2D Gaussian process regression into two-dimensional images created from flux values at each location in wavelength-time space. These ``flux heatmaps" of each supernova detection, along with ``uncertainty heatmaps" of the Gaussian process uncertainty, constitute the dataset for our model. This preprocessing step not only smooths over irregular sampling rates between filters but also allows \scone\ to be independent of the filter set on which it was trained. Our model has achieved impressive performance without redshift on the in-distribution SNIa classification problem: $99.73 \pm 0.26$\% test accuracy with no over/underfitting on a subset of supernovae from PLAsTiCC's unblinded test dataset. We have also achieved $98.18 \pm 0.3$\% test accuracy performing 6-way classification of supernovae by type. The out-of-distribution performance does not fully match the in-distribution results, suggesting that the detailed characteristics of the training sample in comparison to the test sample have a big impact on the performance.  We discuss the implication and directions for future work. All of the data processing and model code developed for this paper can be found in the \href{https://github.com/helenqu/scone\ }{\scone\ software package} located at github.com/helenqu/scone \citep{helen_qu_2021_4660288}.

\section{Introduction}
The discovery of the accelerating expansion of the universe \citep{perlmutter,riess} has led to an era of sky surveys designed to probe the nature of dark energy. Type Ia supernovae (SNe Ia) have been instrumental to this effort due to their standard brightness and light curve profiles. Building a robust dataset of SNe Ia across a wide range of redshifts will allow for the construction of an accurate Hubble diagram that will enrich our understanding of the expansion history of the universe as well as place constraints on the dark energy equation of state.

Modern-scale sky surveys, including SDSS, Pan-STARRS, and the Dark Energy Survey, have identified thousands of supernovae throughout their operational lifetimes \citep{sdss,panstarrs,des}. However, it has been logistically challenging to follow up on most of these detections spectroscopically. The result is a low number of spectroscopically confirmed SNe Ia and a large photometric dataset of SNe Ia candidates. The upcoming Rubin Observatory Legacy Survey of Space and Time (LSST) is projected to discover $10^7$ supernovae \citep{lsstsciencecollaboration2009lsst}, with millions of transient alerts each observing night. As spectroscopic resources are not expected to scale with the size of these surveys, the ratio of spectroscopically confirmed SNe to total detections will continue to shrink. With only photometric data, distinguishing between SNe Ia and other types can be difficult. A reliable photometric SNe Ia classification algorithm will allow us to tap into the vast potential of the photometric dataset and pave the way for confident classification and analysis of the ever-growing library of transients from current and future sky surveys.

Significant progress has been made in the past decade in the development of such an algorithm. Most approaches involve lightcurve template matching \citep{sako}, or feature extraction paired with either sequential cuts \citep{bazin,campbell} or machine learning algorithms \citep{Moller_2016,Lochner_2016,Dai_2018,Boone_2019}. Most recently, the spotlight has been on deep learning techniques since it has been shown that classification based on handcrafted features is not only more time-intensive for the researcher but is outperformed by neural networks trained on raw data \citep{Charnock_2017,moss,Kimura_2017}. Since then, many neural network architectures have been explored for SN photometric classification, such as PELICAN's CNN architecture \citep{pasquet} and SuperNNova's deep recurrent network \citep{supernnova}.

Several photometric classification competitions have been hosted, including the Supernova Photometric Classification Challenge (SPCC) \citep{spcc} and the Photometric LSST Astronomical Time Series Classification Challenge (PLAsTiCC) \citep{plasticc}. These have not only resulted in the development of new techniques, such as PSNID \citep{sako} and Avocado \citep{Boone_2019}, but have also provided representative datasets available to researchers during and after the competition, such as the PLAsTiCC unblinded dataset used in this paper.

In this paper we present \scone, a novel application of deep learning to the photometric classification problem. \scone is a convolutional neural network (CNN), an architecture prized in the deep learning community for its state-of-the-art image recognition capabilities \citep{lecun1989backpropagation,lecun1998gradient,alexnet,zeiler2014visualizing,vgg}. Our model requires raw photometric data only, precluding the necessity for accurate redshift approximations. The dataset is preprocessed using a lightcurve modeling technique via Gaussian processes described in \citet{Boone_2019}, which alleviates the issue of irregular sampling between filters but also allows the CNN to learn from information in all filters simultaneously. The model also has relatively low computational and dataset size requirements without compromising on performance -- 400 epochs of training on our $\sim 10^4$ dataset requires around 15 minutes on a GPU.

We will introduce the datasets used to train and evaluate \scone\, its computational requirements, as well as the algorithm itself in Section 2. Section 3 will focus on the performance of \scone\ on both binary and categorical classification, and Section 4 presents an analysis of misclassified lightcurves and heatmaps for both modes of classification.

\section{Methods}
\subsection{Datasets}
The PLAsTiCC training and test datasets were originally created for the 2018 Photometric LSST Astronomical Time Series Classification Challenge.

The PLAsTiCC training set includes $\sim 8000$ simulated observations of low-redshift, bright astronomical sources, representing objects that are good candidates for spectroscopic follow-up observation. This dataset will be referred to in future sections as the ``spectroscopic dataset". We use this dataset to evaluate the out-of-distribution performance of \scone\ in section 3.1.2.

The PLAsTiCC test set consists of 453 million simulated observations of 3.5 million transient and variable sources, representing 3 years of expected LSST output \citep{plasticc_sims}. The objects in this dataset are generally fainter, higher redshift, and do not have associated spectroscopy. Note that most of the results presented in this paper are produced from this dataset alone and will be referred to as ``the dataset", ``the main dataset", or ``the PLAsTiCC dataset" in future sections.

All observations in both datasets were made in LSST's $ugrizY$ bands and realistic observing conditions were simulated using the LSST Operations Simulator \citep{delgado}. While PLAsTiCC includes data from many other transient sources, we are using only the supernovae in the datasets. We selected all of the type II, Iax, Ibc, Ia-91bg, Ia, and SLSN-1 sources (corresponding to \texttt{true\_target} values of 42, 52, 62, 67, 90, and 95) from this dataset and chose only well-sampled lightcurves by restricting ourselves to observations simulating LSST's deep drilling fields (\texttt{ddf=1}). \texttt{peak\_mjd} values, the modified Julian date of peak flux for each object, were calculated for both the main dataset and the spectroscopic dataset by taking the signal-to-noise weighted average of all observation dates. \texttt{peak\_mjd} is referred to as $t_{\rm peak}$ in future sections. The total source count for the spectroscopic set is 4,556, and the total source count for the main dataset is 32,087. A detailed breakdown by type is provided in Table~\ref{tbl:fulldata}.

The categorical dataset was created using SNANA \citep{snana} with the same models and redshift distribution as the main dataset in order to perform categorical classification with balanced classes. 2,000 examples of each type were randomly selected to constitute a class-balanced dataset of 12,000 examples.

\begin{table}
    \centering
    \caption{Makeup of the PLAsTiCC dataset by type.}
    \begin{tabular}{c c c}
        \hline
        SN type & \multicolumn{2}{c}{number of sources}\\
        \hline
         & spectroscopic & main \\
        \hline
        SNIa & 2,313 & 12,640 \\
        SNII & 1,193 & 15,986 \\
        SNIbc & 484 & 2,194 \\
        SNIa-91bg & 208 & 362 \\
        SNIax & 183 & 807 \\
        SLSN-1 & 175 & 98 \\
        \hline
    \end{tabular}
    \label{tbl:fulldata}
\end{table}

\subsection{Quality Cuts}
In order to ensure that the model is learning only from high-quality information, we have instituted some additional quality-based cuts on all datasets. These cuts are based on lightcurve quality, so all metrics are defined for a single source. The metrics evaluated for these cuts are as follows:
\begin{itemize}
    \item \textbf{number of detection datapoints ($n_{\rm detected}$)}: number of observations where the source was detected. We chose a detection threshold of S/N $>$ 5, based on Fig. 8 of \citet{kessler_2015}
    \item \textbf{cumulative signal-to-noise ratio ($CSNR$)}: cumulative S/N for all points in the lightcurve
    $$CSNR = \sqrt{\sum{\frac{f^2}{\sigma_f^2}}}$$
    \item \textbf{duration}: timespan of detection datapoints
    $$t_{\rm active} = t_{\rm last} - t_{\rm first}$$
\end{itemize}
where $f$ represents the flux measurements from all observations of a given source, $\sigma_f$  represents the corresponding uncertainties, $t$ represents the timestamps of all observations of a given source, $t_{\rm first}$ is the time of initial detection, and $t_{\rm last}$ is the time of final detection. Our established quality thresholds require that:
\begin{itemize}
    \item $n_{\rm detected} \geq 5$
    \item $CSNR > 10$
    \item $t_{\rm active} \geq 30$ days
\end{itemize}
for lightcurve points in the range $t_{\rm peak}-50 \leq t \leq t_{\rm peak}+130$.
1,150 out of 4,556 sources passed these cuts in the spectroscopic dataset and 12,611 out of 32,087 sources passed these cuts in the main dataset. The makeup of these datasets is detailed in Table \ref{tbl:cutdata}. The categorical dataset was created from sources that already passed the cuts, so the makeup is unchanged.

\subsection{Class Balancing}

Maintaining an equal number of examples of each class, or a balanced class distribution, is important for machine learning datasets. Balanced datasets allow for an intuitive interpretation of the accuracy metric as well as provide ample examples of each class for the machine learning model to learn from.

As shown in Table \ref{tbl:cutdata}, the natural distribution of the spectroscopic dataset is more abundant in Ia sources than non-Ia sources. Thus, all non-Ia sources were retained in the class balancing process for binary classification and an equivalent number of Ia sources were randomly chosen. SNIax and SNIa-91bg sources were labeled as non-Ia sources for binary classification. The class-balanced spectroscopic dataset has 496 sources of each class for a total of 992 sources.

In contrast, the natural distribution of the main dataset is more abundant in non-Ia sources than Ia sources. Thus, all Ia sources were retained in the class balancing process for binary classification and an equivalent number of non-Ia sources were randomly chosen. The random selection process does not necessarily preserve the original distribution of non-Ia types. The class-balanced dataset has 6,128 sources of each class for a total of 12,256 sources.

The categorical dataset of 2,000 sources for each of the 6 types was created explicitly for the purpose of retaining balanced classes in categorical classification, as mentioned in Section 2.1.

\begin{table}
    \centering
    \caption{Makeup of the PLAsTiCC dataset by type after applying quality cuts.}
    \label{tbl:cutdata}
    \begin{tabular}{c c c}
        \hline
        SN type & \multicolumn{2}{c}{number of sources}\\
        \hline
         & spectroscopic & main \\
        \hline
        SNIa & 654 & 6,128 \\
        SNII & 262 & 5,252 \\
        SNIbc & 97 & 779 \\
        SNIa-91bg & 41 & 113 \\
        SNIax & 59 & 281 \\
        SLSN-1 & 37 & 58 \\
        \hline
    \end{tabular}
\end{table}

All datasets were split by class into 80\% training, 10\% validation, and 10\% testing. Splitting by class ensures balanced classes in each of the training, validation, and test sets.

\begin{figure}
    \centering
    \includegraphics[scale=0.13,trim={6cm 33cm 0 0}]{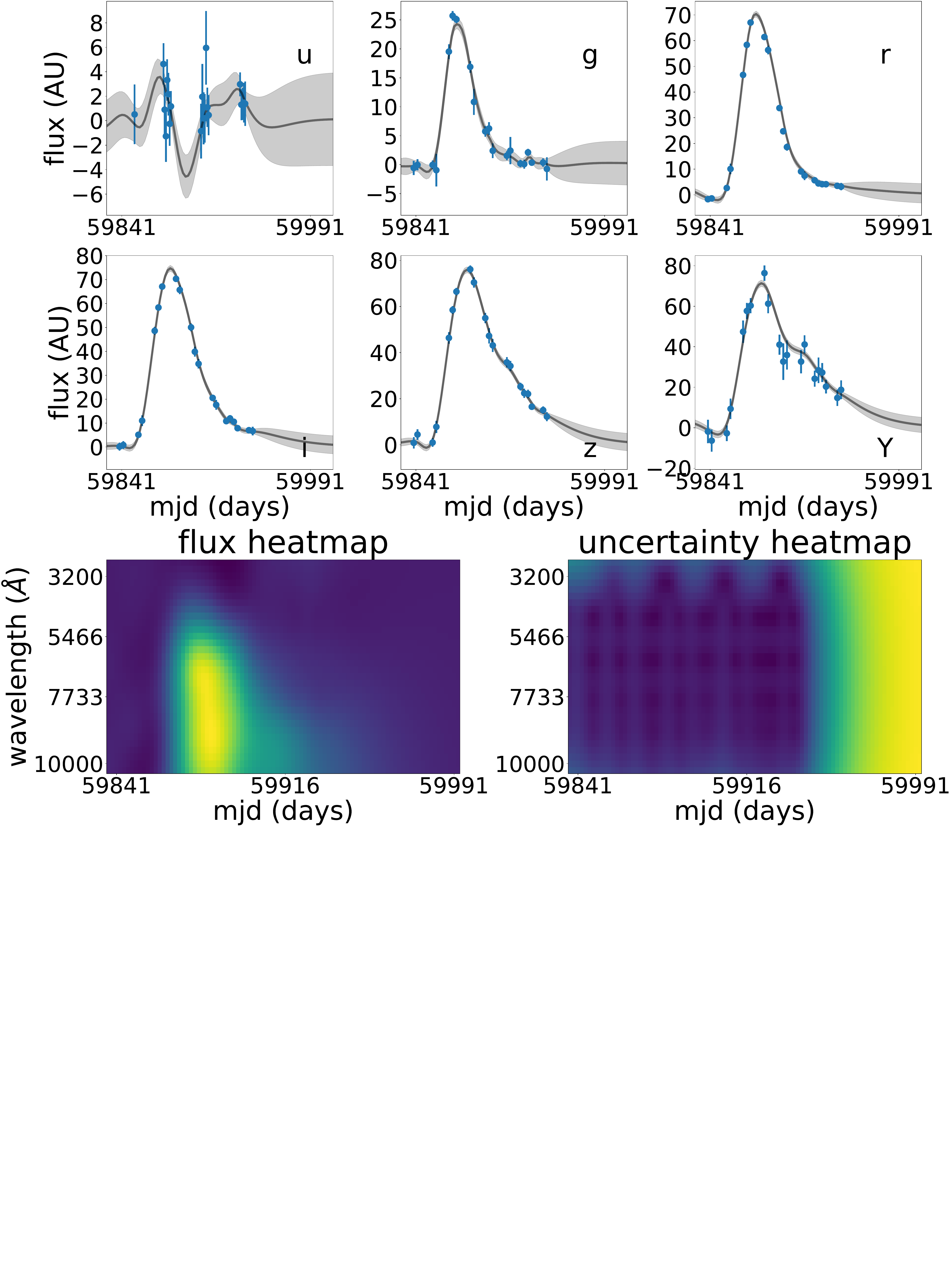}
    \caption{Raw $ugrizY$ lightcurve data with the Gaussian process model at corresponding wavelengths and resulting heatmap and error heatmap for a type Ia SN. The shaded regions in the Gaussian process plots represent the Gaussian process error.}
    \label{fig:lightcurves}
\end{figure}

\subsection{Heatmap Creation}
Prior to training, we preprocess our lightcurve data into heatmap form. First, all observations are labeled with the central wavelength of the observing filter according to Table \ref{tbl:filters}, which was calculated from the filter functions used by \citet{plasticc}. We then use the approach described by \citet{Boone_2019} to apply 2-dimensional Gaussian process regression to the raw lightcurve data to model the event in the wavelength ($\lambda$) and time ($t$) dimensions. We use the Matern 32 kernel with a fixed 6000~\AA\  characteristic length scale in $\lambda$ and fit for the length scale in $t$. Once the Gaussian process model has been trained, we obtain its predictions on a $\lambda,t$ grid and call this our ``heatmap". Our choice for the $\lambda,t$ grid was $t_{\rm peak} - 50 \leq t \leq t_{\rm peak} + 130$ with a 1-day interval and $3000 < \lambda < 10,100$~\AA\ with a $221.875$~\AA\ interval. The significance of this choice is explored further in Section 3.

\begin{figure*}
    \centering
    \includegraphics[scale=0.35,trim={5cm 1cm 0 0}]{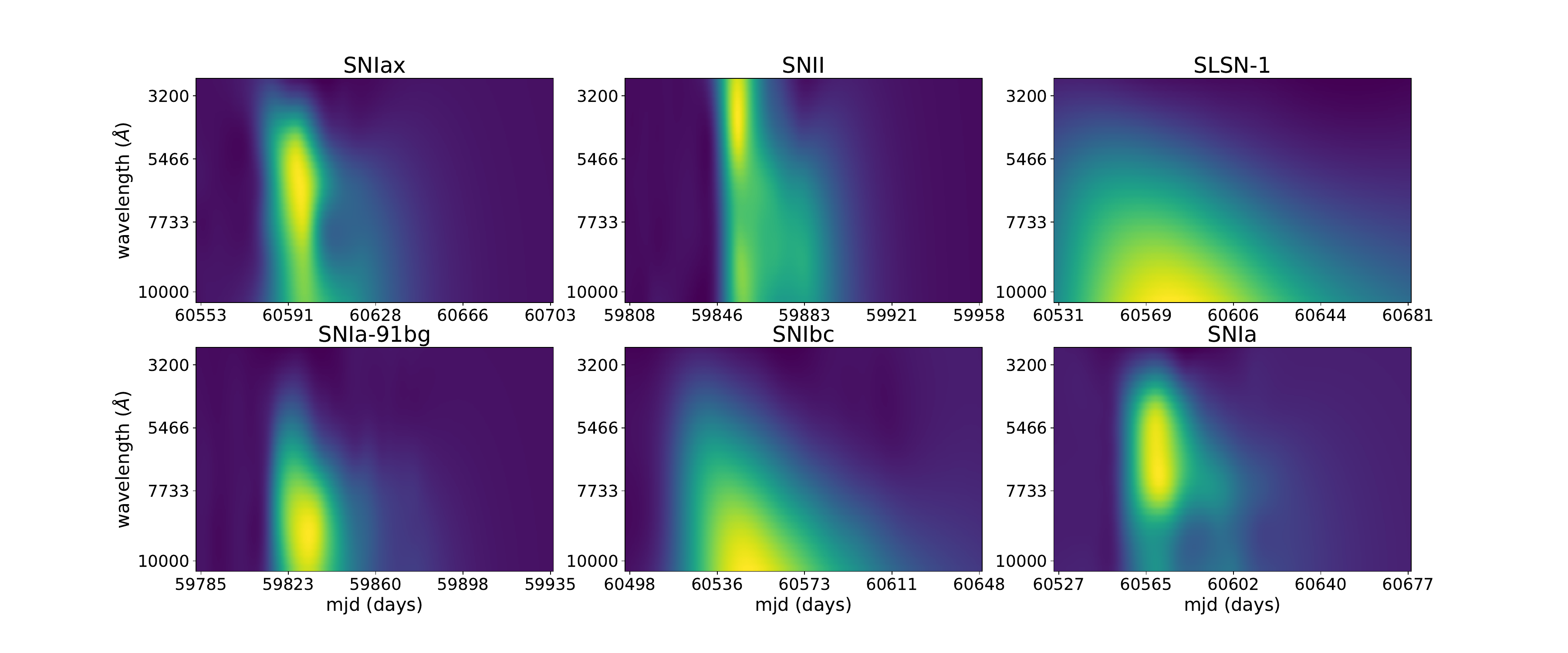}
    \caption{Example flux heatmaps for each supernova type.}
    \label{fig:heatmaps}
\end{figure*}

\begin{table}
    \centering
    \caption{Central wavelength of each filter}
    \label{tbl:filters}
    \begin{tabular}{c c}
        \hline
        Filter & Central Wavelength (\AA) \\
        \hline
        u & 3670.69 \\
        g & 4826.85 \\
        r & 6223.24 \\
        i & 7545.98 \\
        z & 8590.90 \\
        Y & 9710.28 \\
        \hline
    \end{tabular}
\end{table}

The result of this is a $n_{\lambda} \times n_t$ image-like matrix, where $n_{\lambda}$ and $n_t$ are the lengths of the wavelengths array and times array, respectively, given to the Gaussian process. We also take into account the uncertainties on the Gaussian process predictions at each time and wavelength, producing a second image-like matrix. We stack these two matrices depthwise and divide by the maximum flux value to constrain all entries to [0,1]. This matrix is our input to the convolutional neural network.

Figure~\ref{fig:lightcurves} shows the raw lightcurve data in the $t_{\rm peak} - 50 \leq t \leq t_{\rm peak} + 130$ range for each filter in blue, the Gaussian process model in gray, and the resulting flux and uncertainty heatmaps at the bottom. Figure~\ref{fig:heatmaps} shows a representative example of a heatmap for each supernova type.

\subsection{Convolutional Neural Networks}
Convolutional neural networks (CNNs) are a type of artificial neural network that makes use of the convolution operation to learn local, small-scale structures in an image. This is paired with an averaging or sub-sampling layer, often called a \textit{pooling layer}, that reduces the resolution of the image and allows the subsequent convolutional layers to learn hierarchically more complex and less localized structures.

In a \textit{convolutional layer}, each unit receives input from only a small neighborhood of the input image. This use of a restricted receptive field, or \textit{kernel size}, resembles the neural architecture of the animal visual cortex and allows for extraction of local, elementary features such as edges, endpoints, or corners. All units, each corresponding to a different small neighborhood of the image, share the same set of learned weights. This allows them to detect the presence of the same feature in each neighborhood of the image. Each convolutional layer often has several layers of these units, each of which is called a \textit{filter} and extracts a different feature. The output of a convolutional layer is called a \textit{feature map}.

Pooling layers reduce the local precision of a detected feature by sub-sampling the each unit's receptive field, often $2 \times 2$ pixels, based on some rule. Average-pooling, for example, extracts the average of the 4 pixels, and max-pooling extracts the maximum value. Assuming no overlap in the receptive fields, the spatial dimensions of the resulting feature maps will be reduced by half.

Dropout \citep{dropout} is a commonly used regularization technique in fully connected layers. A \textit{dropout layer} chooses a random user-defined percentage of the input weights to set to zero, improving the robustness of the learning process.

A convolutional neural network typically consists of alternating convolutional layers and pooling layers, followed by a series of fully-connected layers that learn a mapping between the result of the convolutions and the desired output.

\subsection{\scone\ Architecture}

\begin{figure}
    \vspace*{0.2cm}
    \includegraphics[scale=0.8,trim={2cm 0 2cm 0}]{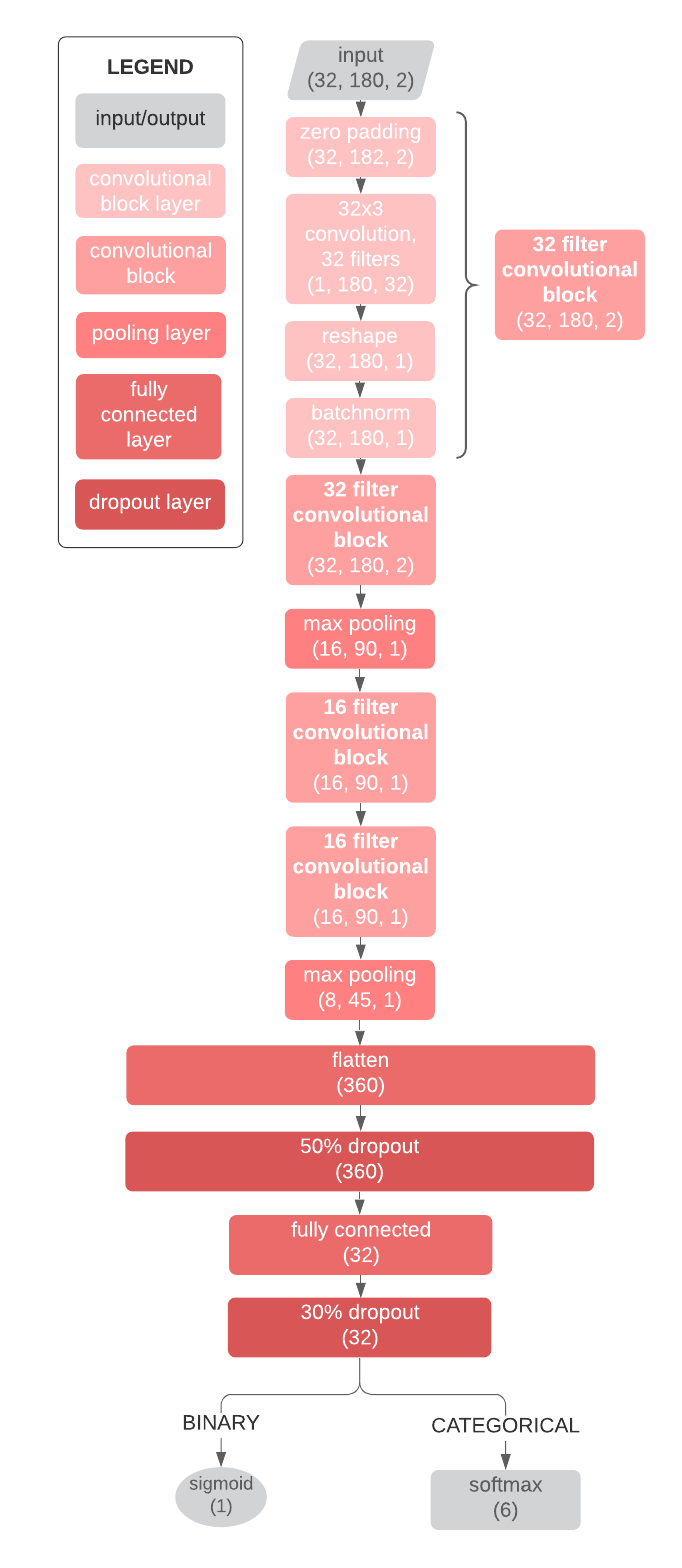}
    \centering
    \caption{\scone\ architecture for binary and categorical classification.}
    \label{fig:scone-architecture}
\end{figure}

The relatively simple architecture of \scone, shown in Figure~\ref{fig:scone-architecture}, allows for a minimal number of trainable parameters, speeding up the training process significantly without compromising on performance. It has a total of 22,606 trainable parameters for categorical classification and 22,441 trainable parameters for binary classification when trained on heatmaps of size $32 \times 180 \times 2$ $(h \times w \times d)$.

As mentioned in Section 2.4, each heatmap is divided by its maximum flux value for normalization. After receiving the normalized heatmap as input, the network pads the heatmap with a column of zeros on both sides, bringing the heatmap size to $32 \times 182 \times 2$. Then, a convolutional layer is applied with $h$ filters and a kernel size of $h \times 3$, which in this case is 32 filters and a $32 \times 3$ kernel, resulting in a feature map of size $1 \times 180 \times 32$. We reshape this feature map to be $32 \times 180 \times 1$, apply batch normalization, and repeat the above process one more time. We have now processed our heatmap through two ``convolutional blocks" with an output feature map of size $32 \times 180 \times 1$.

We apply $2 \times 2$ max pooling to our output, reducing its dimensions to $16 \times 90 \times 1$, and pass it through two more convolutional blocks, but this time $h=16$.

We pass our output through a final $2 \times 2$ max pooling layer, resulting in a $8 \times 45 \times 1$ feature map. This is subsequently flattened into a 360-element array and passed through a 50\% dropout layer. A 32-unit fully connected layer followed by a 30\% dropout layer feeds into the final layer. For binary classification, this is a node with a sigmoid activation that returns the model's predicted Ia probability. For categorical classification, the final layer contains 6 nodes with softmax activations that return the respective probabilities of each of the 6 SN types. Both of these versions of \scone\ are shown in  Figure~\ref{fig:scone-architecture}.

The model is trained with the binary crossentropy loss function for binary classification and the sparse categorical crossentropy loss function for categorical classification. Both classification modes use the Adam optimizer \citep{kingma2017adam} at a constant 1e-3 learning rate for 400 epochs.

\subsection{Evaluation Metrics}

\textit{Accuracy} is defined as the number of correct predictions divided by the number of total predictions. We also evaluated the model on a number of other performance metrics: purity, efficiency, and AUC. 

\textit{Purity} and \textit{efficiency} are defined as: $$\textrm{purity=}\frac{\textrm{TP}}{\textrm{TP+FP}};\; \textrm{efficiency=}\frac{\textrm{TP}}{\textrm{TP+FN}}$$
where TP is true positive, FP is false positive, and FN is false negative.

\textit{AUC}, or \textit{area under the receiver operating characteristic (ROC) curve}, is a common metric used to evaluate binary classifiers. The ROC curve is created by plotting the false positive rate against the true positive rate at various discrimination thresholds, showing the sensitivity of the classifier to the chosen threshold. A perfect classifier would score a 1.0 AUC value while a random classifier would score a 0.5.

\subsection{Computational Requirements}

Due to the minimal number of trainable parameters and dataset size, the time and hardware requirements for training and evaluating with \scone\  are relatively low. The first training epoch on one NVIDIA V100 Volta GPU takes approximately 2 seconds, and subsequent training epochs take approximately 1 second each with TensorFlow's dataset caching. Training epochs on one Haswell node (with Intel Xeon Processor E5-2698 v3), which has 32 cores, take approximately 26 seconds each.

\section{Results}

\subsection{Binary Classification}
Our model achieved $99.93 \pm 0.06$\% training accuracy, $99.71 \pm 0.2$\% validation accuracy, and $99.73 \pm 0.26$\% test accuracy on the Ia vs. non-Ia binary classification problem performed on the class-balanced dataset of 12,256 sources. 

\begin{figure}
    \includegraphics[scale=0.51]{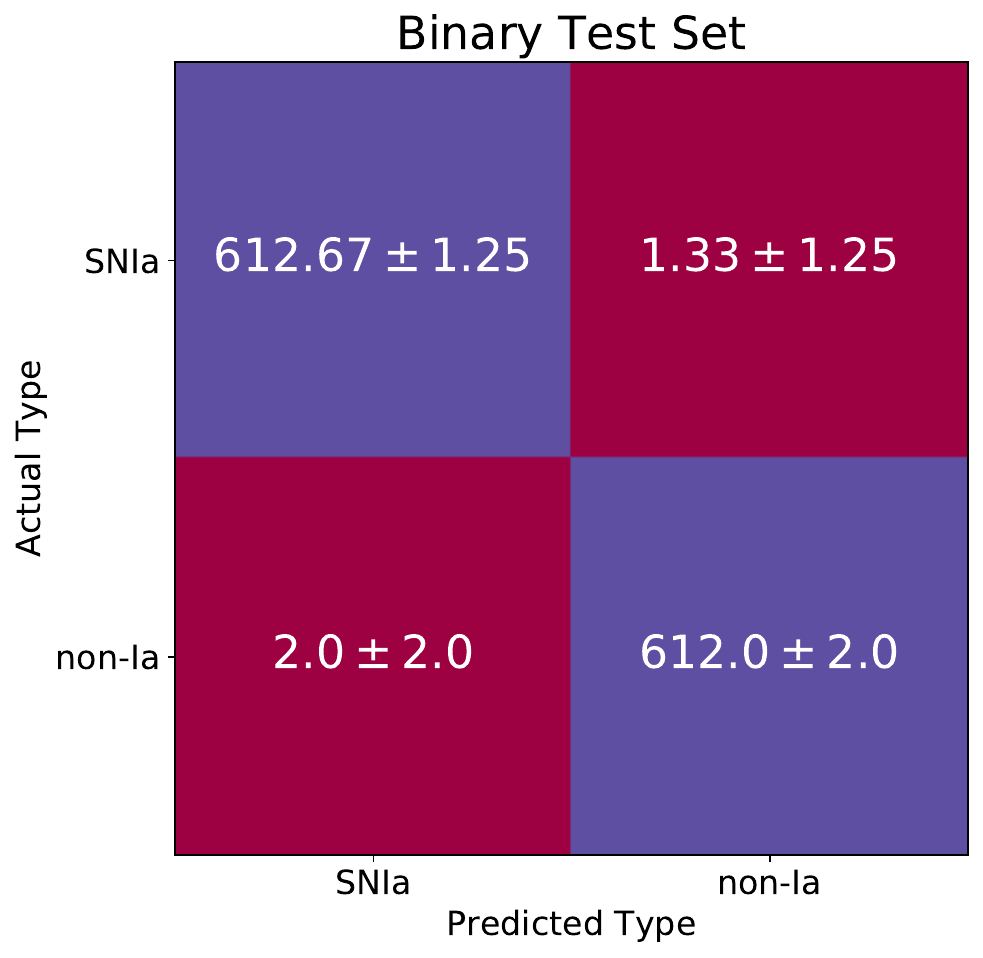}
    \centering
    \caption{Confusion matrix showing average and standard deviation over five runs for binary classification on the test set.}
    \label{fig:binary-confusion}
\end{figure}

\begin{figure}
    \includegraphics[scale=0.35]{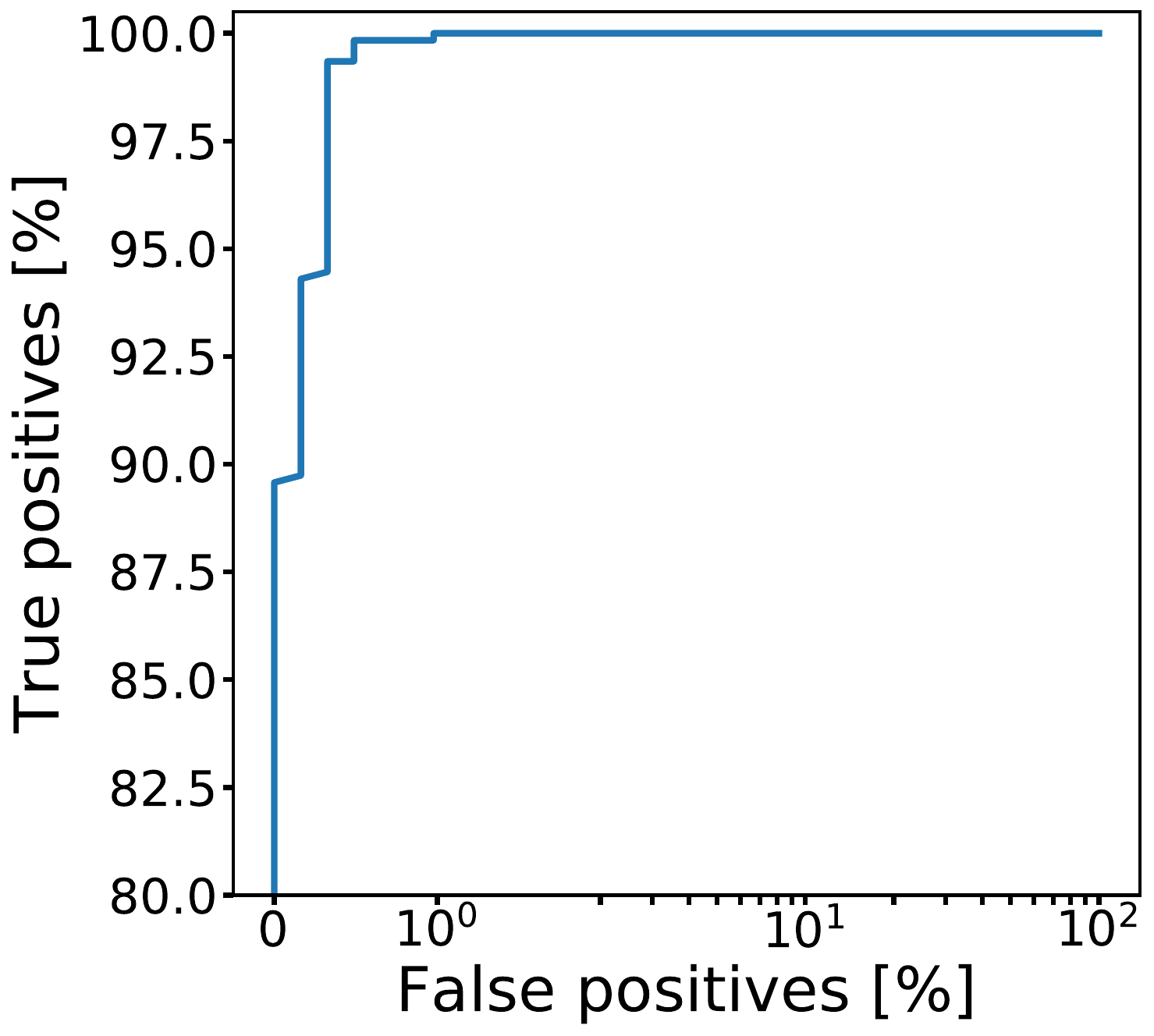}
    \centering
    \caption{Semilog plot of the ROC curve for binary classification on the test set.}
    \label{fig:roc}
\end{figure}

Figure~\ref{fig:binary-confusion} shows the confusion matrices for binary classification, and Table~\ref{tbl:results} shows the model's performance on all the evaluation metrics described in Section 2.7. Both Figure~\ref{fig:binary-confusion} and Table~\ref{tbl:results} were created with data from five independent training, validation, and test runs of the classifier. Unless otherwise noted, the default threshold for binary classification is 0.5, where classifier confidence equal to or exceeding 0.5 counts as an SNIa classification, and vice versa. Altering this threshold produces Figure~\ref{fig:roc}, the ROC curve of one of these test runs.

\begin{table}[t]
    \centering
    \caption{Evaluation metrics for Ia vs. non-Ia classification on cut dataset}
    \label{tbl:results}
    \begin{tabular}{l c c c}
        \hline
        Metric & Training & Validation & Test \\
        \hline
        Accuracy & $99.93 \pm 0.06$\% & $99.71 \pm 0.2$\% & $99.73 \pm 0.26$\%\\
        Purity & $99.93 \pm 0.06$\% & $99.76 \pm 0.25$\% & $99.68 \pm 0.35$\%\\
        Efficiency & $99.93 \pm 0.05$\% & $99.67 \pm 0.23$\% & $99.78 \pm 0.22$\%\\
        AUC & $1.0 \pm 4.1\text{e-}5$ & $0.9991 \pm 1.6\text{e-}3$ & $0.9994 \pm 1\text{e-}3$ \\
        \hline
    \end{tabular}
\end{table}

\begin{figure}
    \includegraphics[scale=0.5]{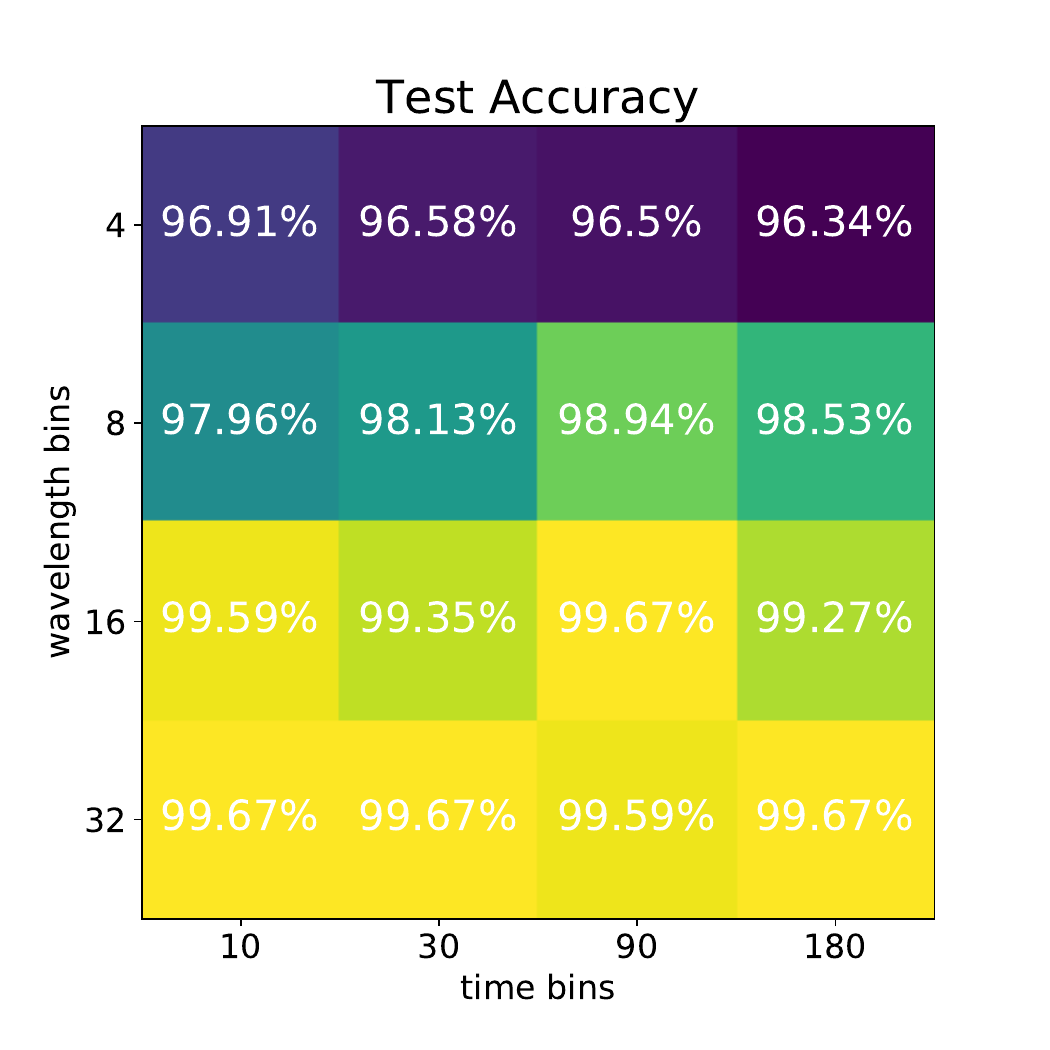}
    \centering
    \caption{Test set accuracies for each choice of wavelength and time bins.}
    \label{fig:grid}
\end{figure}

\subsubsection{Heatmap Dimensions}
We explored the binary classification performance of different heatmap dimensions in both the time (width) and wavelength (height) axes. For our 7100~\AA\ wavelength range ($3000$~\AA $< \lambda < 10,100$~\AA), we chose intervals of 221.875~\AA, 443.8~\AA, 887.5~\AA, and 1775~\AA, resulting in heatmaps with 32, 16, 8, and 4 wavelength "bins", respectively. For our 180 day range ($t_{\rm peak} - 50 \leq t \leq t_{\rm peak} + 130$ ), we chose intervals of 1, 2, 6, and 18 days, resulting in heatmaps with 180, 90, 30, and 10 time bins.

Figure~\ref{fig:grid} shows the training, validation, and test accuracies for each choice of wavelength and time dimensions. Our classifier seems relatively robust to these changes, showing minimal performance impacts for wavelength bins $\geq 16$ and impressive performance for the smaller sizes as well. Test accuracy drops a maximum of 2.67\% between the best- and worst-performing variants, $32 \times 180$ and $4 \times 10$. This is noteworthy as the $32 \times 180$ heatmaps contain 144 times the number of pixels of the $4 \times 10$ heatmaps.

The performance seems to unilaterally improve as expected as the number of wavelength bins increase, but increasing the number of time bins seems to yield varying, though likely not statistically significant, results.

We have reported all of our \scone\ results using one of the best performing variants, the $32 \times 180$ heatmaps.

\subsubsection{Out-of-Distribution Results}
Preliminary exploration into the out-of-distribution task of training on the spectroscopic dataset and testing on the main dataset  yielded 80.6\% test accuracy. The full results of training on 85\% of the spectroscopic dataset, validating on the remaining 15\%, and testing on the full main dataset are shown in Table \ref{tbl:ood}. Since the redshift distribution of the spectroscopic dataset is skewed toward lower redshifts, testing on class-balanced low-redshift subsets of the main dataset yielded 83\% test accuracy for $z < 0.4$ and 87\% test accuracy for $z < 0.3$.  \cite{Boone_2019} introduced redshift augmentation to mitigate this effect.  The mismatch between the test and training sets, however, comes in other forms.  For example, if the training data is generated using models rather than real data, differences in the characteristics of the spectral surfaces can have an impact on classification.  The unknown relative rates of of each type of event also affect the overall performance.  Since out-of-distribution robustness is an integral part of the challenge of photometric SNe classification, improving the performance of \scone\ on these metrics will be the topic of a future paper.\\\\

\begin{table}[h]
    \centering
    \caption{Evaluation metrics for out-of-distribution Ia vs. non-Ia classification}
    \label{tbl:ood}
    \begin{tabular}{l c c c}
        \hline
        Metric & Training & Validation & Test \\
        \hline
        Accuracy & $99.65 \pm 0.36$\% & $98.84 \pm 1.1$\% & $80.61 \pm 1.75$\%\\
        Purity & $99.86 \pm 0.31$\% & $98.74 \pm 1.31$\% & $81.86 \pm 2.22$\%\\
        Efficiency & $99.67 \pm 0.46$\% & $98.75 \pm 1.27$\% & $80.06 \pm 1.31$\%\\
        AUC & $1.0 \pm 8.9\text{e-}5$ & $0.9939 \pm 9.5\text{e-}3$ & $0.8552 \pm 1.4\text{e-}2$ \\
        \hline
    \end{tabular}
\end{table}

\subsection{Categorical Classification}

In addition to binary classification, \scone\  is able to perform categorical classification and discriminate between different types of SNe. We performed 6-way categorical classification with the same PLAsTiCC dataset used for binary classification as well as the class-balanced dataset described in Section 2.1. Our model differentiated between SN types Ia, II, Ibc, Iax, SN-91bg, and SLSN-1. On the PLAsTiCC dataset (not class-balanced), it achieved $99.26 \pm 0.18$\% training accuracy, $99.13 \pm 0.34$\% validation accuracy, and $99.18 \pm 0.18$\% test accuracy. The confusion matrices in Figure~\ref{fig:categorical-plasticc-confusion} show the average by-type breakdown for five independent runs.

\begin{figure}
    \includegraphics[scale=0.35]{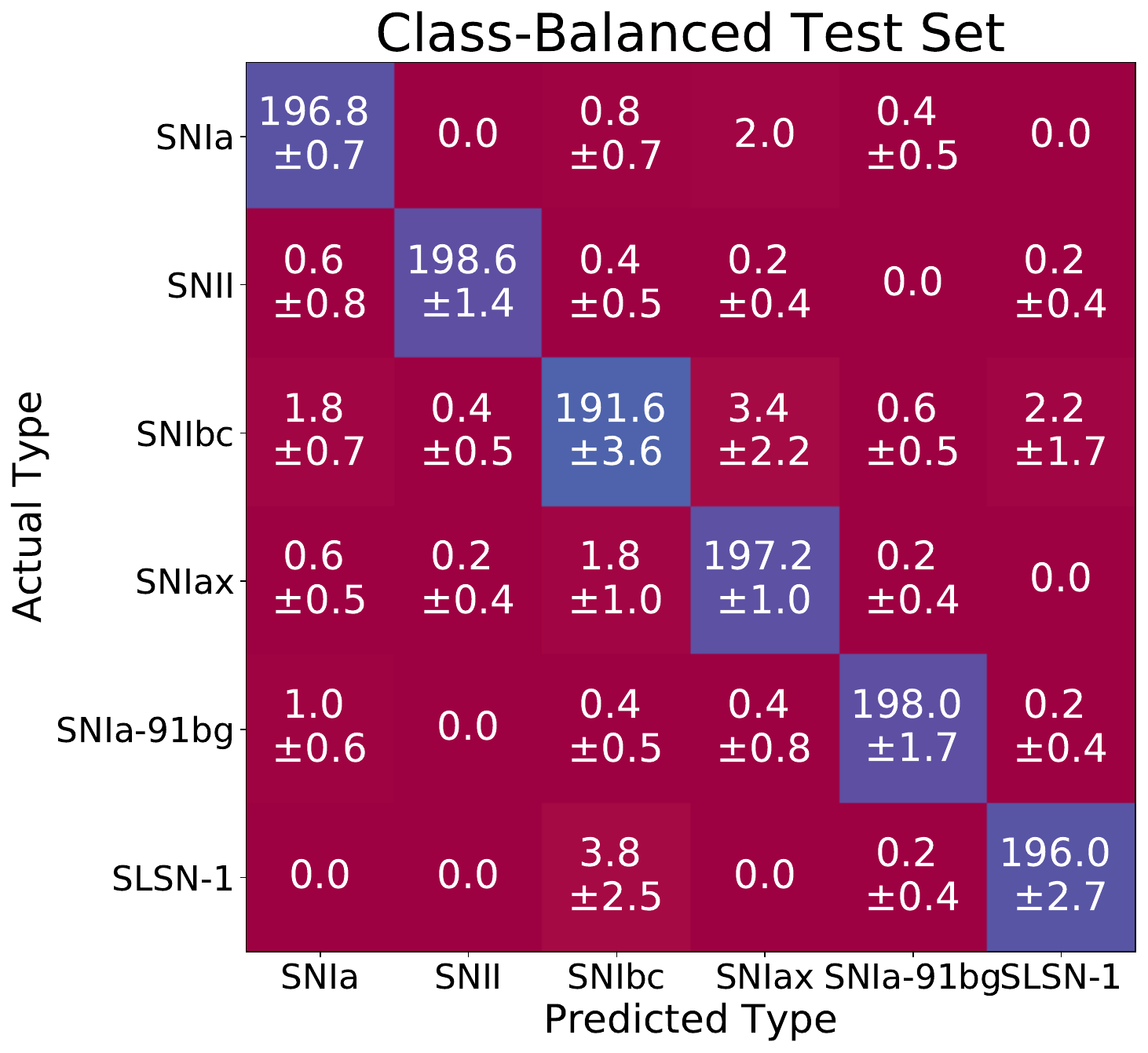}
    \centering
    \caption{Confusion matrix showing average and standard deviation over five runs for categorical classification on the balanced test set.}
    \label{fig:categorical-confusion}
\end{figure}

\begin{figure}
    \includegraphics[scale=0.35]{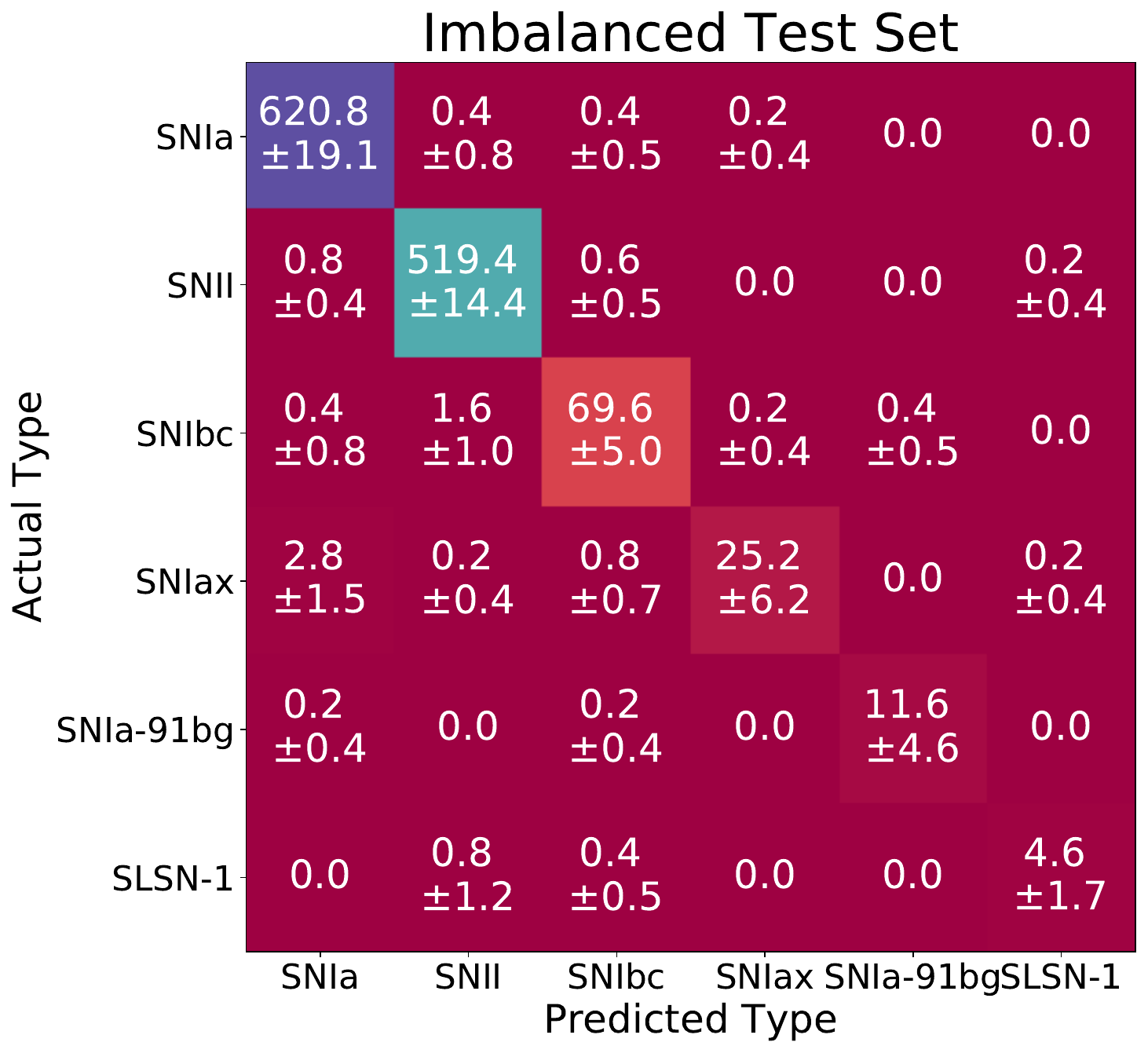}
    \centering
    \caption{Confusion matrix showing average and standard deviation over five runs for categorical classification on the PLAsTiCC (imbalanced) test set.}
    \label{fig:categorical-plasticc-confusion}
\end{figure}

On the balanced dataset, it achieved $97.8 \pm 0.32$\% training accuracy, $98.52 \pm 0.28$\% validation accuracy, and $98.18 \pm 0.3$\% test accuracy. The confusion matrices in Figure~\ref{fig:categorical-confusion} show the average and standard deviations of the by-type breakdown for 5 independent runs.

It is worth noting that we trained and tested with the PLAsTiCC dataset even though it is not class-balanced for this task to try to evaluate the model's performance on a dataset emulating the relative frequencies of these events in nature.

\section{Discussion}
Analysis of misclassified heatmaps was performed for both binary and class-balanced categorical classification. No clear evidence of the effect of redshift on accuracy was found for either mode of classification. The quantity of misclassified SNIa examples per run is not sufficient for us to draw conclusions about the accuracy evolution as a function of redshift.

\subsection{Binary Classification}
According to the data presented in Figure~\ref{fig:binary-confusion}, the model seems to mispredict about the same number of Ia's as non-Ia's. An average of 3.33 Ia's are mispredicted in the training set compared with 3.5 non-Ia's. For the validation set, 2 Ia's are mispredicted compared to 1.5, and 1.33 Ia's and 2 non-Ia's are mislabeled for the test set. 
\begin{table*}
    \centering
    \caption{Misclassified test set heatmaps by true and predicted type for binary classification.}
    \label{tbl:misclassified-binary}
    \begin{tabular}{l l c c c c c}
        \hline
        True Type & Predicted Type & \multicolumn{3}{c}{Confidence} & Total & Percentage \\ 
        \cmidrule(lr){3-5}
        & & $>90$\% & 90-70\% & 70-50\% & & \\
        \hline
        SNIa & non-Ia & 1 & 0 & 0 & 1 & 25\%\\
        \hline
        SNII & \multirow{2}{0em}{SNIa} & 2 & 0 & 0 & 2 & 50\%\\
        SNIax & & 1 & 0 & 0 & 1 & 25\%\\
        \hline
    \end{tabular}
\end{table*}

\begin{table*}
    \centering
    \caption{Misclassified test set heatmaps by true and predicted type for categorical class-balanced classification.}
    \label{tbl:misclassified-categorical}
    \begin{tabular}{l l c c c c c}
        \hline
        True Type & Predicted Type & \multicolumn{3}{c}{Confidence} & Total & Percentage \\ 
        \cmidrule(lr){3-5}
        & & $>90$\% & 90-70\% & 70-50\% & & \\
        \hline
        SNIa & SNIax & 1 & 0 & 0 & 1 & 4.8\%\\
        \hline
        \multirow{2}{0em}{SNII} & SNIa & 1 & 0 & 0 & 1 & 4.8\%\\
        & SNIbc & 0 & 0 & 1 & 1 & 4.8\%\\
        \hline
        \multirow{3}{0em}{SNIbc} & SNIa & 0 & 0 & 1 & 1 & 4.8\%\\
        & SNIax & 1 & 0 & 0 & 1 & 4.8\%\\
        & SLSN-1 & 0 & 0 & 1 & 1 & 4.8\%\\
        \hline
        \multirow{3}{2em}{SNIax}& SNIa & 1 & 2 & 0 & 3 & 14.3\%\\
        & SNIbc & 1 & 2 & 0 & 3 & 14.3\% \\
        & SNIa-91bg & 1 & 0 & 0 & 1 & 4.9\%\\
        \hline
        SLSN-1 & SNIbc & 3 & 1 & 3 & 7 & 33.3\%\\
        \hline
        SNIa-91bg & SNIbc & 0 & 1 & 0 & 1 & 4.8\%\\
        \hline
    \end{tabular}
\end{table*}

The misclassified set summarized in Table~\ref{tbl:misclassified-binary} is the result of one of the five runs represented by the data in Figure~\ref{fig:binary-confusion}. In this example, the model missed 5 examples total during testing -- 1 SNIa, 2 SNII, and 1 SNIax. It was $> 90$\% confident about all of these misclassifications, which is certainly not the case for categorical classification. This could be due to the fact that there are more examples of each type for binary classification than categorical ($\sim 6,000$ and 1,200 per type for training, respectively).


\subsection{Categorical Classification}
The data in Figures~\ref{fig:categorical-confusion} and~\ref{fig:categorical-plasticc-confusion} show some level of symmetry between misclassifications. SNIax and SLSN-1 seem to be easily distinguishable across the board, for example, with 0's in all relevant cells in both figures except one. In Figure~\ref{fig:categorical-confusion}, SNIbc and SLSN-1 are seemingly very similar, as they are misclassified as one another at similarly high rates.

There are notable differences between Figures~\ref{fig:categorical-confusion} and~\ref{fig:categorical-plasticc-confusion}, however. In Figure~\ref{fig:categorical-confusion}, representing the classifier's performance on class-balanced categorical classification, the model mispredicts SNIa's as other types at a similar rate as non-Ia's mispredicted as Ia's. An average of 3.2 Ia's are mispredicted, whereas an average of 4 non-Ia's are misclassified as Ia. In the confusion matrix shown in Figure~\ref{fig:categorical-plasticc-confusion}, significantly more non-Ia's were mispredicted as Ia. An average of 1 Ia was mispredicted compared to an average of 4.2 non-Ia's misclassified as Ia. The rate of SNIbc's mispredicted as SLSN-1 is also significantly lower for the PLAsTiCC dataset than for the balanced dataset. These observations further reinforce the impact of imbalanced classes in classification tasks.

The misclassified set summarized in Table~\ref{tbl:misclassified-categorical} is the result of one of the five class-balanced categorical classification runs represented by the data in Figure~\ref{fig:categorical-confusion}.

One point of interest is the lack of symmetry between misclassifications, in contrast with the analysis of Figures~\ref{fig:categorical-confusion} and~\ref{fig:categorical-plasticc-confusion}. This is clear in the significantly larger number of SLSN-1 misclassified as SNIbc  (7) compared with the number of SNIbc misclassified as SLSN-1 (1). SNIax is also more often misclassified as other types (3 as SNIa, 3 as SNIbc, and 1 as SNIa-91bg) than non-Iax misclassified as Iax (1 SNIa and 1 SNIbc). The more symmetric Figure~\ref{fig:categorical-confusion} suggests that the asymmetry of this table is due to randomness and would be corrected with data from other runs.

The distribution of misclassified examples across the confidence spectrum is non-uniform. In this table, \textit{confidence} refers to the probability assigned to the predicted type by the classifier. Confidence near 100\% for a misclassified example is potentially more insightful than one near 50\%. 9 out of the 21 misclassified heatmaps were misclassified at $> 90$\% confidence, 6 at 90-70\% confidence, and 6 at 70-50\%. Surprisingly, the classifier is confidently wrong almost half the time. One particularly interesting example is a SNIbc "misclassified" as SLSN-1, but the classification probabilities for both SLSN-1 and SNIbc were 50\%.

\subsection{Limitations and Future Work}
As stated in section 2.1, it is important to note that the metrics reported in this paper are in-distribution results since the training, validation, and test sets are mutually exclusive segments of the main dataset. The out-of-distribution performance of \scone, as evaluated in section 3.1.2, is noticeably diminished from the $>99$\% in-distribution test accuracy. The high in-distribution test accuracy shows that \scone\ is robust to previously unseen data, but the lower out-of-distribution test accuracy demonstrates \scone's sensitivity to variations in the parameters of the dataset, such as the redshift distribution, relative rates of different types of SNe, small variations in the SN~Ia model, as well as telescope characteristics. Generalizing \scone\ to become robust to these variations will be the subject of a future paper.

\section{Conclusions}
In this paper we have presented \scone, a novel application of deep learning to the photometric supernova classification problem. We have shown that \scone\  has achieved unprecedented performance on the in-distribution Ia vs. non-Ia classification problem and impressive performance on classifying SNe by type without the need for accurate redshift approximations or handcrafted features.

Using the wavelength-time flux and error heatmaps from the Gaussian process for image recognition also allows the convolutional neural network to learn about the development of the supernova over time in all filter bands simultaneously. This provides the network with far more information than a photograph taken at one moment in time. Our choice of an $h \times 3$ convolutional kernel, where $h$ is the number of wavelength bins, supplements these benefits by allowing the network to learn from data on the full spectrum of wavelengths in a sliding window of 3 days.

As future large-scale sky surveys continue to add to our ever-expanding transients library, we will need an accurate and computationally inexpensive photometric classification algorithm. Such a model can inform the best choice for allocation of our limited spectroscopic resources as well as allow researchers to further cosmological science using minimally contaminated SNIa datasets. \scone\  can be trained on tens of thousands of lightcurves in minutes and confidently classify thousands of lightcurves every second at $> 99$\% accuracy.

Although \scone\  was formulated with supernovae in mind, it can easily be applied to classification problems with other transient sources. The documented source code has been released on Github (github.com/helenqu/scone) to ensure reproducibility and encourage the discovery of new applications.

\section{Acknowledgments}
The authors would like to thank Rick Kessler for his help with SNANA simulations and Michael Xie for his guidance on the model architecture. This research used resources of the National Energy Research Scientific Computing Center (NERSC), a U.S. Department of Energy Office of Science User Facility located at Lawrence Berkeley National Laboratory, operated under Contract No. DE-AC02-05CH11231. This work was supported by DOE grant DE-FOA-0001781 and NASA grant NNH15ZDA001N-WFIRST.

%% file: chapters/scone_early.tex
\section*{Abstract}
 In this work, we present classification results on early supernova lightcurves from \scone, a photometric classifier that uses convolutional neural networks to categorize supernovae (SNe) by type using lightcurve data. \scone\ is able to identify SN types from lightcurves at any stage, from the night of initial alert to the end of their lifetimes. Simulated LSST SNe lightcurves were truncated at 0, 5, 15, 25, and 50 days after the trigger date and used to train Gaussian processes in wavelength and time space to produce wavelength-time heatmaps. \scone\ uses these heatmaps to perform 6-way classification between SN types Ia, II, Ibc, Ia-91bg, Iax, and SLSN-I. \scone\ is able to perform classification with or without redshift, but we show that incorporating redshift information improves performance at each epoch. \scone\ achieved 75\% overall accuracy at the date of trigger (60\% without redshift), and 89\% accuracy 50 days after trigger (82\% without redshift). \scone\ was also tested on bright subsets of SNe ($r<20$ mag) and produced 91\% accuracy at the date of trigger (83\% without redshift) and 95\% 5 days after trigger (94.7\% without redshift). \scone\ is the first application of convolutional neural networks to the early-time  photometric transient classification problem.  All of the data processing and model code developed for this paper can be found in the \href{https://github.com/helenqu/scone\ }{\scone\ software package} located at github.com/helenqu/scone \citep{helen_qu_2021_5602043}.

\section{Introduction}
Observations of transient and supernova phenomena have informed fundamental discoveries about our universe, ranging from its expansion history and current expansion rate \citep{riess, perlmutter, freedman, riess_2019} to the progenitor physics of rare and interesting events \citep{pursiainen, patrick}. In the near future, next generation wide-field sky surveys such as the Vera C. Rubin Observatory Legacy Survey of Space and Time \citep[LSST,][]{lsst} will have the ability to observe larger swaths of sky with higher resolution and certainly uncover even more new and exciting astrophysical phenomena.

These surveys promise to generate ever larger volumes of photometric data at unprecedented rates. However, the availability of spectroscopic resources is not expected to scale nearly as quickly. Thus, the challenge of effectively allocating these limited resources is more important than ever. For type Ia SN cosmology, spectroscopic information is used to minimize contamination in constructing pure and representative samples of SNe Ia to continue to constrain the dark energy equation of state. For supernova physicists, spectra uncover important information about an event's potential progenitor processes \citep{filippenko, perets, federica, sollerman}. Spectra taken near peak brightness of an event are optimal as they include mostly transient information and are not dominated by host galaxy features.

With millions of alerts each night, fast and accurate automatic classification mechanisms will be needed to replace the time-consuming process of manual inspection. More specifically, the ability to perform classification early on in the lifetime of a transient would allow for ample time to take spectra at the peak luminosity of the event or at multiple points over the course of the event's lifetime.

\subsection{Photometric Supernova Classification}
An impressive body of work has emerged over the past decade on photometric classification of supernovae. Since only a small percentage of discovered supernovae have ever been followed up spectroscopically, a reliable photometric classifier is indispensable to the advancement of supernova science. 

The Supernova Photometric Classification Challenge \citep[SNPhotCC,][]{spcc_1,spcc_2} created not only an incentive to invest in photometric SN classification, but also a dataset that would be used to train and evaluate classifiers for years to come. Successful approaches range from empirical template-fitting \citep{sako2008} to making classification decisions based on manually extracted features \citep{richards, karpenka}. The more recent Photometric LSST Astronomical Time-series Classification Challenge \citep[PLAsTiCC,][]{plasticc} diversified the dataset by asking participants to differentiate between 14 different transient and variable object classes, including the 6 common supernova types included in this work.  The top entries made use of feature extraction paired with various machine learning classification methods, such as boosted decision trees and neural networks \citep{plasticc_results}. Ensemble methods, in which the results of multiple classifiers are combined to create the final classification probability, were widely used as well.

Deep learning is a branch of machine learning that seeks to eliminate the necessity of human-designed features, decreasing the computational cost as well as avoiding the introduction of potential biases \citep{charnock_moss, moss, naul, aguirre}. In recent years, many deep learning techniques have been applied to the challenge of photometric SN classification. 

Recurrent neural networks (RNNs) are designed to learn from sequential information, such as time-series data, and have been used with great success on this problem. \cite{charnock_moss} applies a variant of RNNs known as Long Short Term Memory networks \citep[LSTMs,][]{Hochreiter1997LongSM} to achieve impressive performance distinguishing SNIa from core collapse (CC) SNe. \cite{rapid} uses a gated recurrent unit (GRU) RNN architecture to be able to perform real-time and early lightcurve classification. \cite{moller} performs both binary classification and classification by type with full and partial lightcurves using Bayesian RNNs. \cite{superraenn} uses a GRU RNN as an autoencoder to smooth out irregularities in lightcurve data that is then fed into a random forest classifier.

Convolutional neural networks (CNNs), which are used in this work, are a state-of-the-art image recognition architecture \citep{lecun1989backpropagation, lecun1998gradient, alexnet, zeiler2014visualizing}. \cite{pelican} addresses the issue of non-representative training sets by using a CNN as an autoencoder to learn from unlabeled test data. \cite{alerce_stamp} developed an image time-series classifier as part of the ALeRCE alert broker, using a CNN to differentiate between various transient types as well as bogus alerts.

Outside of these traditional models, deep learning is still providing new and creative solutions to the photometric transient classification problem.  Convolutional recurrent neural networks are used to classify a time series of image stamps by \cite{ramanah} to detect gravitationally lensed supernovae. A newer type of deep learning architecture, known as a transformer, achieves a very impressive result when applied to the PLAsTiCC dataset by \cite{transformer}. A variational autoencoder was used by ParSNIP \citep{parsnip} to develop a low-dimensional representation of transient lightcurves that uses redshift-annotated photometric data to perform full lightcurve photometric classification and generate time-varying spectra, among other tasks.

\subsection{Early Photometric Supernova Classification}

Though much progress has been made on the photometric supernova classification problem, most of the solutions tackle classification of full supernova lightcurves retrospectively. However, the earlier an object can be classified, the more opportunities there are for the community to perform follow-up observation. Spectroscopic or photometric follow-up at early stages not only reveals insights into progenitor physics, but can also serve as a benchmark for further observations at later epochs. SN type IIb, for example, exhibit hydrogen features in early spectra that quickly disappear over time \citep{SNIIb}. Shock breakout physics is another use case of follow-up observation. \cite{patrick} was the first to report capturing the complete evolution of a shock cooling lightcurve, a short-lived event preceding peak luminosity that reveals properties of the shock breakout and progenitor star for stripped-envelope supernovae such as the IIb.

Despite the general focus on full lightcurve classification, several notable works have addressed the challenge of early photometric classification. \cite{sullivan} was able to not only differentiate between SNIa and CC SN, but also predict redshift, phase, and lightcurve parameters for SNIa using only two or three epochs of multiband photometry data. \cite{poznanski} also performed binary Ia vs. CC SNe classification, but using a Bayesian template-fitting technique on only single epoch photometry and photometric redshift estimates. {\small{PSNID}} \citep{sako2008, sako2011}, the algorithm that produced the highest overall figure of merit in SNPhotCC, was used by the Sloan Digital Sky Survey \citep{sdss} and the Dark Energy Survey \citep{des} to classify early-time and full supernova lightcurves.

\cite{rapid} is a recent application of deep learning techniques specifically to early-time transient classification. A GRU RNN is trained and tested on a PLAsTiCC-derived dataset of 12 transients, including 7 supernova types, that are labeled at each epoch with ``pre-explosion" prior to the date of explosion and the correct transient type after explosion. Thus, the model is able to produce a classification at each epoch of observation. \cite{moller} has also produced an RNN-based photometric classifier that is capable of classifying partial supernova lightcurves, but primarily achieves good results for Ia vs. CC SN classification. \cite{villar} uses a recurrent variational autoencoder architecture to perform early-time anomaly detection for exotic astrophysical events within the PLAsTiCC dataset, such as active galactic nuclei and superluminous SNe. Finally, LSST alert brokers such as ALeRCE \citep{alerce_lc} specialize in accurate early-time classification of transient alerts.

\subsection{Overview}
Originally introduced in \cite{Qu_2021}, hereafter Q21, as a full lightcurve photometric classification algorithm, \scone\ was able to retrospectively differentiate Ia vs. CC SN with $>99$\% accuracy and categorize SNe into 6 types with $>98$\% accuracy without redshift information. Our approach centers on producing heatmaps from 2-dimensional Gaussian processes fit on each lightcurve in both wavelength and time dimensions. These “flux heatmaps” of each supernova detection, along with
“uncertainty heatmaps” of the Gaussian process uncertainty, constitute the dataset for our model. This preprocessing step smooths over irregular sampling rates between filters, mitigates the effect of flux outliers, and allows the CNN to learn from information in all filters simultaneously.

Section 2 outlines the details of the datasets and models used in this work and we discuss the classifier's performance on the various dataset types in Section 3, including a comparison with existing literature. We state our conclusions and goals for future work in Section 4.

\section{Methods}
\subsection{Simulations}
For this work, \scone\ was trained and tested on a set of LSST deep drilling field (DDF) simulations. The dataset was created with SNANA \citep{snana} using the PLAsTiCC transient class models for supernovae types Ia, II, Ibc, Ia-91bg, Iax, and SLSN \citep{plasticc_data, plasticc-models, SNIa_1, SNIa_2, SNIa_3, SNIax_1, SNII_1, SNII_2, SNII_3, SNII_4, SNIbc_1, SNIbc_2, SNIbc_3, SNIbc_4, SLSN_1, SLSN_2, SLSN_3}. The relative rates and redshift distribution are identical to those of the data produced for the PLAsTiCC challenge. This is the same dataset used to evaluate \scone's categorical classification performance in Q21.  {No cuts on individual low S/N lightcurve points were made, but lightcurves with fewer than two $5\sigma$ detections were removed, as $t_{\rm trigger}$ would be ill-defined in those cases. We note that in observed data, transient light curve samples will contain SNe contaminated by other galactic astrophysical sources, but methods such as \cite{alerce_lc}} are reliably able to distinguish extragalactic and galactic events. Thus, we can assume the feasibility of creating a pure sample of SN lightcurves such as the one used in this work.

\begin{table}
    \centering
    \caption{Training, validation, and test dataset sizes for the $t_{\rm trigger}+N$ datasets.}
    \label{tbl:datasets}
    \begin{tabular}{l c c}
        \hline
        Dataset & Number of Each Type & Total Size\\
        \hline
        Training & 6148 & 36888\\
        Validation & 769 & 4614\\
        Test & 768 & 4608\\
        \hline
        Full & 7685 & 46110\\
    \end{tabular}
\end{table}

\subsection{Trigger Definition}
We define a \textit{detection} as any observation exceeding the 5$\sigma$ signal-to-noise (S/N) threshold. We define the \textit{trigger} as the next detection that occurs at least one night after the first. In this work, the dataset with the least photometric information includes observations up to (and including) the date of trigger. Thus, all SNe in our datasets have at least two epochs of observation. As the date of first detection is also a common choice of trigger date in other transient surveys, the implications of this discrepancy are explored further in Section 3.3. We present results on a dataset where the distinction between these two definitions is small, i.e. $t_{\rm trigger}\leq t_{\rm first\;detection}+5$.

\subsection{Datasets and Heatmap Creation}
\subsubsection{$t_{\mathrm{trigger}}+N$ Datasets}
To evaluate \scone's classification performance on lightcurves at different stages of the supernova lifetime, five sets of heatmaps were created from the simulations described in Section 2.1. All sets of heatmaps take data starting 20 nights prior to the date of trigger ($t_{\mathrm{trigger}}$) and end at $N=$ 0, 5, 15, 25, and 50 days after the date of trigger, respectively. Hereafter, these are collectively referred to as ``$t_{\mathrm{trigger}}+N$ datasets".

\begin{figure*}
    \centering
    \includegraphics[scale=0.22,trim={0cm 2cm 6cm 2cm},clip=true]{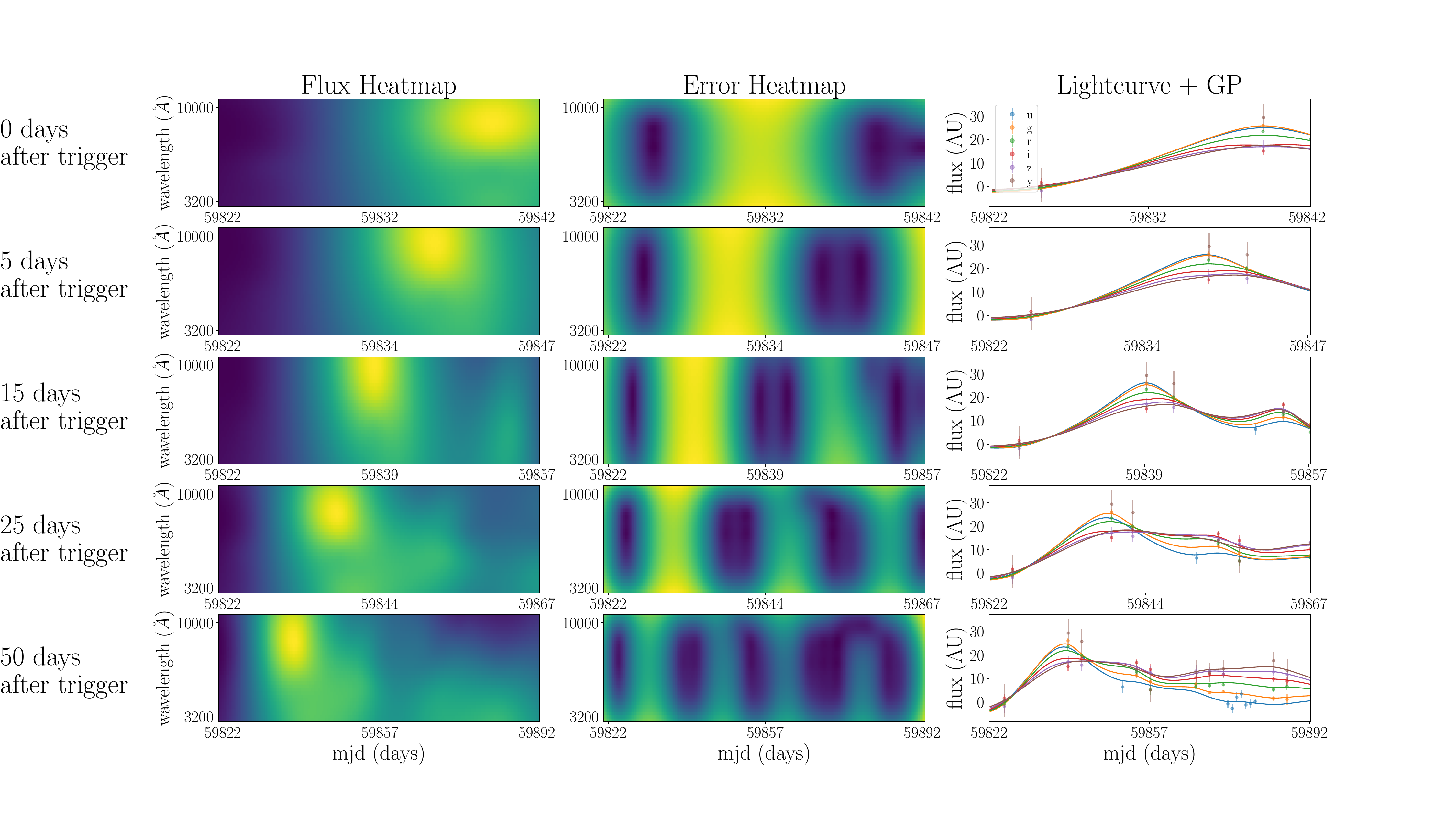}
    \caption{An SNII ($z=0.39$) shown in all five heatmap datasets along with the lightcurves and Gaussian process fits used to create each heatmap. The flux and flux error measurements from the raw photometry are shown as points with error bars, while the Gaussian process fits to each photometry band are shown as curves. The Gaussian process errors, which are used to create the heatmaps in the middle column, are not shown in the lightcurve plots. The $x$-axis limit of the plots in each row are different, as the lightcurve is truncated according to the label on the left for each row in the figure. }
    \label{fig:same-sn}
\end{figure*}

Prior to training, the lightcurve data is processed into heatmaps. We use the approach described by \citet{avocado} to apply 2-dimensional Gaussian process regression to the raw lightcurve data to model the event in the wavelength ($\lambda$) and time ($t$) dimensions. We use the Mat\'ern kernel ($\nu = \frac{3}{2}$) with a fixed 6000~\AA\  characteristic length scale in $\lambda$ and fit for the length scale in $t$. Once the Gaussian process regression model has been trained, we obtain its predictions on a $\lambda,t$ grid and call this our ``flux heatmap".


It is important to note that the Gaussian processes are fit on lightcurves truncated at $N$ days after trigger in each dataset and not given access to lightcurve information past the cutoff date. Thus, though the $\lambda$ axis is not affected by the different choices of $N$, the $t$ range of the input lightcurve data varies for each $t_{\rm trigger} + N$ dataset. For the datasets in this work, the $\lambda,t$ grids were chosen to preserve the shape of the resulting heatmap despite the fact that the number of nights of lightcurve data varies between the $t_{\rm trigger} + N$ datasets. $\lambda$ is chosen to be $3000 < \lambda < 10,100$~\AA\ with a $221.875$~\AA\ interval for all datasets, while the $t$ interval depends on the number of nights of data: $t_{\rm trigger} - 20 \leq t \leq t_{\rm trigger} + N$ with a $\frac{N+20}{180}$ day interval, where $N=0,5,15,25,50$. This ensures that all heatmaps have size $32 \times 180$.

In addition to the flux heatmap, we also take into account the uncertainties on these predictions at each $\lambda_i, t_j$, producing an ``error heatmap". We stack these two heatmaps depthwise for each SN lightcurve and divide by the maximum flux value to constrain all entries to [0,1]. This $32 \times 180 \times 2$ tensor is our input to the convolutional neural network.

An example of the heatmaps and associated lightcurves of a single SN in all 5 datasets is shown in Figure~\ref{fig:same-sn}. Results on the $t_{\rm trigger}+N$ datasets are described in Section 3.2.

\subsubsection{Bright Supernovae}
Our model was also evaluated on the subset of particularly bright supernovae from the $t_{\rm trigger} +0$ and $t_{\rm trigger} +5$ datasets to emulate a real-world use case of \scone\ for spectroscopic targeting, as bright supernovae are better candidates for spectroscopic follow-up. ``Bright SNe" included in these datasets were chosen to be SNe with last included detection $r<20$ mag. With this threshold, there were 907 SNe in the $t_{\rm trigger} +0$ bright dataset and 5,088 SNe in the $t_{\rm trigger} +5$ bright dataset. As described in more detail in Section 2.4, \scone\ was trained with a standard $t_{\rm trigger}+N$ training set combined with 40\% of the $t_{\rm trigger}+N$ bright dataset, and tested on the $t_{\rm trigger}+N$ bright dataset. Results on these datasets are described in Section 3.5.

\subsubsection{Mixed Dataset}
In order to evaluate \scone's ability to classify SNe with any number of nights of photometry, a sixth dataset (the ``mixed" dataset) was created from the same PLAsTiCC simulations. Data is taken starting 20 nights prior to the date of trigger (as with the $t_{\rm trigger}+N$ datasets) but truncated at a random night between 0 and 50 days after trigger. Due to the choice of the $t$ interval described in Section 2.3.1, heatmaps with any number of nights of photometry data are all the same size can thus be mixed in a single dataset in this manner. We train \scone\ on this mixed dataset and evaluate its performance on each of the $t_{\mathrm{trigger}}+N$ datasets in Section 3.6.

\subsection{Dataset Train/Test Split}
Due to the importance of class balancing in machine learning datasets, the same quantity of SNe from each SN type was selected to create the $t_{\rm trigger} +N$ and mixed datasets used to train, validate, and test \scone. 7685 SNe of each of the 6 types were randomly chosen, as this was the quantity of the least abundant type. Thus, the size of each full dataset was 46,110. An 80/10/10 training/validation/test split was used for all results in this work. The sizes of the training, validation, and test subsets of each dataset can be found in Table \ref{tbl:datasets}.

For evaluation on the bright datasets, \scone\ was trained on a hybrid training set of 40\% of the $t_{\rm trigger} +N$ bright dataset combined with a $t_{\rm trigger} +N$ training set, prepared as described in Section 2.3.1. Thus, the training set was not quite class-balanced, as the bright dataset is not class-balanced but the $t_{\rm trigger}+N$ training set is. The trained model was then evaluated on the full bright dataset to produce the results shown in Figure~\ref{fig:bright}. Due to the imbalanced nature of the bright datasets, the confusion matrices in this figure take the place of an accuracy metric, which could be misleading. We chose to include 40\% of the bright dataset in the training process to ensure that the model has seen enough of these particularly bright objects to make reasonable predictions.

\subsection{Model}
\begin{figure*}
\centering
    \includegraphics[scale=0.45, trim={2.25cm 0cm 0cm 0cm}]{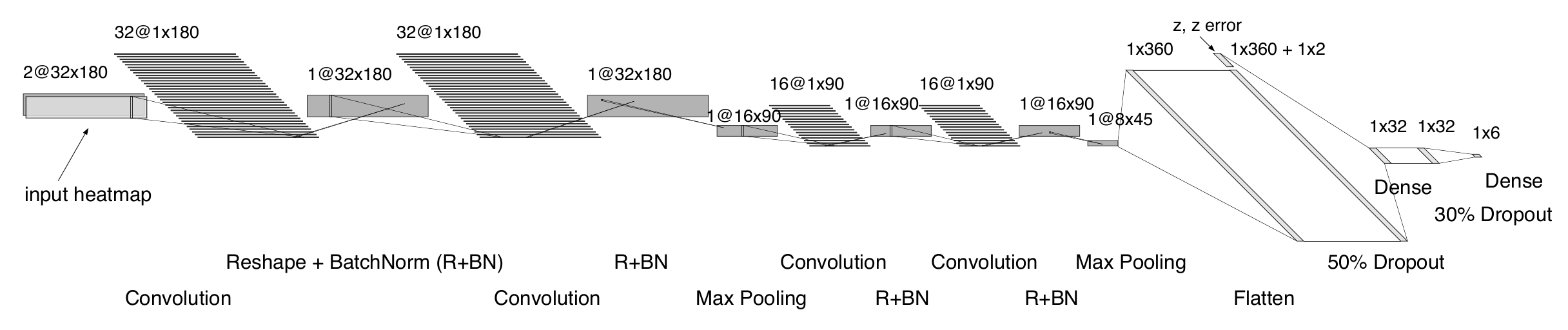}
    \centering
    \caption{\scone\ architecture with redshift information for categorical early lightcurve classification.}
    \label{fig:early-architecture}
\end{figure*}

In this work, we report early lightcurve classification results using the vanilla \scone\ model developed in Q21 as well as a variant of \scone\ that incorporates redshift information. The architecture of \scone\ with redshift is shown in Figure~\ref{fig:early-architecture}. Both redshift and redshift error are concatenated with the output of the first dropout layer and used as inputs to the fully connected classifier. The model uses spectroscopic redshift information when available and photometric redshift estimates if not.

Prior to training and testing, the input flux and error heatmaps are divided by the maximum flux value of each heatmap for normalization. This means that absolute brightness information is not used for classification. All results in this work, with and without redshift, used the sparse categorical crossentropy loss function, the Adam optimizer \citep{adam}, and trained for 400 epochs with a batch size of 32. \scone\ without redshift used a constant 1e-3 learning rate, whereas \scone\ with redshift used a constant 5e-4 learning rate.


\subsection{Computational Performance}
The time required for the heatmap creation process was measured using a sample of 100 heatmaps on a single 32-core NERSC Cori Haswell compute node (with Intel Xeon Processor E5-2698 v3). The time required to create one heatmap was $0.03 \pm 0.01$ seconds. When producing larger-scale datasets, this process is also easily parallelizable over multiple cores or nodes to further decrease heatmap creation time.

\scone\ without redshift has 22,606 trainable parameters and \scone\ with redshift has 22,670 trainable parameters, while other photometric classification models require at least hundreds of thousands. The performance gains of this simple but effective model compounded with a small training set make \scone\ lightweight and fast to train. The first training epoch on a NVIDIA V100
Volta GPU takes approximately 17 seconds (4 milliseconds per batch with a batch size of 32), and subsequent training epochs take approximately 5 seconds each with TensorFlow’s dataset caching. The first training epoch on
a Haswell node takes approximately 12 minutes (625 milliseconds per batch), and subsequent epochs take approximately 6 minutes each. Test time per batch of 32 examples is 3 milliseconds on GPU and 10 milliseconds on a Haswell CPU.

\section{Results and Discussion}
\subsection{Evaluation Metrics}
The \textit{accuracy} of a set of predictions describes the frequency with which the predictions match the true labels. In this case, we define our prediction for each SN example as the class with highest probability output by the model, and compare this to the true label to obtain an accuracy.

The \textit{confusion matrix} is a convenient visualization of the correct and missed predictions by class, providing a bit more insight into the model's performance. The confusion matrices shown in Figure~\ref{fig:cm} are normalized such that the $(i,j)$ entry describes the fraction of the true class, $i$, classified as class $j$. The confusion matrices in Figure~\ref{fig:bright} are colored by the normalized values, just like Figure~\ref{fig:cm}, but overlaid with absolute (non-normalized) values. For both figures, the $(i,i)$ entries, or those on the diagonal, describe correct classifications.

The \textit{receiver operating characteristic (ROC) curve} makes use of the output probabilities for each class rather than simply taking the highest probability class, as the previous two metrics have done. We consider an example to be classified as class $i$ if the output probability for class $i$, or $p_i$, exceeds some threshold $p$ ($p_i > p$). The ROC curve sweeps values of $p$ between 0 and 1 and plots the true positive rate (TPR) at each value of $p$ against the false positive rate (FPR). 

TPR is the percentage of correctly classified objects in a particular class, or true positives (TP), as a fraction of all examples in that class, true positives and false negatives (TP+FN). Other names for TPR include \textit{recall} and \textit{efficiency}. The values along the diagonal of the normalized confusion matrices in Figure~\ref{fig:cm} are efficiency values.
$$\mathrm{Efficiency=TPR = \frac{TP}{TP+FN}}$$  
FPR is the percentage of objects incorrectly classified as a particular class, or false positives (FP), as a fraction of all examples not in that class, false positives and true negatives (FP+TN).
$$\mathrm{FPR = \frac{FP}{FP+TN}}$$

The \textit{area under the ROC curve}, or \textit{AUC}, is used to evaluate the classifier from its ROC curve. A perfect classifier would have an AUC of 1, while a random classifier would score (on average) a 0.5.

The \textit{precision} or \textit{purity} of a set of predictions is the percentage of correctly classified objects in a particular predicted class.
$$\mathrm{Precision=\frac{TP}{TP+FP}}$$

\begin{table*}
    \centering
    \caption{Training, validation, and test accuracies \textit{without} redshift information for each early lightcurve dataset. These averages and standard deviations were computed from 5 independent runs of \scone.}
    \label{tbl:no-z-acc}
    \begin{tabular}{l c c c c c}
        \hline
        Accuracy & \multicolumn{5}{c}{Days After Trigger}\\
        \hline
        & 0 days & 5 days & 15 days & 25 days & 50 days \\
        \hline
        Training & $58.36 \pm 0.14$\% & $68.92 \pm 0.21$\% & $73.99 \pm 0.14\%$& $76.89 \pm 0.29$\% & $80.93 \pm 0.14$\%\\
        Validation & $59.57 \pm 0.51$\% & $70.74 \pm 0.59$\% & $73.31 \pm 3.01\%$ & $79 \pm 0.84$\% & $82.5 \pm 2.35$\%\\
        Test & $59.66 \pm 0.43$\% & $70.05 \pm 0.63$\% & $73.66 \pm 2.36$\%& $79 \pm 0.86$\% & $82.2 \pm 1.8$\%\\
    \end{tabular}
\end{table*}

\begin{table*}
    \centering
    \caption{Training, validation, and test accuracies \textit{with} redshift information for each early lightcurve dataset. These averages and standard deviations were computed from 5 independent runs of \scone.}
    \label{tbl:with-z-acc}
    \begin{tabular}{l c c c c c}
        \hline
        Accuracy & \multicolumn{5}{c}{Days After Trigger}\\
        \hline
        & 0 days & 5 days & 15 days & 25 days & 50 days \\
        \hline
        Training & $72.73 \pm 0.27$\% & $79.61 \pm 0.3$\% & $83.07 \pm 0.2\%$& $84.68 \pm 0.2$\% & $87.17 \pm 0.26$\%\\
        Validation & $74.78 \pm 0.18$\% & $80.52 \pm 1.42$\% & $83.98 \pm 1.15\%$& $86.75 \pm 0.5$\% & $89.2 \pm 0.85$\%\\
        Test & $74.27 \pm 0.51$\% & $80.2 \pm 0.93$\% & $84.14 \pm 1.37$\%& $86.71 \pm 1$\% & $89.04 \pm 0.39$\%\\
    \end{tabular}
\end{table*}

\begin{figure*}
    \centering
    \includegraphics[scale=0.35,trim={0 0 0 0}]{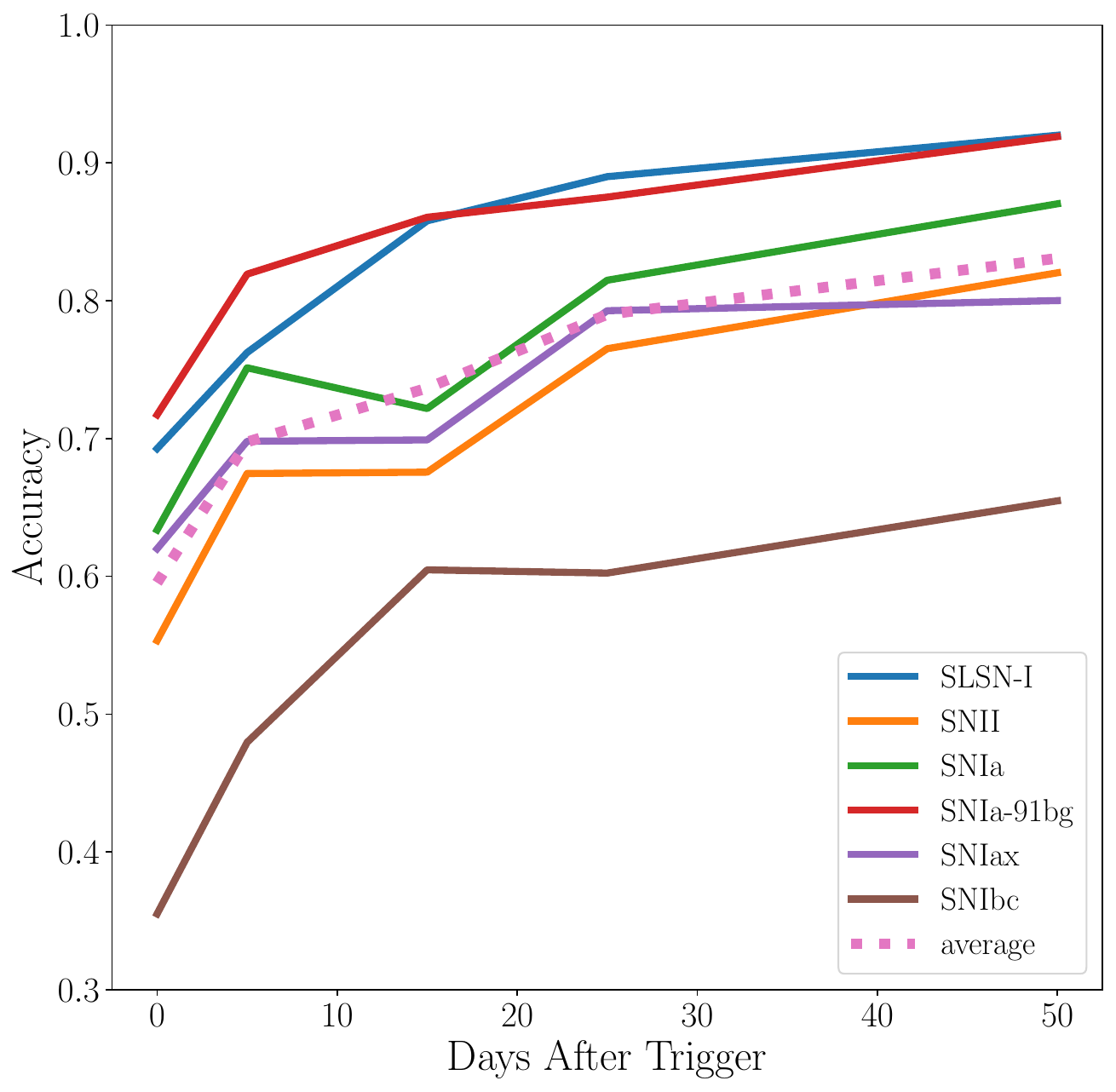}
    \includegraphics[scale=0.35,trim={0 0 0 0}]{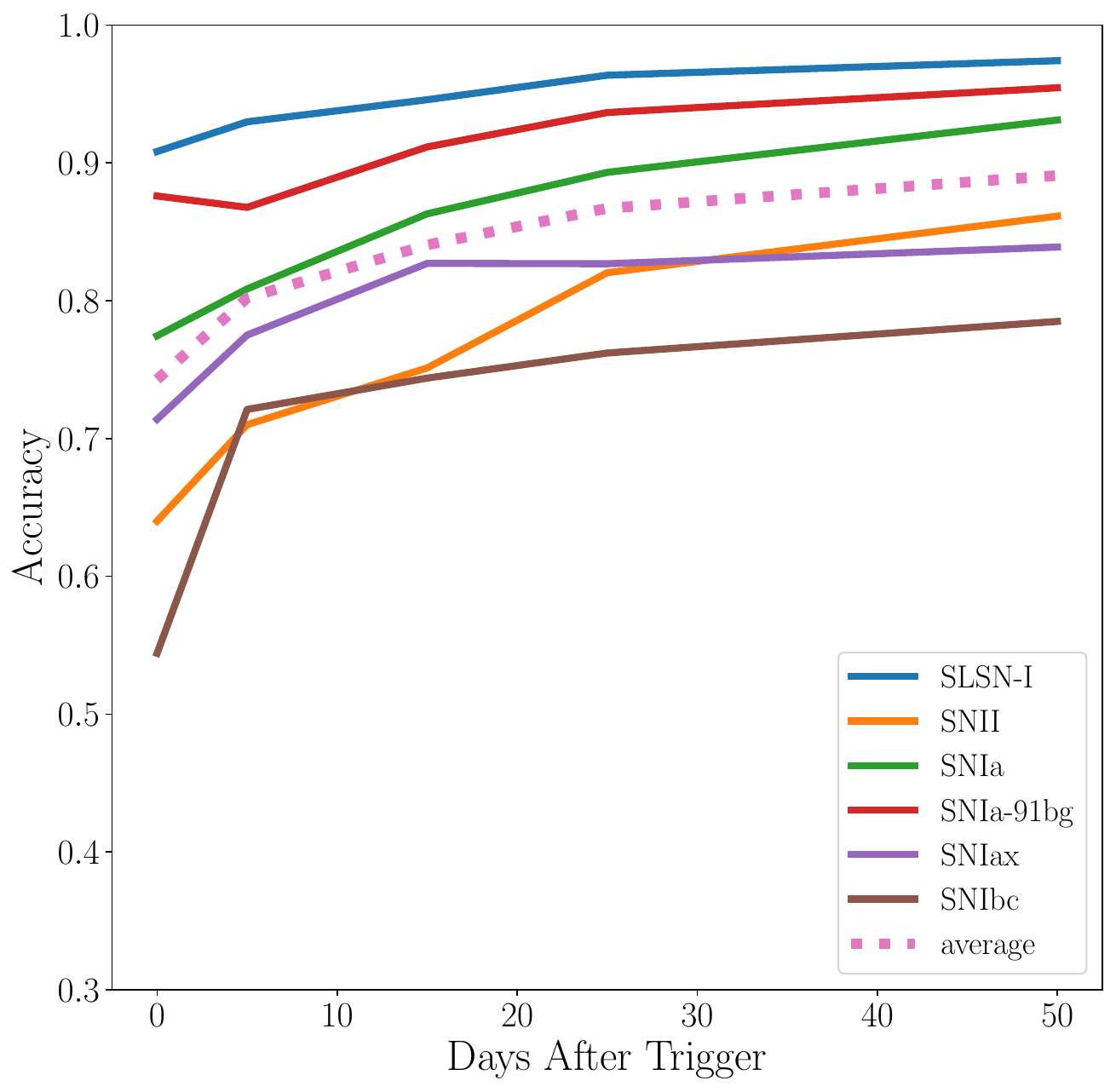}

    \centering
    \caption{Accuracy/efficiency over time for each supernova type without redshift (left) and with redshift (right) for the $t_{\mathrm{trigger}}+N$ test datasets. The values used in this plot correspond with the diagonals on each normalized confusion matrix in Figure~\ref{fig:cm}.}
    \label{fig:accs}
\end{figure*}

\subsection{$t_{\mathrm{trigger}}+N$ Datasets}
The accuracies our model achieved without redshift on each $t_{\mathrm{trigger}}+N$ dataset are described in Table~\ref{tbl:no-z-acc}, and the accuracies with redshift are described in Table~\ref{tbl:with-z-acc}. These tables show that redshift unequivocally improves classification performance, especially at early times when there is little photometric data to learn from. The inclusion of redshift information not only increases the average accuracies for each dataset but also improves the model's generalizability, as the standard deviations for the validation and test accuracies are lower overall in Table~\ref{tbl:with-z-acc}.

The largest improvement in accuracy between $t_{\mathrm{trigger}}+N$ datasets occurred between 0 and 5 days after trigger for all datasets. Since the explosion likely reached peak brightness during this period, the lightcurves truncated at 5 days after trigger includes much more information necessary for differentiating between the SN types.

Figure~\ref{fig:accs} shows the accuracy evolution over time for each supernova type in the test sets. From the test sets with redshift plot on the right, it is clear that the jump in overall accuracy between 0 and 5 days after trigger can be attributed to the sharp accuracy boost experienced by SNIbc at 5 days after trigger. Overall, SNIbc benefited the most from the inclusion of redshift, though classification performance on all types saw improvement. Note that, as described in Section 2.3, all heatmaps are normalized to values between 0 and 1 so absolute flux values are not used to differentiate between types. Thus, the model cannot rely on relative luminosity information.

\begin{figure*}
    \centering
    \includegraphics[scale=0.35,trim={1cm 0 0 0}]{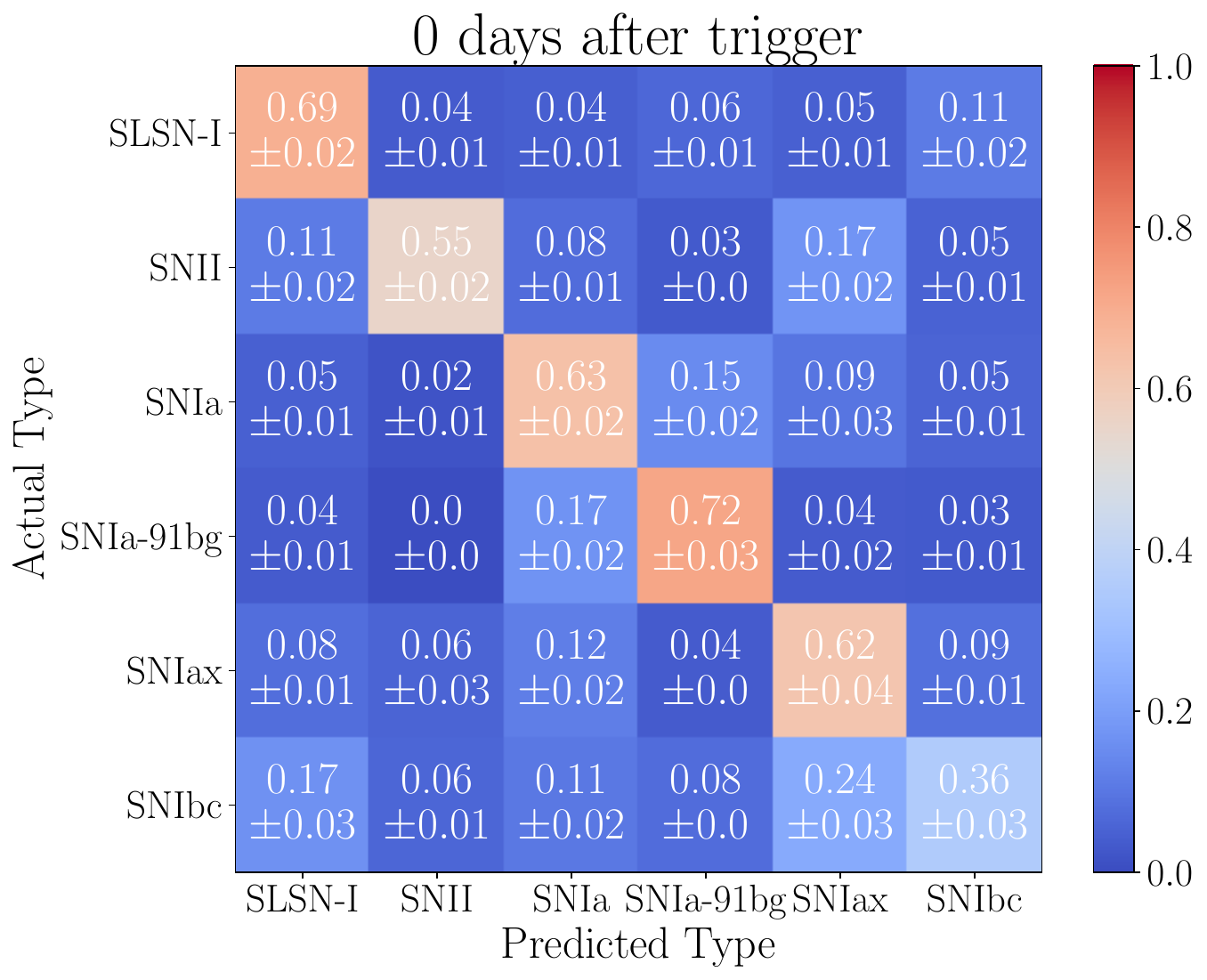}
    \includegraphics[scale=0.35,trim={1cm 0 0 0}]{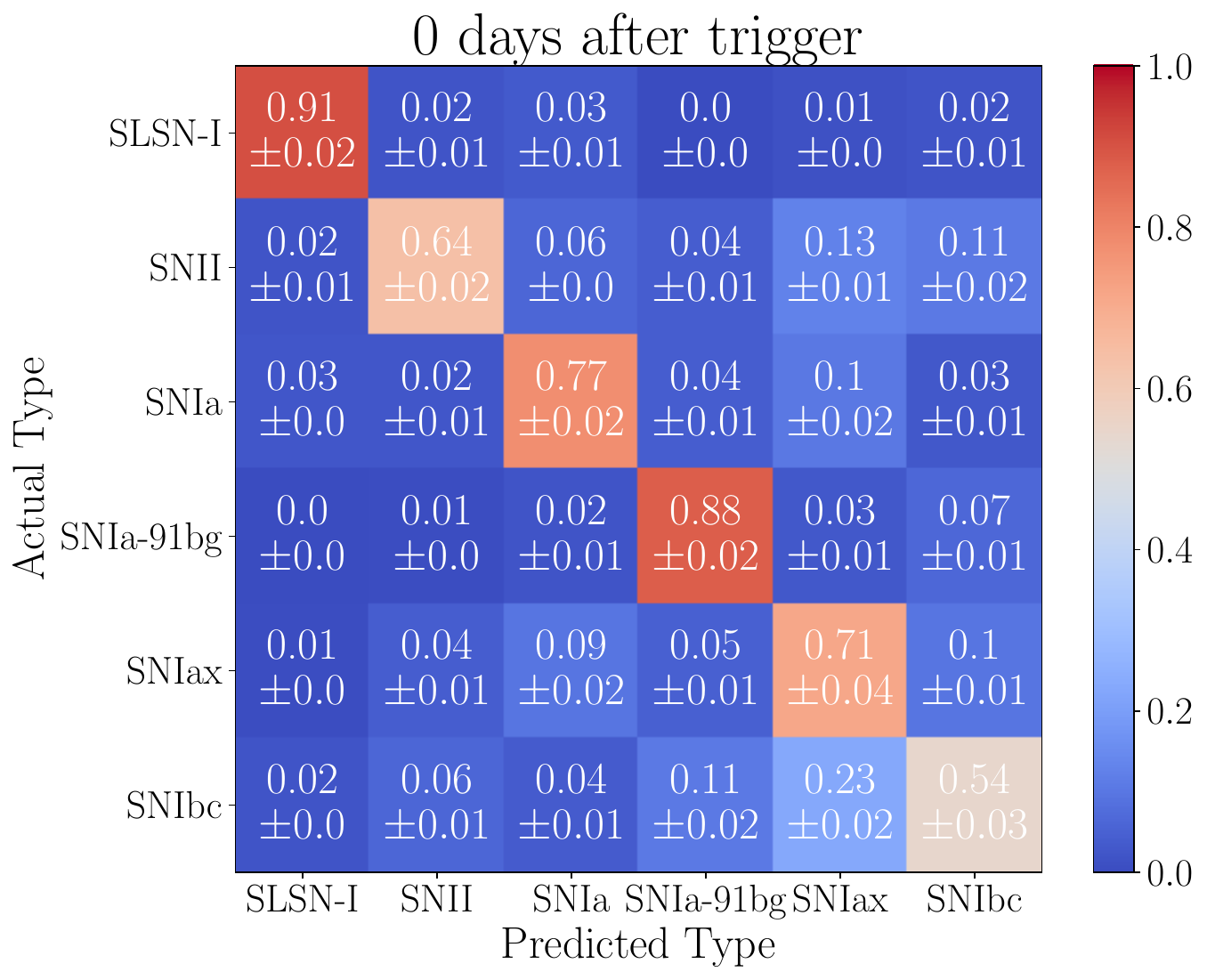}
    \includegraphics[scale=0.35,trim={1cm 0 0 0}]{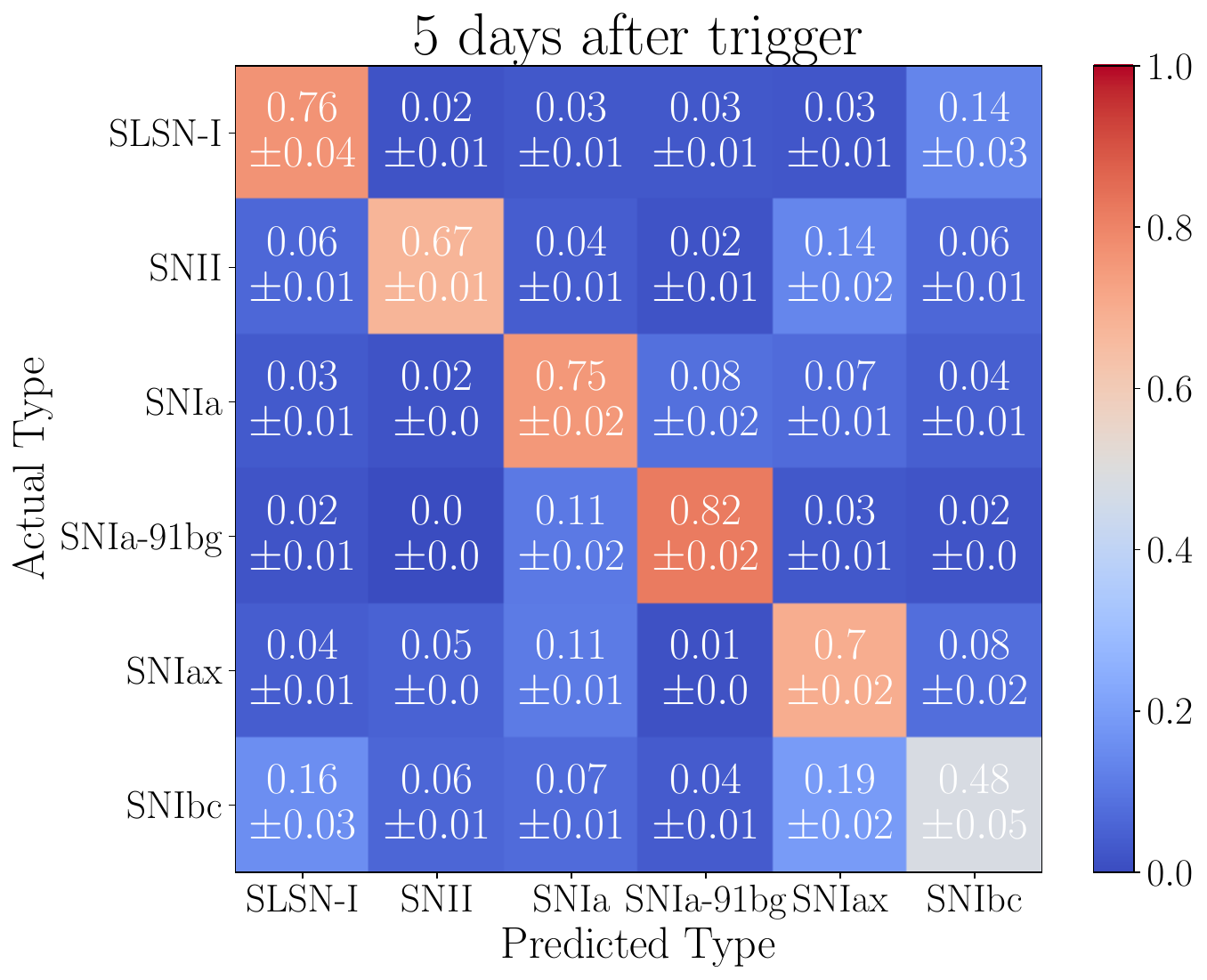}
    \includegraphics[scale=0.35,trim={1cm 0 0 0}]{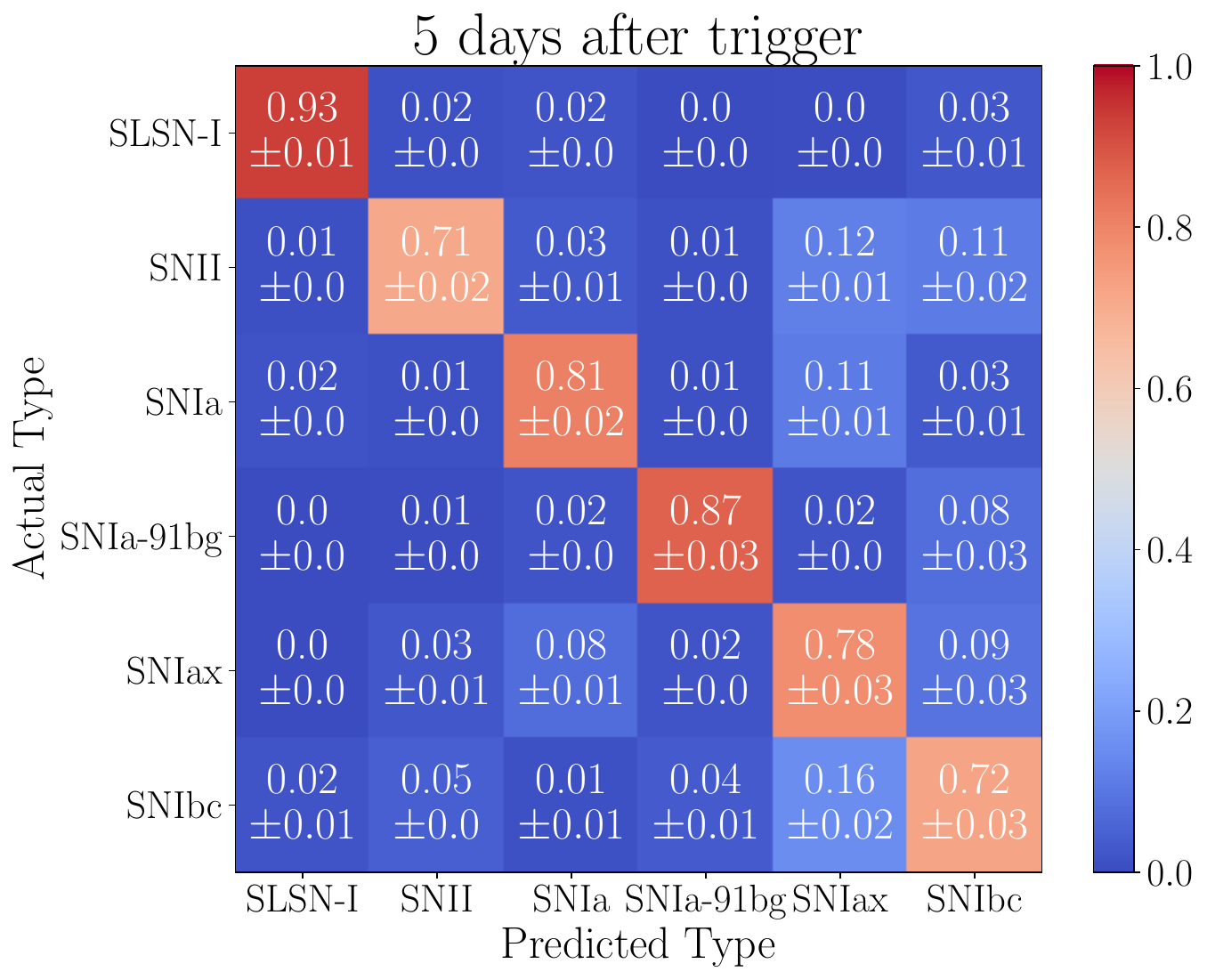}
    \includegraphics[scale=0.35,trim={1cm 0 0 0}]{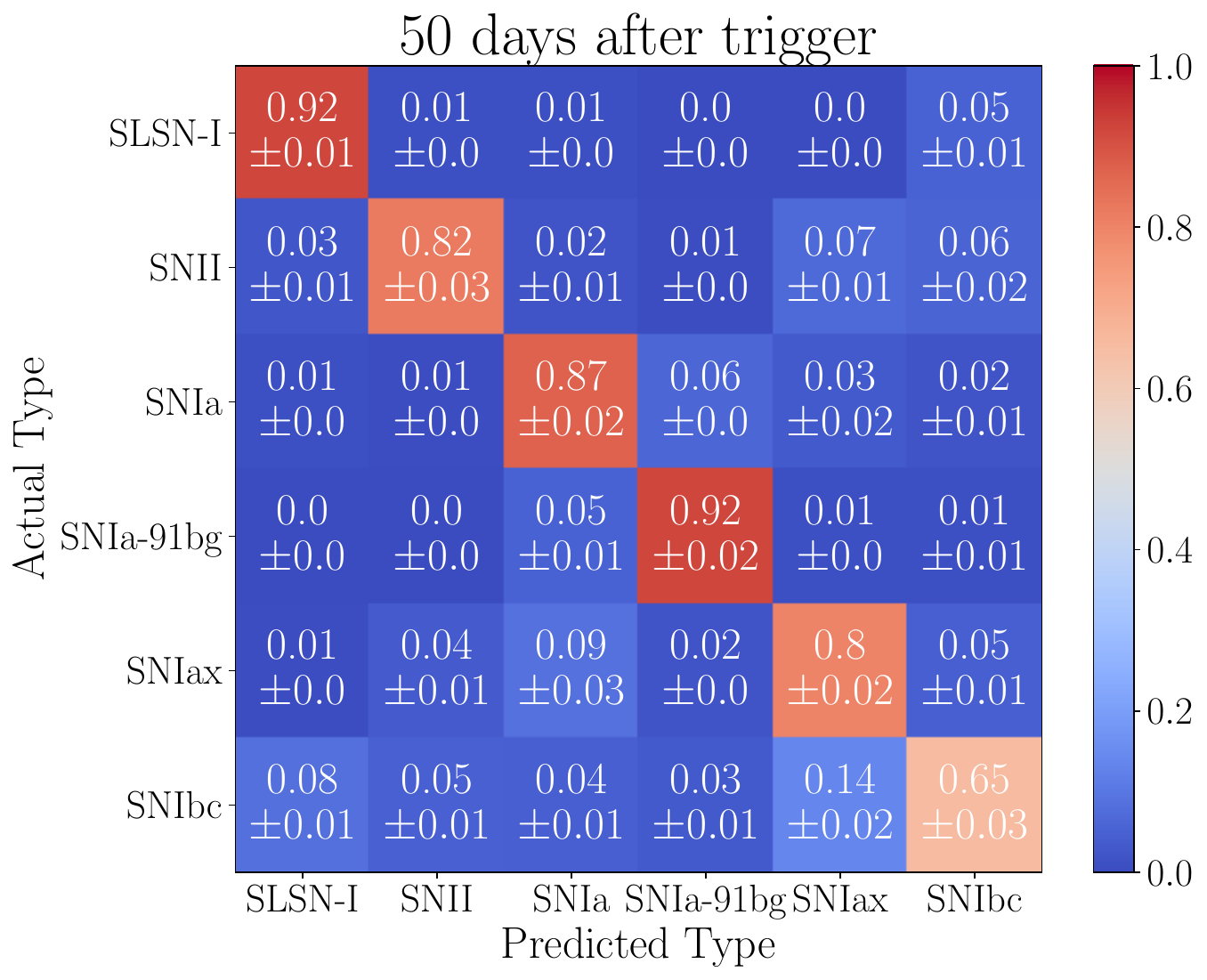}
    \includegraphics[scale=0.35,trim={0cm 0 0 0}]{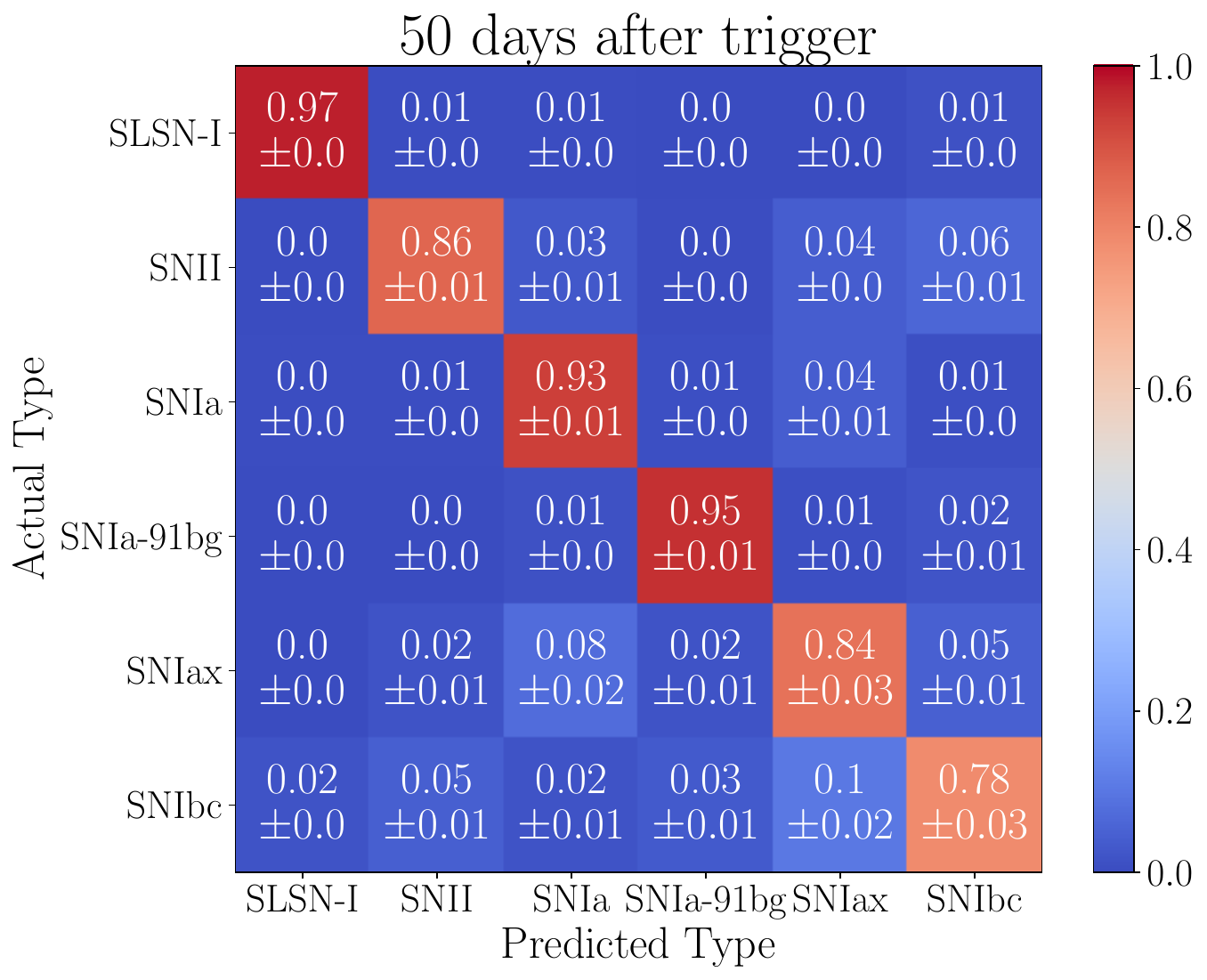}

    \caption{Normalized confusion matrices produced by \scone\ without (left) and with (right) redshift for the $t_{\mathrm{trigger}}+\{0,5,50\}$ test sets (heatmaps created from lightcurves truncated at 0, 5, and 50 days after the date of trigger). These matrices were made with test set classification performance from 5 independent runs of \scone.}
    \label{fig:cm}
\end{figure*}

The confusion matrices for $t_{\mathrm{trigger}}+\{0,5,50\}$ test sets with and without redshift information are shown in Figure~\ref{fig:cm}. The top two panels are early epoch classification results (0 and 5 days after trigger) and the bottom panel shows late epoch results. The confusion matrices from intermediate epochs (15 and 25 days after trigger) were omitted for brevity. 

At the date of trigger (top panel of Figure~\ref{fig:cm}), the incorporation of redshift information primarily prevents confusion between SLSN-I and SNIbc. True SLSN-I events misclassified as SNIbc decreased from 11\% on average to 2\% with redshift. True SNIbc misclassified as SLSN-I decreased from 16\% on average to 2\% with redshift. Overall, SLSN-I were classified with 91\% accuracy with redshift compared to 69\% without redshift, and SNIbc were classified with 54\% accuracy with redshift compared to 36\% without redshift. All types saw marked improvement in classification performance without redshift from 0 to 5 days after trigger, while classification with redshift saw drastic improvement in SNIbc accuracy but only minor improvement for other types. Finally, the effect of added redshift becomes less noticeable by late epochs, where classification accuracy (along the diagonal) is only mildly improved in the bottom panel of Figure~\ref{fig:cm}.

The confusion matrices in Figure~\ref{fig:cm} are normalized by true type, meaning that the values in each row sum to 1. Thus, the values along the diagonal are \textit{efficiency} scores. Normalizing by predicted type, such that the values in each column sum to 1, would result in \textit{purity} scores along the diagonal. However, since all datasets used in Figure~\ref{fig:cm} are class-balanced, the purity scores can be reconstructed from these confusion matrices by dividing each main diagonal value by the sum of the values in its column.

\begin{figure*}
    \centering
    \includegraphics[scale=0.3,trim={0 0 0 1cm}]{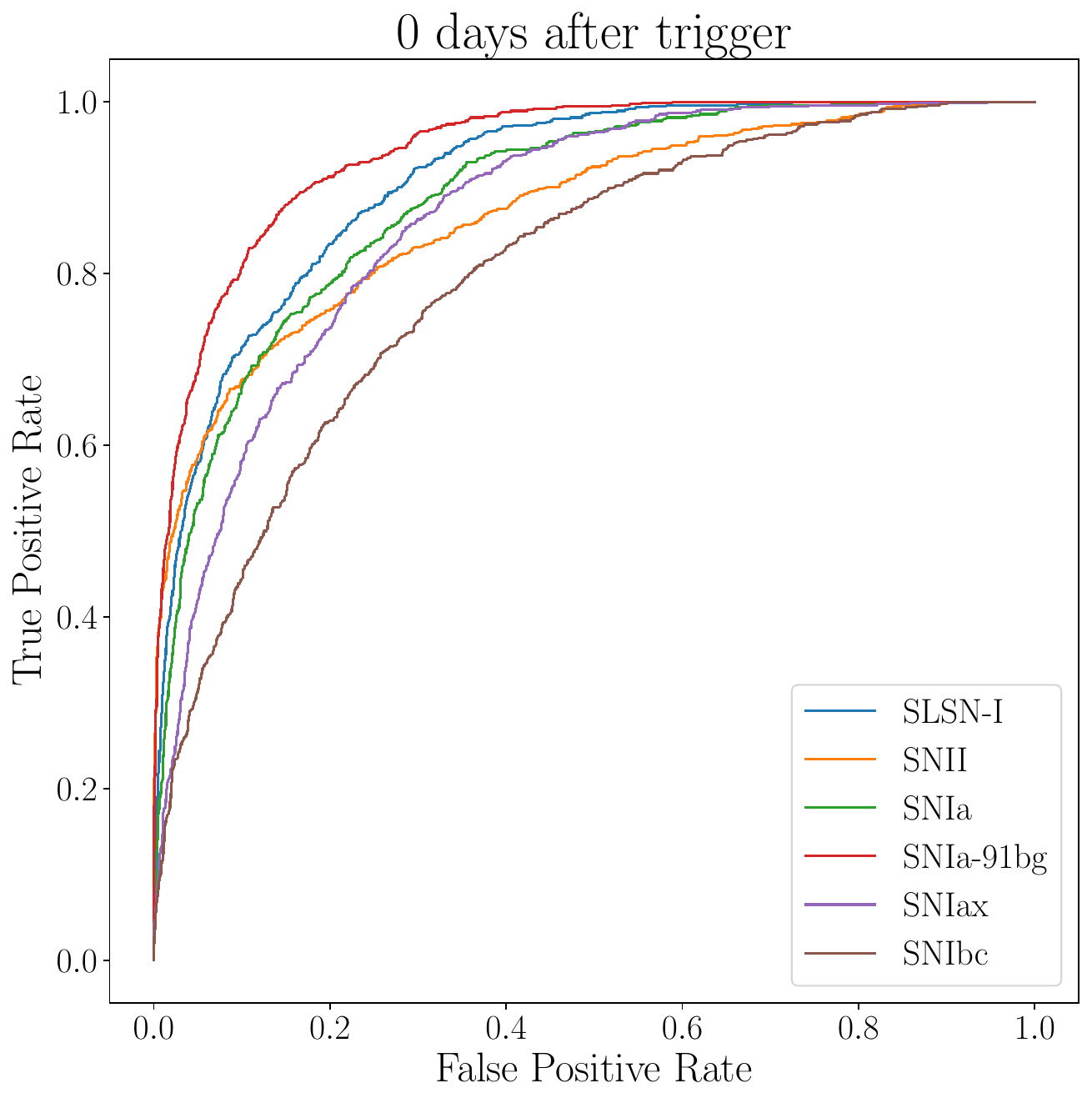}
    \includegraphics[scale=0.3,trim={0 0 0 1cm}]{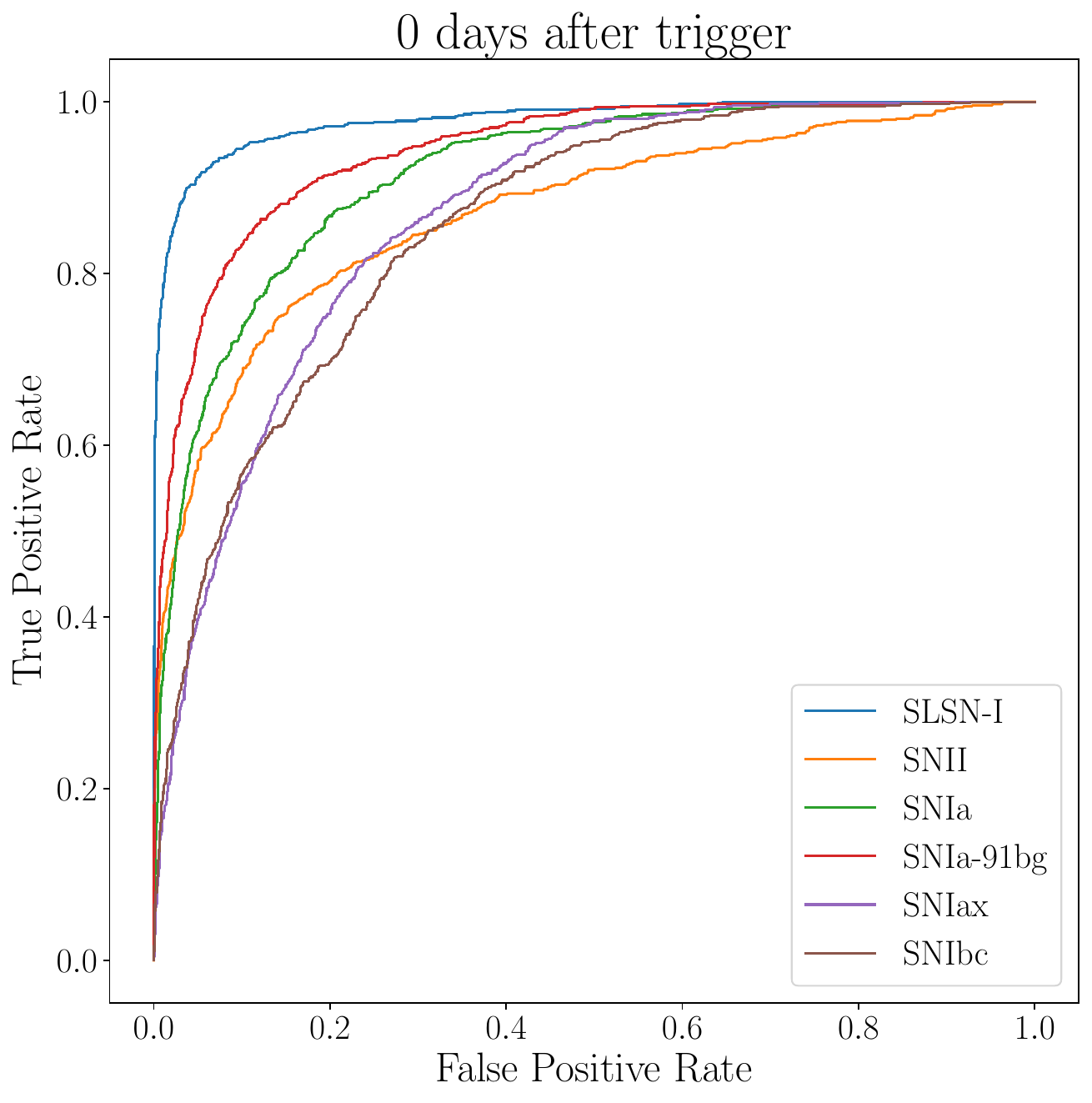}
    \includegraphics[scale=0.3,trim={0 0 0 0}]{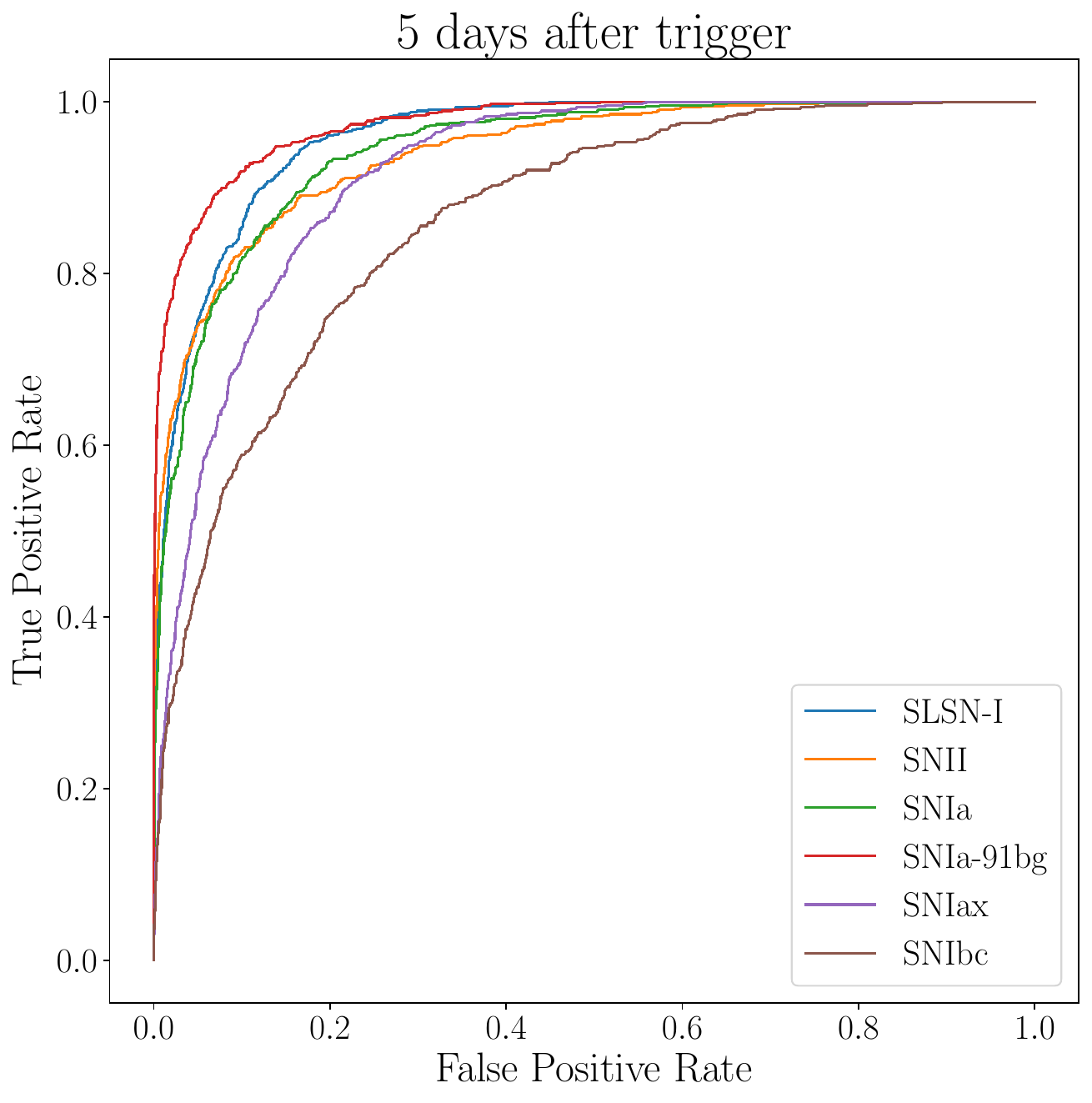}
    \includegraphics[scale=0.3,trim={0 0 0 0}]{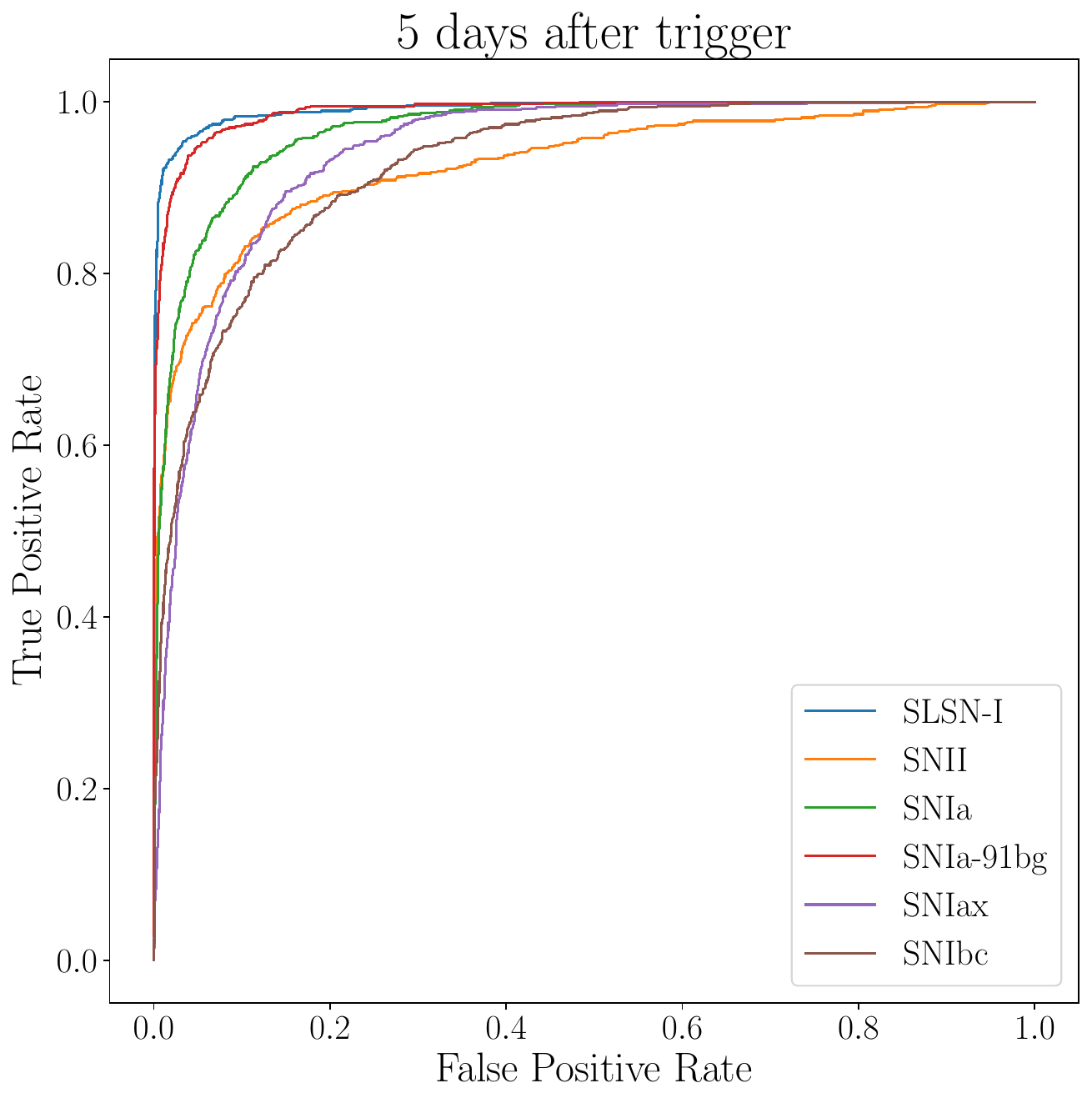}
    \includegraphics[scale=0.3,trim={0 0 0 0}]{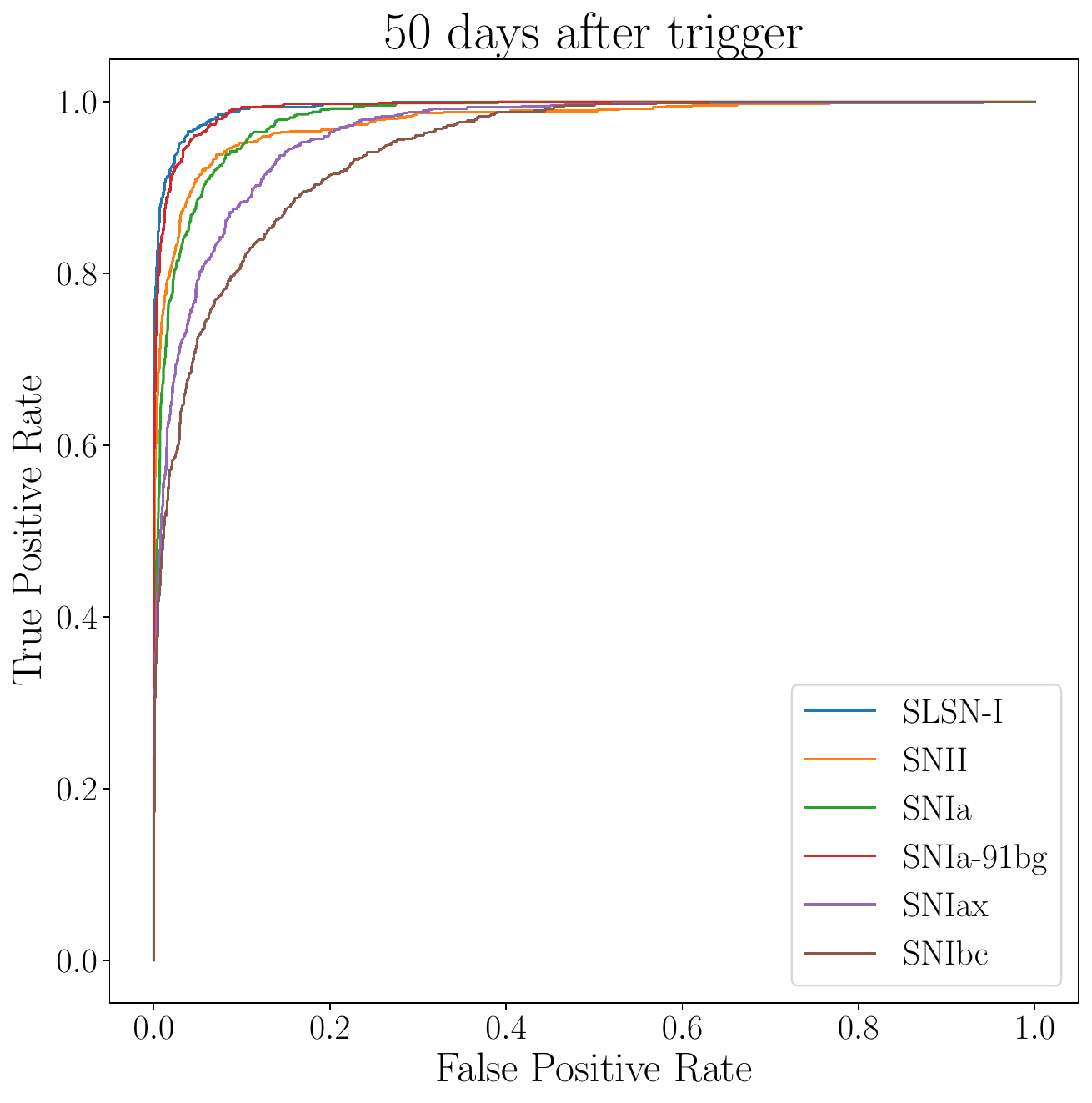}
    \includegraphics[scale=0.3,trim={0cm 0 0 0}]{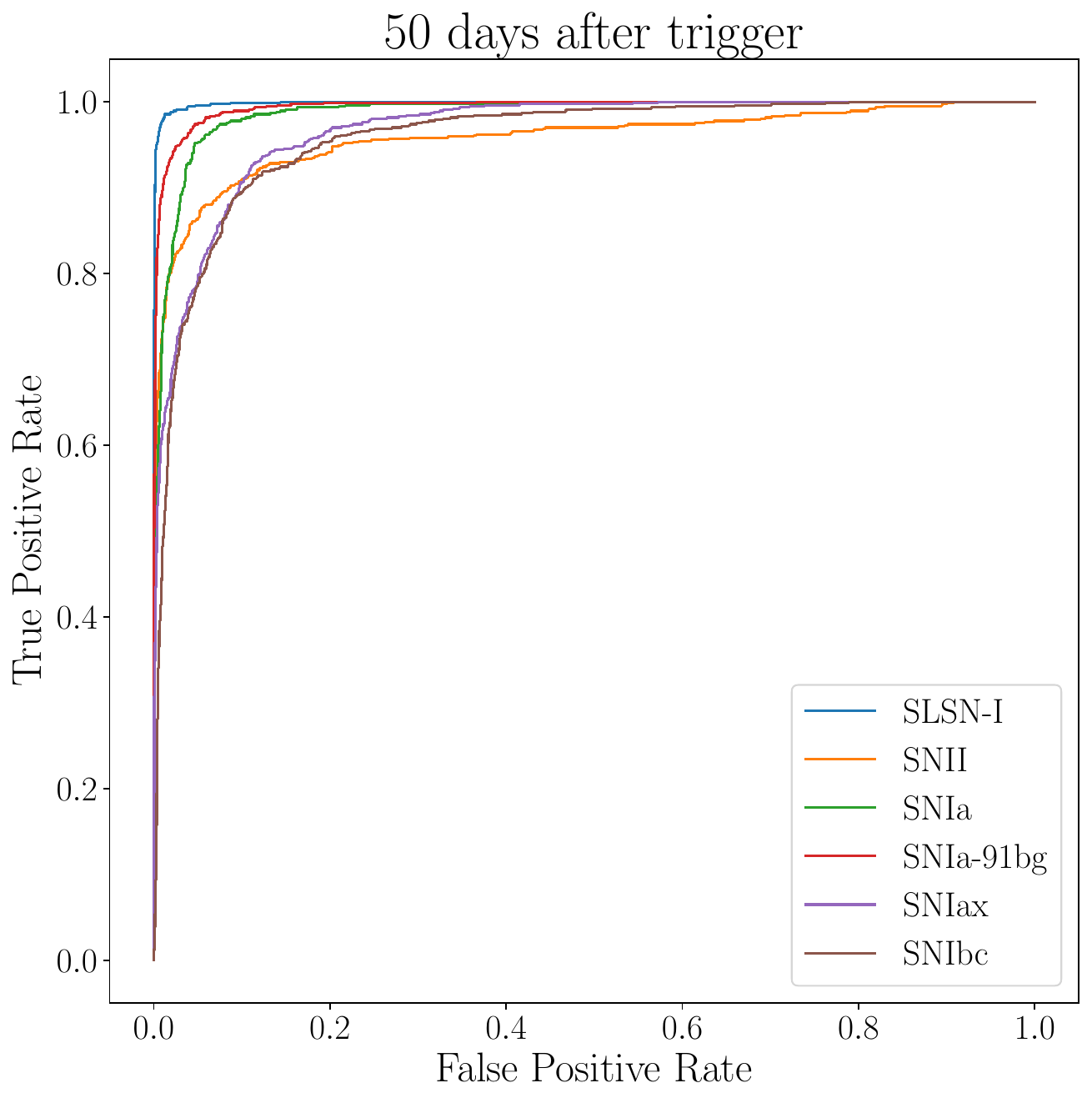}

    \caption{Receiver operating characteristic (ROC) curves produced by \scone\ without (left) and with (right) redshift for the $t_{\mathrm{trigger}}+\{0,5,50\}$ test sets (heatmaps created from lightcurves truncated at 0, 5, and 50 days after the date of trigger).}
    \label{fig:early-roc}
\end{figure*}

The datasets used for the confusion matrices in Figure~\ref{fig:cm} were also used to create ROC curves for each SN type. ROC curves for test sets without redshift are shown on the left side of Figure~\ref{fig:early-roc} and ROC curves for test sets with redshift are shown on the right. The addition of redshift information seems to most notably improve the model's ability to classify SLSN-I -- all three panels on the right show SLSN-I as the highest AUC curve whereas all three panels on the left show SNIa-91bg with a higher AUC curve than SLSN-I. This is consistent with our earlier observations from the confusion matrices and accuracy plots.

\begin{figure*}
    \centering
    \includegraphics[scale=0.35,trim={0 0 0 0cm}]{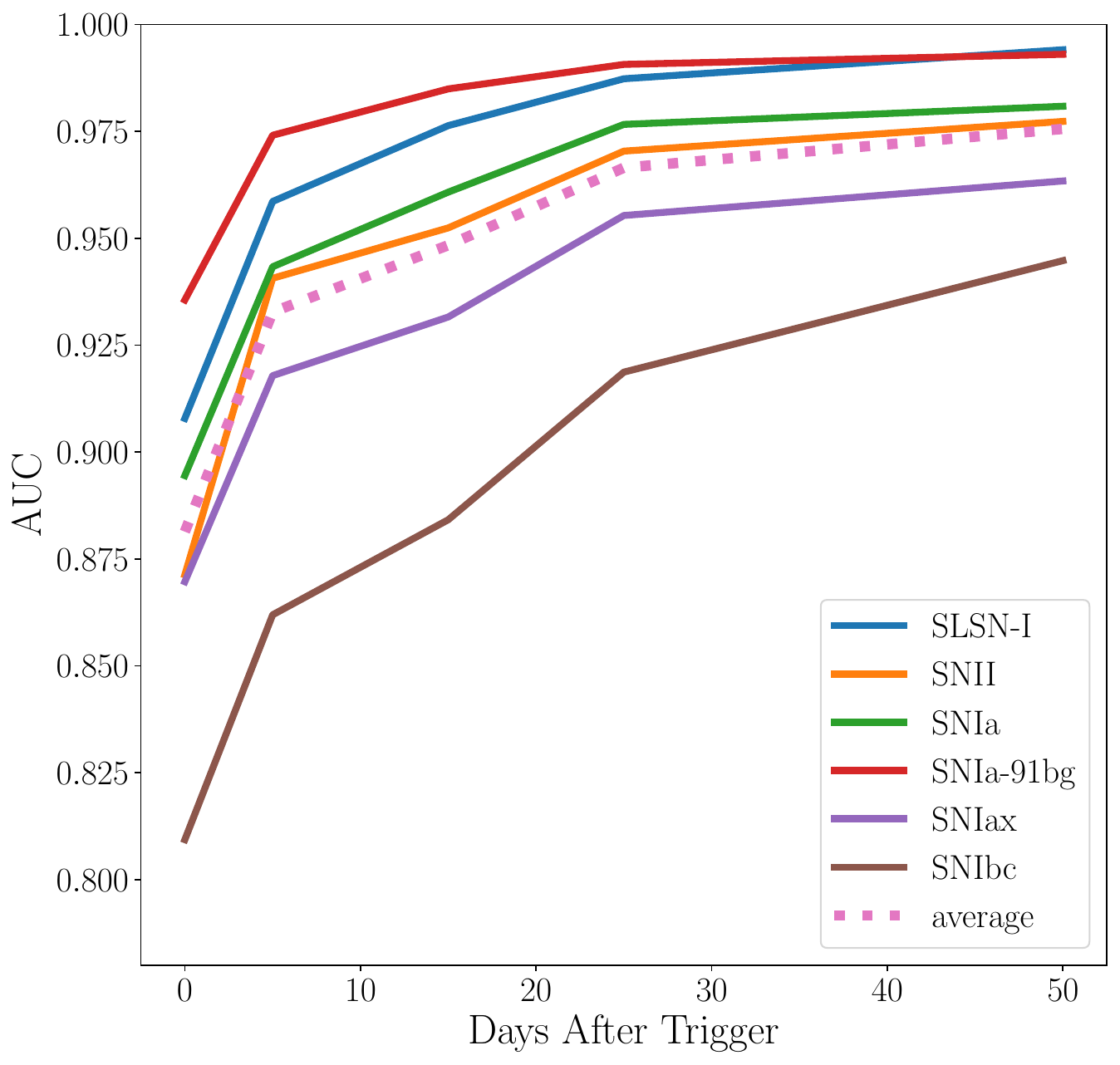}
    \includegraphics[scale=0.35,trim={0 0 0 0cm}]{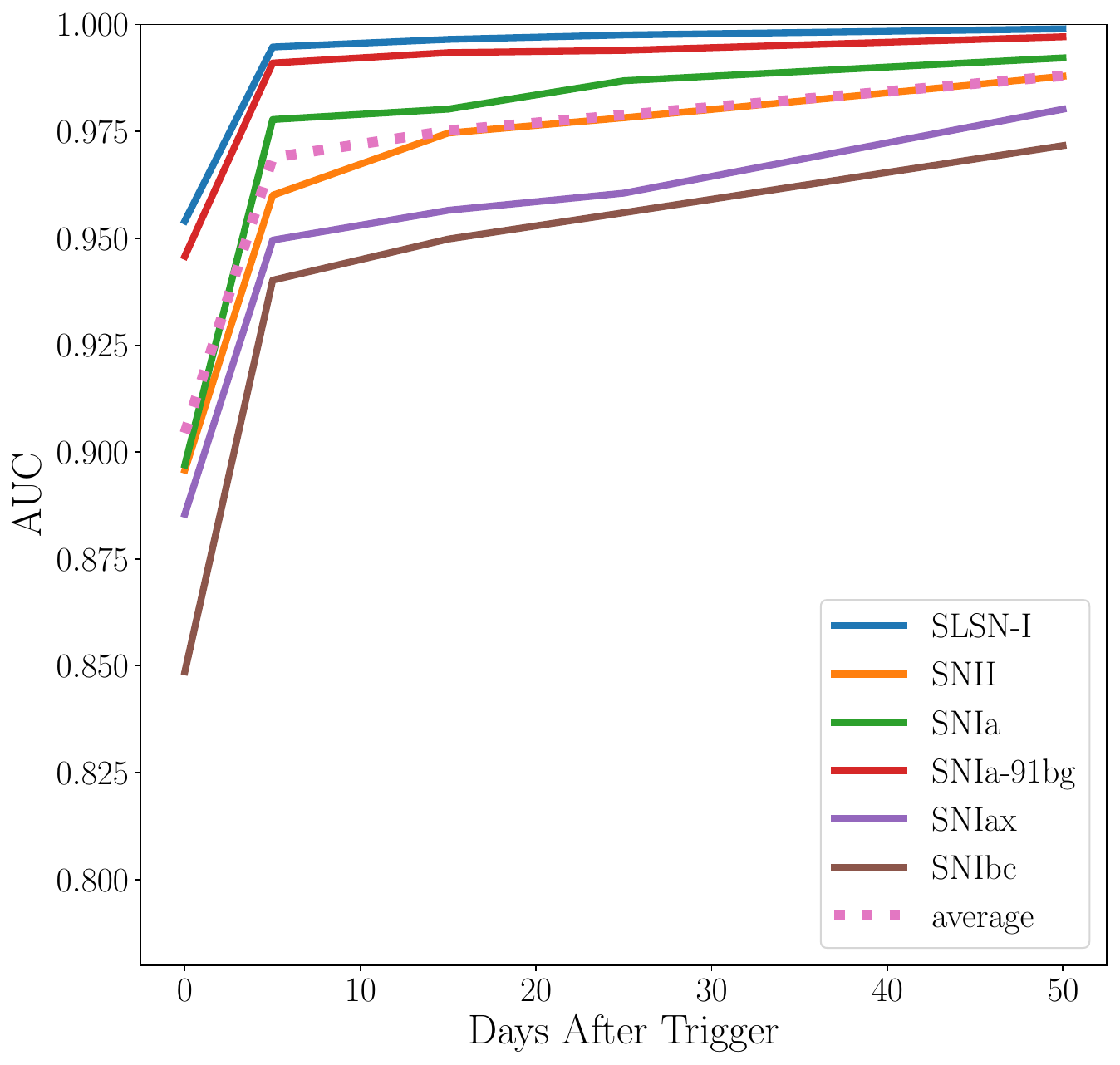}

    \caption{Area under the ROC curve (AUC) without (left) and with (right) redshift over time for each supernova type.}
    \label{fig:auc}
\end{figure*}

The information in the ROC curves for all $t_{\mathrm{trigger}}+N$ datasets is summarized in Figure~\ref{fig:auc}, AUC over time plots with and without redshift. The performance looks quite impressive, starting at an average AUC of above 0.9 with redshift at the date of trigger and increasing to 0.975 by 50 days after trigger. Without redshift, average AUC is still respectable, starting at 0.88 and increasing to 0.97.

\subsection{Approximating a First-Detection Trigger Definition}
Another common trigger definition used in transient surveys places the trigger at the date of the first detection ($t_{\rm first\;detection}$)rather than the second, which is the definition followed in this work. In order to more directly compare \scone's results with those of other classifiers following the first detection trigger definition, the distribution of $t_{\rm trigger}- t_{\rm first\;detection}$ was examined as well as \scone's performance on the subset of the $t_{\rm trigger}+0$ dataset with date of second detection ($t_{\rm trigger}$) at most 5 days after the date of first detection (i.e. $t_{\rm trigger}\leq t_{\rm first\;detection}+5$).

\begin{figure}
    \centering
    \includegraphics[scale=0.6,trim={0 0 0 0cm}]{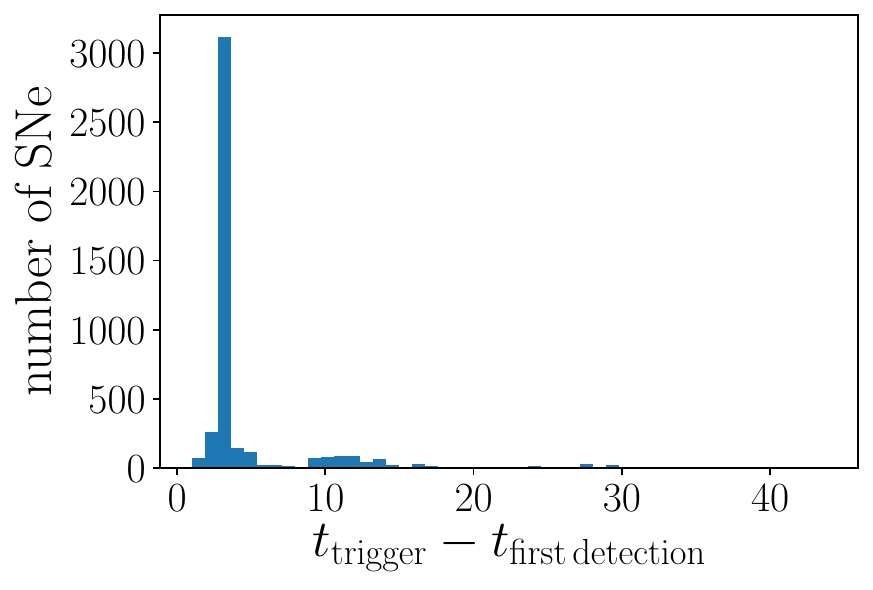}

    \caption{Distribution of $t_{\mathrm{trigger}}-t_{\mathrm{first\, detection}}$ in a \scone\ test dataset of 4608 SNe.}
    \label{fig:trigger-dist}
\end{figure}

\begin{figure*}
    \centering
    \includegraphics[scale=0.35,trim={0 0 0 0cm}]{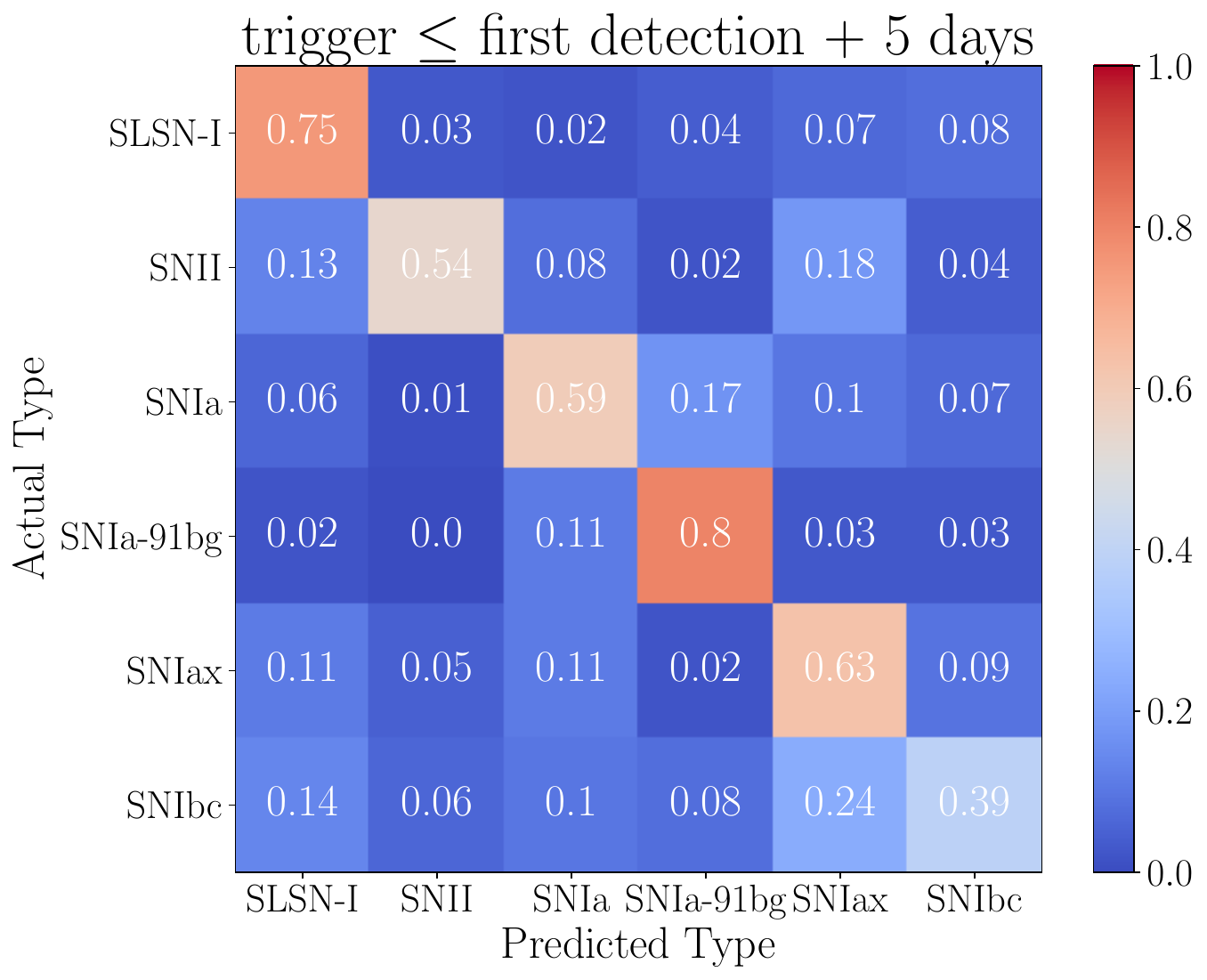}
    \includegraphics[scale=0.35,trim={0 0 0 0cm}]{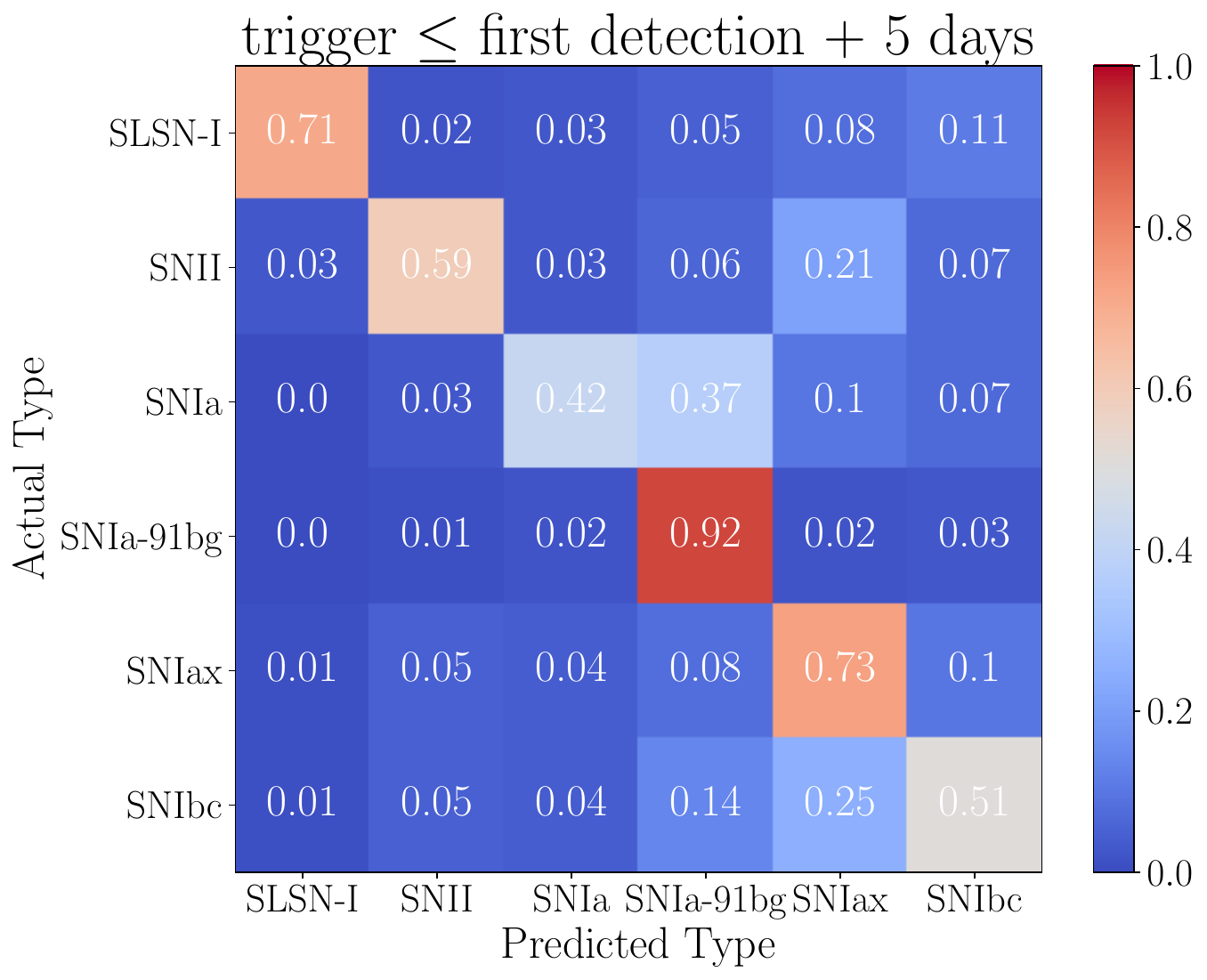}

    \caption{Normalized confusion matrices produced by \scone\ without (left) and with (right) redshift for the $t_{\rm trigger}\leq t_{\rm first\;detection}+5$ subset of the $t_{\mathrm{trigger}}+0$ test set. This cut ensures that the lightcurves used for performance evaluation are not given substantially more information than those created with the first detection trigger definition.}
    \label{fig:trigger}
\end{figure*}

Figure~\ref{fig:trigger-dist} shows that $>65$\% of $t_{\rm trigger}$ dates are no more than 5 days after the date of first detection. To further understand the direct impact of this choice of trigger definition, \scone\ was tested on the subset of the $t_{\rm trigger}+0$ dataset with date of second detection ($t_{\rm trigger}$) at most 5 days after the date of first detection. This cut ensures that the lightcurves used for classification are not given substantially more information than those created with the first detection trigger definition. The normalized confusion matrices for the $t_{\rm trigger}\leq t_{\rm first\;detection}+5$ dataset are shown with and without redshift in Figure~\ref{fig:trigger}.

With redshift, \scone's performance primarily suffers on SLSN-I and SNIa classification. SLSN-I appears to more strongly resemble SNIax and SNIbc at early times, as the SNIax confusion rose to 8\% from 1\% and the SNIbc confusion rose to 11\% from 2\%. SNIa were commonly misclassified as SNIa-91bg at early times, which is not reflected in the $t_{\rm trigger}+0$ confusion matrices in Figure~\ref{fig:cm}. Surprisingly, true SNIa-91bg were not misclassified as SNIa despite the prevalence of SNIa misclassified as SNIa-91bg. Without redshift, however, \scone's performance on the $t_{\rm trigger}\leq t_{\rm first\;detection}+5$ subset very closely resembles the $t_{\rm trigger}+0$ results shown in Figure~\ref{fig:cm}.

\subsection{Baseline Model}
A multi-layer perceptron model \citep[MLP,][]{HORNIK1989359} was developed as a baseline for direct comparison to \scone. MLP architectures are a simple type of feedforward neural network with at least 3 layers (input, hidden, output) in which each node in a particular layer is connected to every node in the subsequent layer. They have been successfully used in many general as well as image classification tasks \citep{MLPMixer,gMLP}.

The $32 \times 180 \times 2$ input heatmap is split into 180 non-overlapping ``patches" of size $32 \times 1$. The patches were chosen to be full height in the wavelength dimension to remain consistent with the full height convolutional kernels used in \scone. A $180 \times 64$-dimensional hidden layer is then computed via $h_{1,ij} = \mathrm{relu}(x^j_{i} W_{1,ji}+b_{1,j})$, where $\mathrm{relu}(x)=\mathrm{max}(0,x)$ is the rectified linear unit, $x^j$ is the $j^{\rm th}$ input heatmap patch, $W_1$ is the weight matrix learned by the network, and $b_1$ is the learned bias vector. The dimensionality of the hidden layer is then squashed to a single 64-dimensional vector with global average pooling: $h_{2,i} = \mathrm{average}(h_{1,ij})$. Finally, the output class is computed via $y_{k} = \sigma(h_{2,i} W_{2,ji}+b_{2,j})_k$, where $\sigma(\vec{x})_k=\frac{e^{x_k}}{\sum_j{e^{x_j}}}$ is the softmax function, $W_2$ is the learned weight matrix, and $b_2$ is the learned bias vector.

\begin{figure*}
    \centering
    \includegraphics[scale=0.35,trim={0 0 0 0cm}]{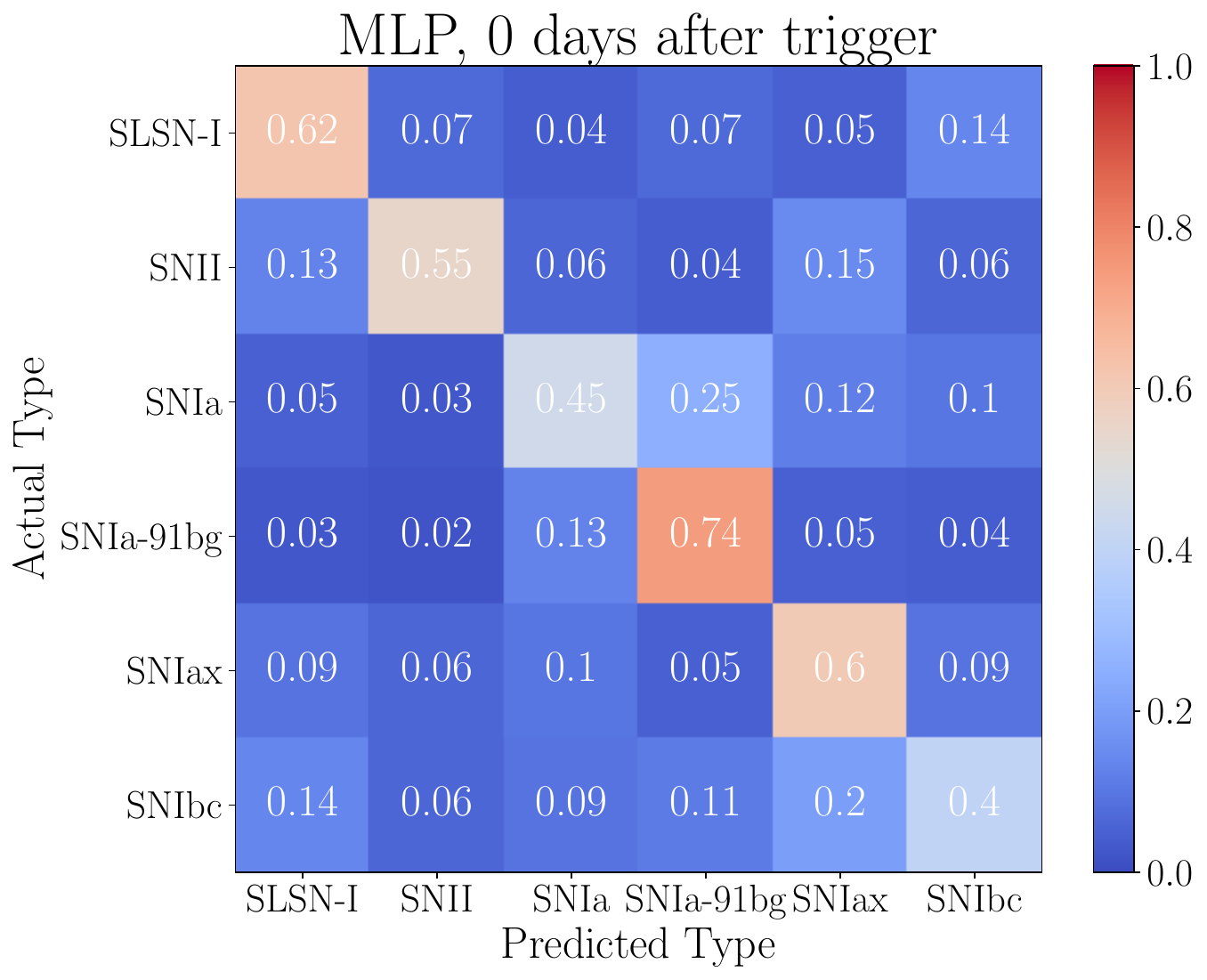}
    \includegraphics[scale=0.35,trim={0 0 0 0cm}]{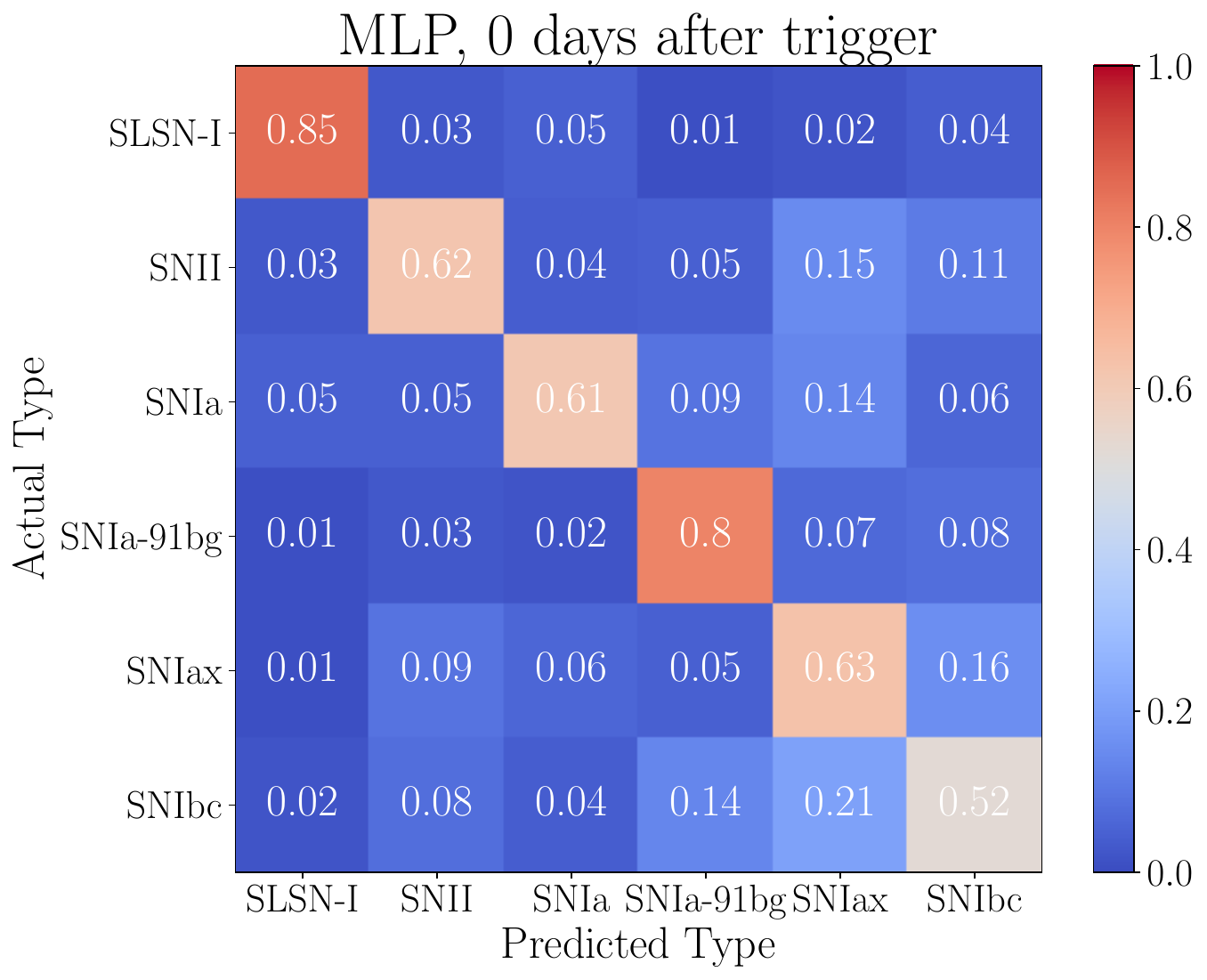}

    \caption{Normalized confusion matrices produced by the baseline MLP model without (left) and with (right) redshift for the $t_{\mathrm{trigger}}+0$ test set (heatmaps created from lightcurves truncated at the date of trigger).}
    \label{fig:baseline}
\end{figure*}

Without redshift, our model achieved a test accuracy of 56\%. With redshift, the test accuracy improved to 67.19\%. The performance of the MLP on the $t_{\mathrm{trigger}}+0$ dataset with and without redshift is summarized in the confusion matrices in Figure~\ref{fig:baseline}. Compared to the performance of \scone\ on the $t_{\mathrm{trigger}}+0$ dataset in the top panel of Figure~\ref{fig:cm}, the MLP is less accurate at classifying most SN types, most noticeably with redshift. The degraded but still respectable performance of the MLP on classification both with and without redshift shows that these supernova types can indeed be differentiated in some hyperdimensional space by a neural network, and that \scone\ in particular possesses the required discriminatory power for this task.

\subsection{Bright Supernovae}

Bright supernovae, defined as supernovae with last included $r$-band observation $r<20$~mag, were identified from both the $t_{\mathrm{trigger}}+0$ and $t_{\mathrm{trigger}}+5$ datasets. Since fewer (and likely dimmer) observations were included for each supernova in the $t_{\mathrm{trigger}}+0$ dataset, there are much fewer examples of bright supernovae than in the $t_{\mathrm{trigger}}+5$ dataset. The bright supernovae subsets of these datasets are referred to as the ``bright $t_{\mathrm{trigger}}+N$ datasets".

\begin{figure*}
    \centering
    \includegraphics[scale=0.35,trim={0 0 0 0}]{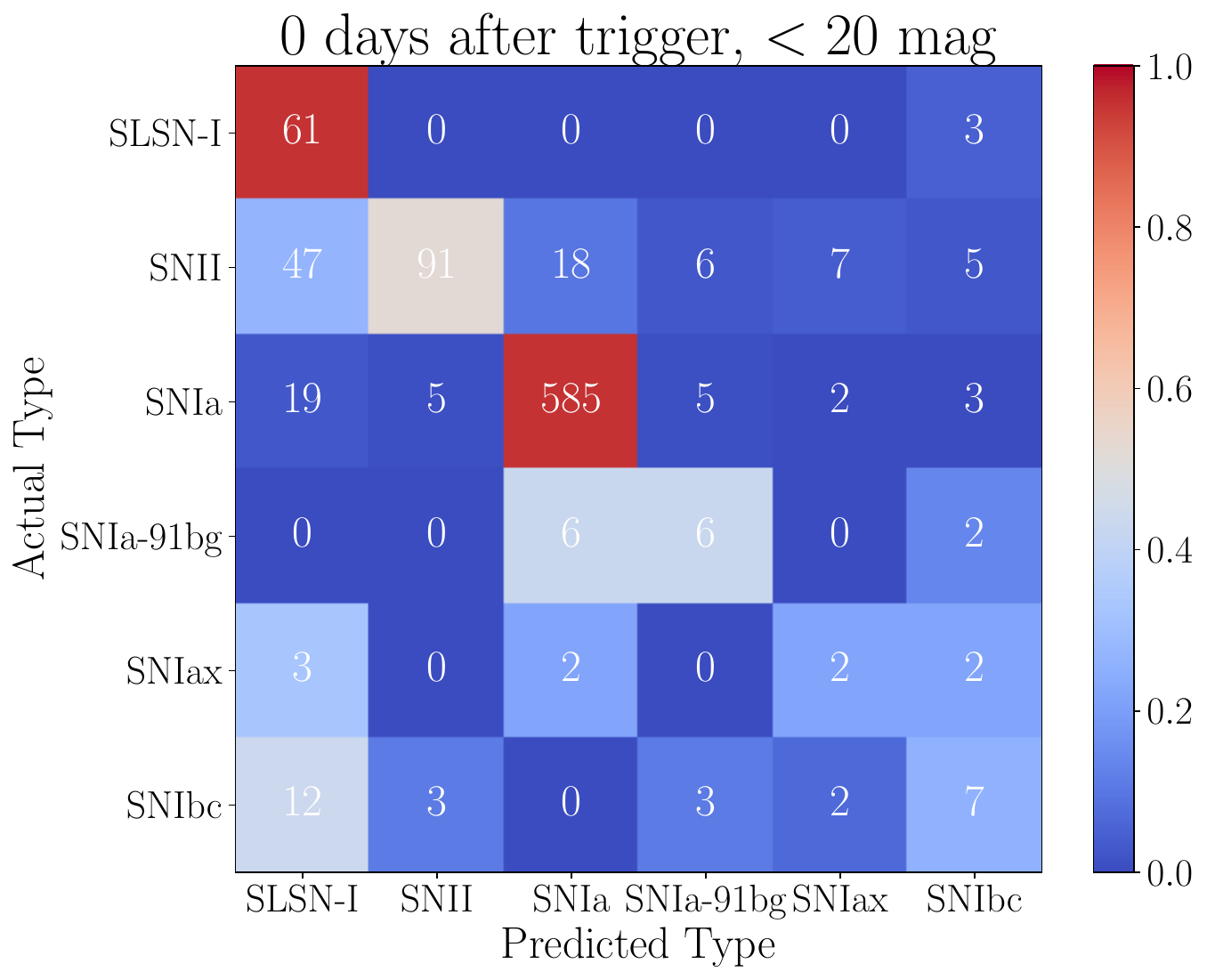}
    \includegraphics[scale=0.35,trim={0 0 0 0}]{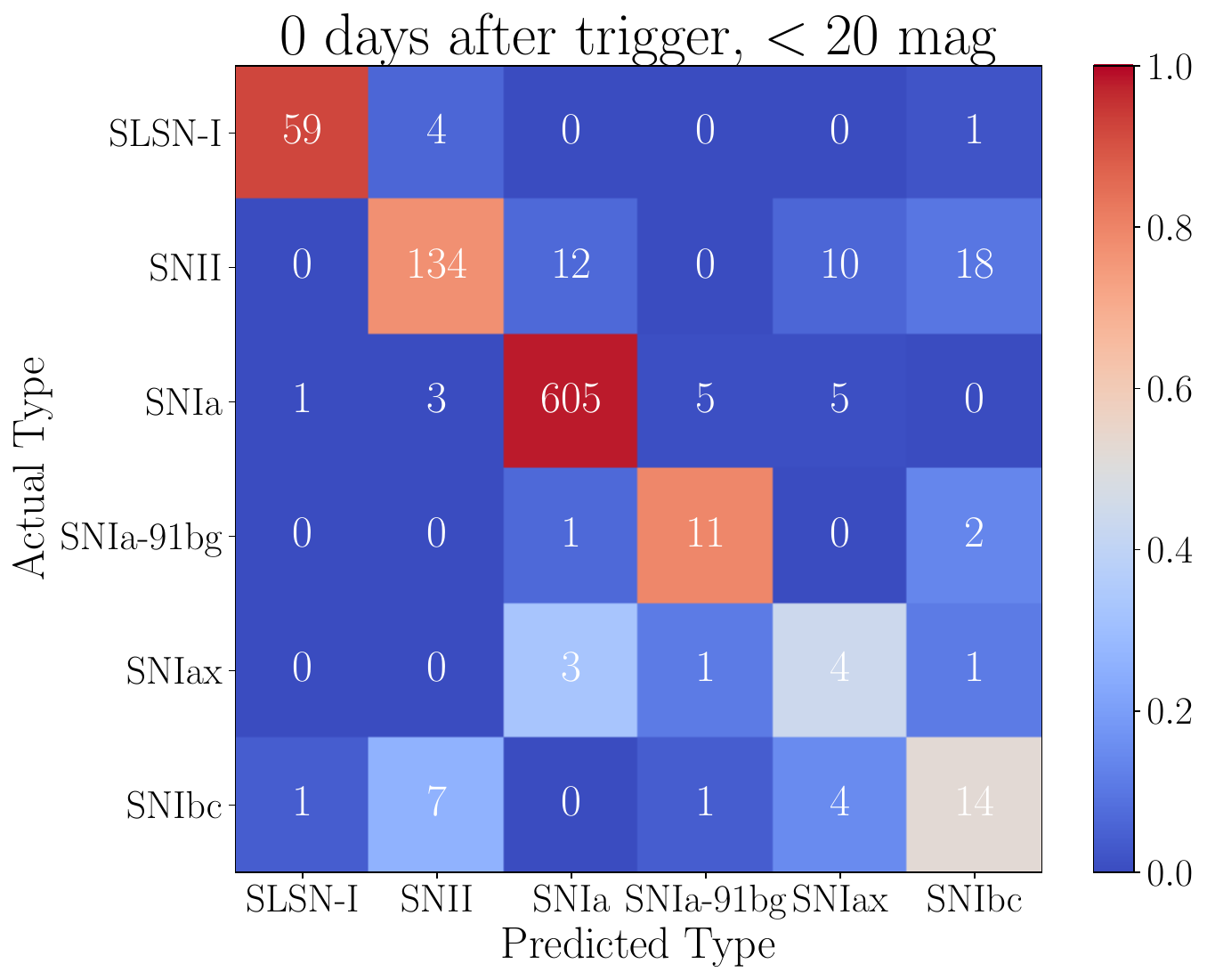}
    \includegraphics[scale=0.35,trim={0 0 0 0}]{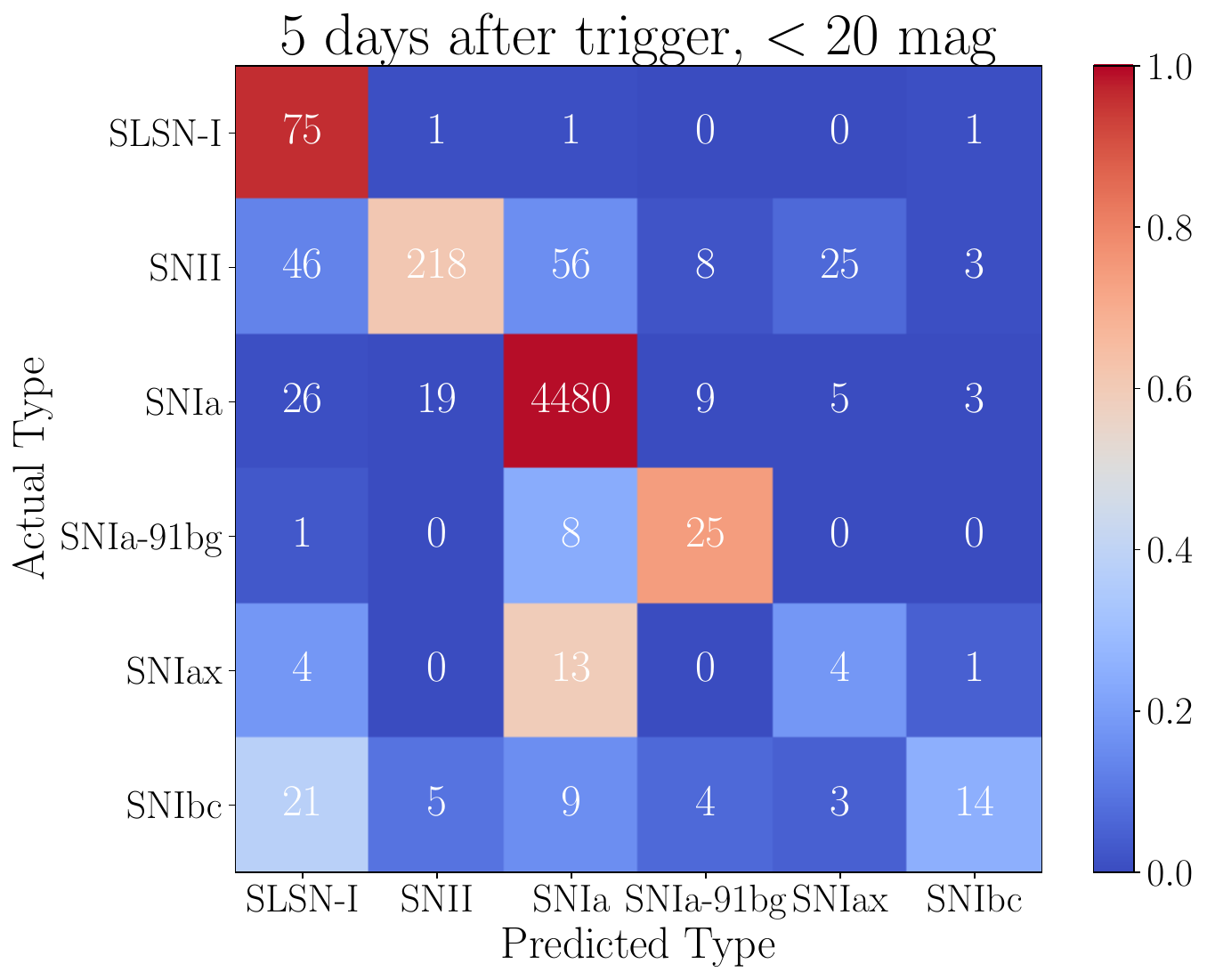}
    \includegraphics[scale=0.35,trim={0 0 0 0}]{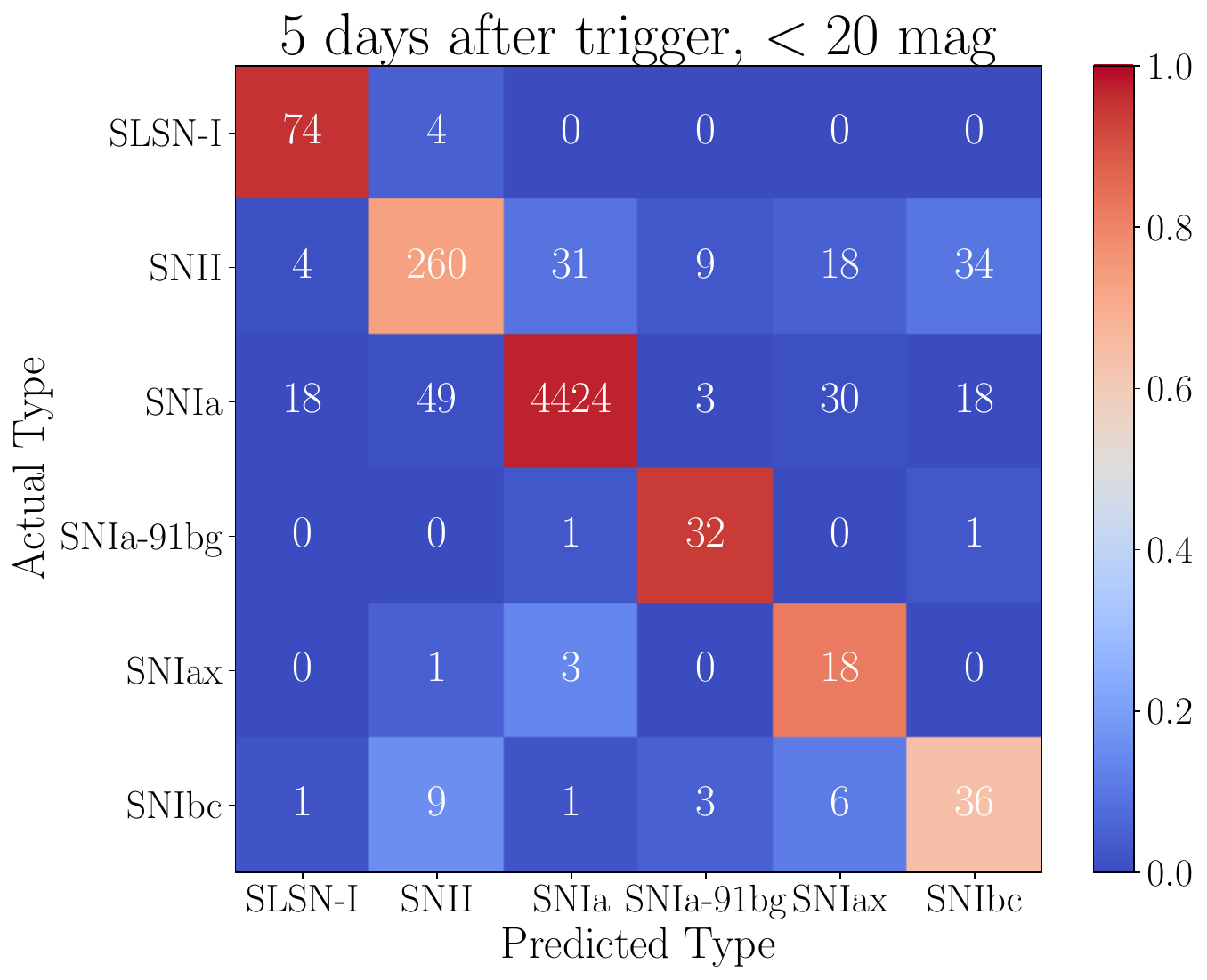}

    \caption{Early epoch confusion matrices with (right) and without (left) redshift for the bright supernovae ($<20$ magnitude) in each $t_{\mathrm{trigger}}+N$ dataset. \scone\ was trained with a class-balanced $t_{\mathrm{trigger}}+N$ training set combined with 40\% of $<20$ magnitude supernovae. These confusion matrices were created by testing the trained \scone\ model on the full $<20$ magnitude supernovae dataset. The confusion matrices are colored according to normalized accuracies, as in Figure~\ref{fig:cm}, and are overlaid with absolute (non-normalized) values since the dataset is imbalanced.}
    \label{fig:bright}
\end{figure*}

To evaluate the performance of \scone\ on identifying bright supernovae at early epochs, the model was trained on a regular class-balanced $t_{\mathrm{trigger}}+N$ training set, prepared as described in Section 2.4, combined with 40\% of the bright $t_{\mathrm{trigger}}+N$ dataset. The results of testing the trained \scone\ model on the bright $t_{\mathrm{trigger}}+N$ datasets are shown in Figure~\ref{fig:bright}. These confusion matrices, like the ones in Figure~\ref{fig:cm}, are colored by efficiency score. However, since the dataset is not class-balanced, the overlaid values are absolute (non-normalized) to preserve information on the relative abundance of each type. Thus, an efficiency (purity) score for each type can be calculated by dividing each main diagonal value by the sum of the values in its row (column). The overall accuracies as well as the total number of SNe in each dataset are summarized in Table~\ref{tbl:bright-acc}.

The benefits of redshift information are much more pronounced for certain types than others. As also noted in analyses of Figures~\ref{fig:cm} and ~\ref{fig:auc}, the quantity of SNIbc misclassified as SLSN-I was significantly reduced in results from \scone\ with redshift information. At the date of trigger, 44.4\% of SNIbc were misclassified as SLSN-I without redshift. This contamination rate was reduced to only 3.7\% with redshift. However, classification of bright SNIa seems relatively unaffected by the presence of redshift information. 5 days after trigger, SNIa were classified with an efficiency/accuracy of 98.6\% and a purity score of 98.1\% without redshift, and 97.4\% efficiency/accuracy and 99.1\% purity with redshift.

\begin{table}
    \centering
    \caption{Test accuracies with and without redshift information for the bright datasets.}
    \label{tbl:bright-acc}
    \begin{tabular}{l c c c}
        \hline
        & Total & Accuracy no $z$ & Accuracy with $z$\\
        \hline
        bright $t_{\mathrm{trigger}}+0$ & 907 & 82.91\% & 91.18\%\\
        bright $t_{\mathrm{trigger}}+5$ & 5088 & 94.65\% & 95.2\% \\
    \end{tabular}
\end{table}

\subsection{Mixed Dataset}

Training on the $t_{\mathrm{trigger}}+N$ datasets represents one way of deploying \scone\ for real-world transient alert applications, while training on a mixed dataset is a much less computationally expensive alternative. On one hand, testing a $t_{\mathrm{trigger}}+N$-trained model on a $t_{\mathrm{trigger}}+N$ test set yields the best classification accuracies. However, this approach requires the creation of separate datasets for each choice of $N$, which could be an expensive initial time investment depending on the number of datasets and size of each dataset (see Section 2.6 for computational requirements for heatmap creation). In this work, only five datasets ($N=0,5,15,25,50$) were created, but perhaps $N=0,1,...,50$ will be needed to accurately classify real-world transient alerts with any number of nights of photometry. Training on a mixed dataset, where each heatmap is created with a random number of nights of photometry after trigger, is a viable alternative for resource- or time-constrained applications.

\begin{figure*}
    \centering
    \includegraphics[scale=0.35,trim={0 0 0 0}]{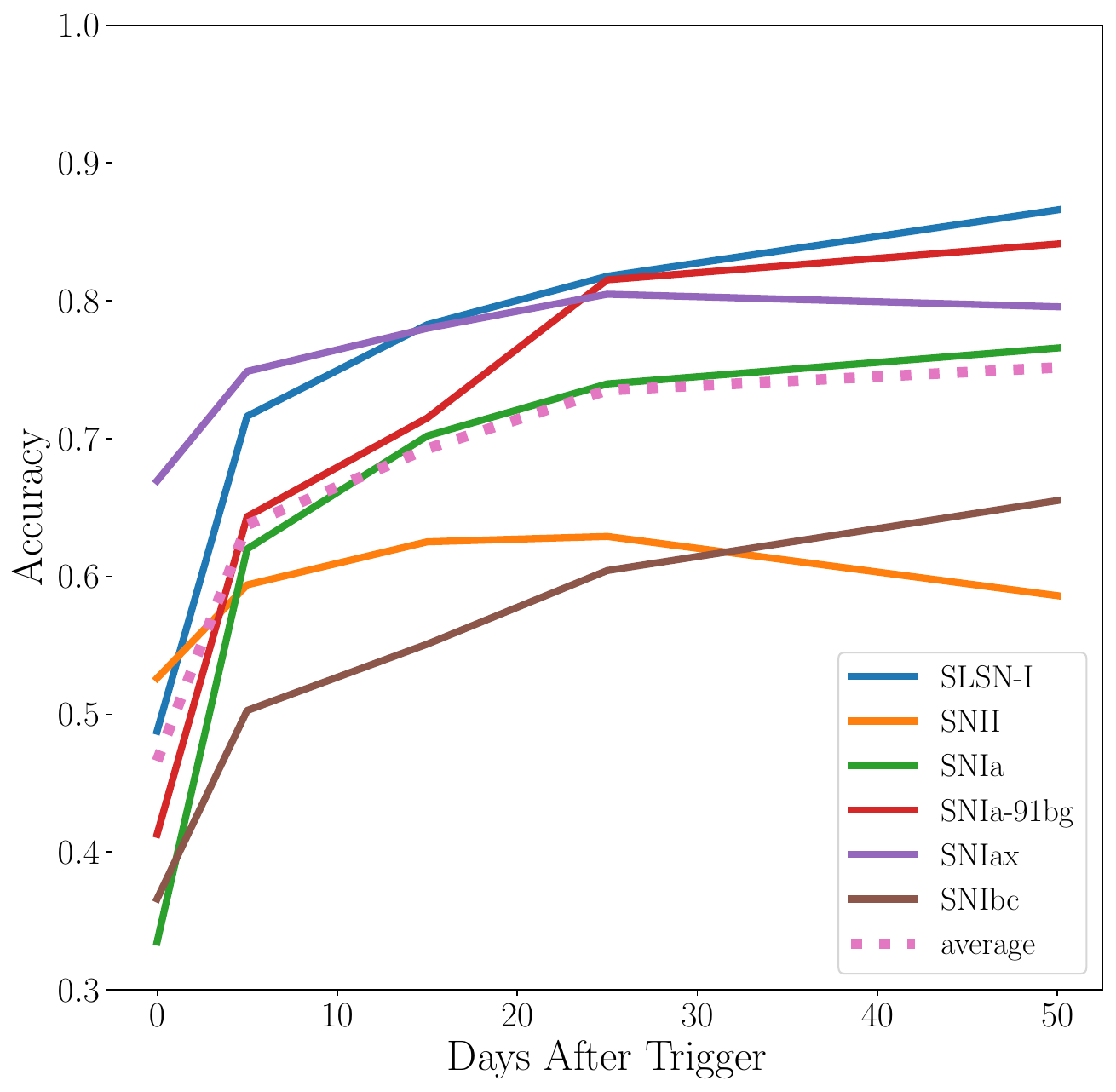}
    \includegraphics[scale=0.35,trim={0 0 0 0}]{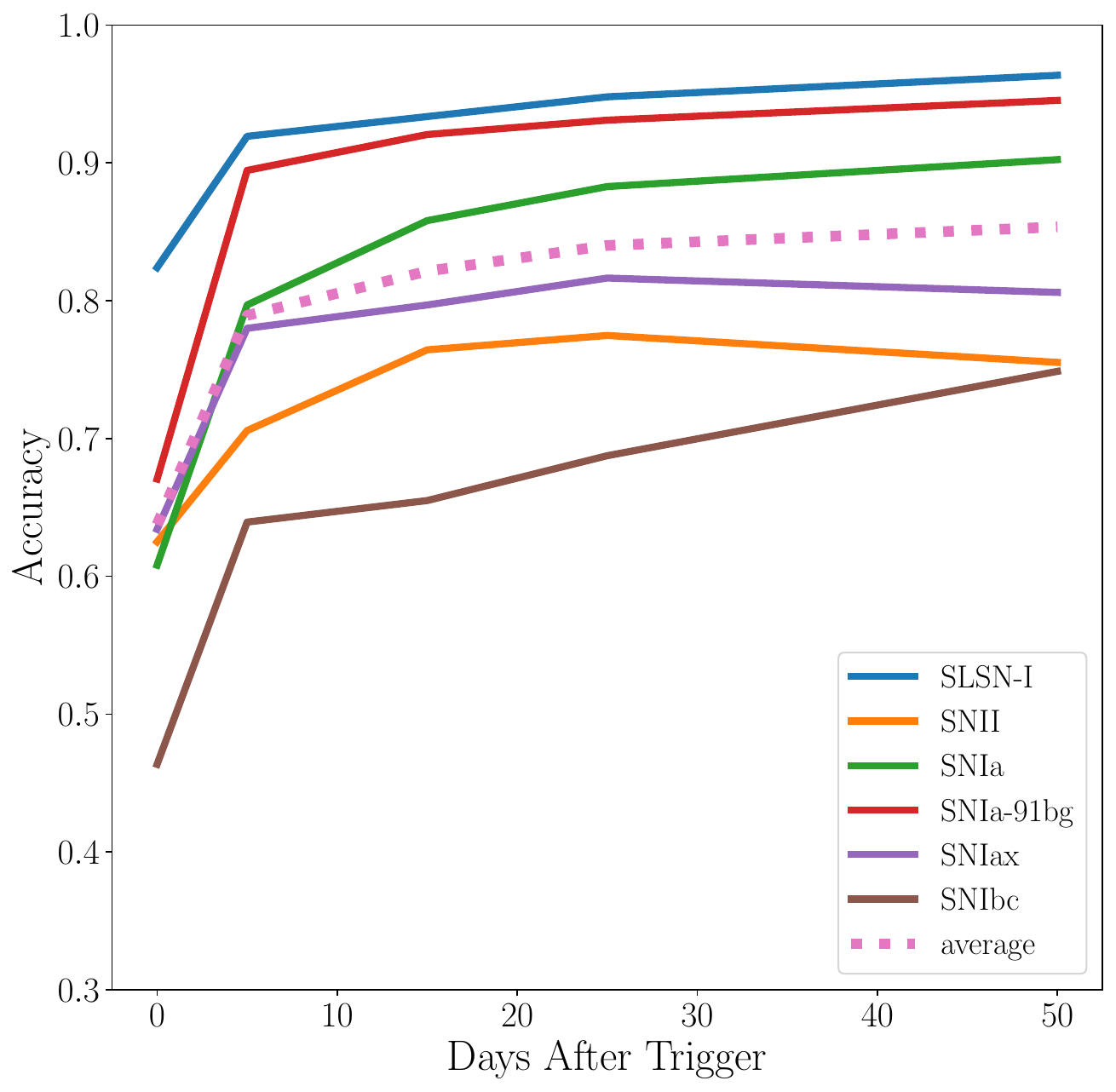}

    \caption{Test set accuracy/efficiency without (left) and with (right) redshift over time for \scone\ trained on the mixed dataset and tested on each individual $t_{\mathrm{trigger}}+N$ dataset. The values used in these plots correspond with the diagonals on a normalized confusion matrix.}
    \label{fig:mixed-accs}
\end{figure*}

\begin{figure*}
    \centering
    \includegraphics[scale=0.35,trim={0 0 0 0cm}]{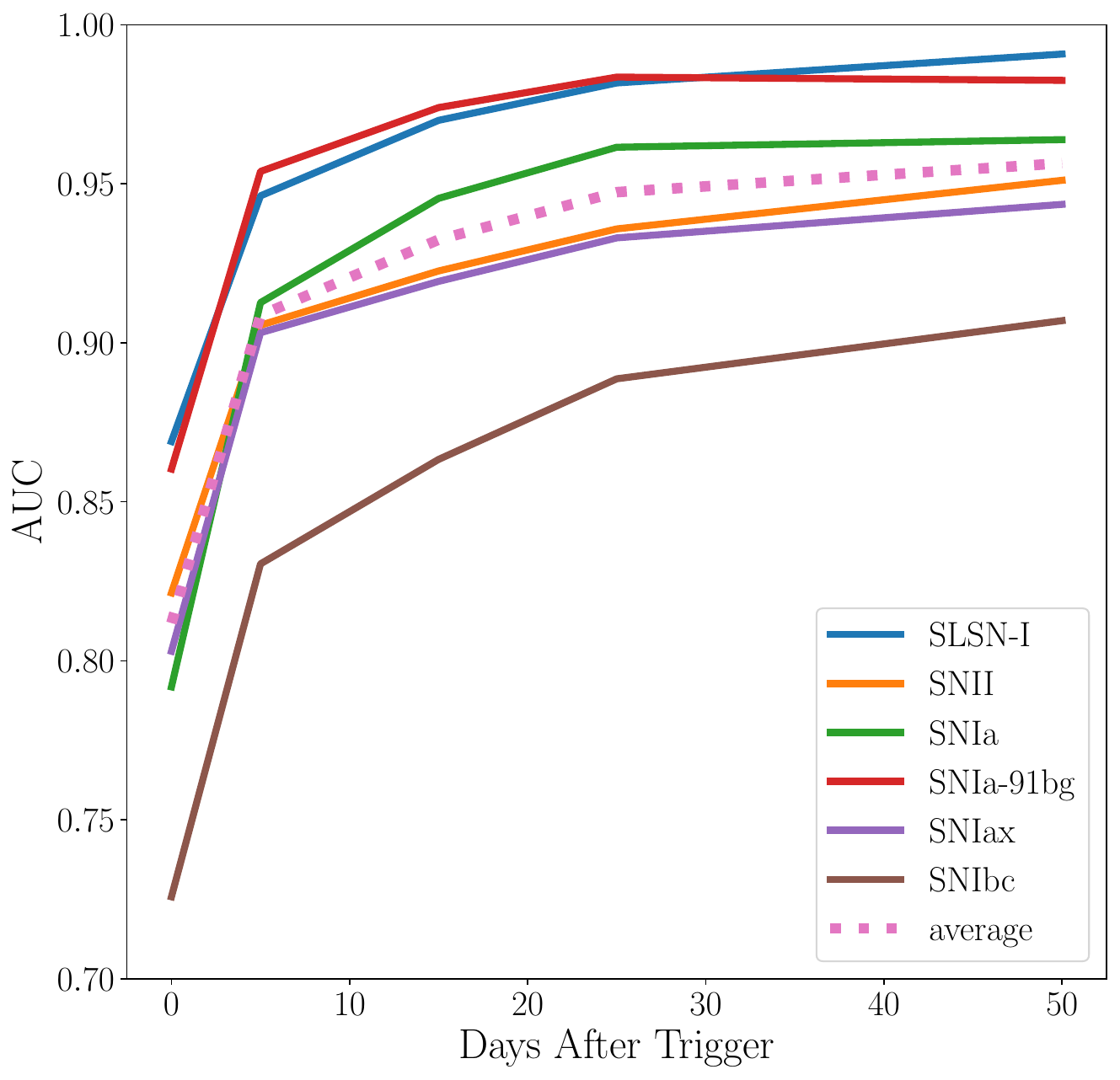}
    \includegraphics[scale=0.35,trim={0 0 0 0cm}]{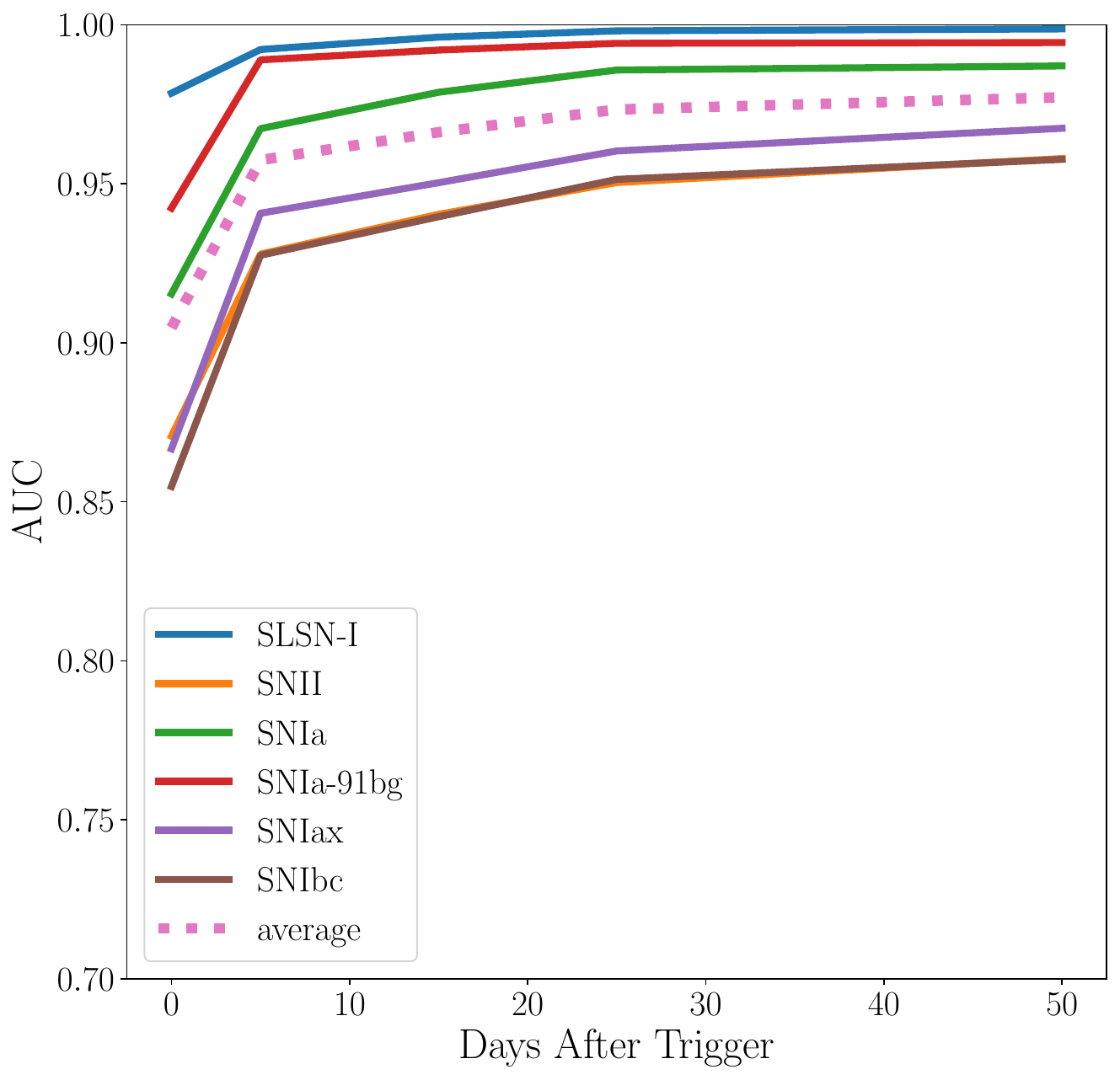}

    \caption{Area under the ROC curve (AUC) without (left) and with (right) redshift over time for \scone\ trained on the mixed dataset and tested on each individual $t_{\mathrm{trigger}}+N$ dataset.}
    \label{fig:mixed-auc}
\end{figure*}

To directly compare the performance of \scone\ trained on the mixed dataset and the $t_{\mathrm{trigger}}+N$ datasets, the mixed-dataset-trained model was tested on each individual $t_{\mathrm{trigger}}+N$ dataset. The accuracies over time split by SN type are summarized in Figure~\ref{fig:mixed-accs}. Compared to the results of \scone\ trained and tested on each individual $t_{\mathrm{trigger}}+N$ dataset (Figure~\ref{fig:accs}), the accuracies are lower but still respectable. The performance at the date of trigger is the most dissimilar, with average accuracy ~74\% with $z$ for a model trained on $t_{\mathrm{trigger}}+0$ and ~64\% with mixed. The performance of the mixed-trained model performs similarly to the $t_{\mathrm{trigger}}+N$-trained model by 5 days after trigger, however, both averaging just under 80\% with $z$. The AUCs over time split by SN type are shown in Figure~\ref{fig:mixed-auc}. These AUC plots are comparable to the $t_{\mathrm{trigger}}+N$ AUCs in Figure~\ref{fig:auc}, indicating that the performances of both models are comparable when averaged over all values of the prediction threshold $p$. However, the predicted class for categorical classification is not typically calculated with respect to a threshold; rather, it is defined as the class with the highest prediction confidence for each example. Thus, the AUCs are analogous to analyzing the performance on each type as its own binary classification problem, resulting in slight discrepancies from the accuracies.

\subsection{Comparison with Existing Literature}
At the time of this writing, the only work in existing literature with a similarly strong focus on early photometric classification of supernovae is RAPID \citep[][hereafter M19]{rapid}, a GRU RNN approach to photometric transient classification that differentiates between 12 transient types, including 7 supernova types. RAPID differs in several significant ways from the data, model, and results presented in this work. We highlight some of the differences in these two works below.

\subsubsection{Comparison of Methods}

The most obvious difference is in the type of neural network architecture used for classification. RAPID uses a uni-directional RNN architecture, which is designed to learn from time-series data chronologically. \scone\ employs a convolutional neural network architecture, which is most commonly used for image recognition tasks. In this instance, however, \scone\ is designed to read in data chronologically. Convolutional layers in a CNN work by computing functions on a ``sliding window" of the input image, thereby allowing the model to learn small-scale structures in the image. These windows, or the convolutional kernel, is typically a small square chosen to match the characteristic length scale of structures in the images. \scone's convolutional kernel, however, is chosen to span the full height of the input heatmap, resulting in a window that slides chronologically along the horizontal, or time, axis.

M19 trained and tested on a dataset of simulated Zwicky Transient Facility (ZTF) lightcurves, which have $g$- and $r$- band photometry, compared to the LSST lightcurves used in this work, with $ugrizY$ photometry bands. In addition to the 6 supernova types that this work focuses on, M19 includes 4 rare transient classes (pair instability supernovae (PISN), intermediate luminosity transients (ILOT), calcium-rich gap transients (CART), tidal disruption events (TDE)) as well as point-Ia and KN.

Other differences include the addition of a ``pre-explosion" class, rest- vs. observer-frame time intervals, and the choice of trigger definition. M19 chooses to include an additional class, ``pre-explosion", to describe examples at time-steps prior to the occurrence of the transient event. M19 also converts time intervals out of observer-frame by dividing by $1+z$, which is not done in this work to ensure that mistakes in redshift estimates will not be propagated to affect the lightcurve data. Finally, M19 uses the first-detection trigger date definition, while this work defines $t_{\rm trigger}$ to be the date of the second detection.

\subsubsection{Comparison of Results}
The results of \scone\ classification with redshift (right side panels of Figures~\ref{fig:cm}-\ref{fig:auc}) is used to compare with RAPID's results, as RAPID also incorporates redshift information. As described in the previous section, this work differs in many ways from M19 and the following comparison does not account for these differences; a rigorous comparison of the two models against a single dataset is left to a future work. 

Most notably, \scone\ improves upon RAPID's SNIbc and SNII classification accuracy, while RAPID performs very well at classifying early-time SNIa. From Figure 7 of M19, 12\% of SNIbc are correctly classified 2 days after detection, compared to \scone's 54\% accuracy at the date of trigger. In RAPID's results, 30\% of true SNIbc are misclassified as CART, which is not included in the datasets in this work. The second and third largest contaminants (SNIax at 19\%, then SNIa and SNIa-91bg at 8\% each), are both part of this analysis. From Figure~\ref{fig:cm}, we find that SNIax and SNIa-91bg are also major contaminants for \scone\ at 23\% and 11\%, respectively, at the date of trigger and 16\% and 4\%, respectively, 5 days after trigger. However, there is no significant contamination from SNIa, with contamination rates at 4\% on the date of trigger and 1\% 5 days after trigger.

2 days after detection, SNII is classified at 7\% accuracy by RAPID compared to 64\% accuracy at the date of trigger by \scone. The primary contaminant of SNII for RAPID 2 days after detection is SNIa at 21\%, which is not reflected in \scone's results, where the contamination rate is 6\% at the date of trigger and 3\% 5 days after trigger. The second largest contaminant, SLSN-I, is also not an issue in \scone's SNII classification. Surprisingly, the improvement over time of RAPID's SNII classification accuracy outpaces its SNIbc classification accuracy, as it is able to achieve 49\% accuracy on SNII 40 days after detection compared to 31\% accuracy on SNIbc.

While \scone's SNIa classification accuracy slowly climbs from 77\% at the date of trigger to 93\% 50 days after trigger, RAPID is able to classify SNIa at 88\% accuracy almost immediately after detection. A future direct comparison will aid in concluding whether this discrepancy is due to differences in the datasets, such as M19's exclusion of $z \geq 0.5$ objects, or something more fundamental to the model architectures.

\section{Conclusions}

Our ability to observe the universe has improved in leaps and bounds over the past century, allowing us to find new and rare transient phenomena, enrich our understanding of transient physics, and even make cosmological discoveries aided by observational data. Our photometric observing capabilities greatly outpace the rate at which we can gather the associated spectroscopic information, resulting in a vast trove of photometric data sparsely annotated by spectroscopy. In the era of large-scale sky surveys, with millions of transient alerts per night, an accurate and efficient photometric classifier is essential not only to make use of the photometric data for science analysis, but also to determine the most effective spectroscopic follow-up program early on in the life of the transient.

In this work, we presented \scone's performance classifying simulated LSST early-time supernova lightcurves for SN types Ia, II, Ibc, Ia-91bg, Iax, and SLSN-I. As a neural network-based approach, \scone\ avoids the time-intensive manual process of feature selection and engineering, and requires only raw photometric data as input. We showed that the incorporation of redshift estimates as well as errors on those estimates significantly improved classification accuracy across the board, and was especially noticeable at very early times. Notably, this is the first application of convolutional neural networks to this problem.

\scone\ was tested on 3 types of datasets: datasets of lightcurves that were truncated at 0, 5, 15, 25, and 50 days after trigger ($t_{\mathrm{trigger}}+N$ datasets); bright ($< 20$ magnitude) subsets of the $t_{\mathrm{trigger}}+\{0,5\}$ datasets; and a dataset of lightcurves truncated at a random number of nights between 0 and 50 (``mixed"). Without redshift, \scone\ was able to classify $t_{\mathrm{trigger}}+0$ lightcurves with 60\% overall accuracy, which increases to 82\% at 50 days after trigger. \scone\ with redshift information starts at 74\% overall accuracy at the date of trigger and improves to 89\% 50 days after trigger. Confusion matrices, ROC plots, and accuracy over time as well as AUC over time plots of results with and without redshift were presented to better understand classification performance and identify areas of improvement. For the bright subsets, overall accuracy is $>90$\% at the date of trigger with redshift and over 80\% without. These results improve to around 95\% accuracy both with and without redshift by 5 days after trigger. The overall accuracy over time of a mixed-dataset-trained model tested on the $t_{\mathrm{trigger}}+N$ datasets shows some degradation in accuracy at very early epochs, but may be a worthwhile lightweight alternative to the more resource-intensive process of creating many $t_{\mathrm{trigger}}+N$ datasets.

We showed that \scone's performance with redshift is competitive with existing work on early classification, such as M19, while improving on computational time requirements. \scone\ has a lightweight pre-processing step and can achieve impressive performance with a small training set. It requires only hundredths of a second to preprocess each lightcurve into a heatmap and seconds for each training epoch on GPU. This makes \scone\ a great candidate for incorporation into alert brokers for LSST and future wide-field sky surveys.

In future work, we plan to apply this model to real data to further validate the approach. We also plan to extend \scone\ to classify both full-duration and early lightcurves for more transient and variable classes in the PLAsTiCC simulations.

\section{Acknowledgments}
This work was supported by DOE grant DE-FOA-0002424, NASA Grant NNH15ZDA001N-WFIRST, and NSF grant AST-2108094. This research
used resources of the National Energy Research Scientific Computing Center (NERSC), a U.S. Department
of Energy Office of Science User Facility located at
Lawrence Berkeley National Laboratory, operated under Contract No. DE-AC02-05CH11231.

%% file: chapters/host_mismatch.tex
\newcommand{\ddlr}{$d_{\mathrm{DLR}}$ }
\newcommand{\snnz}{$\text{SNN}_{+Z}$}
\newcommand{\snnnoz}{$\text{SNN}_{NoZ}$}
\newcommand{\diffimg}{\texttt{DiffImg}}
\newcommand{\ythreegold}{\texttt{Y3GOLD}}

\section*{Abstract}
Redshift measurements, primarily obtained from host galaxies, are essential for inferring cosmological parameters from type Ia supernovae (SNe Ia). Matching SNe to host galaxies using images is non-trivial, resulting in a subset of SNe with mismatched hosts and thus incorrect redshifts. We evaluate the host galaxy mismatch rate and resulting biases on cosmological parameters from simulations modeled after the Dark Energy Survey 5-Year (DES-SN5YR) photometric sample. For both DES-SN5YR data and simulations, we employ the directional light radius method for host galaxy matching. In our SN Ia simulations, we find that 1.7\% of SNe are matched to the wrong host galaxy, with redshift difference between the true and matched host of up to 0.6. Using our analysis pipeline, we determine the shift in  the dark energy equation of state parameter ($\Delta w$) due to including SNe with incorrect host galaxy matches.  For SN Ia-only simulations, we find $\Delta w = 0.0013 \pm 0.0026$ with constraints from the cosmic microwave background (CMB). Including core-collapse SNe and peculiar SNe Ia in the simulation, we find that $\Delta w$ ranges from 0.0009 to 0.0032 depending on the photometric classifier used. This bias is an order of magnitude smaller than the expected total uncertainty on $w$ from the DES-SN5YR sample of $\sim 0.03$.  We conclude that the bias on $w$ from host galaxy mismatch is much smaller than the uncertainties expected from the DES-SN5YR sample, but we encourage further studies to reduce this bias through better host-matching algorithms or selection cuts.

\section{Introduction}
Type Ia supernovae (SNe Ia) enabled the discovery of accelerating cosmic expansion \citep{perlmutter, riess} and have since been an important probe of the dark energy thought to cause it. To constrain cosmology, each SN Ia must have an accurate estimate of cosmological redshift as well as physical distance. Distance estimates are obtained from the standardized luminosities of SNe Ia, earning them the title of \textit{standardizable candles}, but redshift is difficult to determine without spectroscopy. The depth of modern photometric surveys has resulted in orders of magnitude more photometrically observed SNe than can be followed up spectroscopically. Thus, the vast majority of SNe Ia used for estimation of cosmological parameters are assigned the redshift of their matched host galaxies. Moreover, galaxy redshifts are more precise than redshifts measured from SN spectroscopy due to the broadened features and phase-dependent nature of SN spectra.

\begin{figure}
    \includegraphics[scale=0.32,trim={2cm 0cm 0cm 0cm}]{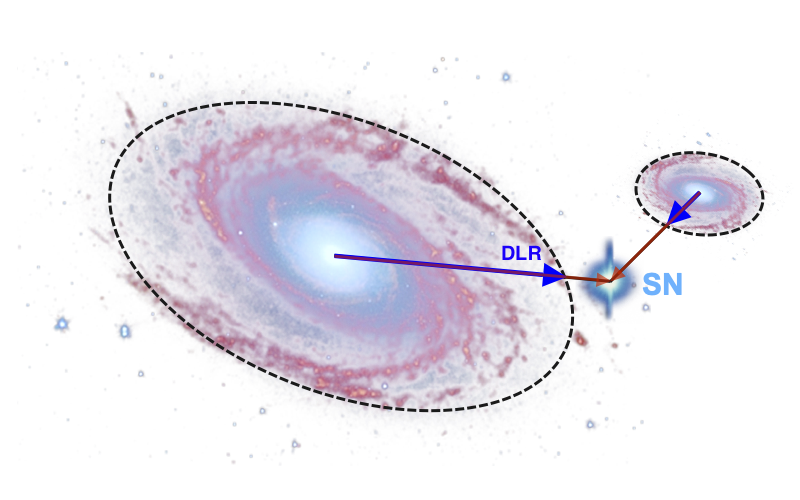}
    \centering
    \caption{An illustration adapted from \cite{gupta} showing an example of a challenging host galaxy matching problem. The supernova (labeled ``SN") is closer in angular separation (red arrows) to the smaller galaxy on the right, but it is closer to the edge of the larger galaxy on the left. The directional light radius (DLR) of each galaxy is shown in the blue arrows. According to the DLR method, the most likely host galaxy is the one with minimal $d_{\mathrm{DLR}}$ value, or ratio of the angular separation to the DLR, which would correctly identify the larger galaxy on the left as the more likely host.}
    \label{fig:host-confusion}
\end{figure}

This work investigates the impact of mismatched host galaxies and the resulting incorrect SN redshifts on the measurement of cosmological parameters. Using images alone, host galaxy matching is a nontrivial problem when there are multiple galaxies near the SN location. Two-dimensional images have little distance information, and galaxies in a crowded field can be difficult to disentangle without distance measurements to each. In addition, the large scale structure of the universe dictates that many galaxies occur in pairs, groups, or clusters, making it difficult to determine which galaxy is the host. Finally, some ``hostless" SNe explode in extremely faint or distant galaxies that fall below the threshold of detection, even in the deep coadded images created for this work described in Section~\ref{subsec:hostgal-catalog}. These SNe could be incorrectly matched to brighter, nearby galaxies that are close in projected distance and thus assigned an incorrect redshift. Figure~\ref{fig:host-confusion} illustrates a challenging example of host galaxy matching, where the larger and more likely host galaxy is further in terms of SN-galaxy separation than the smaller galaxy on the right.

Though automated methods for host galaxy matching have been in use since the SuperNova Legacy Survey (SNLS) analysis \citep[][S06]{sullivan_2006}, accurate characterization of systematic uncertainties such as the effect of mismatched host galaxies has become more pressing with the advent of wider and deeper SN surveys. Future surveys such as the Legacy Survey of Space and Time (LSST) at the Vera Rubin Observatory will observe hundreds of thousands more SNe in the coming decade \citep{lsst_book}, further shrinking statistical uncertainties and necessitating accurate measurements of systematic uncertainties. This work represents the first thorough exploration of systematics related to the host galaxy mismatch problem and its effect on cosmological parameter estimates as part of an ongoing cosmology analysis.


A commonly used method for matching SNe with a host galaxy is the directional light radius (DLR) method, initially developed to characterize host properties for the SNLS survey by S06 and tested extensively on simulations by \citet{gupta}. 
More details about the DLR method can be found in Section~\ref{subsec:dlr}. Recently, \citet[][P20]{popovic2020} performed the first retrospective estimate of cosmological biases resulting from host galaxy mismatches by applying the DLR method on simulations based on SDSS data. P20 found a mismatch rate of 0.6\% with a resulting bias on $w$ of $\Delta w = 0.0007$. 

Recently, a number of promising alternative approaches for host matching have emerged, including DELIGHT \citep{delight} and GHOST \citep{ghost}. DELIGHT introduces a deep learning-based approach to host galaxy identification in which a convolutional neural network is trained on transient images to predict a 2-dimensional vector connecting the transient position with the position of its host galaxy center. GHOST uses a novel gradient ascent method for host galaxy identification that was shown to produce accurate matches in situations where DLR was unable to identify a match. Though this work did not explore these novel approaches, they certainly merit consideration for future SNe Ia cosmological analyses.

In this work, we focus on understanding the cosmological biases arising from host galaxy mismatch using the DLR method on the full Dark Energy Survey 5-Year photometric SNe Ia sample~\citep[DES-SN5YR;][]{vincenzi2021, photo_Ia_des}. Section~\ref{sec:methods} details the DLR host matching algorithm and an overview of our analysis strategy. Section~\ref{sec:data} describes the DES-SN5YR data sample along with the host galaxy catalog used in this analysis. Section~\ref{sec:sims} reviews the simulations used to model the host mismatch rate and characterize its effects on the cosmological parameters, and Section~\ref{subsec:params} describes the methods and results used to ensure consistency between the DES-SN5YR data and our simulations. Section~\ref{sec:cosmo} reviews the framework used for cosmological parameter estimation in this work, and Section~\ref{sec:hostmatch-results} describes the cosmological biases resulting from host galaxy mismatch.

\section{Host Galaxy Matching}
\label{sec:methods}

\subsection{The DLR Method}
\label{subsec:dlr}
The directional light radius (DLR) method for host galaxy matching, developed in S06, involves computing a dimensionless distance ($d_{\mathrm{DLR}}$) for each potential host galaxy measured between the SN position and the centroid of the galaxy normalized by the galaxy size in the direction of the SN. Explicitly, $d_{\mathrm{DLR}}$ is defined as
\begin{equation}
d_{\mathrm{DLR}} = \frac{\mathrm{\Delta \theta}}{\mathrm{DLR}}
\end{equation}
where $\Delta \theta$ is the angular separation (arcsec) between the galaxy centroid and the SN, and DLR (arcsec) is the radius of the galaxy in the direction of the SN. We require all galaxies in our catalog (see Section~\ref{subsec:hostgal-catalog}) to be well modeled by ellipses parameterized by a semi-major axis $a$, a semi-minor axis $b$, and a position angle $\phi$, which is defined relative to the positive RA axis. Thus, the DLR value for each SN-galaxy pair is computed as follows:

\begin{equation}
    \textrm{DLR} = \frac{ab}{\sqrt{(a \,\textrm{sin} \, \phi)^2 + (b \,\textrm{cos} \, \phi)^2}}.
\end{equation}

For our DLR calculations, the values for the $a, b, \phi$ parameters are the \texttt{A\_{IMAGE}}, \texttt{B\_{IMAGE}} (converted into arcsec), and \texttt{THETA\_{IMAGE}} parameters output by Source Extractor \citep[][\texttt{sextractor}]{psfex} using the coadded $r+i+z$ detection image. The galaxy with the lowest $d_{\mathrm{DLR}}$ value is chosen as the host galaxy.

\subsection{Analysis Overview}
\label{}
We quantify the cosmological biases resulting from host galaxy mismatch by generating two sets of simulated SNe, one with and one without host galaxy mismatches, and comparing the fitted cosmological parameters. Details about the DES data can be found in Section~\ref{sec:data} and the simulations are described in Section~\ref{sec:sims}. 

To ensure that our results from simulations are applicable to real DES data, we show consistency between simulations and data across relevant parameter distributions, following \cite{popovic2020}: 
\begin{itemize}
    \item angular SN-galaxy separation ($\Delta \theta$),
    \item directional light radius (DLR),
    \item the DLR-normalized SN-galaxy separation, $d_{\mathrm{DLR}} = \frac{\Delta \theta}{\mathrm{DLR}}$
    \item $r$-band host galaxy magnitude ($m_{r,\mathrm{gal}}$),
    \item the ratio between smallest and second smallest $d_{\mathrm{DLR}}$ values, $r_{\mathrm{DLR}} = \frac{d_{\mathrm{DLR, HOSTGAL1}}}{d_{\mathrm{DLR, HOSTGAL2}}}$, where $d_{\mathrm{DLR, HOSTGAL}\{i\}}$ is the $i^{\text{th}}$ smallest $d_{\mathrm{DLR}}$ value, i.e. the $d_{\mathrm{DLR}}$ value of the $i^{\text{th}}$ most likely host galaxy.
\end{itemize}
 Further description of these parameters and the results of our consistency checks can be found in Section~\ref{subsec:params}. Host matching is performed on one set of simulated SNe to model mismatches as well as the DES-SN5YR data with the DLR method using the catalog of galaxies generated from deep DES imaging stacks, as described in Section~\ref{subsec:hostgal-catalog}. Finally, we estimate biases by comparing the fitted cosmological parameters from simulations with matched hosts and an identical set of simulations with true hosts, i.e. those assigned by the simulation.

\section{Data and host galaxy catalog}
\label{sec:data}

\subsection{Dark Energy Survey 5-Year Photometric Sample}
DES is an optical imaging survey designed to deliver precision cosmological results and constraints on dark energy by combining the probing power of weak gravitational lensing, baryon acoustic oscillations, galaxy clusters, and SNe Ia \citep{des_combined_probes}. DES imaged 5000 $\text{deg}^2$ of the southern sky for 6 years using the Dark Energy Camera \citep{decam} mounted on the 4m Blanco Telescope at the Cerro Tololo Inter-American Observatory. The time-domain survey component of the DES survey strategy covers a smaller area on the sky (10 supernova fields covering 27 $\text{deg}^2$), but exposures were repeated approximately weekly over the course of the survey. Eight of the ten fields were surveyed to a depth of $\sim 23.5$ mag per visit (‘shallow fields’), and the remaining two to a deeper limit of $\sim 24.5$ mag per visit (‘deep fields’), thus extending the SN detection limit to $z \sim 1.2$. Transient identification from images was performed using difference imaging \citep[\diffimg;][hereafter K15]{diffimg}. Spectroscopic follow-up of SN Ia candidates as well as their host galaxies was performed as described in \citet{smith2020,lidman2020}.

\subsection{Host Galaxy Catalog}
\label{subsec:hostgal-catalog}
To ensure sufficient depth and density of potential host galaxies as well as consistency between the galaxy catalogs used for real data matching and simulation, we produce a deep galaxy catalog by coadding DES images, identifying sources, and estimating photometric redshifts and galaxy parameters for each source.

The coadd procedure is similar to those described in \citet{wiseman2020} with several updates.  First, we apply stricter selection requirements (cuts) on the quality of the single-epoch images.  Images with effective exposure time ratio 
\footnote{$\tau_{\mathrm{eff}} = \Big( \frac{FWHM_{fid}}{FWHM} \Big) ^2 \Big( \frac{B_{fid}}{B} \Big) F_{trans}$, where $FWHM$ and $FWHM_{fid}$ are the measured and fiducial PSF full width half max, respectively; $B$ and $B_{fid}$ are the measured and fiducial sky background; $F_{trans}$ is the atmospheric transmission relative to a nearly clear night. See \citet{morganson} for details.}

$\tau_{\mathrm{eff}}$ $\le 0.3$ and those with point spread function full width at half maximum (PSF FWHM) $\ge 1.3", 1.2", 1.1", 1.0"$ in {\it griz}, respectively, are excluded to ensure higher quality images across the focal plane and to mitigate source confusion.  
Second, each image is scaled to a common zeropoint using the same framework adopted in \diffimg\ (see K15).  Third, rather than excluding images from the season in which the supernova candidate was discovered, we instead use all images from the 5-year survey and perform a median coaddition with \texttt{swarp} \citep{psfex}. Although median coaddition is not statistically optimal, it loses only $\sim 0.1$ mag in depth compared to other weighted average methods and is a more robust way to exclude light from transients and image artifacts when adding hundreds of images.  Finally, we determine the PSF model of the coadded images with \texttt{psfex} \citep{psfex} using our tertiary standard stars. 
The PSFs are used by \texttt{sextractor} \citep{psfex} to fit for the true (unblurred) Sérsic profiles and derive their parameters, which are necessary for the placement of simulated SNe in their host galaxies.


Sources from all 10 DES SN fields are identified using \texttt{sextractor}. Magnitudes are corrected for Milky Way dust using $E(B-V)$ values from \citet{sfd} and extinction coefficients for DECam filters from \citet{schlafly2011} and assuming $R_V = 3.1$.  These coefficients are $A/E(B-V) = 3.237, 2.176, 1.595,$ and $1.217$ in DECam $griz$, respectively.

The galaxies were matched to a spectroscopic redshift catalog compiled from multiple surveys including OzDES \citep{lidman2020}, SDSS \citep{sdss_followup}, 6dF \citep{6df}, ATLAS \citep{atlas}, GAMA \citep{gama}, VVDS \citep{vvds}, VIPERS \citep{vipers}, ACES \citep{aces}, DEEP2 \citep{deep2}, 2dFGRS \citep{2dfgrs}, UDS/VANDELS \citep{uds}, and PRIMUS \citep{primus}. Among the 7.8 million galaxies, a total of 124,824 have secure spectroscopic redshifts. 
Although we require a spectroscopic redshift for the DES data, the simulation needs a complete galaxy catalog to accurately model mis-matches. We therefore use photometric redshifts when spectroscopic redshifts are not available. 

\begin{figure}
    \includegraphics[scale=0.35,trim={0cm 0cm 0cm 0cm}]{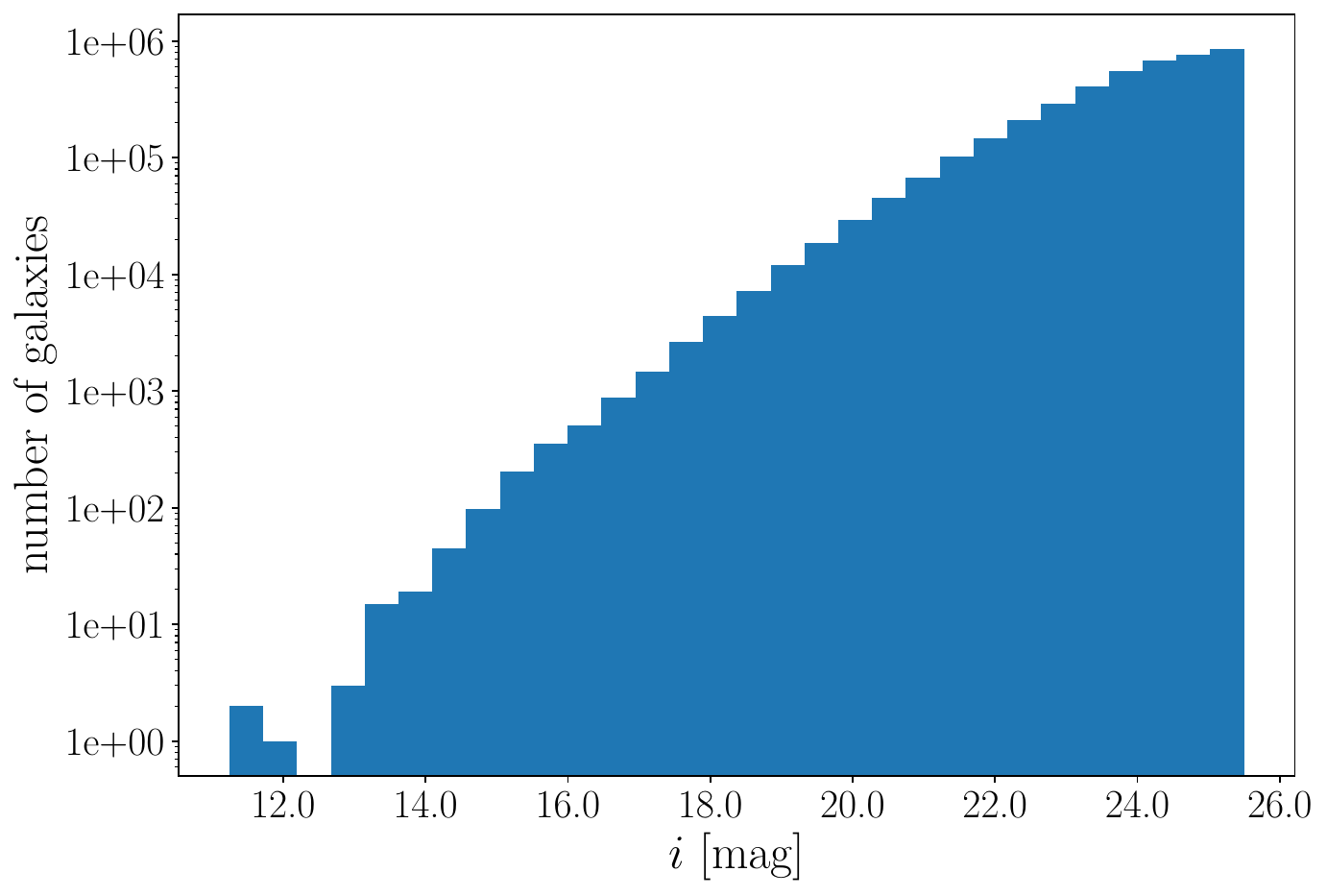}
    \centering
    \caption{$i$-band \texttt{MAG\_MODEL} magnitude distribution of all galaxies in our deep DES galaxy catalog with a spectroscopic or photometric redshift estimate. The catalog cuts off at $i \sim 26$ because photometric redshifts are unavailable for galaxies fainter than $i=25.5$ mag.}
    \label{fig:mag_dist}
\end{figure}

For the vast majority of galaxies without a spectroscopic redshift, we estimate photometric redshifts of galaxies that are brighter than $i=25.5$~mag using the method described in Section~\ref{subsec:photo-z}.  Sources fainter than this limit do not have reliable photo-$z$ estimates and are not included in the galaxy catalog. Figure~\ref{fig:mag_dist}  shows the magnitude distributions of the remaining sources.


\subsubsection{Host Galaxy Photometric Redshifts}
\label{subsec:photo-z}

Host galaxy photometric redshift estimation is performed independently for the deep and shallow SN fields, consisting of two and eight fields, respectively, as described in Section~\ref{subsec:hostgal-catalog}. For each galaxy in each field, we have a set of $griz$ photometric flux measurements and their corresponding uncertainties. Using this data, we train a \emph{Self-Organizing Map} (SOM) to characterize and discretize the photometric space of host galaxies, using the SOM algorithm described in the Appendix of \citet{Sanchez2020}. This SOM algorithm uses unsupervised learning to project the 4-dimensional photometric data ($griz$) onto a lower-dimensional grid, in our case a 2-dimensional grid, while attempting to preserve the topology of the 4-dimensional space. This means that similar objects in the 4-D space will be grouped together in the SOM, enabling a visual understanding of features. 

The particular SOM algorithm used in this work differs from the SOM algorithm used in previous DES analyses \citep[such as][]{y3-sompz, y3-2x2ptaltlenssompz} in order to improve the classification of galaxies with low- and modest-S/N photometry, which is relevant given the faint nature of many SN host galaxies. First, we alter the distance metric used by the SOM algorithm to incorporate flux uncertainties. Next, we include flux information, not just colors, and we do not impose periodic boundary conditions on the map. For the application in this work, we build a SOM of size $22 \times 22$, with a total of 484 cells. These alterations are described in detail in \citet{Sanchez2020}.

After the SOM is constructed, we use the subset of SN host galaxies with spectroscopic redshifts to populate the SOM and compute the redshift distribution of each SOM cell. The spectroscopic subset for deep SN fields has a total of 45,937 galaxies, while the one for shallow SN fields has a total of 78,887 galaxies. Once we populate the SOMs with redshift information, we assign to each galaxy the redshift distribution of the cell it is assigned to. Even if these spectrosopic subsets provide a good coverage of the corresponding photometric spaces, some SOM cells do not contain redshift information and therefore we do not estimate redshifts for galaxies in them. This is the case for 3.9\% of galaxies in the deep SN fields, and 6.7\% of galaxies in the shallow SN fields. 

To validate this procedure, we split the spectroscopic samples into separate training and validation samples, with a random 90\% of galaxies going into training and the remaining 10\% into validation. Next, we compare the estimated and spectroscopic redshifts for galaxies in the validation sample to assess the quality of the reconstruction. For this purpose, and to enable comparisons with previous DES estimates, we use photo-$z$ metrics from \citet{Sanchez2014}.  We construct the $\Delta z \equiv z_{phot} - z_{spec}$ distribution, and we compute two different metrics: 
\begin{enumerate}
    \item We estimate the photo-$z$ precision by calculating the 68-percentile width $\sigma_{68}$ of the $\Delta z$ distribution around its median value. This estimator, $\sigma_{68}$, measures the width of the core of the $\Delta z$ distribution. In particular, it is defined as half of the width of the distribution, measured with respect to the median, where 68\% of the data are enclosed. 
    \item We estimate the fraction of photo-$z$ outliers by calculating the fraction of objects with 3$\sigma$ deviations in $\Delta z$, $\mathrm{out}_{3\sigma} \equiv |\Delta z| > 3\sigma_z$, where $\sigma_z$ is the standard deviation of the $\Delta z$ distribution. 
\end{enumerate}
For the photo-$z$ estimation procedure described above, we find $\sigma_{68} = 0.124$ and $\mathrm{out}_{3\sigma} = 0.017$ for the deep SN fields, and $\sigma_{68} = 0.124$ and $\mathrm{out}_{3\sigma} = 0.015$ for the shallow SN fields. For the wide-field DES survey, the photo-$z$ requirements set prior to the start of the survey specified $\sigma_{68} < 0.12$ and $\mathrm{out}_{3\sigma} < 0.015$ for 90\% of the sample of galaxies. The values we find are slightly above these requirements, but it is important to point out that galaxy samples from the SN fields reach significantly fainter magnitudes than wide-field DES galaxies, and hence it is more difficult to satisfy the wide-field requirements. However, the SOM model is constructed such that it is primarily sensitive to color, and the color-redshift relation should be agnostic to galaxy brightness. In addition, the numbers we find are similar to those reported by several photo-$z$ codes in \citet{Sanchez2014}, demonstrating the comparable performance of our method when applied to fainter galaxies. 

\subsubsection{Profile Fitting}
\label{subsec:profile-fit}
Both the measured profile, which was used to calculate DLR values and described in Section~\ref{subsec:dlr}, and the \textit{intrinsic} galaxy light profile are used in this work.

The intrinsic light profile is used to determine the location of each simulated SN within its assigned host galaxy (see Section~\ref{subsec:host-matching}). To calculate the intrinsic light profile, each galaxy in the catalog is fit with a Sérsic profile \citep{sersic} that describes the variation of galaxy intensity $I$ with radius $R$: 
\begin{equation}
\label{eq:sersic}
    I(R) = I_e \; \mathrm{exp} \Bigl\{ -b_n \Bigl[ \Bigl(\frac{R}{R_e}\Bigr)^{1/n} - 1 \Bigr] \Bigr\}
\end{equation}
$I_e$ is the galaxy intensity at the half-light radius $R_e$, $n$ is the Sérsic index describing the cuspiness of the profile, and $b_n$ is a known function of $n$. The fitting was performed by \texttt{sextractor}, which fits for $R_e$ as well as additional parameters \texttt{theta} and \texttt{aspect}, which describe the angle relative to the positive RA axis and the ellipticity of the galaxy, respectively. These additional parameters enable us to model each galaxy as an ellipse and determine its semimajor and semiminor axes, which we denote $a, b$. These quantites are calculated with $R_e = \sqrt{ab},\; \texttt{aspect} = \frac{a}{b}$. The fitted Sérsic half-light radii were scaled by a factor of 0.8 to obtain better data-simulation agreement (see Section~\ref{subsec:data-sim-agreement} for details on the data-simulation matching procedure and Section~\ref{subsec:sersic-scale} for comparisons with and without this scaling factor). This is notably the same scaling factor found to best match the DES3YR data when comparing distributions of host galaxy surface brightness at the SN position \citep[see Figure 6 of][]{des3yr}.

\subsubsection{Catalog Cuts and Parameter Fitting}
\label{subsec:catalog-cuts-params}

\begin{table}
    \centering
    \caption{Cuts applied to the host galaxy library and resulting galaxy counts.}
    \label{tbl:catalog-cuts}
    \begin{tabular}{l c}
        \hline
        Cut Requirement & Galaxies Remaining \\
        \hline
        Full catalog & 8,401,139 \\
        Removed duplicate galaxies & 7,860,305 \\ 
        Has photo-$z$ or spec-$z$ & 5,118,585 \\
        Has CIGALE galaxy parameter fit & 5,108,812 \\
        $5 < \mathrm{log}(M_{\star} / M_{\odot}) < 14$ & 4,938,372 \\
        Has reasonable Sérsic fit & 4,221,001 \\
        
        \hline
        \textbf{Final} & \textbf{4,221,001}\\
    \end{tabular}
\end{table}

Several selection criteria are applied to select \texttt{sextractor} sources that balance the trade-off between preserving realistic catalog depth/density, and maintaining the quality of the galaxy photometry. The list of cuts as well as the number of galaxies remaining after each is shown in Table~\ref{tbl:catalog-cuts}.

First, duplicate observations of galaxies are removed, prioritizing deep field observations when possible. The duplication is due to the slight overlap of certain DES SN fields, causing galaxies in the overlapping regions to appear in multiple coadded images. Next, galaxies without a spectroscopic redshift determination or photometric redshift estimate are removed, because simulated SNe cannot be associated with such galaxies. As stated in Section~\ref{subsec:hostgal-catalog}, galaxies fainter than $i=25.5$ mag do not have reliable photometric redshift estimates and are removed. In addition to the $i=25.5$ mag cut, 3.9\% of galaxies in the deep SN fields and 6.7\% of galaxies in the shallow SN fields were not successfully assigned a redshift estimate by the SOM and are removed as well.

Certain galaxy properties are known to be correlated with SN Ia rate \citep[S06;][]{mannucci,graur2017,wiseman2021}, and we use these known correlations to assign suitable host galaxies in the simulation. We estimate the total stellar mass $(M_{\star})$ and the star formation rate (SFR) for galaxies in our catalog using CIGALE \citep{cigale}. CIGALE uses grid search to find the best-fit (lowest-$\chi^2$) combination of user-specified model parameters given galaxy photometry and redshift estimates. For the galaxy parameter fits, we assume a delayed star formation history, where SFR is defined by
\begin{equation}
    \mathrm{SFR}(t) \propto \frac{t}{\tau^2} \mathrm{exp}(-t/\tau),\; 0 \leq t \leq t_0
\end{equation}
with $t_0$ the age of the onset of star formation and $\tau$ the time at which the SFR peaks. The \texttt{bc03} \citep{bc03} library of single stellar populations with a Salpeter initial mass function is used to compute the intrinsic stellar spectrum. Attenuation from dust and other sources is parameterized by the \citet{calzetti_2000} starburst attenuation curve extended with the \citet{leitherer} curve. Nebular emission is modeled by templates from \citet{inoue_2011}.
We remove a small subset of galaxies with poorly constrained CIGALE parameter fits by restricting our library to galaxies with $5 < \mathrm{log}(M_{\star}/M_{\odot}) < 14$.

We found by manual inspection that in some cases, such as very diffuse galaxies with low Sérsic index, the Sérsic profile fit fails catastrophically and produces greatly exaggerated estimates of $R_e$. This will result in SNe placed very far from the galaxy center in simulations, potentially creating a clear mismatch with SN-galaxy separations measured from the DES data. To remove galaxies with these pathological fits, we calculate ellipse areas from \texttt{sextractor} parameters ($A_{\texttt{sextractor}} = \texttt{A\_IMAGE}*\texttt{B\_IMAGE}$) as well as from Sérsic ellipse parameters ($A_{\text{Sérsic}} = ab$). We select galaxies with a ``reasonable" Sérsic fit, which we define to be $\frac{A_{\mathtt{sextractor}}}{A_{\text{Sérsic}}} \geq 0.25$.

\subsection{DES Data Host Matching}
\label{subsec:des-matching}

\begin{table}
    \centering
    \caption{Summary of host galaxy matching for the DES Y5 sample.}
    \label{tbl:des-matching}
    \begin{tabular}{l c c}
        \hline
         & Number of SNe & Percent of Total \\
        \hline
        Total sample size & 2,186 & 100\%\\ 
        Has $\geq 1$ host match & 2,047 & 94\%\\ 
        Has $\geq 2$ host matches & 126 & 6\%\\
        \hline
    \end{tabular}
\end{table}

Host galaxy matching for the DES Y5 sample is performed using the DLR method with the host galaxy catalog described in Section~\ref{subsec:hostgal-catalog}. \ddlr values are computed for all galaxies in the catalog within 15" of the SN position. Galaxies with \ddlr $> 4$ are discarded, following the SDSS convention \citep{sako2014}, and those with lowest and second lowest \ddlr values are identified as the most likely and next most likely host match (\texttt{HOSTGAL1} and \texttt{HOSTGAL2}). If fewer than two galaxies have \ddlr $\leq 4$, those SNe are considered to be missing a host match or (if one galaxy has \ddlr $\leq 4$) missing a second host match. Summary statistics for the DES data with host matches is shown in Table~\ref{tbl:des-matching}.

Ideally, matching should be done with the full catalog, including galaxies without a known redshift. This would allow for us to remove SNe hosted by galaxies without known redshifts, rather than match them incorrectly to a galaxy with known redshift. However, we find that $\sim 98\%$ of DES SNe have the same host match when matched with the full catalog, and those that do not mostly become ``hostless" when the cuts are employed. Thus, we conclude that matching with the cut catalog is a valid method, as these SNe are ineligible for the analysis regardless of the cuts. 

\section{Simulations and event selection}
\label{sec:sims}

\subsection{Simulations}
All simulations for this work are produced with the SuperNova ANAlysis (SNANA) software \citep{snana}. The SNANA simulation starts with a SN spectral template and generates survey-specific photometry under realistic observing conditions by utilizing a cadence library containing zero points, sky noise, and PSF information for each telescope pointing on each observing night.

We note that overlaying simulated SNe on actual galaxy images is the ideal method of performing a host matching analysis; however, it would be very computationally expensive to run the volume of simulations needed to develop and constrain the host mismatch systematic. For improved computational efficiency, we use the catalog-level simulation from SNANA.

\subsubsection{SN Models}
\label{subsec:sn-models}

SN Ia simulations for this work are created using the SALT2 model \citep{salt2} with training parameters determined from the Joint Lightcurve Analysis \citep[JLA,][]{jla}. The SALT2 model defines several restframe parameters for SN lightcurves: the time of SN peak brightness $t_0$, a stretch-like parameter $x_1$, a color parameter $c$ and the lightcurve normalization parameter $x_0$. Nuisance parameters $\alpha, \beta$ are determined according to \citet{scolnic2016}, while the color ($c$) and stretch ($x_1$) populations follow \citet{popovic2021}. The SNe Ia are simulated with a redshift-dependent volumetric rate, using measured rates from \citet{dilday} and \citet{perrett} and recomputed by \cite{frohmaier}. To model empirically measured Hubble scatter after the SN Ia standardization procedure, we use the spectral-variation intrinsic scatter model in \citet{k13} that is based on the model uncertainties determined in \cite{g10}. The simulations include two separate populations: the DES-like photometric sample and a spectroscopically confirmed low-$z$ anchor. For simplicity, the low-$z$ anchor is an ad-hoc DES simulation applied to $0 < z < 0.1$ with an inflated rate to match the true number of low-$z$ events in the DES-SN5YR sample.

Two types of core-collapse SNe as well as two types of peculiar SNe Ia are simulated alongside the SNe Ia to evaluate the effects of host galaxy mismatch on redshift-dependent photometric classification: SNII, SNIbc, SNIax, and SNIa-91bg. These simulations use the SED templates introduced in \cite{v19} with luminosity functions and rates following \cite{vincenzi2021}. We use relative rates as measured by \citet{shivvers} anchored by an overall rate following the cosmic star formation history presented in \cite{madau} normalized by the local SN rate from \cite{frohmaier20}.

All simulations are generated assuming a flat $\Lambda$CDM cosmology with $H_0 = 70\; \mathrm{km\;s^{-1}\;Mpc^{-1}}$ and $\Omega_{\mathrm{M}} =0.311$.

\begin{figure}
    \includegraphics[scale=0.38,trim={0cm 0cm 0cm 0cm}]{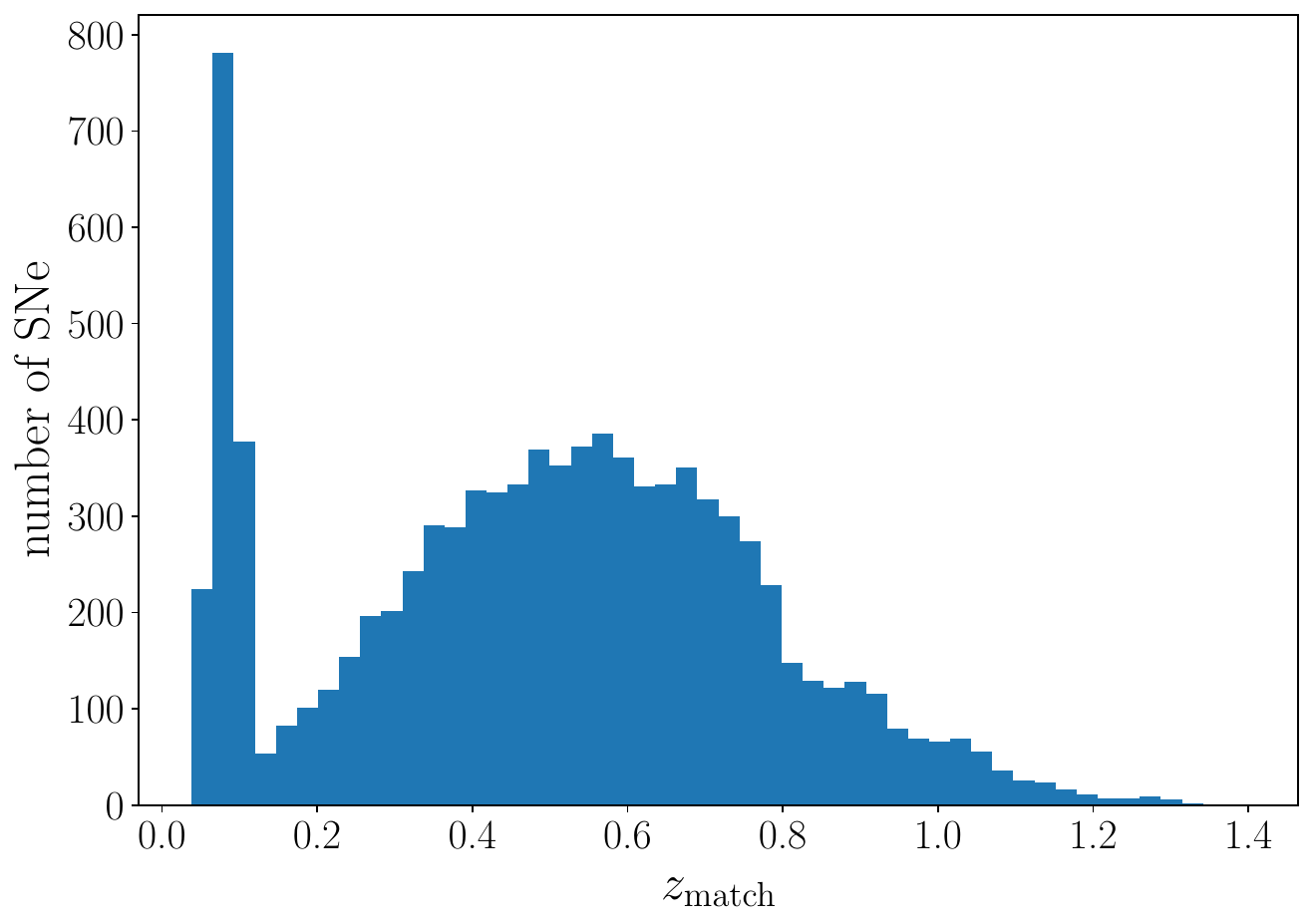}
    \centering
    \caption{Redshift distribution of matched host galaxies in one realization of the SNIa-only simulations. Since we find the mismatch rate resulting from the DLR method to be quite low (see Table~\ref{tbl:sim-cuts-Ia-cc}), the true redshift distribution is not visibly different and was not included in this plot.}
    \label{fig:Ia_z_dist}
\end{figure}

\begin{figure*}
    \includegraphics[scale=0.38,trim={0cm 0cm 0cm 0cm}]{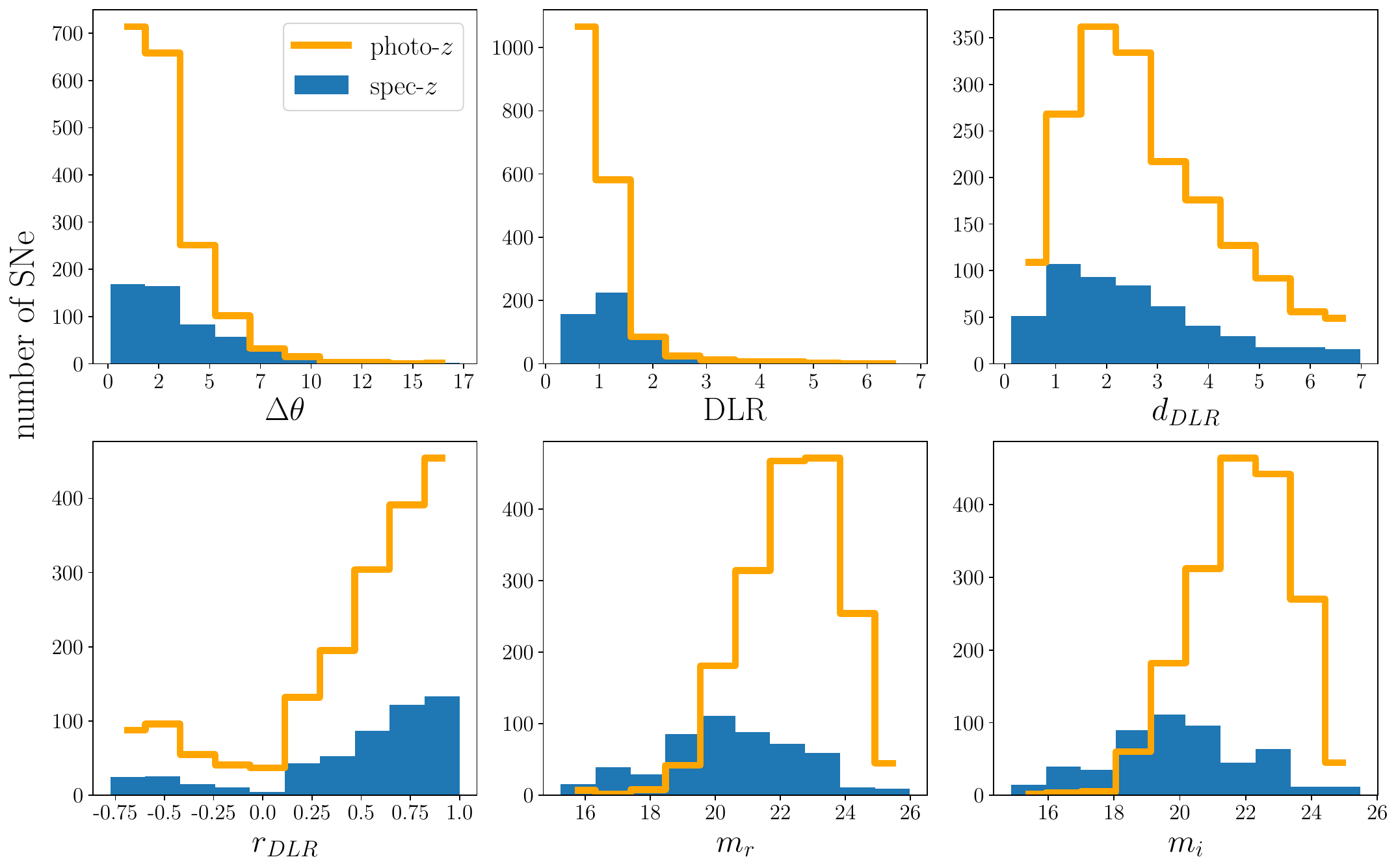}
    \centering
    \caption{Parameter distributions of matched host galaxies in one realization of the SNIa-only simulations split into two populations: galaxies with spectroscopic redshifts and those with photometric redshift estimates only. Details and definitions of these parameters can be found in Section~\ref{subsec:data-sim-agreement}. Since we find the mismatch rate resulting from the DLR method to be quite low (see Table~\ref{tbl:sim-cuts-Ia-cc}), the true distributions are not visibly different and were not included in this plot.}
    \label{fig:spec_vs_phot_dists}
\end{figure*}

\subsubsection{Host Association and Matching}
\label{subsec:host-matching}
To associate a simulated SN with a host galaxy, the simulation first generates properties for each SN, including color, stretch, redshift, and sky location (RA, DEC). Subsequently, all galaxies whose redshifts match the true SN redshift within a small tolerance ($dz_{\mathrm{tol}}= \text{max} |z_{\mathrm{SN}} - z_{\mathrm{GAL}}| = 0.002 + 0.04z$) are selected. The \textit{assigned} host galaxy is chosen from this subset using the host mass-dependent weighting shown in \cite{wiseman2021}. The simulated host association and matching use the same deep coadded DES galaxy catalog used for the data analysis.

Note that the process of assigning a host for a simulated SN does not take into account the locations of the host or the SN, only the redshifts of the SN and galaxy as well as the galaxy stellar mass. The host and its neighboring galaxies are moved near the SN such that it satisfies the simulated SN-host separation as well as models host confusion in the analysis. This strategy is efficient for modeling SN-host correlation and incorrect host matches, but it does not model large scale structure.

The SN-host separation is determined by placing the SN at a radial distance $R$ from the center of the assigned host galaxy according to the probability distribution $p(R) \sim I(R)$, where $I(R)$ is the galaxy intensity at radius $R$ described by the galaxy's fitted Sérsic profile (see Equation~\ref{eq:sersic}). This approach follows past work using SNANA such as \citet{des3yr}.

To replicate the host matching procedure used for real data, we apply host galaxy matching with the DLR method to the simulated SNe. The simulation computes $d_{\mathrm{DLR}}$ values for up to 10 galaxies within a 10"\ radius of the assigned host and saves the galaxies with the lowest and second lowest $d_{\mathrm{DLR}}$ values as the most and next most likely host match (\texttt{HOSTGAL1} and \texttt{HOSTGAL2}), as long as $d_{\mathrm{DLR}}\leq 4$. The redshift distribution of the matched host galaxies for one realization of a SNIa-only simulation is shown in Figure~\ref{fig:Ia_z_dist} and relevant population parameter distributions (e.g. $d_{\mathrm{DLR}}$) are shown for host galaxies with spectroscopic and photometric redshifts in Figure~\ref{fig:spec_vs_phot_dists}.
The host matching procedure is run for the full set of simulations, including the low-$z$ population, though low-$z$ mismatches are rare. 

\subsection{Event Selection}
\subsubsection{Lightcurve Fitting}
\label{subsec:lcfit}

All DES-SN observed and simulated SN lightcurves are fit with the same SALT2 model with JLA parameters that was used to simulate the SNe Ia. The fit is performed with a $\chi^2$-minimization program included in SNANA and determines several parameters under the assumption that the event is a SN Ia: the time of SN peak brightness $t_0$, a stretch-like parameter $x_1$, a color parameter $c$ and the lightcurve normalization parameter $x_0$, as well as their uncertainties and covariances (i.e., $\sigma_{t_0}$, etc.). These parameters are used to calculate the distance modulus $\mu$, allowing the SNe to be placed on the Hubble diagram.

We apply selection cuts on these fitted parameters and select SN lightcurves well described by the SALT2 model. Specifically, we restrict our sample to SNe satisfying the following criteria:

\begin{itemize}
  \item $|x_1| < 3$,
  \item $|c| < 0.3$,
  \item $\sigma_{x_1} < 1$, and
  \item $\sigma_{t_0} < 2$ days.
\end{itemize}

\begin{figure*}
    \includegraphics[scale=0.28,trim={0cm 0cm 0cm 0cm}]{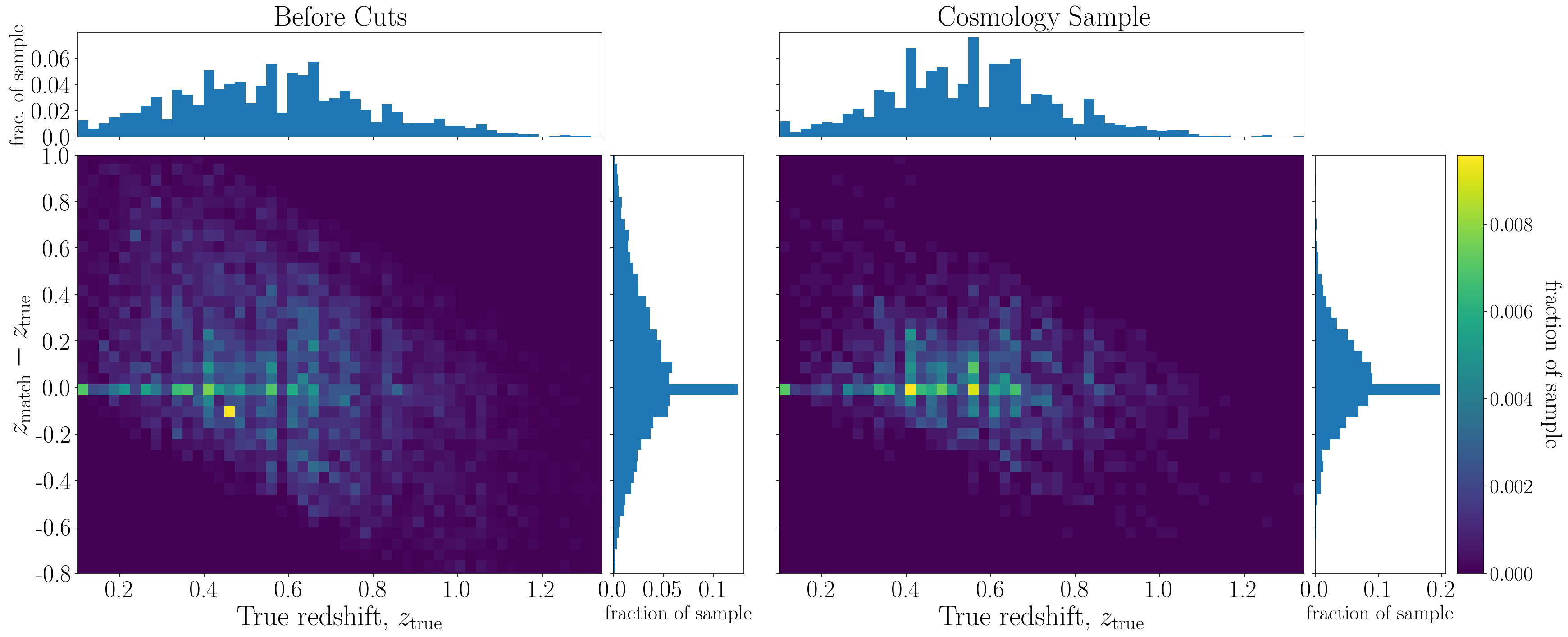}
    \centering
    \caption{Redshift differences ($\Delta z \equiv z_{\mathrm{match}} - z_{\mathrm{true}}$) for simulated SNe Ia with mismatched hosts. (left) $\Delta z$ as a function of $z_{\mathrm{true}}$ for the full simulated sample prior to applying any selection cuts. (right) $\Delta z$ as a function of $z_{\mathrm{true}}$ after all selection cuts (described in Section~\ref{subsec:lcfit}) and removal of SNe without a valid bias correction (described in Section~\ref{subsec:bcor}). Our selection process not only reduces the mismatch rate (see Table~\ref{tbl:sim-cuts-Ia-cc}) but also removes SNe with extremely biased redshift estimates from mismatched hosts, reducing the spread in $\Delta z$. Quantitatively, the spread in $z_{\mathrm{match}} - z_{\mathrm{true}}$ characterized by the middle 90\% of the distribution is reduced from 1.1 before cuts to 0.6 after cuts.}
    \label{fig:Ia_delta_z}
\end{figure*}

A summary of these cuts on our simulations can be found in Table~\ref{tbl:sim-cuts-Ia-cc}. The right panel of Figure~\ref{fig:Ia_delta_z} shows the effect of these cuts on the redshift error distribution between the matched and true hosts. These data show that the cuts not only reduce the average mismatch rate over 25 realizations from 2.5\% to 1.7\%, but also significantly reduce the spread in redshift error. Quantitatively, the cuts reduced the middle 90\% of the $z_{\text{match}}-z_{\text{true}}$ distribution for mismatched pairs from 1.1 (left panel of Figure~\ref{fig:Ia_delta_z}) to 0.6 after cuts (right panel).

\begin{table*}
    \centering
    \caption{Summary of the mismatch rate averaged over 25 realizations after each selection cut on the SNIa+CC simulated dataset, split by Ia vs. non-Ia SNe. The non-Ia SNe include SNIax, SNIa-91bg, SNII, and SNIbc. All results using SNe Ia only are using the SN Ia subset of this dataset.}
    \label{tbl:sim-cuts-Ia-cc}
    \begin{tabular}{l c c c c c c}
        \toprule
        Cut & \multicolumn{3}{c}{SN Ia} & \multicolumn{3}{c}{non-Ia SNe}\\
        \cmidrule(lr){2-4} \cmidrule(lr){5-7} 
         & Mismatches & Total & Mismatch Rate & Mismatches & Total & Mismatch Rate\\
        \midrule
        No cuts & 233 & 9,356 & 2.5\% & 151 & 5,957 & 2.5\%\\ 
        $|x_1| < 3$, $|c| < 0.3$ & 95 & 5,562 & 1.7\% & 22 & 800 & 2.8\%\\ 
        $\sigma_{x_1} < 1$ & 83 & 4,886 & 1.7\% & 15 & 545 & 2.8\%\\
        $\sigma_{t_0} < 2$ & 83 & 4,870 & 1.7\% & 15 & 543 & 2.8\%\\
        \bottomrule
    \end{tabular}
\end{table*}

\subsubsection{Photometric Classification}
\label{subsec:classification}

In the absence of SN spectra, we rely on photometric classifiers to remove non-Ia contaminants from the cosmological sample. We choose two recent neural-network-based classifiers with high demonstrated accuracies for this work: SuperNNova \citep[SNN,][]{supernnova} and SCONE \citep{scone}. High quality SN redshifts have been shown to improve SNN accuracy classifying SN Ia vs. non-Ia, so we test SNN in both redshift-dependent and redshift-independent configurations to evaluate the impact of misidentified redshifts. SNN models were trained in both the redshift-independent and dependent configurations following \citet{Vincenzi_2022}. SCONE is redshift-independent and performs classification based on SN lightcurves alone, so these results should not be affected by host misidentification. Each classifier outputs $P_{\mathrm{Ia}}$ values, the predicted probability of each SN to be a type Ia. The SNIa-only simulations are not run through photometric classification; all objects are simply labeled as SNe Ia.

\section{Comparing Data with Simulations}
\label{subsec:params}

\subsection{Host Matching Rates and Hostless SNe}

First, we compare the fraction of simulated and observed DES SNe  that pass our selection cuts with one or more matched hosts. Since we have reliable SN Ia photometric classifiers, we compare simulated true SNe Ia with DES SNe Ia as predicted by SCONE \citep{scone}. Since most of the redshifts for the DES SNe come from host galaxies, we chose to use SCONE for this comparison as it does not require redshift information.

After applying the cuts specified in Table~\ref{tbl:sim-cuts-Ia-cc} to both data and simulations, the DLR algorithm is able to find at least one host galaxy match in 98.3\% of SNe in our SN Ia-only simulations compared to 98.2\% of SCONE-classified SNe Ia in the DES data. The remaining $\sim 2$\% of SNe in our data and simulations is the rate of ``hostless" SNe. SNe can appear hostless because their host galaxies are too faint to be detected and our imposed \ddlr $\leq 4$ cut on potential host matches avoids matching to unrealistically faraway galaxies.

SNe in galaxy-dense regions have multiple potential host matches with \ddlr $\leq 4$. We find that ($9.35\pm 0.09$)\% of simulated SN Ia have more than one host match, compared to ($8.71\pm 1.06$)\% of SCONE-classified DES SNe Ia. The agreement in the fraction of hostless and multiple-host events
provides confidence in our simulated data sample.

\begin{figure*}
    \includegraphics[scale=0.33,trim={0cm 0cm 0cm 0cm}]{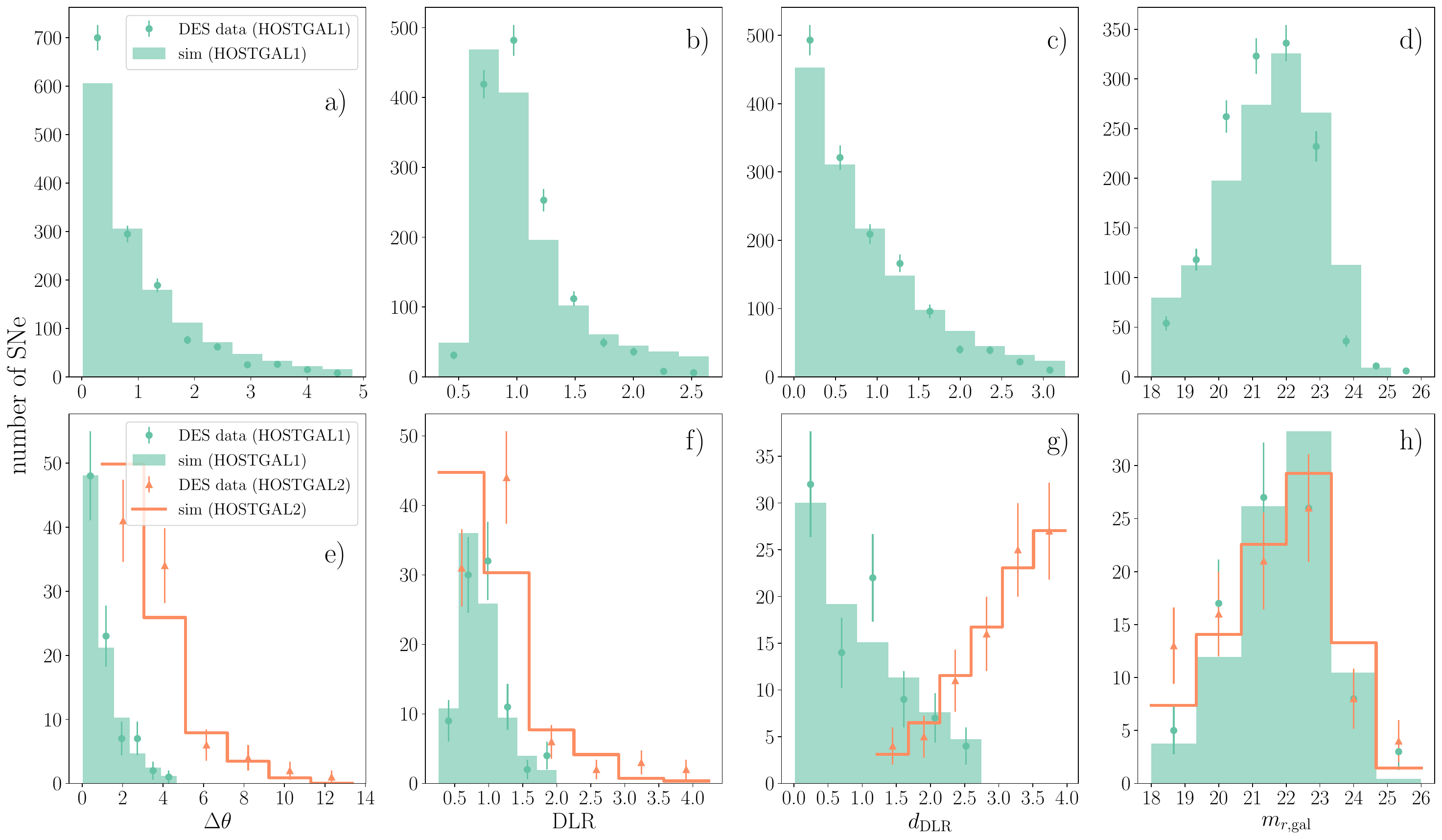}
    \centering
    \caption{Histograms comparing our simulated SNe Ia against DES Y5 photometrically confirmed SNe Ia for angular SN-galaxy separation ($\Delta \theta$, arcsec), directional light radius (DLR), the DLR-normalized SN-galaxy separation ($d_{\mathrm{DLR}} = \frac{\Delta \theta}{\mathrm{DLR}}$), and $r$-band host galaxy magnitude ($m_{r,\mathrm{gal}}$) (see Section~\ref{subsec:params} for explanations of each). For ease of comparison, the simulation histograms are normalized to match the integral of the data histogram and the $x$ axis limits of each histogram are determined by the middle 90\% of the data distribution to remove outliers. In both rows, points with error bars represent parameter distributions measured from the DES data and filled or unfilled histograms represent the analogous quantities for simulations. (top row) Histograms of parameter values for closest host galaxy match (\texttt{HOSTGAL1}) for SNe with only one host galaxy match. (bottom row) Histograms of parameter values for closest (\texttt{HOSTGAL1}) and second closest (\texttt{HOSTGAL2}) host galaxy match for SNe with 2 or more host galaxy matches.}
    \label{fig:param-dists}
\end{figure*}

\subsection{Comparing Parameter Distributions}
\label{subsec:data-sim-agreement}

To further verify that the host mismatch rate estimates and resulting cosmological biases derived from our simulations are representative of the DES sample, we compare simulations and data over five relevant parameter distributions: SN-galaxy separation ($\Delta \theta$), DLR, $d_{\mathrm{DLR}}$, $r$-band host galaxy magnitude ($m_{r, \mathrm{gal}}$), and the \texttt{HOSTGAL1}  to \texttt{HOSTGAL2}  \ddlr ratio ($r_{\mathrm{DLR}}$), following \cite{popovic2020}. A comparison of these distributions for the DES-SNY5YR sample (shown in points) and the simulations (shown in filled/unfilled bars) used for this analysis is shown in Figure~\ref{fig:param-dists}.

The top row of Figure~\ref{fig:param-dists} shows the parameter distributions of data and simulations for SNe with a single host galaxy match (i.e. only one galaxy with \ddlr $\leq 4$). The second row shows the same parameter distributions for SNe with at least two host galaxy matches, where green show distributions for the closest host galaxy match (smallest \ddlr value, labeled \texttt{HOSTGAL1}) and orange denotes the second closest host galaxy match (labeled \texttt{HOSTGAL2}).

The angular separation between the center of the galaxy and the SN position, {$\Delta \theta$, is shown in panels a) and e) of Figure~\ref{fig:param-dists}. Good agreement in the data vs. simulation $\Delta \theta$ distribution validates that (1) the algorithm used by the simulation to place SNe within their host galaxies is representative of real observations, and (2) the Sérsic ellipse parameters $a,b$ used to place SNe within their host galaxies are well estimated. Note that, as described in Section~\ref{subsec:profile-fit}, the fitted Sérsic parameters were scaled by a factor of 0.8. The distributions from data and simulations match very well overall, but the simulations slightly underestimate \texttt{HOSTGAL1} matches at the very low end of the $\Delta \theta$ distribution.

DLR, shown in panels b) and f), corresponds to the size of the matched host and is measured as in Section~\ref{subsec:dlr}. Agreement in this parameter verifies that matched hosts are similar in size between data and simulations. The data and simulations agree well for both \texttt{HOSTGAL1} and \texttt{HOSTGAL2}.

The distribution of \ddlr values is shown in panels c) and g). \ddlr agreement is an important quantity, since it is used to determine which galaxies are host matches, and shows that simulated SNe are placed at reasonable distances from the host center. These distributions largely show the same trend as the $\Delta \theta$ distributions, where the \texttt{HOSTGAL2} distributions match very well but the \texttt{HOSTGAL1} distributions from simulations appear to be skewed slightly higher than those of the DES data.

$m_{r,\text{gal}}$, the host galaxy $r$-band magnitude, is shown in panels d) and h). Agreement in this parameter is an additional validation of similarity in the overall populations of host galaxies between data and simulations. The slight discrepancy between these distributions can likely be attributed to the fact that the spectroscopic efficiency is defined using \texttt{MAG\_AUTO}, but we applied the efficiency to \texttt{MAG\_MODEL} magnitudes, which tend to be slightly fainter.

Finally, we define
\begin{equation}
r_{\mathrm{DLR}} = \frac{d_{\mathrm{DLR, HOSTGAL1}}}{d_{\mathrm{DLR, HOSTGAL2}}}.
\end{equation}
Figure~\ref{fig:rddlr_cigale} shows the distributions of this parameter for DES data and our simulations. Since the definition requires $d_{\mathrm{DLR, HOSTGAL2}}$, only SNe with at least 2 host galaxy matches are included. Consistency in $r_{\mathrm{DLR}}$ and the \texttt{HOSTGAL2} distributions indicate that the galaxy catalog is sufficiently dense, since the spacing between galaxies will affect the \texttt{HOSTGAL2} parameter distributions much more than \texttt{HOSTGAL1}.

\begin{figure}
    \centering
    \includegraphics[scale=0.5]{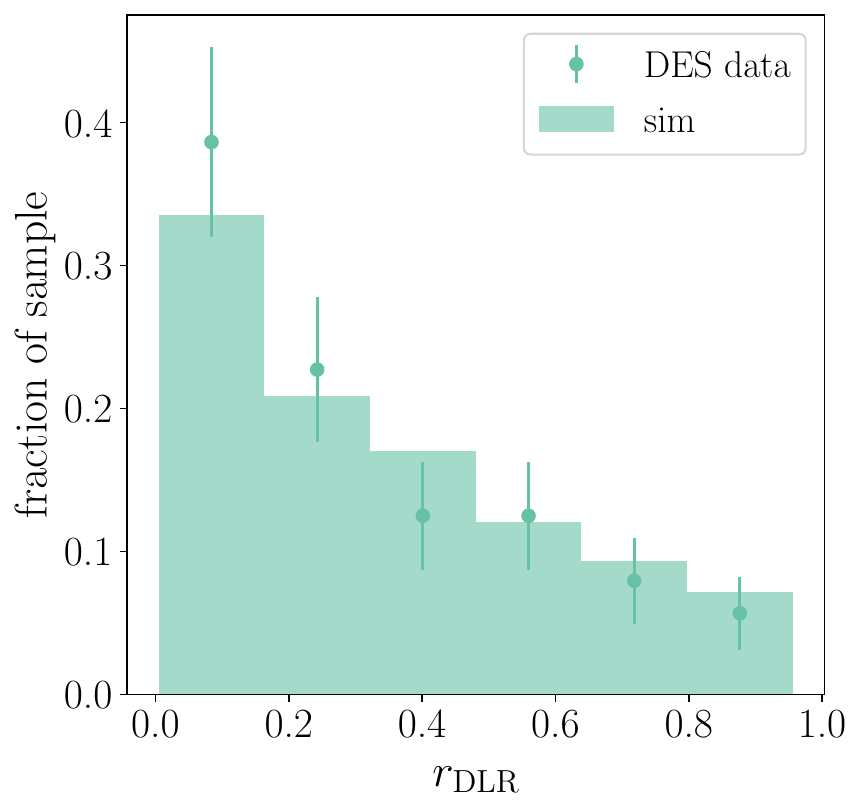}
    \caption{Distributions of the ratio $r_{\mathrm{DLR}} = \frac{d_{\mathrm{DLR, HOSTGAL1}}}{d_{\mathrm{DLR, HOSTGAL2}}}$} for DES Y5 photometrically confirmed SNe Ia and our simulated SNe Ia.
    \label{fig:rddlr_cigale}
\end{figure}


\section{Cosmological Parameter Estimation}
\label{sec:cosmo}

\subsection{Cosmology Analysis Overview}
The cosmology analysis pipeline is orchestrated end-to-end by the \texttt{Pippin} framework \citep{pippin}, beginning with simulations or data as input and concluding with cosmological parameter estimates. After SNe are simulated, the resulting lightcurves are fit with an empirical model that outputs characteristics for each SN, such as color and stretch (see Section~\ref{subsec:lcfit}). SNe that are fit successfully are then assigned a SN Ia probability by a photometric classifier (see Section~\ref{subsec:classification}) and bias corrections on distance moduli are computed for each SN based on its fitted parameters as well as SN Ia probability. Finally, the bias corrected distance moduli are used to determine constraints on cosmological parameters.

\subsection{Bias Corrections}
\label{subsec:bcor}

Biases due to core collapse contamination and survey selection are modeled and corrected for with the BEAMS with Bias Corrections framework \citep[BBC,][]{bbc}, an extension of BEAMS \citep{beams}, which allows photometrically classified SNe Ia to be used for cosmology. The primary output of the BBC framework applied on a SN sample is a redshift-binned Hubble diagram corrected for biases from selection effects and non-SNIa contamination.

First, systematic biases due to selection effects are modeled by a large simulation of SNe Ia ($\sim 800,000$ SNe Ia). We found empirically that our current framework for modeling and correcting for biases, typically used for e.g. the Malmquist bias, is not suited for including wrong hosts in our bias correction simulations. Thus, our bias correction simulation includes only assigned hosts for our primary results. In Section~\ref{subsec:bcor-mismatch}, we explore incorporating host matching (and thus mismatched hosts) into the bias correction simulations. 

Using this large SN sample, distance moduli are calculated using the Tripp formula \citep{tripp}, 
\begin{equation}
\label{eq:tripp}
\mu_\mathrm{obs} = m_B + \alpha x_1 - \beta c + M_B+\Delta \mu_{\mathrm{bias}}
\end{equation}
for each SN. $m_B=-2.5 \mathrm{log_{10}}(x_0)$ and $M_B$ is the absolute magnitude of a SN Ia with $x_1 = c = 0$ and $\alpha,\beta$ are nuisance parameters determined according to \citet{scolnic2016}. Biases from a reference cosmology, $\Delta \mu_{\mathrm{bias}}$, are calculated in a 3-dimensional grid of $\{z, x_1, c\}$ bins using the method described above. We use this grid of estimated biases to correct all of our distance moduli prior to cosmology fitting. A small percentage of SNe with parameter values that do not fit into the grid are discarded. 
Finally, the BEAMS method is used to estimate binned distance moduli from the bias-corrected distance moduli from the previous step in the presence of core-collapse contamination. This is done by minimizing the BEAMS likelihood, which models the SNe Ia population and a population of contaminants separately. These terms are weighted by $P_{\mathrm{Ia}}$, the probability of each SN to be a type Ia as output by a photometric classifier. We omit the mass step correction in Equation~\ref{eq:tripp}; details on the mass step and the impact of host mismatches can be found in Section~\ref{subsec:mass-step}.

\subsection{Cosmological Parameters}
\label{subsec:cosmo-params}
 We fit for $w$ and $\Omega_m$ using \texttt{wfit}, a fast cosmology grid-search program in SNANA,  assuming a diagonal covariance matrix $\mathcal{C}_{\text{stat}}$ and an approximate CMB prior computed with the $R$-shift parameter (see e.g. Equation 69 in \citet{WMAP:2008lyn}) from the same cosmological parameters used to generate the SNe Ia. The $R$ uncertainty is  $\sigma_R = 0.006$, tuned to have the same constraining power as \citet{planck}. As we are only interested in the impact of host mismatches on cosmology, the approximation of a diagonal covariance matrix is sufficient for our purposes.
\texttt{wfit} calculates the $\chi^2$ of the SN likelihood to compute our final cosmological fit,
\begin{equation}
\label{eqn:chi2}
    \chi^2 = \Delta \mu_{\text{model}}^T \cdot \mathcal{C}_{\text{stat}}^{-1}\cdot\Delta \mu_{\text{model}}
\end{equation}
where 
\begin{equation}
    \Delta \mu_{\text{model}} = \mu - \mu_{\text{model}}(\Omega_m, w).
\end{equation}

\section{Results and Discussion}
\label{sec:hostmatch-results}

We use the cosmological parameter estimation framework described in Section~\ref{sec:cosmo} to evaluate the effects of host galaxy mismatch on the resulting best fit cosmology. We focus primarily on shifts in the best fit value for $w$, the dark energy equation of state parameter. We quantify this shift by creating two sets of identical simulations that differ only in whether or not the DLR method is run to determine the matched hosts using the procedure described in Section~\ref{subsec:host-matching}. We define $S_{\mathrm{match}}$ as simulations with matched hosts and $S_{\mathrm{truehost}}$ as simulations with perfect host matching. In $S_{\mathrm{truehost}}$, we do not calculate matched hosts using the DLR method; we instead force a match to the true hosts assigned by the simulation. This ensures that there will be no mismatches and serves as a baseline for comparison.

We define the $w$ shift as the difference between the inferred $w$ values from $S_{\mathrm{match}}$ ($w_{\text{match}}$) and $S_{\mathrm{truehost}}$ ($w_{\text{truehost}}$). We average over 25 realizations with the same simulation parameters. Explicitly, we define the $w$ shift as
\begin{equation}
    \Delta w = \langle w_{\textrm{match}} - w_{\textrm{truehost}} \rangle_{\mathrm{(25\;realizations)}}
\end{equation}
where $\langle \rangle$ denotes the inverse-variance weighted average. 
We calculate the associated uncertainty on $\Delta w$ as follows:
\begin{equation}
\sigma_{\Delta w} = \sqrt{\frac{\sum_i (w_{\textrm{match},i} - w_{\textrm{truehost}, i})^2}{25}}.
\end{equation}

\subsection{Cosmological Biases with SNe Ia Only}
\label{subsec:cosmo-Ia}

For simulations with SNe Ia only, we find $\Delta w = 0.0013 \pm 0.0026$. \textbf{In the flat $w_0 w_a$ CDM model, we find biases of $\Delta w_0=0.037 \pm 0.041, \Delta w_a = -0.22 \pm 0.25$. Biases in both cosmological models are consistent with zero.} In Figure~\ref{fig:Ia_delta-mu}, we show the biases on the binned Hubble diagram comparing distance moduli $\mu$ from simulations with DLR matched hosts ($\mu_{\text{match}}$) and true hosts ($\mu_{\text{truehost}}$). We define 
\begin{equation}
\label{eq:delta-mu}
    \Delta \mu = \mu_{\textrm{match}} - \mu_{\textrm{truehost}}.
\end{equation}
We see that the bias is consistent with 0 until $z \sim 1$, where the sample becomes very sparse (see Figure~\ref{fig:Ia_z_dist} for the redshift distribution of the sample).

Figure~\ref{fig:Ia_HD} shows the Hubble diagram and Hubble residuals for a single realization of simulations with mismatches. In this particular realization, 79 SNe were matched to the wrong host out of 5,811 total simulated SNe, which translates to a 1.4\% mismatch rate. The mismatched SNe, shown in red circles, show similar Hubble residuals compared to the correctly matched SNe, which is consistent with the small recovered bias on $w$. The observed similarity in Hubble residuals is likely due to the fact that catastrophic outliers in redshift are removed by the selection criteria described in Table~\ref{tbl:sim-cuts-Ia-cc} and shown in Figure~\ref{fig:Ia_delta_z}.

The binned Hubble residuals for all 25 simulations with mismatch are shown in Figure~\ref{fig:Ia_residuals}. This plot shows $\Delta \mu = \mu_{\mathrm{match}} - \mu_{\mathrm{model}}$, as opposed to Figure~\ref{fig:Ia_delta-mu}, which shows $\mu_{\mathrm{match}} - \mu_{\mathrm{truehost}}$. This allows us to compare residuals from the subpopulations of SNe with wrong and correct host match with respect to a fiducial cosmology. Aggregated over all 25 simulations, the bias from SNe with the wrong host match (shown in orange) is clearly distinct from the nearly unbiased subset of SNe with the correct host match (teal). Although wrong hosts clearly lead to biases on the Hubble diagram, SNe with the wrong host match make up $<2\%$ of the sample, resulting in the small $\Delta w$ value we observe.

\begin{figure}
    \includegraphics[scale=0.45,trim={0cm 0cm 0cm 0cm}]{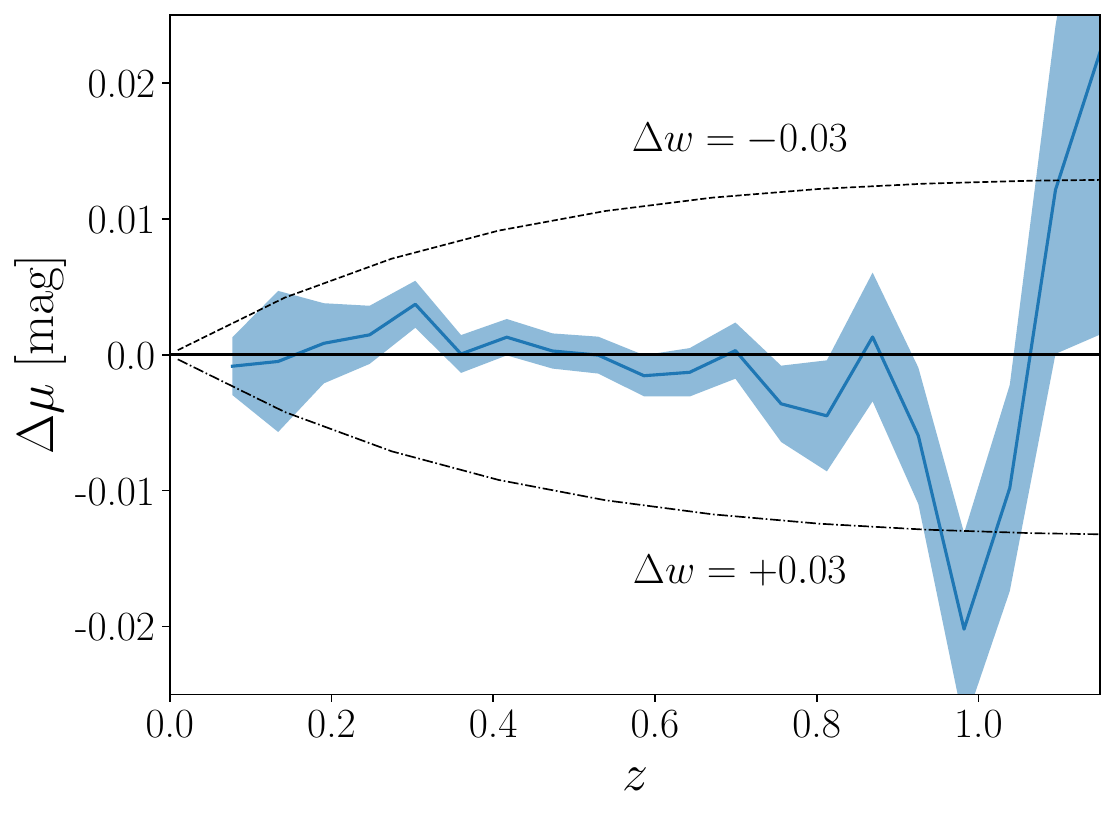}
    \centering
    \caption{Biases in binned Hubble residuals for the SN Ia-only sample between samples with and without mismatched host galaxies, $\Delta \mu = \mu_{\text{match}} - \mu_{\text{truehost}}$, as a function of redshift. Uncertainties are shown as the shaded region and calculated from the binned standard deviations of $\mu_{\text{match}}$ and $\mu_{\text{truehost}}$. Lines showing $\Delta w = \pm 0.03$ are also plotted for reference.}
    \label{fig:Ia_delta-mu}
\end{figure}

\begin{figure}
    \includegraphics[scale=0.37,trim={0cm 0cm 0cm 0cm}]{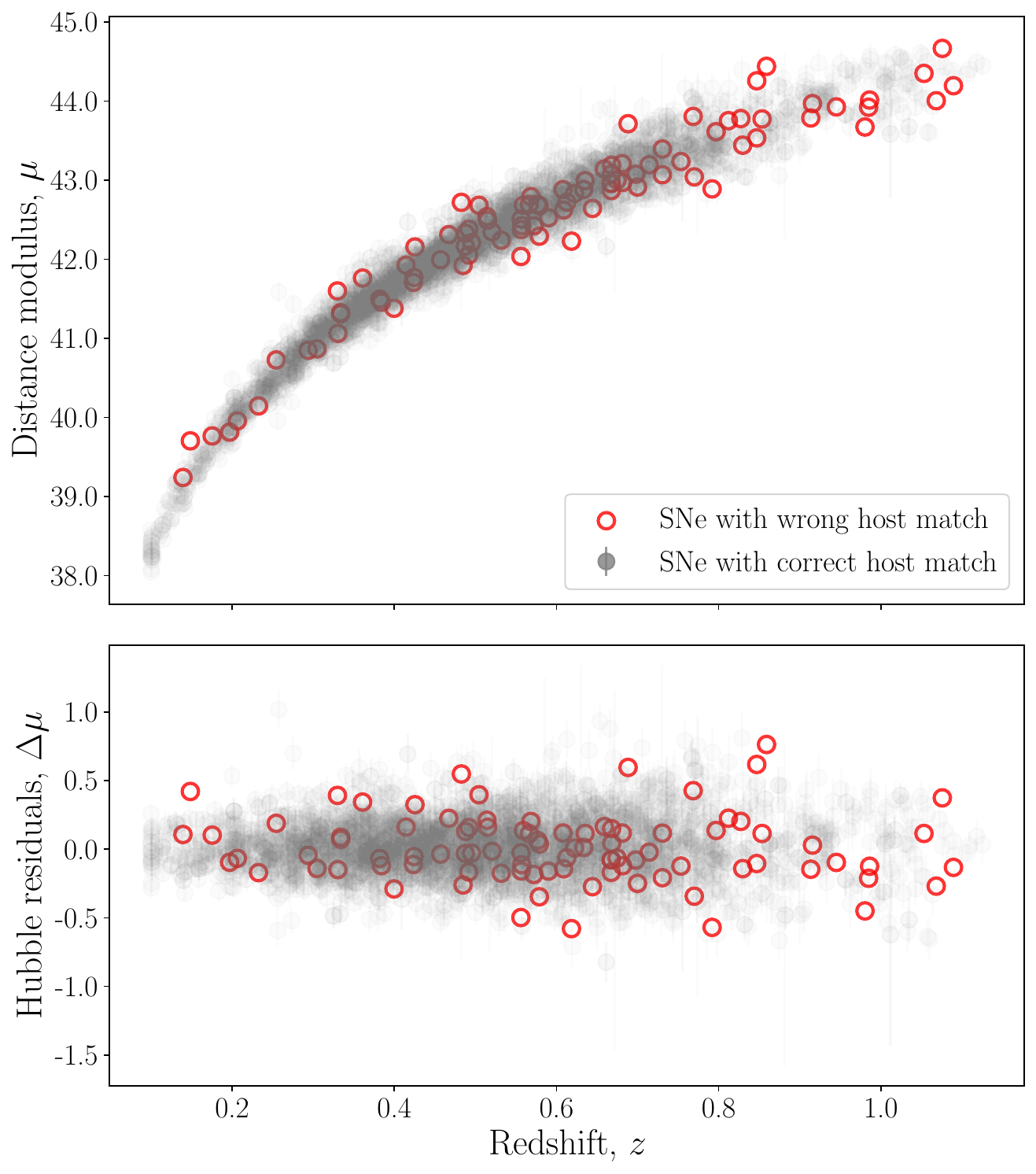}
    \centering
    \caption{Hubble diagram for a single realization of our SNIa-only simulation with matched hosts. This realization has 79 SNe matched to an incorrect host out of 5,811 total SNe, a 1.4\% mismatch rate. The low-$z$ sample is omitted from this plot, since it is not part of the main DES sample.}
    \label{fig:Ia_HD}
\end{figure}

\begin{figure}
    \includegraphics[scale=0.45,trim={0cm 0cm 0cm 0cm}]{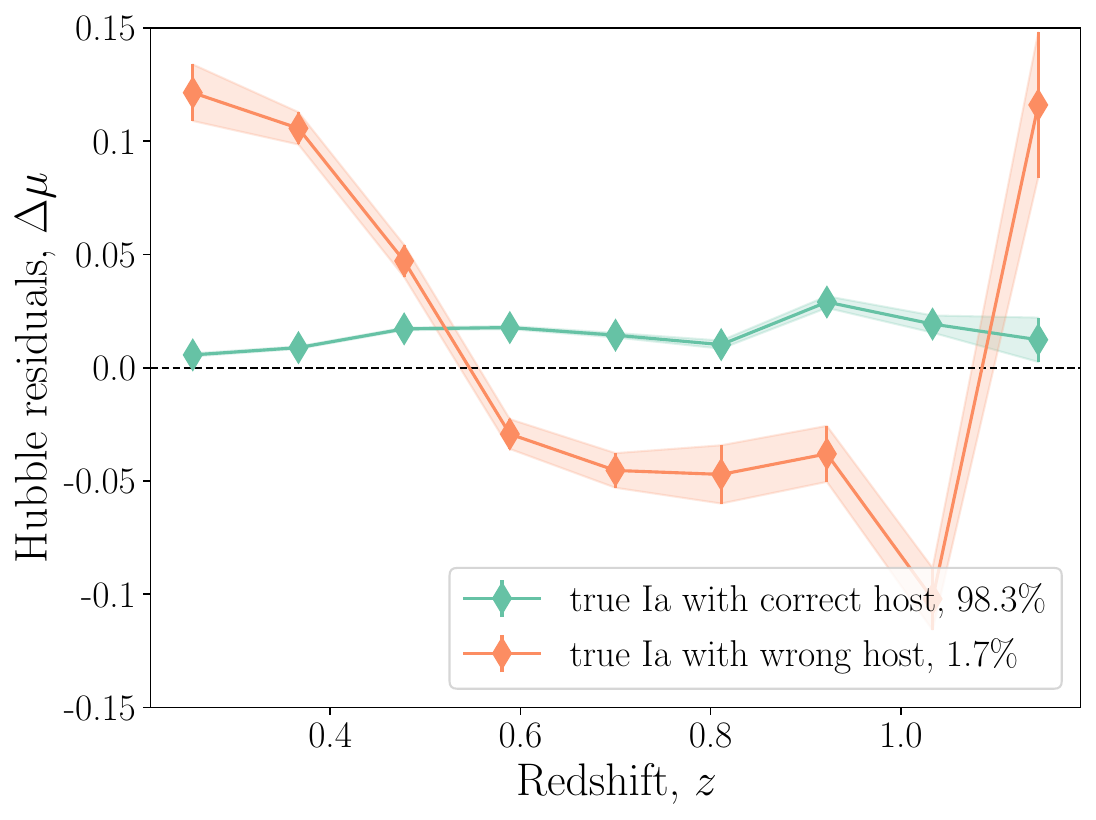}
    \centering
    \caption{Binned Hubble residuals ($\mu_{\text{match}}-\mu_{\text{model}}$) for all 25 realizations of our SNIa-only simulation with mismatch. The percentages in the legend show the fraction of the sample represented by each population in the plot.}
    \label{fig:Ia_residuals}
\end{figure}


\begin{table*}
    \centering
    \begin{tabular}{l l c c c c}
        \toprule
        Classifier & SNe & $\Delta w$ & \multicolumn{3}{c}{Ia vs. non-Ia Classification Accuracy} \\
        \cmidrule(lr){4-6} 
         & & & with host mismatch & no host mismatch & change\\
        \midrule
        Perfect & Ia only & $0.0013 \pm 0.0026$ & 100\% & 100\% & 0\% \\ 
        Perfect & Ia+CC & $0.0011 \pm 0.0027$ & 100\% & 100\% & 0\% \\ 
        \snnz & Ia+CC & $0.0032 \pm 0.0040$ & 97.94\% & 98.06\% & -0.1\% \\ 
        \snnnoz & Ia+CC & $0.0009 \pm 0.0028$ & 97.05\% & 97.05\% & 0\% \\
        SCONE & Ia+CC & $0.0016 \pm 0.0032$ & 96.13\% & 96.13\% & 0\%\\
        \bottomrule
    \end{tabular}
    \caption{$\Delta w$ and classification accuracies for each classifier. Accuracy change is only expected for the \snnz\ classifier, as it is the only classifier tested that requires redshift information. Accuracy change is defined as $\textrm{Accuracy (with mismatch)} - \textrm{Accuracy (no mismatch)}$.}
    \label{tbl:Ia-cc-results}
\end{table*}

\subsubsection{Host Mismatch and the Mass Step}
\label{subsec:mass-step}

The `mass step' is the observed correlation between SNe Ia intrinsic luminosity and host galaxy stellar mass, $M_{\star}$. Specifically, SNe Ia in more massive galaxies are more luminous after lightcurve corrections than their counterparts occurring in galaxies with lower stellar mass, with the average corrected luminosity distribution following a two-part step function with a break at log($M_{\star} / M_{\odot}) \sim 10$ \citep{kelly2010, mass_step, jla, smith2020, kelsey2021, kelsey2023}. Though the underlying astrophysical cause is unknown, recent cosmological analyses have incorporated a correction in which SN luminosities in hosts with log($M_{\star} / M_{\odot}$) $\geq 10$ and those in hosts with log($M_{\star} / M_{\odot}$) $< 10$ are fit for separately. When this two-part fit is employed, incorrect host matches will produce additional bias through incorrect host mass estimates, leading to different best fit values for SN luminosities. We observe that 34.3\% of our simulated SNe Ia with the wrong host match ``switch sides", i.e. the correct host is on one side of the mass step but the matched host is on the other. Given that the rate of mismatches averaged over 50 realizations is 1.7\% after selection cuts for our Ia-only sample, the overall fraction of SNe Ia that switch sides of the mass step is $\sim 0.6\%$. We assume that such a small percentage of the sample switching sides makes little impact and we do not pursue this aspect of the analysis further.

\subsection{Cosmological Biases with Photometric Classification}
\label{subsec:cosmo-Ia+CC}

Some photometric classifiers, such as SuperNNova (SNN), rely on SN redshift information to improve classification accuracy. To evaluate the impact of incorrect redshifts from host galaxy mismatches on the predictions from photometric classifiers, we jointly simulate SNe Ia, two types of peculiar SNe Ia, and two types of core collapse SNe: SNII, SNIbc, SNIax, and SNIa-91bg. Details on these simulations can be found in Section~\ref{subsec:sn-models} and Table~\ref{tbl:sim-cuts-Ia-cc}.

We tested 4 different photometric classifiers on our SNIa+CC simulations: the baseline perfect classification, SNN with redshift information (\snnz), SNN without redshift information (\snnnoz), and SCONE. SNN is typically used with SN redshift information and has been shown to produce highly accurate Ia vs. non-Ia classification results in this paradigm, so testing both redshift-dependent and redshift-independent configurations will show which effect is more detrimental to performance: incorrect redshift information or lack of redshift information altogether. SCONE uses SN lightcurves alone without the need for redshift information, so its predictions are not affected by host matching. For this analysis, we choose to define an SN Ia classification as $P_{\text{Ia}} \geq 0.5$.

Following the same approach as the Ia-only analysis, $w$ shifts were calculated for each photometric classifier by comparing two sets of identical simulations with and without mismatches. The results are shown in Table~\ref{tbl:Ia-cc-results}.


\begin{figure}
    \includegraphics[scale=0.38,trim={0cm 0cm 0cm 0cm}]{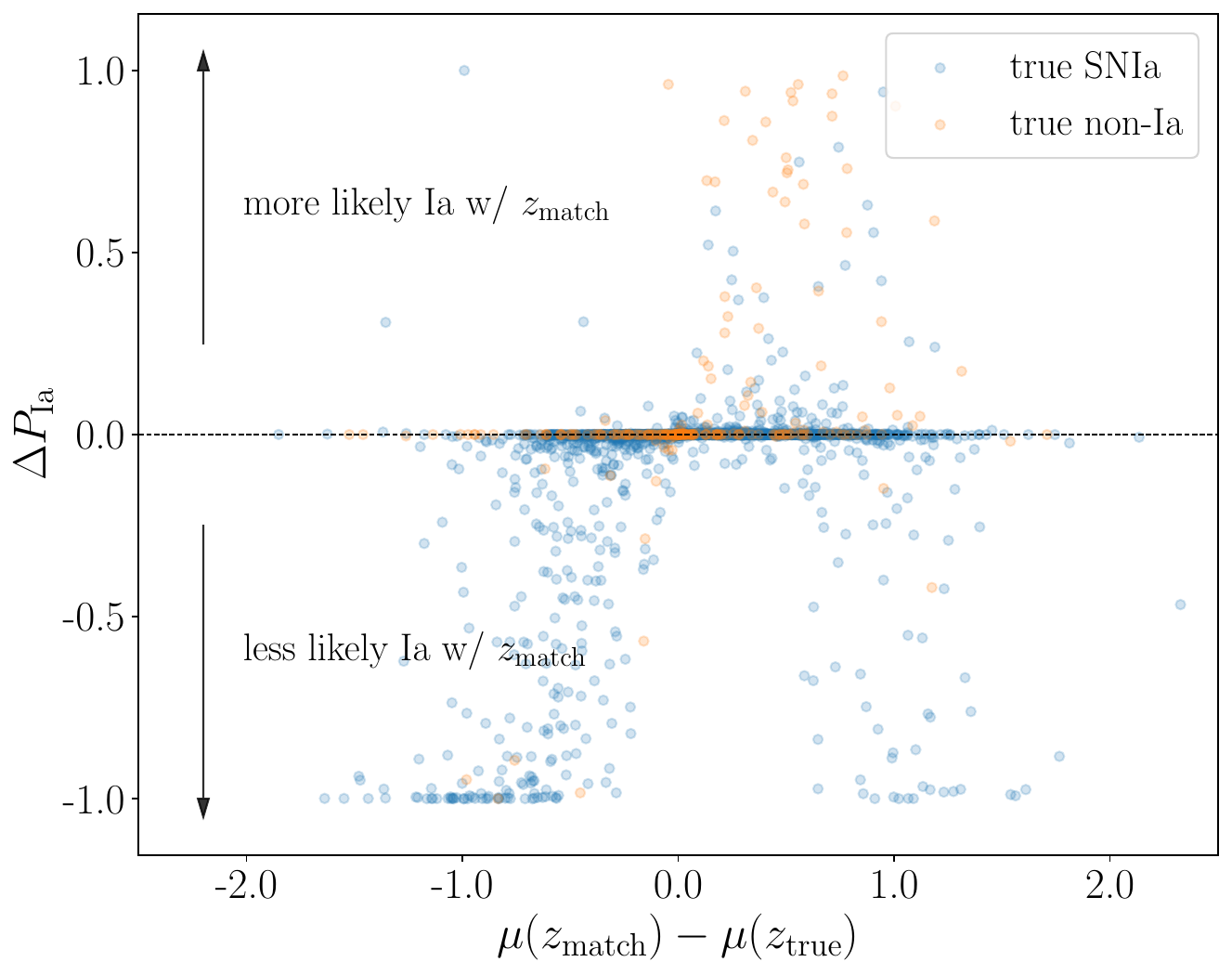}
    \centering
    \caption{Visualization of the impact of incorrect redshifts on \snnz\ predictions. $\Delta P_{\mathrm{Ia}}=P_{\mathrm{Ia, wronghost}} - P_{\mathrm{Ia, correcthost}}$ is the difference between $P_{\mathrm{Ia}}$ values output by the \snnz\ classifier given the wrong host redshift ($P_{\mathrm{Ia, wronghost}}$) and the correct host redshift ($P_{\mathrm{Ia, correcthost}}$) for SNe with mismatched hosts. $\Delta P_{\mathrm{Ia}}$ values are plotted against the difference between the distance modulus $\mu$ calculated at the wrong ($\mu(z_{\text{match}})$) and correct host redshifts ($\mu(z_{\text{true}})$). As $\mu(z_{\text{match}})-\mu(z_{\text{true}})$ deviate from 0, we would expect larger deviations in $P_{\mathrm{Ia}}$ values, i.e. $|\Delta P_{\mathrm{Ia}}| > 0$. Non-Ia SNe (orange points) more likely to be misclassified as Ia with the wrong redshift will appear in the upper half of the plot ($\Delta P_{\text{Ia}} > 0$), whereas SNe Ia (blue points) more likely to be misclassified as non-Ia with the wrong redshift will appear in the lower half. 7\% of mismatched SNe are incorrectly classified as a result of wrong host redshifts, leading to a overall 0.1\% reduction in classification accuracy compared to a simulation with correct host redshifts.}
    \label{fig:SNN_diffs}

\end{figure}

\begin{figure}
    \includegraphics[scale=0.38,trim={0cm 0cm 0cm 0cm}]{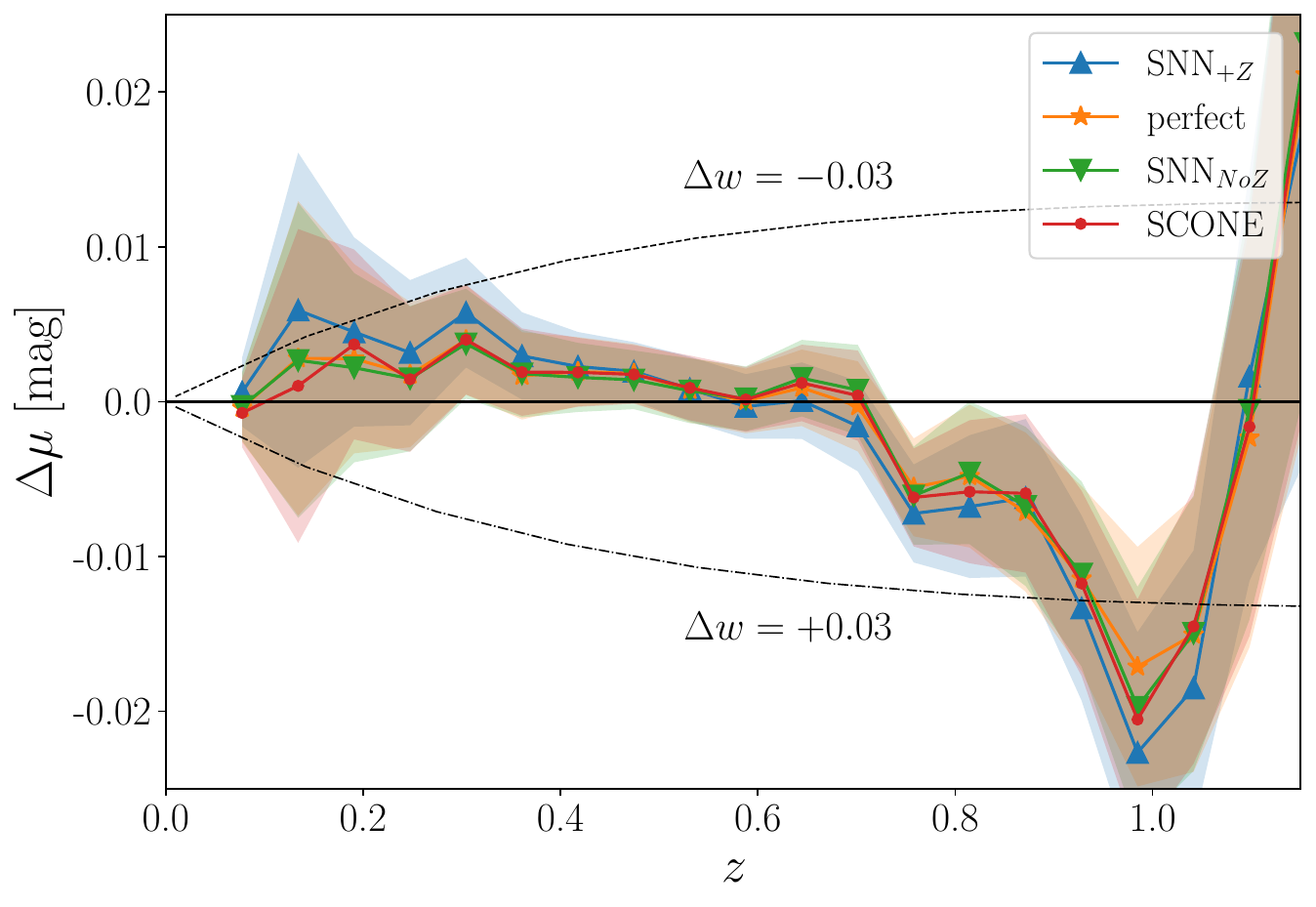}
    \centering
    \caption{Biases in binned Hubble residuals for all 4 photometric classifiers between samples with and without host galaxy mismatch. Uncertainties are shown as the shaded region and calculated from the binned standard deviations of $\mu_{\mathrm{match}}$ and $\mu_{\mathrm{truehost}}$. Lines showing $\Delta w = \pm 0.03$ are also plotted for reference.}
    \label{fig:Ia+CC_delta-mu}
\end{figure}

\begin{figure*}
    \centering
    \includegraphics[scale=0.255,trim={0cm 0cm 0cm 0cm}]{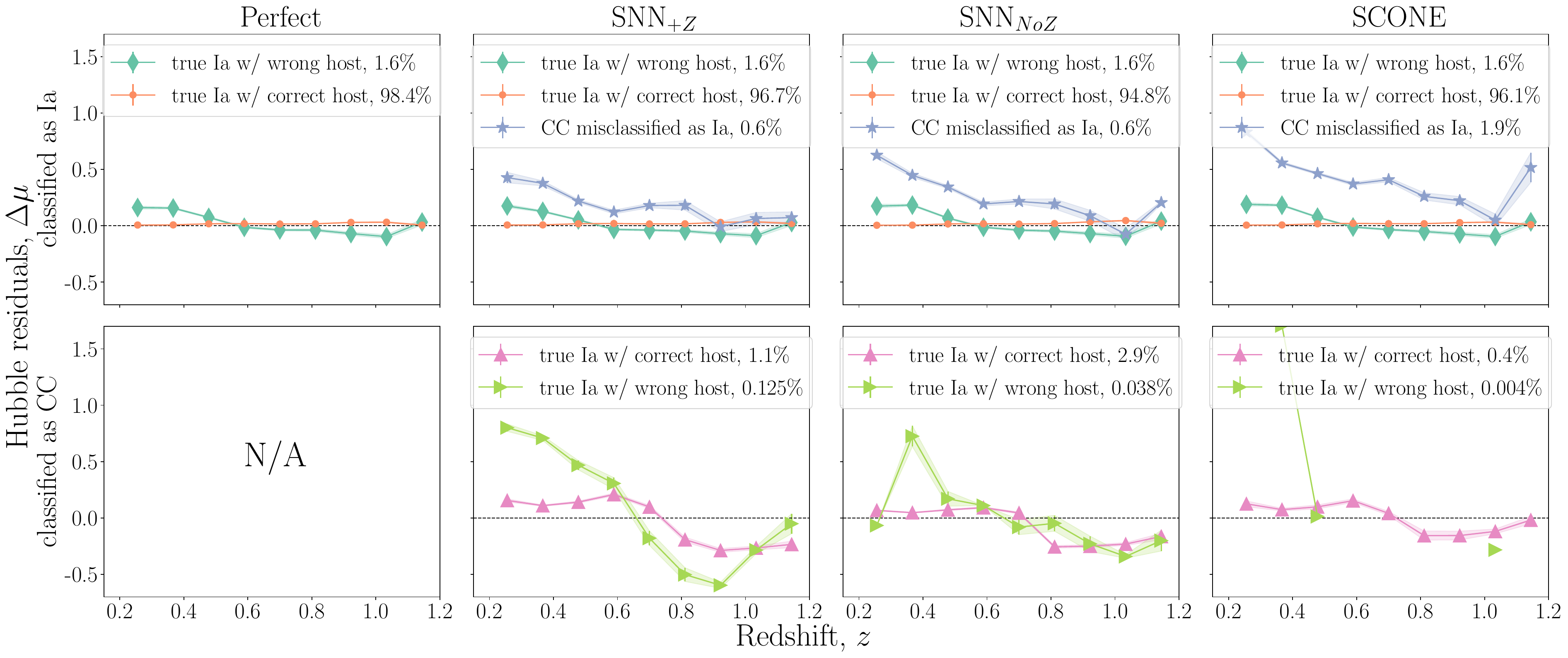}
    \caption{Binned Hubble residuals, $\Delta \mu = \mu_{\mathrm{match}} - \mu_{\mathrm{model}}$, for the 4 photometric classifiers for our SNIa+CC simulation with mismatch. (top) SN populations classified as SNe Ia by each photometric classifier, including true SNe Ia with mismatched hosts as well as core collapse contamination. (bottom) SN populations classified as CC SNe, including SNe with mismatched hosts.}
    \label{fig:Ia+CC_residuals}
\end{figure*}
\snnz\ is the only classifier that should be impacted by host galaxy mismatches, and this relationship is observed in the larger $\Delta w$ value and lower classification accuracy for simulations with mismatched host galaxies. Figure~\ref{fig:SNN_diffs} shows the difference between predicted SN Ia probability output by the \snnz\ classifier given the wrong host redshift ($P_{\mathrm{Ia,wronghost}}$) as opposed to the correct host redshift  ($P_{\mathrm{Ia,correcthost}}$)}. Wrong redshifts indeed cause SNN to produce incorrect predictions for both Ia and non-Ia SNe, leading to the observed drop in accuracy with wrong redshifts. SCONE and \snnnoz\ are oblivious to any host matching changes and produce the same predictions for both sets of simulations, as expected. 

Figure~\ref{fig:Ia+CC_delta-mu} shows the biases on the binned Hubble diagram for each of the photometric classifiers with $\Delta \mu$ defined as in Equation~\ref{eq:delta-mu}. Overall, the classifiers perform quite similarly and exhibit very small differences in $\Delta \mu$ over the full redshift range. As expected from the small $\Delta w$ values, the $\Delta \mu$ curves for the Ia+CC simulations exhibit a slight redshift-dependent bias, though still mostly consistent with 0 up to high redshifts. Further validating the observed $\Delta w$ values for each classifier, we see that the two classifiers with most similar $w$ shifts, the perfect classifier (shown in orange) and \snnnoz\ (green), have the most similar $\Delta \mu$ values. \snnz\ (shown in blue), which has the largest $w$ shift, also consistently appears furthest from $\Delta \mu=0$ across all redshift bins.

The binned Hubble residuals of SNe Ia in the Ia+CC simulations as predicted by the 4 photometric classifiers are shown in Figure~\ref{fig:Ia+CC_residuals}. This plot shows $\Delta \mu = \mu_{\mathrm{match}} - \mu_{\mathrm{model}}$, as opposed to Figure~\ref{fig:Ia+CC_delta-mu}, which shows $\mu_{\mathrm{match}} - \mu_{\mathrm{truehost}}$. This allows us to compare residuals from the subpopulations of SNe with wrong and correct host match with respect to a fiducial cosmology. 
The biases on Hubble residuals from SNe Ia with wrong hosts (shown in teal diamonds on the top row) appear similar between the four classifiers, reflecting the small recovered $\Delta w$ values shown in Table~\ref{tbl:Ia-cc-results}. We also observe that the bias from wrong hosts is much more pronounced in SNe Ia misclassified as CC (shown in green triangles on the bottom row), particularly in the \snnz\ panel, indicating that \snnz\ was able to identify severe redshift outliers and rejected them from the SN Ia sample.

\subsection{Robustness of Cosmological Biases}
\subsubsection{Impact of CMB Prior}
The $w - \Omega_m$ contour estimated from measurements of the CMB exhibits a nearly orthogonal direction of degeneracy to the SN-only contour for a flat $w\text{CDM}$ model, providing strong constraints and drastically reducing the impact of systematics that act along the SN degeneracy direction. All $\Delta w$ values reported in Sections~\ref{subsec:cosmo-Ia} and~\ref{subsec:cosmo-Ia+CC} were calculated with a CMB prior. In this section, we evaluate the impact of the CMB prior on cosmological biases.

\begin{figure}
     \centering
     \includegraphics[scale=0.6,trim={1cm 0cm 0cm 0cm}]{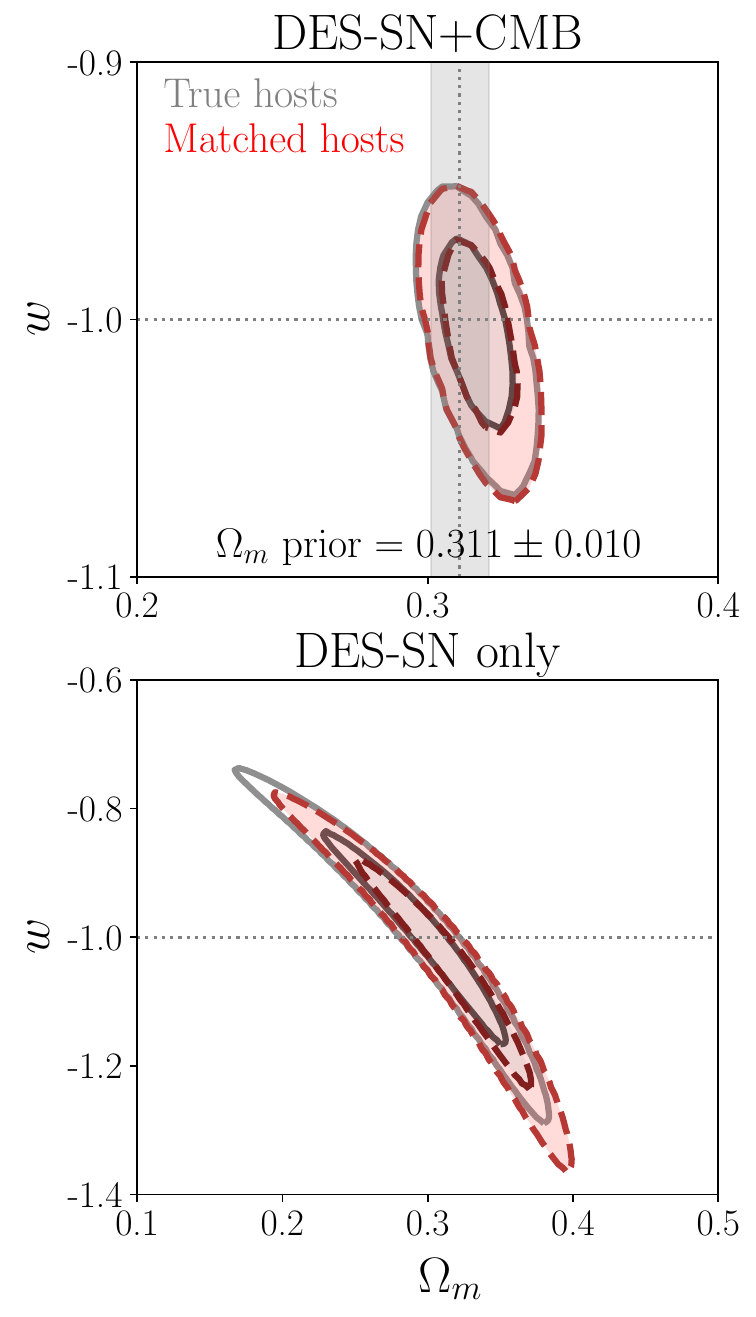}
    \caption{Cosmological contours computed with one realization of simulated DES Y5 SNe Ia with (red) and without host galaxy mismatch (gray). (top) SN+CMB contours, in which an approximate CMB prior $\Omega_m=0.311\pm 0.010$ prior is applied (see Section~\ref{subsec:cosmo-params} for details). This results in good agreement between the two contours. (bottom) SN-only contours showing a substantial shift in best fit $w, \Omega_m$ between the two contours. }
     \label{fig:contour}
     
 \end{figure}

The cosmological biases on both $w$ and $\Omega_m$ with and without the CMB prior are shown in Table~\ref{tbl:shifts-cmb} for the SNIa-only sample. The associated cosmological contours are shown in Figure~\ref{fig:contour}. Both $\Delta w$ and $\Delta \Omega_m$ are significantly inflated without the CMB prior, though still within their uncertainties (right column). The larger shift in $ w$ and $\Omega_m$ are visible when comparing the contours in Figure~\ref{fig:contour}. The contours computed with a CMB prior (top panel) are very nearly identical, whereas with a flat $\Omega_m$ prior, the contour with matched hosts is visibly shifted from the true hosts contour. However, these results are still consistent with 0 for the analysis we performed for the DES data, but should be studied further in future surveys.

\begin{table}
    \centering
    
    \begin{tabular}{c c c }
        \toprule
       Parameter & with CMB prior & no CMB prior\\
        \midrule
        $\Delta w$ & $0.0013 \pm 0.0026$ & $-0.062 \pm 0.072$ \\ 
        $\Delta \Omega_m$ & $0.0014 \pm 0.0017$ & $0.028 \pm 0.029$ \\
        \bottomrule
    \end{tabular}
    \caption{$\Delta w$ and $\Delta \Omega_m$ values for the Ia-only SN population with and without a CMB prior. The values in the $\Delta w$ with CMB prior cell are reproduced from Table~\ref{tbl:Ia-cc-results}.}
    \label{tbl:shifts-cmb}
\end{table}

    
\subsubsection{Bias Correction Simulations with Mismatch}
\label{subsec:bcor-mismatch}
In the earlier sections, distance moduli from both $S_{\mathrm{match}}$ and $S_{\mathrm{truehost}}$ are bias corrected using a large simulation of SNe Ia generated with the same parameters as $S_{\mathrm{truehost}}$, i.e. each SN is matched to its true host. We define this set of perfectly matched bias correction simulations as $B_{\mathrm{truehost}}$. An alternative bias correction strategy is to ``correct like with like", i.e. generating two separate bias correction simulations, one identical to $S_{\mathrm{match}}$ as well as the existing one identical to $S_{\mathrm{truehost}}$. We define this new set of DLR-matched bias correction simulations as $B_{\mathrm{match}}$. $B_{\mathrm{match}}$ models the biases arising from mismatched hosts in the bias correction simulations in order to correct for these biases in the simulated data. The reference $w$ value, $w_{\mathrm{truehost}}$ is still computed from $S_{\mathrm{truehost}}$ corrected with the baseline bias correction simulations with correct hosts only ($B_{\mathrm{truehost}}$). In this scheme, we define $w(S,B)$ as the best-fit $w$ value computed from simulations $S$ corrected with bias correction simulations $B$. The resulting $\Delta w$ equation then becomes
\begin{multline}
    \Delta w, \mathrm{bcor+mismatch}= \\\langle w(S_{\mathrm{match}}, B_{\mathrm{match}}) - w(S_{\mathrm{truehost}}, B_{\mathrm{truehost}}) \rangle_{\mathrm{(25\;realizations)}}.
\end{multline}

The $\Delta w$ values following this approach are shown in Table~\ref{tbl:w-shifts-bcor}. While the $\Delta w$ values using this bias correction scheme are still consistent with zero, the uncertainties are larger than those using bias correction simulations with perfect host matching; this may arise from comparing $w$ values corrected with two statistically independent sets of bias correction simulations, i.e. $S_\mathrm{match}$ is corrected with $B_\mathrm{match}$, whereas $S_{\mathrm{truehost}}$ is corrected with $B_\mathrm{truehost}$. Further investigation of the interplay of bias correction simulations with different sources of bias, including incorrect redshifts from host mismatch, will be addressed in a future work.

\begin{table}
    \centering
    \caption{$\Delta w$ values for the Ia-only and Ia+CC simulated SN populations with an alternative bias correction scheme (``bcor+mismatch") that includes mismatched hosts in the bias correction simulations. The values in the $\Delta w$, baseline column are reproduced from Table~\ref{tbl:Ia-cc-results}.}
    \label{tbl:w-shifts-bcor}
    \begin{tabular}{l c c c}
        \toprule
        Classifier & SNe  & $\Delta w$, baseline & $\Delta w$, bcor+mismatch \\
        \midrule
        Perfect & Ia only & $0.0013 \pm 0.0026$ & $-0.0094 \pm 0.0099$\\ 
        Perfect & Ia+CC & $0.0011 \pm 0.0027$ & $-0.0110 \pm 0.0120$ \\ 
        \snnz & Ia+CC & $0.0032 \pm 0.0040$ & $-0.0081 \pm 0.0086$ \\ 
        \snnnoz & Ia+CC & $0.0009 \pm 0.0028$ & $-0.0100 \pm 0.0012$ \\
        SCONE & Ia+CC & $0.0016 \pm 0.0032$ & $-0.0099 \pm 0.0100$ \\
        \bottomrule
    \end{tabular}
\end{table}


\subsubsection{Results from Varying Sérsic Scale}
\label{subsec:sersic-scale}
We vary the scaling of the fitted Sérsic $a$ and $b$ parameters, which describe each galaxy's semi-major and semi-minor axes, respectively. Simulated SNe are placed according to the intrinsic light profile described by these parameters. The scaled parameters $a'$ and $b'$ are calculated as $a'=ka, b'=kb$, where $k$ is the scaling parameter we vary. We tested $k \in [0.5, 1.2]$ with an interval of 0.1 and evaluated $\chi^2$ values on the histograms in Figure~\ref{fig:param-dists} to find the best match between data and simulations. We found that $k=0.8$ minimized the $\chi^2$ and was thus chosen for the main analysis, as described in Section~\ref{subsec:catalog-cuts-params}. We note that our conclusion runs contrary to that of \citet{Li_2016}, which found that the Sérsic effective radius, $R_e = \sqrt{ab}$, is underestimated for stacked galaxy images rather than overestimated, as we discovered.

We performed cosmological parameter estimation using simulations with and without Sérsic scaling ($k=0.8$ and $k=1$) and found a modest benefit of using the $k=0.8$ scaling for redshift-independent photometric classifiers as well as perfectly classified SNe Ia, but a noticeable improvement for SNe classified using the redshift-dependent classifier, \snnz. This seems to indicate that the Sérsic scaling improves the host matching efficiency significantly, since \snnz\ is most affected by incorrect redshifts. This is notably the same scaling factor found to best match the DES3YR data when comparing distributions of host galaxy surface brightness at the SN position \citep[see Figure 6 of][]{des3yr}. The recovered biases on $w$ from mismatched hosts ($\Delta w$) with and without Sérsic scaling for all simulations and classifiers are shown in Table~\ref{tbl:w-shifts-sersic}.

\begin{table}
    \centering
    \caption{$\Delta w$ values for the Ia-only and Ia+CC simulated SN populations with and without Sérsic scaling ($k=0.8$ and $k=1$). The primary results presented in this work (Sections~\ref{subsec:cosmo-Ia} and~\ref{subsec:cosmo-Ia+CC}) use simulations scaled with $k=0.8$. The \snnz\ results with $k=1$ have an inflated $\Delta w$ as well as $\sigma_{\Delta w}$ due to a realization with poor $\chi^2$ fit from \texttt{wfit}.}
    \label{tbl:w-shifts-sersic}
    \begin{tabular}{l c c c}
        \toprule
        Classifier & SNe & $\Delta w(k=0.8)$ & $\Delta w(k=1)$ \\
        \midrule
        Perfect & Ia only & $0.0013 \pm 0.0026$ & $0.0030 \pm 0.0042$\\ 
        Perfect & Ia+CC & $0.0011 \pm 0.0027$ & $0.0016 \pm 0.0026$ \\ 
        \snnz & Ia+CC & $0.0032 \pm 0.0040$ & $0.0120 \pm 0.0640$ \\ 
        \snnnoz & Ia+CC & $0.0009 \pm 0.0028$ & $0.0021 \pm 0.0031$ \\
        SCONE & Ia+CC & $0.0016 \pm 0.0032$ & $0.0025 \pm 0.0034$ \\
        \bottomrule
    \end{tabular}
\end{table}

\subsubsection{Host Confusion Parameter}
\begin{figure}
    \includegraphics[scale=0.36]{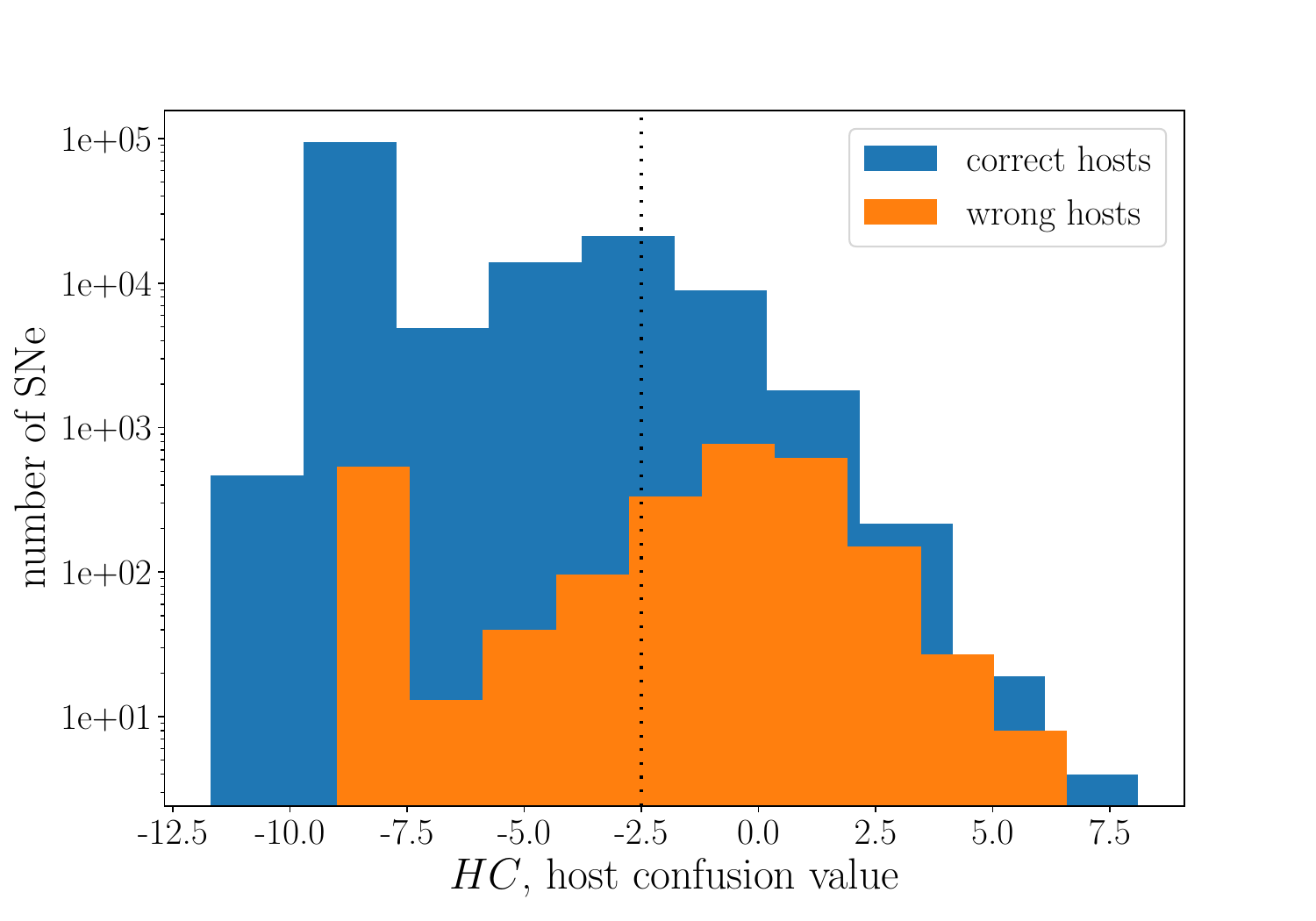}
    \centering
    \caption{Distributions of $HC$ values for SNe with wrong and correct host matches. The dotted line is drawn at our chosen threshold of $HC = -2.5$.}
    \label{fig:host-confusion-dists}
\end{figure}

Equation 3 in \cite{gupta} defines a quantity characterizing the likelihood of a wrong match: the host confusion parameter, or $HC$. This parameter is a function of the \ddlr values of each galaxy in the search radius around a SN, and aims to distinguish situations with a clear correct host from those without. In this analysis, we attempt to remove wrong hosts by cutting out SNe with high host confusion ($HC \geq -2.5$). The distribution of $HC$ values for wrong and correct hosts in our simulations is shown in Figure~\ref{fig:host-confusion-dists}. The $-2.5$ threshold was chosen by visual inspection of this distribution. The $HC$ distributions for both populations look quite similar, but 72\% of SNe with an incorrect host match are removed by the $HC \geq -2.5$ cut, while 12\% of SNe with correct host matches are removed. 

We performed a full cosmology analysis with perfectly classified Ia-only simulations and find $\Delta w = -0.0043 \pm 0.0047$ by comparing Ia populations with and without wrong hosts, both subject to the $HC < -2.5$ selection requirement. In this cosmology analysis, we use bias correction simulations with host matching (i.e. $B_{\mathrm{match}}$ from Section~\ref{subsec:bcor-mismatch}) and select only the subset of the bias correction simulations with the same $HC < -2.5$ requirement for consistency. Both simulations (with perfectly matched hosts and with wrong hosts) are bias corrected with $B_{\mathrm{match}}$ in order to apply the $HC$ selection requirement globally, but it may not be suitable to correct simulations without wrong hosts in this way. The magnitude of this $\Delta w$ value is larger than that of the baseline (i.e. no $HC$ selection requirement), likely due to our inclusion of wrong hosts in the bias correction simulations, but is still consistent with zero.

\begin{figure}
    \includegraphics[scale=0.55]{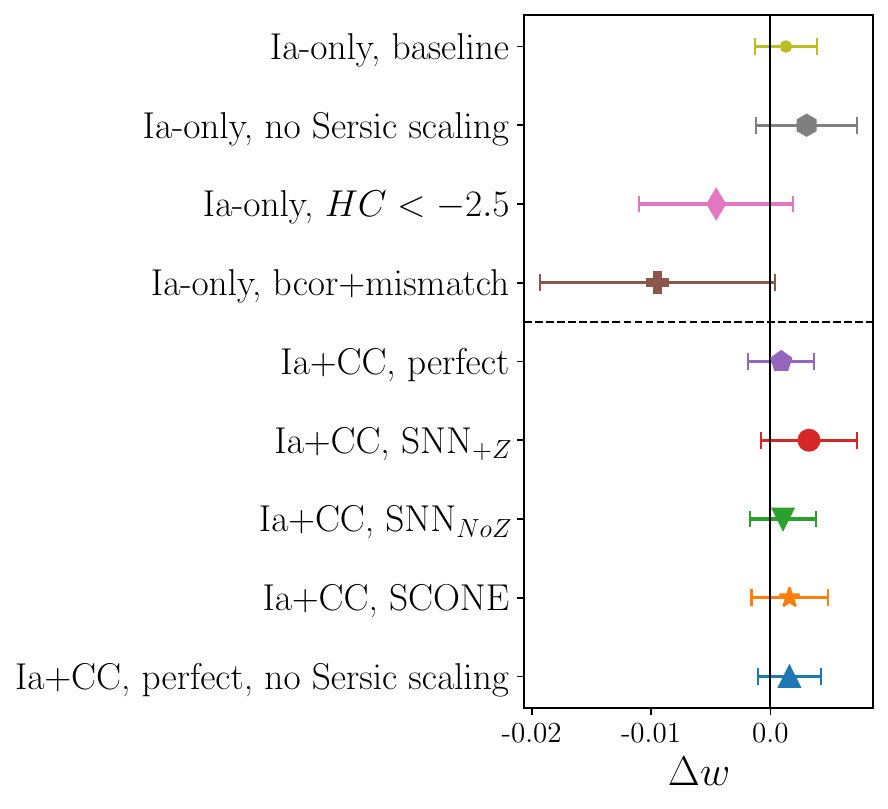}
    \centering
    \caption{$\Delta w$ values from all tested variants of simulated DES SNe with CMB prior. Note that the total uncertainty on $w$ from the DES3YR analysis \citep{des3yr_cosmo}, $\delta w = 0.059$, is much larger than the bounds of this plot. Details about each variant can be found in Section~\ref{sec:hostmatch-results}.}
    \label{fig:w_shifts}
\end{figure}

\section{Conclusions}

Matching SNe to their host galaxies is a non-trivial problem when using 2-dimensional images of 3-dimensional space. Accurate host galaxy matches are important for cosmology because host galaxies are used to measure the vast majority of SN redshifts. Identifying incorrect host galaxies alters the shape of the Hubble diagram and can impact the resulting fitted cosmological parameters.

In this work, we investigated the impact of mismatched host galaxies for the DES Y5 SN cosmology analysis. To this end, we created a galaxy catalog from DES deep field images that was sufficiently deep and dense, and calculated photometric redshifts, galaxy parameters, as well as Sérsic parameter fits for these 4 million galaxies. Simulations using this galaxy catalog were verified to be sufficiently similar to the DES Y5 photometric SN Ia sample using distributions such as Figure~\ref{fig:param-dists}. This simulation enabled us to predict the prevalence of mismatched host galaxies in DES data and, in turn, characterize the effect of host mismatches on cosmological parameter estimates.

Host matching was performed using the directional light radius (DLR) method, a  technique that identifies the host galaxy by minimizing SN-galaxy distance normalized by galaxy radius. This analysis aims to characterize the systematic error due to host misidentification by the DLR method for the DES Y5 cosmology analysis.

The main findings of this work are summarized in Figure~\ref{fig:w_shifts}, which shows the observed shifts in the dark energy equation of state parameter $w$ as a result of host galaxy mismatches with different approaches to the analysis. We defined $\Delta w$ by producing two identical sets of simulations: one with perfect host matching and one with host matches from the DLR method. We also probed the interplay between host galaxy mismatches and photometric classification, as classifiers such as SuperNNova use SN redshift estimates for more accurate SN type predictions. Finally, we explored the impact of variations to our analysis, such as the choice of $\Omega_m$ prior, bias correction simulations, S{\'e}rsic scaling factor, and additional selection cuts to remove incorrect host matches.

We found that the baseline $w$ shift with perfectly classified type Ia SNe is $\Delta w = 0.0013 \pm 0.0026$, and changes in certain properties can increase this to $| \Delta w | \sim 0.004$. We also find that the choice of photometric classifier makes an impact on the $w$ shift: classifiers requiring redshift estimates for prediction tend to misclassify SNe with the wrong redshift, leading to a larger $w$-bias. When the CMB prior is replaced with a flat $\Omega_m$ prior, the $\Delta w$ value changes to $-0.062 \pm 0.072$. Though the $\Delta w$ uncertainty is larger with the flat $\Omega_m$ prior, the shift is still consistent with 0 given the associated inflation in uncertainty. 

In conclusion, we find that our current estimate on the systematic error associated with host galaxy mismatch is substantially smaller than the statistical error for DES Y5, but as future surveys continue to discover more SNe Ia, we encourage continued study and improvement of the accuracy of host galaxy matching and prevalence of catastrophic errors in redshift.

\section*{Acknowledgements}
H.Q., J.L., and M.S. were supported by DOE grant DE-FOA-0002424 and NSF grant AST-2108094.
L.G. acknowledges financial support from the Spanish Ministerio de Ciencia e Innovaci\'on (MCIN), the Agencia Estatal de Investigaci\'on (AEI) 10.13039/501100011033, and the European Social Fund (ESF) ``Investing in your future" under the 2019 Ram\'on y Cajal program RYC2019-027683-I and the PID2020-115253GA-I00 HOSTFLOWS project, from Centro Superior de Investigaciones Cient\'ificas (CSIC) under the PIE project 20215AT016, and the program Unidad de Excelencia Mar\'ia de Maeztu CEX2020-001058-M.
PW acknowledges support from the Science and Technology Facilities Council (STFC) grant ST/R000506/1.
L.K. thanks the UKRI Future Leaders Fellowship for support through the grant MR/T01881X/1. 

This work was completed in part with resources provided by the University of Chicago’s Research Computing Center, as well as resources of the National Energy Research Scientific Computing
Center (NERSC), a DOE Office of Science User Facility
supported by the Office of Science of the U.S. Department of
Energy under Contract No. DE-AC02-05CH11231. 

Funding for the DES Projects has been provided by the U.S. Department of Energy, the U.S. National Science Foundation, the Ministry of Science and Education of Spain, 
the Science and Technology Facilities Council of the United Kingdom, the Higher Education Funding Council for England, the National Center for Supercomputing 
Applications at the University of Illinois at Urbana-Champaign, the Kavli Institute of Cosmological Physics at the University of Chicago, 
the Center for Cosmology and Astro-Particle Physics at the Ohio State University,
the Mitchell Institute for Fundamental Physics and Astronomy at Texas A\&M University, Financiadora de Estudos e Projetos, 
Funda{\c c}{\~a}o Carlos Chagas Filho de Amparo {\`a} Pesquisa do Estado do Rio de Janeiro, Conselho Nacional de Desenvolvimento Cient{\'i}fico e Tecnol{\'o}gico and 
the Minist{\'e}rio da Ci{\^e}ncia, Tecnologia e Inova{\c c}{\~a}o, the Deutsche Forschungsgemeinschaft and the Collaborating Institutions in the Dark Energy Survey. 

The Collaborating Institutions are Argonne National Laboratory, the University of California at Santa Cruz, the University of Cambridge, Centro de Investigaciones Energ{\'e}ticas, 
Medioambientales y Tecnol{\'o}gicas-Madrid, the University of Chicago, University College London, the DES-Brazil Consortium, the University of Edinburgh, 
the Eidgen{\"o}ssische Technische Hochschule (ETH) Z{\"u}rich, 
Fermi National Accelerator Laboratory, the University of Illinois at Urbana-Champaign, the Institut de Ci{\`e}ncies de l'Espai (IEEC/CSIC), 
the Institut de F{\'i}sica d'Altes Energies, Lawrence Berkeley National Laboratory, the Ludwig-Maximilians Universit{\"a}t M{\"u}nchen and the associated Excellence Cluster Universe, 
the University of Michigan, NSF's NOIRLab, the University of Nottingham, The Ohio State University, the University of Pennsylvania, the University of Portsmouth, 
SLAC National Accelerator Laboratory, Stanford University, the University of Sussex, Texas A\&M University, and the OzDES Membership Consortium.

Based in part on observations at Cerro Tololo Inter-American Observatory at NSF's NOIRLab (NOIRLab Prop. ID 2012B-0001; PI: J. Frieman), which is managed by the Association of Universities for Research in Astronomy (AURA) under a cooperative agreement with the National Science Foundation.

The DES data management system is supported by the National Science Foundation under Grant Numbers AST-1138766 and AST-1536171.
The DES participants from Spanish institutions are partially supported by MICINN under grants ESP2017-89838, PGC2018-094773, PGC2018-102021, SEV-2016-0588, SEV-2016-0597, and MDM-2015-0509, some of which include ERDF funds from the European Union. IFAE is partially funded by the CERCA program of the Generalitat de Catalunya.
Research leading to these results has received funding from the European Research
Council under the European Union's Seventh Framework Program (FP7/2007-2013) including ERC grant agreements 240672, 291329, and 306478.
We  acknowledge support from the Brazilian Instituto Nacional de Ci\^encia
e Tecnologia (INCT) do e-Universo (CNPq grant 465376/2014-2).

This manuscript has been authored by Fermi Research Alliance, LLC under Contract No. DE-AC02-07CH11359 with the U.S. Department of Energy, Office of Science, Office of High Energy Physics.

%% file: chapters/photoz.tex
\newcommand{\name}{Photo-$z$SNthesis}

\section*{Abstract}
Upcoming photometric surveys will discover tens of thousands of Type Ia supernovae (SNe Ia), vastly outpacing the capacity of our spectroscopic resources. In order to maximize the science return of these observations in the absence of spectroscopic information, we must accurately extract key parameters, such as SN redshifts, with photometric information alone. We present \name, a convolutional neural network-based method for predicting full redshift probability distributions from multi-band supernova lightcurves, tested on both simulated Sloan Digital Sky Survey (SDSS) and Vera C. Rubin Legacy Survey of Space and Time (LSST) data as well as observed SDSS SNe. We show major improvements over predictions from existing methods on both simulations and real observations as well as minimal redshift-dependent bias, which is a challenge due to selection effects, e.g. Malmquist bias. Specifically, we show a $61\times$ improvement in prediction bias $\langle \Delta z \rangle$ on PLAsTiCC simulations and $5\times$ improvement on real SDSS data compared to results from a widely used photometric redshift estimator, LCFIT+Z. The PDFs produced by this method are well-constrained and will maximize the cosmological constraining power of photometric SNe Ia samples. 

\section{Introduction}
\input{chapters/photoz/intro}

\section{Data}
\input{chapters/photoz/data}

\section{Model}
\input{chapters/photoz/model}
 
\section{Results and Discussion}
\input{chapters/photoz/results}
\section{Conclusions}
\input{chapters/photoz/conclusion}

\section*{Acknowledgments}
H.Q. and M.S. were supported by DOE grant DE-FOA-0002424 and NSF grant AST-2108094. The authors would like to thank Sang Michael Xie for insightful discussions on the network architecture and training process, Brodie Popovic for providing SDSS simulations, Rebecca Chen for helpful discussions and LCFIT+Z setup, Rick Kessler for assisting with LCFIT+Z, SNANA, and LSST simulations, and Carles S\'{a}nchez and Jaemyoung Lee for helpful discussions.

This research used resources of the National Energy Research Scientific Computing Center (NERSC), a U.S. Department of Energy Office of Science User Facility located at Lawrence Berkeley National Laboratory, operated under Contract No. DE-AC02-05CH11231. This work was completed in part with resources provided by the University of Chicago’s Research Computing Center.
\section*{Appendix A: Survey-Agnostic Model Performance}
\input{chapters/photoz/appendix}

%% file: chapters/photoz/intro.tex
The study of Type Ia supernovae (SNe Ia) has proven to be a crucial tool in modern cosmology, providing insight into the expansion rate of the universe and the properties of dark energy \citep{riess, perlmutter}. Measuring the cosmological redshift of each SN, a proxy quantity for recessional velocity, is essential for accurate estimation of the distance-redshift relation and resulting cosmological analyses. However, traditional methods of measuring redshifts are time-consuming and resource-intensive, primarily relying on spectroscopic observations of the SNe themselves or their host galaxies. Using host galaxy redshifts can also lead to cosmological biases if the host is incorrectly identified (Qu et al., in prep).
\begin{figure*}
    \centering
    \includegraphics[scale=0.45]{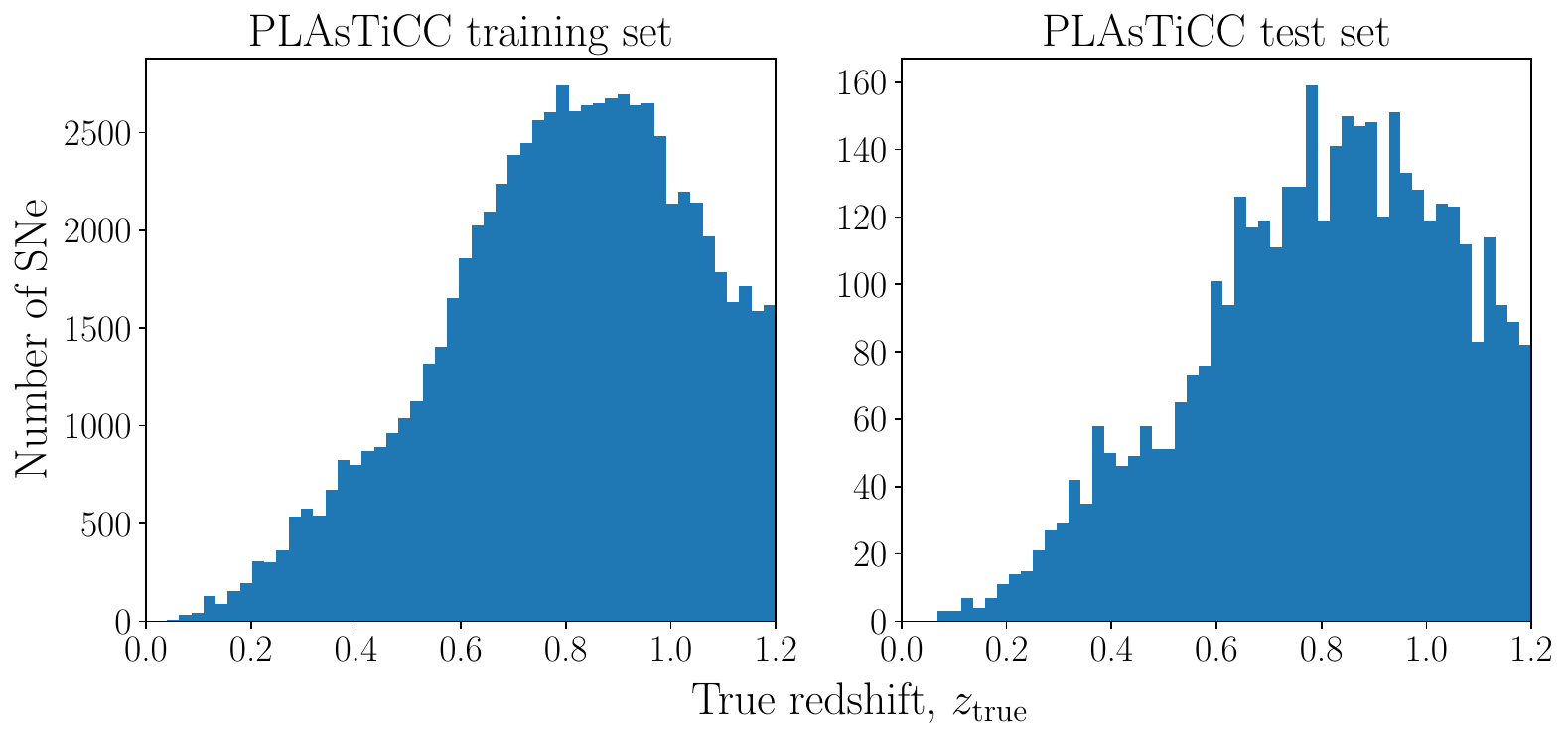}
    \caption{True redshift distribution of the 73,620 simulated SNe Ia in the LSST (PLAsTiCC) training dataset (left) and 4,057 simulated SNe Ia in the test dataset (right).}
    \label{fig:plasticc-z-dist}
\end{figure*}

Most SNe Ia cosmological analyses so far have relied on SN spectra for SN type confirmation as well as redshift information, but only with samples of up to $\sim 1,500$ SNe \citep[e.g.][]{Abbott_2019, pantheonplus}. Spectroscopic follow-up of all SNe Ia candidates or their host galaxies will become infeasible with future sky surveys such as the Legacy Survey of Space and Time at the Vera C. Rubin Observatory (LSST), which will discover tens of thousands of SNe Ia over the course of their observational lifetimes \citep{lsst_book}. Recent improvements in photometric SN classification \citep[e.g.][]{supernnova, scone} have drastically reduced the likelihood of non-Ia contamination in photometric SNe Ia samples and enabled cosmological analyses with photometrically classified samples \citep{vincenzi_contamination}. However, spectroscopic redshifts were still available for the host galaxies of these photometrically confirmed SNe Ia and were used as the SN redshifts. SN redshift estimates independent of host galaxy redshift can also serve as an independent cross-check for host galaxy association, ensuring accurate studies of host galaxy correlations and corrections for the mass step \citep[e.g.][]{Rigault_2020}. Accurate photometric redshift estimates for SNe that are independent of host galaxy spectroscopic redshifts are thus the final building block required to enable SN Ia cosmology for the LSST era.



Cosmological inference frameworks that account for the inflated uncertainties from photometric redshifts are currently being developed. \citet{mitra2022, Dai_2018} show promising results with simulated LSST samples and SN photometric redshifts fitted using host galaxy redshift priors. In particular, \citet{Dai_2018} recovers a fitted $\Omega_m$ value consistent with the input cosmology when using SN photo-$z$s fitted with a host galaxy photo-$z$ prior. \citet{mitra2022} shows a 2\% effect on fitted cosmological parameters of an assumed systematic uncertainty due to the use of SN photo-$z$s of 0.01. Using observed data from the Dark Energy Survey, \citet{chen2022} performed a cosmological analysis using a subset of $\sim 100$ SNe Ia hosted by galaxies in the redMaGiC catalog, which have both photometric and spectroscopic redshifts. The difference in best-fit cosmological parameters between using spectroscopic and photometric redshifts was found to be minimal, $\Delta w \sim 0.005$. redMaGiC galaxies were chosen for this analysis due to their particularly well-constrained photometric redshift estimates, which is not representative of the full population of SN host galaxies. However, even with a sound cosmological inference framework, galaxy photometric redshifts are often inaccurate or plagued with large uncertainties, and requiring host galaxy information to produce SN photometric redshift estimates may introduce additional biases. 

These issues with traditional redshift determination along with the development of a cosmological framework for photometric redshifts has led to a growing interest in alternative techniques for predicting redshifts for Type Ia supernovae. In this work, we intoduce a novel machine learning algorithm to predict full redshift probability distributions for Type Ia supernovae based solely on lightcurve data. Our estimator harnesses the constraining power on redshift of the SN lightcurves and can additionally provide an independent cross-check on host galaxy matches, minimizing mismatch rates. We present a detailed analysis of our model, including its accuracy and limitations, and discuss the potential implications of our findings for future cosmological studies.

\subsection{Photo-z Estimation}

Most of the existing literature on photometric redshift estimation is on galaxy photo-$z$s. These approaches generally use either (1) a training set of galaxy photometric observables, such as colors and magnitudes, to determine a mapping to spectroscopic redshifts \citep[e.g.][]{Brunner_1997}; or (2) template fitting, in which observed properties are compared with redshifted template spectra to determine the best fit redshift value \citep[e.g.,][]{Benitez_2000}. There have also been successful machine learning-based galaxy photo-$z$ models developed, e.g. \citet{disanto, Pasquet_2018} using convolutional neural networks on galaxy images, and \citet{SOM} using a self-organizing map to relate color-magnitude space and redshifts.

\begin{figure*}
    \centering
    \includegraphics[scale=0.45]{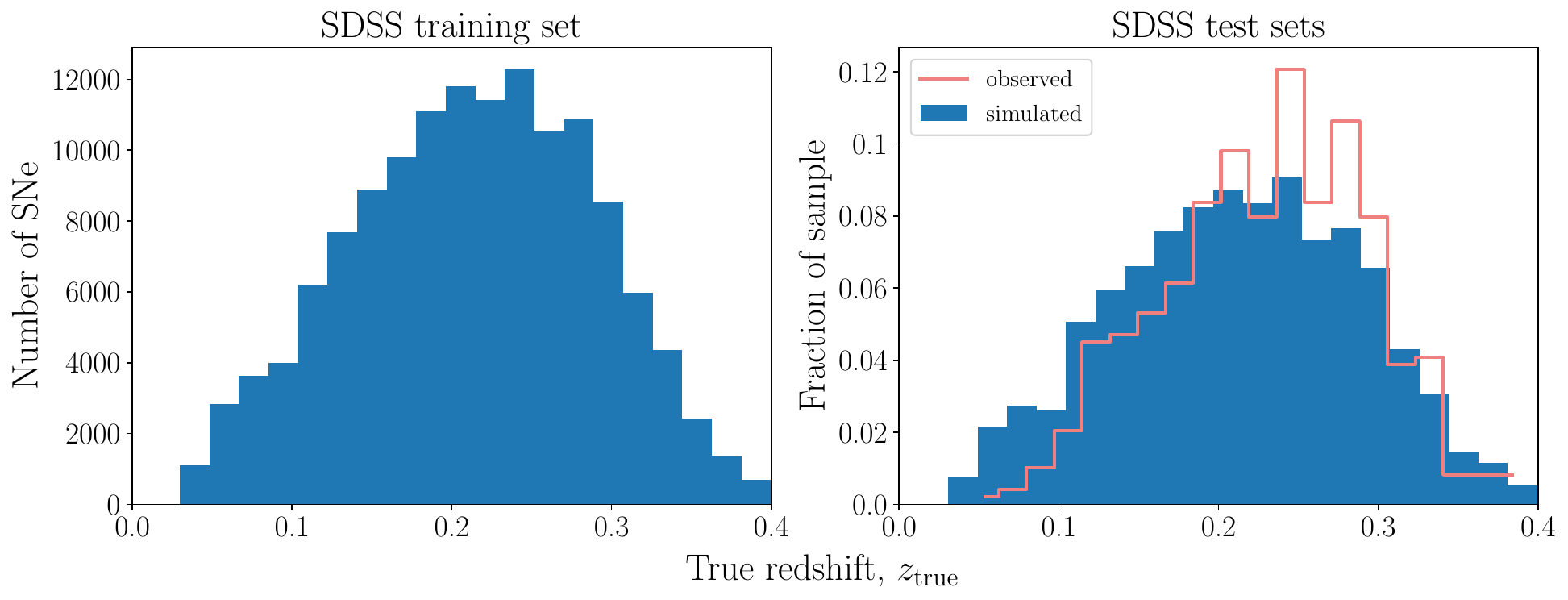}
    \caption{True redshift distribution of the 135,521 simulated SNe Ia in the SDSS training dataset (left) and the 5,274 simulated SNe Ia and 489 observed photometrically confirmed SNe Ia in the SDSS test sets (right).}
    \label{fig:sdss-z-dist}
\end{figure*}
Photometric redshift estimation for SNe uses many of the same techniques. \citet{Kim_2007} used the SALT2 lightcurve model \citep{salt2} to determine photometric redshifts and distances as well as uncertainties using the Fisher information matrix. \citet{Palanque_Delabrouille_2010} and \citet[][LCFIT+Z]{Kessler_2010} extended the SALT2 lightcurve fit to include redshift as a fitted parameter, and incorporating the host galaxy redshift as a prior. These approaches find the best fit redshift by comparing photometric observables to those expected from the SALT2 model via $\chi^2$ minimization. An analytic approach was presented in \citet{wang2015} which assumed a functional form for the redshifts as a function of multi-band photometric fluxes and fit for free parameters using a training set of SNe Ia with spectroscopic redshifts. \citet{deOliveira:2022tvw} applied machine learning techniques to this problem, performing regression using features obtained from principal component analysis. However, many of these results suffer from redshift-dependent bias, in which photo-$z$s for high-redshift SNe are systematically underestimated (see e.g. Figure 6 of \citet{Kessler_2010}, Figure 7 of \citet{deOliveira:2022tvw}). Most of these studies also estimate a redshift value and uncertainty rather than the full probability density function (PDF).

\subsection{Overview}
In this work we present \name, a convolutional neural network model that uses multi-band photometry to predict a full redshift PDF. Our approach uses the raw photometric data and does not require any manual feature engineering to determine the most predictive features. We evaluated \name\ using LSST and SDSS simulated SNe Ia as well as the SDSS photometric SN Ia sample, and show that our redshift predictions have low scatter and suffer from minimal redshift-dependent bias on all tested SN samples. Our results are a promising step towards precision photometric SN Ia cosmology.

We introduce the simulated and observed data samples used for evaluation in \S 2, as well as the preprocessing step used to transform multi-band lightcurves into convolutional neural network inputs. In \S 3, we describe the model architecture and training strategy. We present results, comparisons with LCFIT+Z, and further experiments in \S 4, and conclude in \S 5.

%% file: chapters/photoz/data.tex
\subsection{Data Sources}
\label{subsec:data}
We present results on simulated SNe Ia from two surveys: LSST \citep{ivezic_lsst} and the SDSS-II SN survey \citep{sdss}. We also demonstrate that our model generalizes well to an observed photometric SN Ia dataset by showing photo-$z$ predictions on SDSS SNe classified as type Ia by SuperNNova \citep{supernnova}.
We use the SuperNova ANAlysis software \citep[SNANA,][]{snana} to produce all simulated SN lightcurves used in this work.

\subsubsection{Simulated LSST SNe Ia (PLAsTiCC)}
LSST is a ground-based dark energy survey program that will discover millions of SNe over the 10 year survey duration. The 8.4m Simonyi Survey optical telescope at the Rubin Observatory uses a state-of-the-art 3200 megapixel camera with a 9.6 $\text{deg}^2$ field of view that will provide deeper and wider views of the universe with unprecedented quality. LSST will observe nearly half the night sky each week to a depth of $24^{\text{th}}$ magnitude in $ugrizY$ photometric bands spanning wavelengths from ultraviolet to near-infrared.

We simulate LSST-like observations of SNe Ia in $ugrizY$ photometric bands following the model, rates, and LSST observing conditions developed for the PLAsTiCC dataset \citep{plasticc_data, plasticc_models} for $0.05 < z < 1.2$. We use the SALT2 lightcurve model \citep{salt2} with training parameters derived from the Joint Lightcurve Analysis \citep{jla} extended into the ultraviolet and near-infrared by \citet{hounsell2018} following the procedure described in \citet{pierel2018}. While PLAsTiCC included two LSST observing strategies, the Deep Drilling Fields (DDF) as well as the Wide-Fast-Deep (WFD), we simulate only the DDF subsample. We coadd all observations within the same night, following PLAsTiCC, resulting in observations that are $\sim 2.5$x more frequent and $\sim 1.5$ mag deeper than the WFD sample. Though we focused on the DDF sample for this work, evaluating the performance of \name\ on WFD simulations is an important direction for future studies.

As PLAsTiCC is the name of the dataset we emulated while LSST is the survey we simulate, we will use both interchangeably to denote ``simulated LSST SN lightcurves following the PLAsTiCC dataset".

\subsubsection{Simulated SDSS SNe Ia}
\label{subsec:sdss-sims}
The SDSS-II Supernova Survey identified and measured light curves for intermediate-redshift ($0.05 < z < 0.4$) SNe Ia using repeated five-band ($ugriz$) imaging of Stripe 82, a stripe $2.5^{\circ}$ wide centered on the celestial equator in the Southern Galactic Cap. The primary instrument for this survey is the SDSS CCD camera mounted on a dedicated 2.5m telescope at Apache Point Observatory, New Mexico. Over the three observing seasons between 2005 and 2007, SDSS discovered 10,258 transient and variable objects, with 536 spectroscopically confirmed SNe Ia and an additional 907 photometrically classified SNe Ia candidates \citep{sako2018}. \citet{sdss} provides an in-depth review of the SDSS-II SN survey.

We simulate SDSS SNe Ia in $ugriz$ photometric bands using the SALT3 model described by \citet{Kenworthy_2021} due to its improved wavelength range coverage in the near infrared, allowing $z$ band observations to be simulated for SNe at low redshifts. We additionally extend this model to $500$\AA\ to simulate $u$ band observations at high redshifts. We use simulated observing conditions from \citet{sdss_simlib} and host galaxy spectroscopic detection efficiency from \cite{snana}. We simulate a \textit{photometric} SDSS SNe Ia dataset using the host spectroscopic detection efficiency rather than a SN spectroscopic detection efficiency to demonstrate the performance of \name\ on the practical use case of a photometric sample.

\subsubsection{Observed SDSS SNe Ia}
In addition to simulated SNe, we test \name\ on SDSS lightcurves from \citet{sako2018} classified by SuperNNova \citep{supernnova} as likely SNe Ia, defined as SNe with SNIa probability $P_{\text{Ia}} \geq 0.5$. ``True" redshifts $z_{\text{true}}$ for the sample are from spectra of the SNe themselves or their host galaxies. Details on the training data and training procedure for the SuperNNova model used here can be found in Popovic et al., in prep. 489 lightcurves remain after the selection cuts described in Sections~\ref{subsec:cuts} and~\ref{subsec:preprocess}. The number of SDSS lightcurves remaining after each set of cuts is shown in Table~\ref{tbl:sdss-cuts}. 

\subsubsection{Lightcurve Selection}
\label{subsec:cuts}
We apply basic selection cuts to the simulated LSST sample. We require

\begin{itemize}
    \item at least 5 observations of each SN
    \item signal-to-noise ratio (SNR) $> 3$ for at least one observation each in 2 of the $griz$ filters
\end{itemize}

We apply selection cuts, following \citet{popovic2020}, to both the simulated and observed SDSS data to remove poor quality lightcurves. We define a rest-frame age, $T_{\text{rest}} = (t - t_{\text{peak}})/(1+z)$, where $t$ is the observation date, $t_{\text{peak}}$ is the estimated epoch of SN peak brightness from SNANA, and $z$ is the redshift of the event. We require

\begin{itemize}
    \item $0 < T_{\text{rest}} < 10$ 
    \item SNR $> 5$ for at least one observation each in 2 of the $griz$ filters
\end{itemize}

We additionally require SNe in the SDSS data to have a spectroscopic redshift from either the SN spectrum or the SN host galaxy. 

\begin{table}
    \centering
    \begin{tabular}{l c}
        \toprule
        Cut & Number of SNe \\
        \midrule
        Full sample & 10,258 \\
        Lightcurve selection (\S\ref{subsec:cuts}) & 2,044 \\
        Likely SNe Ia ($P_{\text{Ia}} \geq 0.5$) & 1,037 \\
        SALT fit cuts (\S\ref{subsec:preprocess}) & 555\\
        Successful LCFIT+Z fit & 489 \\
        \bottomrule
    \end{tabular}
    \caption{Number of SDSS lightcurves remaining after each selection cut. We evaluate \name\ on the remaining 489 SNe.}
    \label{tbl:sdss-cuts}
\end{table}

\begin{figure*}
    \centering
    \includegraphics[scale=0.45]{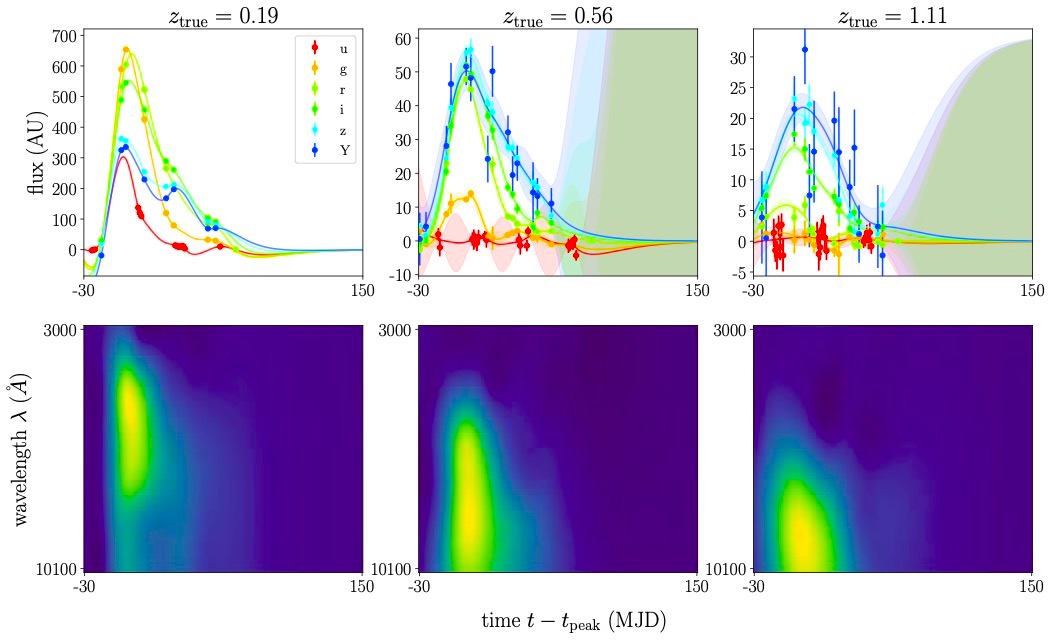}
    \caption{Examples of a low (left), medium (middle), and high (right) redshift simulated PLAsTiCC SN. \textbf{(top panel)} The multi-band lightcurves are shown in colored points, with Gaussian process predictions and uncertainties for this lightcurve shown in colored lines and shaded regions, respectively. \textbf{(bottom panel)} The flux matrix created from the lightcurve in the top panel (see \S\ref{subsec:lc-preprocess} for details). Large flux values are colored in yellow while lower flux values are in dark blue. The physical effects of redshifting, including longer wavelengths and longer transient durations, are observed in the flux matrices in the downward (towards higher wavelengths) shift of the yellow high flux region and the increased width of the yellow region with increasing redshift.}
    \label{fig:plasticc-lc-examples}
\end{figure*}
\subsection{Data Preprocessing}
\label{subsec:preprocess}

All SNe in all datasets are fit with a $\chi^2$-minimization program included in SNANA to determine several restframe parameters under the assumption that the event is a SN Ia: the time of SN peak brightness $t_\mathrm{peak}$, a stretch-like parameter $x_1$, a color parameter $c$, and the lightcurve normalization parameter $x_0$, as well as their uncertainties and covariances (i.e., $\sigma_{x_1}$, etc.). These parameters are used to calculate the distance modulus $\mu$, allowing the SNe to be placed on the Hubble diagram. SDSS datasets are fit with the SALT3 model while the LSST dataset is fit with the same extended SALT2 model used for the simulations.

We additionally select only SNe in our datasets that are well-described by the SALT model. These criteria were chosen following those used in past cosmology analyses, e.g. \citet{jla}:

\begin{itemize}
    \item $|c| < 0.3$
    \item $|x_1| < 3$
    \item $\sigma_{x_1} \leq 1$
    \item $\sigma_{t_\mathrm{peak}} \leq 2$
\end{itemize}

Finally, we choose SNe within the redshift range $0.01 \leq z \leq 0.4$ for SDSS and $0.01 \leq z \leq 1.2$ for LSST that have a successful LCFIT+Z \citep{Kessler_2010} fit, in order to draw direct comparisons between the performance of \name\ and LCFIT+Z. A more detailed description of LCFIT+Z can be found in \S\ref{subsec:lcfit+z}.

146,069 simulated SDSS SNe Ia and 81,734 simulated LSST SNe Ia remain after all cuts, as well as 489 observed SDSS SNe photometrically classified as SNe Ia. The number of observed SDSS lightcurves remaining after each cut is described in Table~\ref{tbl:sdss-cuts}. The redshift distributions of the PLAsTiCC training and test datasets after selection cuts are shown in Figure~\ref{fig:plasticc-z-dist}, while the redshift distributions of the SDSS simulated and observed datasets are shown in Figure~\ref{fig:sdss-z-dist}. We note that, while most machine learning classifiers perform better when trained on a dataset with balanced classes (i.e. flat redshift distribution), we found that this was not the case for \name. While our simulations are able to generate any artificial redshift distribution, an unphysical one such as a flat distribution could lead to strange artifacts in the dataset, confusing the classifier.

\subsubsection{Lightcurve Preprocessing}
\label{subsec:lc-preprocess}
To preprocess each lightcurve, we first take observations within the range $t_\mathrm{peak}-30 \leq t \leq t_\mathrm{peak}+150$, where $t_\mathrm{peak}$ is the epoch of SN peak brightness estimated by SNANA. This is to ensure the lightcurves contain just the SN and no extraneous information.

We use 2-dimensional Gaussian process (GP) regression to model each SN lightcurve in time ($t$) and wavelength ($\lambda$) space, then use the predictions of the fitted GP on a fixed $(\lambda, t)$ grid as the input to our neural network model. Modelling lightcurves of astronomical transients using 2D Gaussian processes was originally introduced in \citet{Boone_2019}; and \citet[][Q21]{scone} used the GP models to create ``images" from lightcurve data, which is the technique used for this work. Following Q21, we use the Mat\'{e}rn kernel ($\nu=3/2$) with a fixed $6000$\AA\ length scale in wavelength space and fit for the time length scale as well as the amplitude using maximum likelihood estimation. This GP model is fit separately to each lightcurve and used to produce a smooth 2D representation of the lightcurve by predicting flux values at each point in a $(\lambda, t)$ grid. We choose 32 equally spaced points in the range $3,000$\AA\ $\leq \lambda \leq 10,100$\AA\ ($\delta \lambda = 221.875$\AA) and 180 points in the range $t_\mathrm{peak}-30 \leq t \leq t_\mathrm{peak}+150$ ($\delta t=1$ day). This produces a $32 \times 180$ matrix of predicted flux values. We also produce a matrix of prediction uncertainties at each $\lambda_i, t_j$ of equal size.

We stack the flux and uncertainty matrices depthwise to produce a $32 \times 180 \times 2$ tensor and divide elementwise by the maximum flux value to constrain all entries to [0,1].

We show low, medium, and high redshift examples of PLAsTiCC lightcurves, the fitted GP models and their uncertainties, and the resulting flux matrices in Figure~\ref{fig:plasticc-lc-examples}. Redshifting increases the wavelength of light, which we see in the figure as the yellow (high flux) region moving down (toward $\lambda = 10,100$\AA) with increasing redshift. We also expect to observe a longer duration for higher redshift transients, which is evident in the increase in width of the yellow region. This data format is particularly well suited to the redshift prediction task, as we are able to visibly see expected physical results of redshifting in the flux matrices.

\subsubsection{Redshift Preprocessing}
The redshift range for each survey is discretized into $n_z$ discrete and non-overlapping bins ($0.01 \leq z \leq 0.4, n_z = 50$ for SDSS and $0.01 \leq z \leq 1.2, n_z = 150$ for LSST). We chose these $n_z$ values to preserve the approximate width of each redshift bin across surveys ($\delta z \sim 0.0078$). The bin corresponding to the true redshift of each SN is one-hot encoded and passed into the model as the training label.

%% file: chapters/photoz/model.tex
\name\ is a convolutional neural network model \citep[e.g. ][]{lecun,  zeiler2014visualizing, Simonyan2014VeryDC, krizhevsky2017imagenet} that takes in GP-interpolated lightcurves as well as the GP prediction uncertainties, prepared as described in \S\ref{subsec:preprocess}, and predicts a probability distribution over fine-grained redshift bins. This approach allows us to produce a discretized full PDF over redshift space for each individual SN without any assumptions on the underlying distribution, and has been used in a variety of contexts including image generation \citep{pixelrnn} and prediction of precipitation probabilities \citep{metnet}. Treating this as a categorical classification problem using the cross-entropy loss function has also been shown to accurately approximate Bayesian posterior probabilities \citep{lippman}. 

\subsection{Convolutional Neural Networks}
The convolutional neural network (CNN) is a class of artificial neural network with properties particularly suited to object and image recognition. It requires fewer trainable parameters than the standard feedforward network due to the convolution operation, learning a single weight matrix for small neighborhoods of the input image. This property is not only parameter-efficient but also imparts CNNs with translation-equivariance, i.e. the same feature shifted by $n$ pixels will produce the same response shifted by $n$ pixels. These convolutional layers are paired with pooling layers, which downsample the input to allow for the next set of convolutional layers to learn hierarchically more complex features. CNNs are prized in the machine learning community for being simple yet performant on image recognition benchmark datasets such as ImageNet \citep{imagenet}.

\subsection{Residual Learning}
\label{subsec:resnet}
We implement residual learning \citep{he2016deep} due to its state-of-the-art performance on the ImageNet benchmark at the time of publication as well as its widespread adoption over vanilla CNNs. Residual learning was presented as a solution to the degradation problem in CNNs, in which performance degraded past a certain threshold of network depth.  Since the additional layers could simply act as identity mappings and not affect the network performance, \citet{he2016deep} decreased the difficulty by not only feeding inputs sequentially through a series of layers, but also adding the inputs back in to the outputs of those layers. This allows the layers to learn the zero mapping rather than the identity mapping.
%
Since these layers are tasked with learning the \textit{residual} with respect to the input, this is known as \textit{residual learning} and the stack of layers is known as a \textit{residual block}. The residual connections in the \name\ architecture are shown as lines curving around each residual block in Figure~\ref{fig:architecture}.

\subsection{Architecture}
\label{subsec:architecture}
\begin{figure}
    \centering
    \includegraphics[scale=0.61, trim={16cm 11cm 9.5cm 26cm},clip]{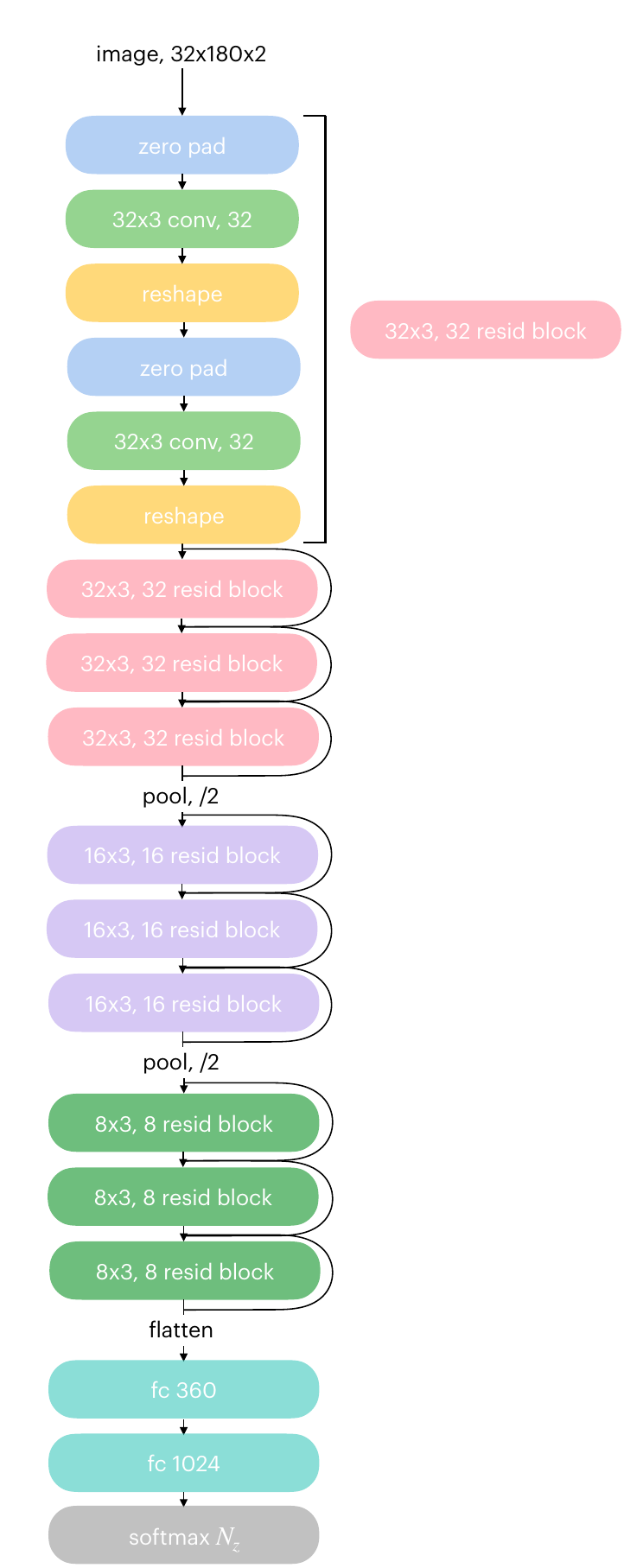}
    \caption{The model architecture developed for this work, described in the text in \S\ref{subsec:architecture}. A layer shown in the figure as ``$h \times d$ conv, $n_{\mathrm{filters}}$" is a 2D convolutional layer with a kernel size of $h \times d$ and $n_{\mathrm{filters}}$ filters. Similarly, ``$h \times d, n_{\mathrm{filters}}$ resid block" defines a residual block containing two $h \times d$ conv, $n_f$ convolutional layers. ``fc $n_{\mathrm{nodes}}$" denotes a fully connected layer with $n_{\mathrm{nodes}}$ nodes. The pooling layers apply max pooling, downsampling both the height and width by 2. The output softmax layer has $N_z$ nodes, $N_z=50$ for SDSS and $N_z=150$ for LSST.}
    \label{fig:architecture}
\end{figure}
Our model (Figure~\ref{fig:architecture}) takes as input a $32 \times 180 \times 2$ tensor containing the GP-interpolated lightcurve data on a $32 \times 180$ grid of wavelength and time values, and the GP uncertainties at each of those points. Table~\ref{tbl:layers} shows input and output dimensions of example layers in the model architecture. We also provide the model with the one-hot encoded vector specifying the correct redshift bin as the training label. 

\begin{table}
    \centering
    \begin{tabular}{l c c}
        \toprule
        Layer & Input Shape & Output Shape \\
        \midrule
        Zero Padding & $32 \times 180 \times 2$ & $32 \times 182 \times 2$ \\
        $32\times 3$ Convolutional & $32 \times 182 \times 2$ & $180 \times 32 \times 1$ \\
        Reshape & $180 \times 32 \times 1$ & $32 \times 180 \times 1$\\
        Max Pooling & $32 \times 180 \times 1$ & $16 \times 90 \times 1$\\
        \bottomrule
    \end{tabular}
    \caption{Description of example layers in the model architecture.}
    \label{tbl:layers}
\end{table}

The input is zero-padded on both sides to $32 \times 182 \times 2$ to ensure that feature maps output by the convolutional layers retain the original shape, then passed through a convolutional layer with 32 filters and kernel size $32 \times 3$. Since the convolutional kernel determines the receptive field of each unit in the layer, we choose a convolutional kernel that spans the wavelength space to allow each unit to learn from all wavelengths simultaneously while preserving the linearity of time. The output of this convolutional layer is $1 \times 180 \times 32$, which we then reshape back to $32 \times 180 \times 1$. This first feature map is now passed through a series of residual blocks.

Each residual block contains two convolutional blocks, each consisting of a ReLU nonlinearity, batch normalization, and the zero-padding, convolutional layer, and reshaping layer identical to the ones described above. The input to each residual block is then added to the output as described in \S\ref{subsec:resnet}. After each series of three residual blocks, the output is passed through a $2 \times 2$ max-pooling layer, downsampling the height and width of the output by a factor of 2.

Finally, the output is flattened and passed through a fully connected layer with 1,024 hidden nodes, which connects to the final softmax layer. The nodes in this layer correspond to redshift bins, thus it has 50 nodes for processing SDSS data and 150 for LSST data. The array of output probabilities is interpreted as the probability density over redshifts for our input SN.

\subsection{Calibration}

We also performed temperature scaling \citep{guo2017} to ensure that the probabilities output by \name\ are properly calibrated. In this process, we learn a single ``temperature" parameter used to scale the output probabilities. 

Before scaling, the output probabilities $\textbf{p}_i$ for input SN $i$ are derived from the softmax function 
\begin{equation}
    \textbf{p}_i^{(k)} = \sigma_{\text{SM}}(\textbf{q}_i)^{(k)} = \frac{\text{exp}(\textbf{q}_i^{(k)})}{\sum_{j=1}^{K} \text{exp}(\textbf{q}_i^{(j)})}
\end{equation}
where $\textbf{q}_i$ is the vector of network logits corresponding to SN $i$, i.e. the output of the final hidden layer of the network. The temperature parameter $T$ is learned by minimizing the cross-entropy loss between the one-hot encoded labels and the scaled probabilities, $\textbf{p}'_i$,
\begin{equation}
\textbf{p}'_i = \sigma_{\text{SM}}(\textbf{q}_i / T)    
\end{equation}

\begin{figure}
    \centering
    \includegraphics[scale=0.6, trim={0 0 0 0},clip]{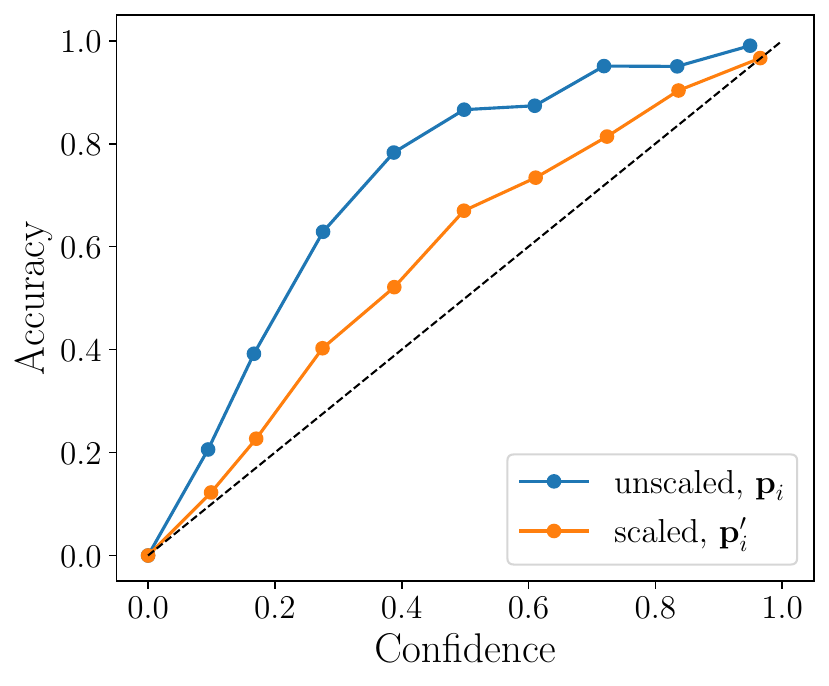}
    \caption{Reliability diagram showing PLAsTiCC model calibration before and after temperature scaling.}
    \label{fig:calibration}
\end{figure}

Reliability diagrams \citep{degroot, niculescu} and the expected calibration error statistic \citep[ECE,][]{naeini} are common evaluation methods for calibration. Reliability diagrams show the prediction accuracy as a function of \textit{confidence}, which is defined as the probability associated with the predicted class: $\text{max}(\textbf{p}_i)$. A reliability diagram for a perfectly calibrated classifier will show the identity function, and any deviation from a perfect diagonal is a sign of miscalibration. The reliability diagram before and after temperature scaling for the PLAsTiCC model is shown in Figure~\ref{fig:calibration}. The scaled probabilities are much closer to the diagonal, representing a significant improvement in calibration. 

ECE is a weighted average of the difference between the accuracy and the confidence in bins of confidence values. ECE is more precisely defined as
\begin{equation}
    \text{ECE} = \sum_{m=1}^M \frac{|B_m|}{N} |\text{acc}(B_m) - \text{conf}(B_m)|
\end{equation}
where $B_m$ is the $m^{\text{th}}$ confidence bin and $N$ is the total number of samples in the dataset.
Prior to temperature scaling, the ECE for the PLAsTiCC model probabilities was 0.24, which improved to 0.08 after scaling.

\subsection{Implementation Details}
We split both the SDSS and LSST datasets into 90\% training, 5\% validation, and 5\% test datasets and trained both models with batch sizes of 2048 for 750 epochs. We minimize the cross-entropy loss function with the Adam optimizer \citep{adam} with an initial learning rate of 1e-3 that is halved after 25 epochs of no improvement in the validation loss. The total number of trainable parameters in the SDSS model is 451,654 and 554,154 for the LSST model due to the difference in number of redshift bins and resulting difference in output layer size. To prevent overfitting, we use a weight decay of 1e-3 as well as dropout layers \citep{dropout}. Calibration was performed with the validation set of both datasets and with an initialization of $T=1$. We minimize a cross-entropy loss function using Adam and find $T_{\text{SDSS}}=0.82$ and $T_{\text{LSST}}=0.64$.

\begin{figure*}
    \centering
    \includegraphics[scale=0.55]{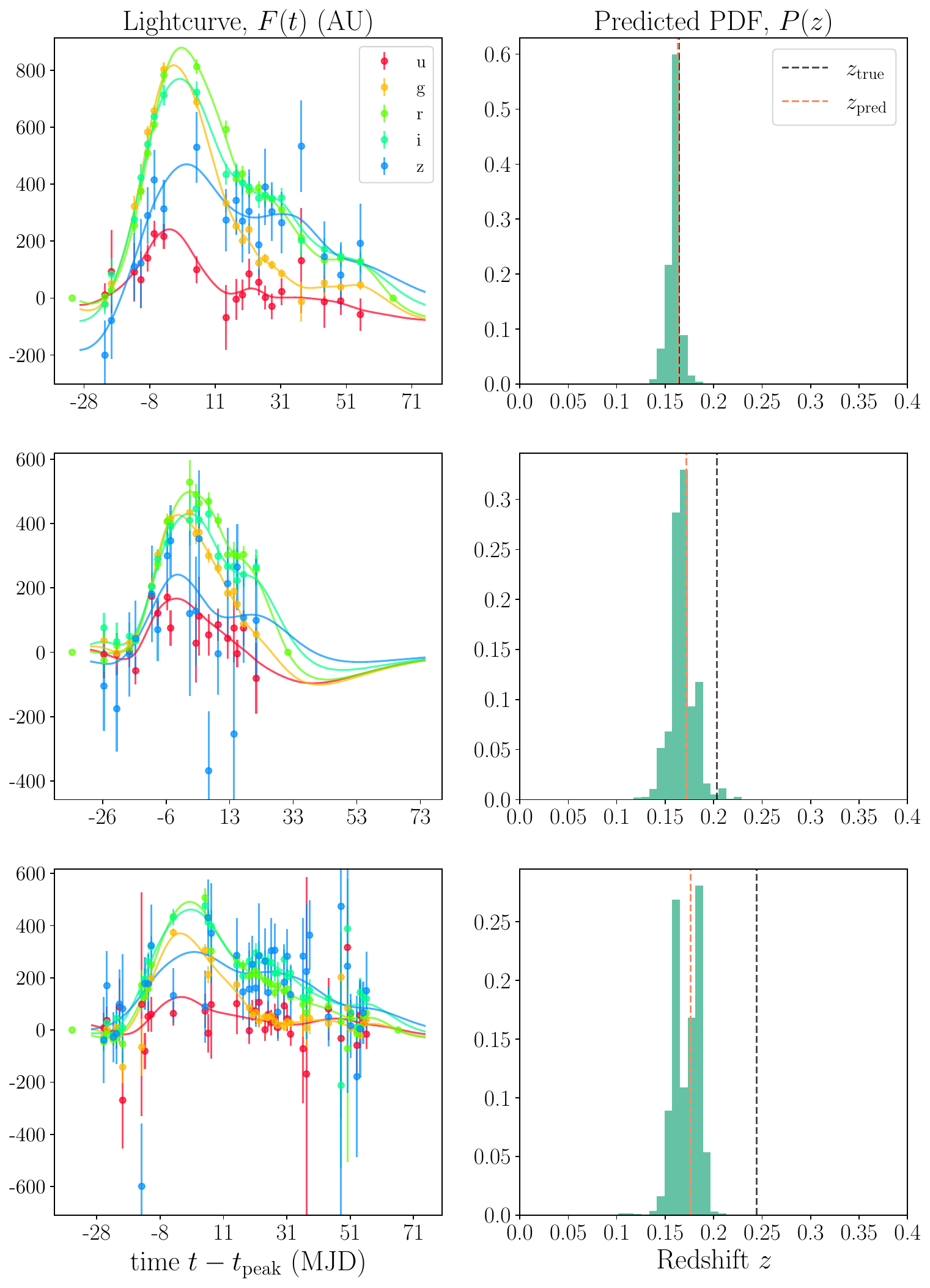}
    \caption{Examples of a high accuracy ($\Delta z < 0.005$, top), medium accuracy ($0.01 \leq \Delta z \leq 0.05$, middle), and outlier ($\Delta z > 0.05$, bottom) lightcurve and Gaussian process fit from the SDSS spectroscopic sample and their predicted PDFs. The true (spectroscopic) redshift is shown in the black dotted line, while the predicted redshift (mean(PDF)) is shown in orange. The medium and poor accuracy lightcurves have larger errors and scatter than the high accuracy lightcurve, and the low accuracy lightcurve appears to have much fewer points.}
    \label{fig:sdss-lc-examples}
\end{figure*}

%% file: chapters/photoz/results.tex
Examples of lightcurves and their corresponding predicted redshift PDFs are shown in Figure~\ref{fig:sdss-lc-examples}. We chose spectroscopically confirmed SDSS SNe Ia lightcurves, and selected a high accuracy ($\Delta z < 0.005, \Delta z \equiv \frac{z_{\mathrm{pred}}-z_{\mathrm{true}}}{1+z_{\mathrm{true}}}$), medium accuracy ($0.01 \leq \Delta z \leq 0.05$), and an outlier ($\Delta z > 0.05$) example.

\subsection{Evaluation Metrics and Basis for Comparison}
\subsubsection{Point Estimates and Metrics}
\label{subsec:metrics}
\begin{figure}
    \centering
    \includegraphics[scale=0.45]{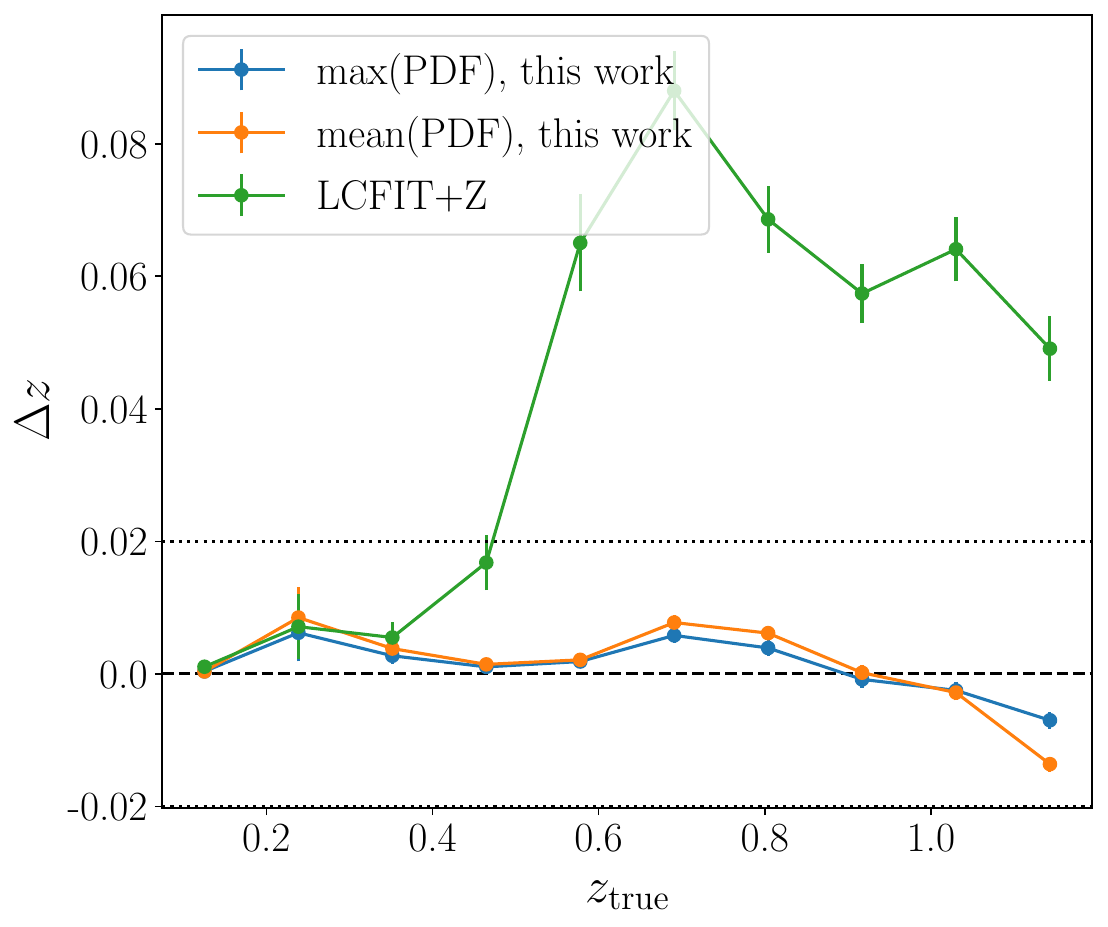}
    \caption{Mean binned residuals $\Delta z$ as a function of true redshift $z_{\mathrm{true}}$ for the PLAsTiCC simulated SNe Ia test set. Predictions from our model have much lower biases as well as scatter compared to predictions from LCFIT+Z. The max(PDF) and mean(PDF) point estimates for our model also agree quite well, resulting in similar $\Delta z$ values. Dotted lines are plotted at $\Delta z = \pm 0.02$ for reference.}
    \label{fig:plasticc-resids}
\end{figure}

\begin{figure*}
    \centering
    \includegraphics[scale=0.55]{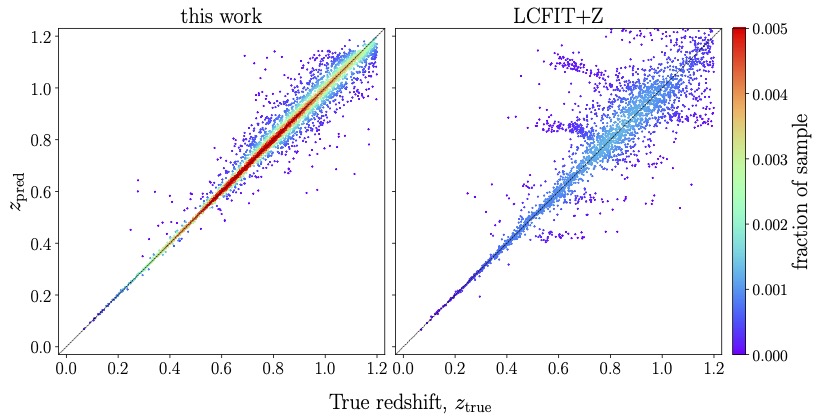}
    \caption{Predicted vs. true redshifts for the PLAsTiCC simulated SNe Ia sample colored by the fraction of each sample represented by each point. \textbf{(left)} Predictions from \name\ described in this work, \textbf{(right)} predictions from LCFIT+Z.}
    \label{fig:plasticc-scatter}
\end{figure*}

Although full photometric redshift PDFs are preferred for further statistical analyses (i.e. cosmological analyses), point estimates can also be computed from each PDF. We require these point estimates to evaluate our model performance against the true redshift values. We compare two possible methods to condense a PDF into a point estimate:
\begin{equation}
     \text{mean(PDF)} = \frac{1}{2} \sum_i p(Z_i) \frac{\lceil Z_{i} \rceil ^2 - \lfloor Z_{i} \rfloor ^2}{\lceil Z_{i} \rceil - \lfloor Z_{i} \rfloor}
\end{equation}
where $Z$ represents the vector of redshift bins, $\lceil Z_i \rceil $ the right edge of bin $i$, $\lfloor Z_i \rfloor $ the left edge of bin $i$, and $p(Z_i)$ the output probability assigned by the model to bin $i$; and
\begin{equation}
    \text{max(PDF)} = \tilde{Z}_{\text{argmax}_i(p(Z_i))}\\
\end{equation}
where $\tilde{Z}$ represents the array of midpoints of the redshift bins $Z$. Taking the weighted mean is a common summary statistic for PDFs, while taking the bin with maximum probability as the predicted output is typical for classification tasks. As shown in Tables~\ref{tab:plasticc-results} and~\ref{tab:sdss-results} and Figures~\ref{fig:plasticc-resids} and~\ref{fig:sdss-delta-z}, the two methods give similar results with mean(PDF) performing slightly better on the real SDSS dataset, and thus we use the mean point estimate for Figures~\ref{fig:plasticc-scatter},~\ref{fig:sdss-scatter},~\ref{fig:sdss-delta-z-corrected},~\ref{fig:hubble}, and~\ref{fig:sdss-delta-z-prior}.

We compute the following metrics for both of our point estimates, following e.g. \citet{Pasquet_2018}:
\begin{itemize}
    \item the residuals $\Delta z \equiv \frac{z_{\mathrm{pred}}-z_{\mathrm{true}}}{1+z_{\mathrm{true}}}$,
    \item the bias $\langle \Delta z \rangle$,
    \item the mean absolute deviation $\sigma_{\mathrm{MAD}} = 1.4826 \times \text{median}(|\Delta z - \text{median}(\Delta z)|)$,
    \item the fraction of outliers $\eta$ with $|\Delta z|>0.05$.
\end{itemize}

We evaluate \name\ using these metrics on the PLAsTiCC dataset as well as the simulated and observed SDSS datasets, and show the results in Tables~\ref{tab:plasticc-results} and~\ref{tab:sdss-results}.

\subsubsection{Comparison with LCFIT+Z}
\label{subsec:lcfit+z}
LCFIT+Z \citep{Kessler_2010} is an extension of the lightcurve fitting code described in \S\ref{subsec:data} that treats redshift as an additional free parameter. It determines the best fit lightcurve parameters by minimizing $\chi^2$ values of the observed and model fluxes, which are computed as a function of the free parameters (4 lightcurve parameters $t_{\mathrm{peak}}$, stretch $x_1$, color $c$, flux normalization $x_0$; and redshift $z_{\mathrm{phot}}$). LCFIT+Z is actively used for recent and ongoing experiments performing full cosmological analyses with photometric redshifts \citep[e.g.][]{Dai_2018, mitra2022}. LCFIT+Z produces point estimates with uncertainties as opposed to a full PDF, so we perform comparisons using our point estimates. 

We compare our results with LCFIT+Z as opposed to other SN photometric redshift estimators \citep[e.g., ][]{wang2015,deOliveira:2022tvw} due to its demonstrated performance, widespread use, and integration into SNANA. We run LCFIT+Z on the same datasets used to evaluate our model, enabling a direct comparison. Note that most documented uses of LCFIT+Z place a prior on the redshift fit using the redshift of the SN host galaxy, resulting in much more constrained fits. However, since (1) host spectroscopic redshifts are unavailable in future survey environments, and (2) to draw direct comparisons with \name, which does not require any host galaxy information, we omit the host galaxy redshift prior when running LCFIT+Z. In order to compare how these methods perform using all available information in the photometric survey era, we test \name\ and LCFIT+Z with host galaxy \textit{photometric} redshift priors in Section~\ref{subsec:with-prior}.

\subsection{LSST (PLAsTiCC) Results}
\label{subsec:plasticc}

We evaluate our model and our baseline for comparison, LCFIT+Z, on a test set of 4,057 simulated PLAsTiCC-like lightcurves with true redshift distribution shown in the right panel of Figure~\ref{fig:plasticc-z-dist}. 
The values of the evaluation metrics for results from this work as well as LCFIT+Z are shown in Table~\ref{tab:plasticc-results}. The two methods for obtaining point estimates from \name\ PDFs, mean and max, give similar results, though the mean point estimates have a degraded $\sigma_{\mathrm{MAD}}$ value compared to the max point estimates but a smaller outlier rate. However, the differences between mean and max point estimates are orders of magnitude smaller than the improvement we see relative to LCFIT+Z. The mean gives a $\sim 3.2\times$ larger result than max on $\sigma_{\mathrm{MAD}}$, compared with a $\sim 180\times$ larger result from LCFIT+Z.

\begin{table}
    \centering
    \begin{tabular}{l c c c}
        \toprule
        Metric & \multicolumn{2}{c}{this work} & LCFIT+Z\\
        \cmidrule(lr){2-3}
        & mean & max & \\
        \midrule
         bias $\langle \Delta z \rangle$ & 0.00095 & \textbf{0.00075 }& 0.058\\
         $\sigma_{\mathrm{MAD}}$ & 0.0081 & \textbf{0.0025 }& 0.0450\\
         outlier rate $\eta$ & \textbf{3.87\% }& 4.24\% & 32.3\%\\
         \bottomrule
    \end{tabular}
    \caption{Evaluation metrics computed for the PLAsTiCC test dataset for both the mean(PDF) and max(PDF) point estimates for our model as well as LCFIT+Z. The best result for each metric is shown in bold.}
    \label{tab:plasticc-results}
\end{table}

We show the mean binned residuals, $\Delta z$, as a function of true redshift for our model and LCFIT+Z in Figure~\ref{fig:plasticc-resids}. We see that while the residuals of our model and LCFIT+Z match quite well up to $z_{\mathrm{true}} \sim 0.4$, LCFIT+Z has a tendency to dramatically overestimate at higher redshifts. We also see that the spread within each residual bin in LCFIT+Z results is much larger, as shown in the size of the error bars on each point. In contrast, our model shows a relatively flat $\Delta z$ over the full redshift range with minimal redshift-dependent bias, which has not been achieved by other SN photometric redshift estimators.

Figure~\ref{fig:plasticc-scatter} more clearly shows the spread of predicted redshifts, where $z_{\mathrm{pred}}$ is calculated as the mean(PDF) for our model. The predictions produced by our model (left panel) lie much closer to the $z_{\mathrm{pred}} = z_{\mathrm{true}}$ line with approximately equal amounts of scatter on either side, reinforcing the minimal redshift-dependent bias shown in Figure~\ref{fig:plasticc-resids}. The dark red area along the $z_{\mathrm{pred}} = z_{\mathrm{true}}$ line also indicates that most of the sample is localized there. The LCFIT+Z results (right panel) have much larger spread and little localization, as evidenced by the lack of red in the plot.

\subsection{SDSS Results}
\label{subsec:sdss}
\begin{table*}
    \centering
        \begin{tabular}{l c c c c c c}
        \toprule
        Metric & \multicolumn{3}{c}{SDSS simulated} & \multicolumn{3}{c}{SDSS real}\\
        \cmidrule(lr){2-4} \cmidrule(lr){5-7} 
         & this work (mean) & this work (max) & LCFIT+Z & this work (mean) & this work (max) &  LCFIT+Z \\
        \midrule
        bias $\langle \Delta z \rangle $ & \textbf{7.8e-5} & 0.00073 &  0.023 & \textbf{0.0050 }& 0.0052 & 0.027\\
         $\sigma_{\mathrm{MAD}}$ & 0.011 & \textbf{0.010 }& 0.028 & \textbf{0.018 }& 0.020 & 0.047 \\
         outlier rate $\eta$ & \textbf{0.85\% }& 1.16\% & 19.9\% & \textbf{5.14\%} & 5.78\% & 26.1\%\\ 
        \bottomrule
    \end{tabular}
         
    \caption{Evaluation metrics computed for the SDSS simulated and real test datasets for both the mean(PDF) and max(PDF) point estimates for our model as well as LCFIT+Z. The best result for each metric and dataset is shown in bold.}
    \label{tab:sdss-results}
\end{table*}
\begin{figure*}
    \centering
    \includegraphics[scale=0.5]{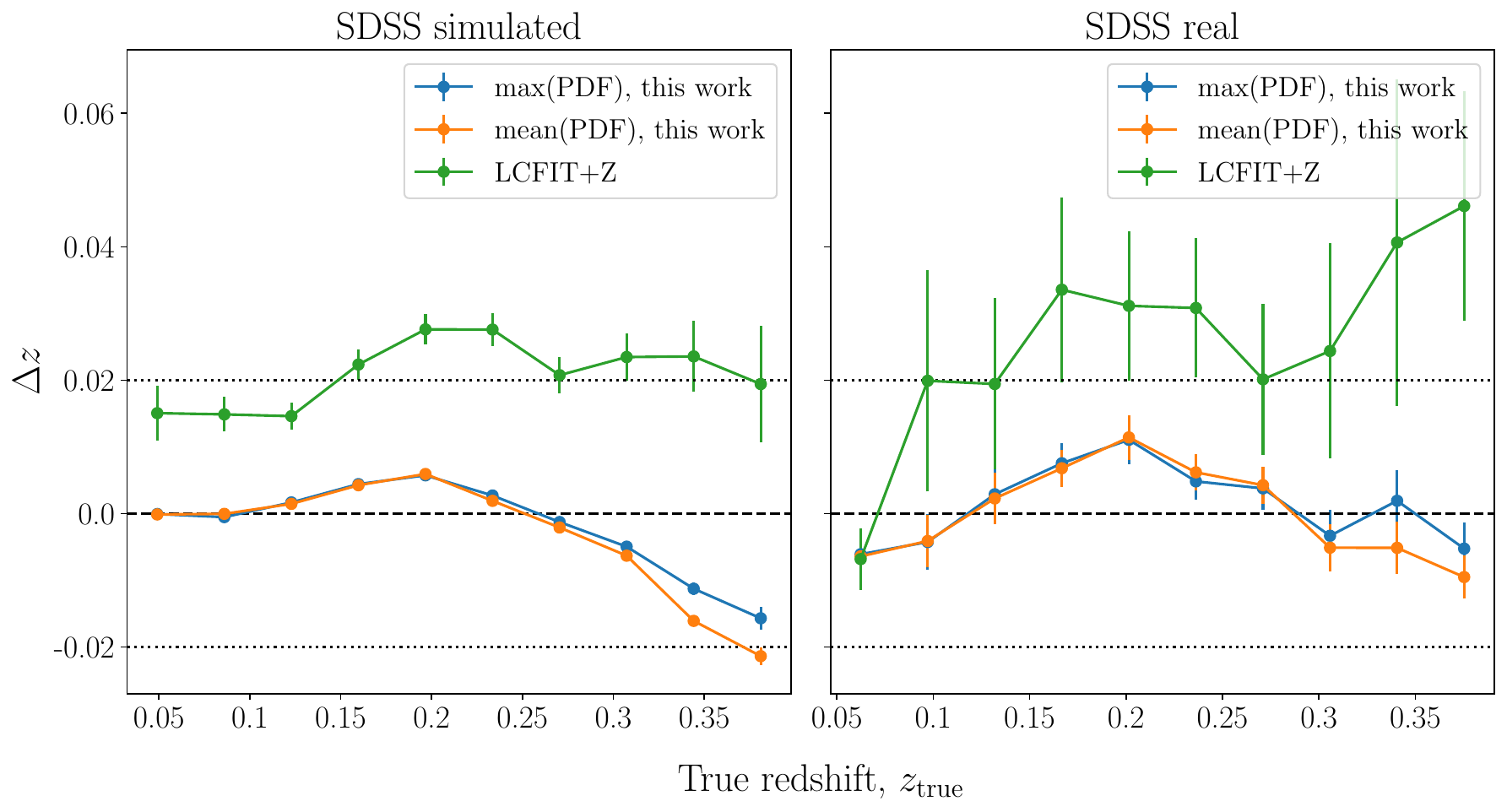}
    \caption{Mean binned residuals $\Delta z$ as a function of true redshift $z_{\mathrm{true}}$ for the SDSS simulated and real SNe Ia samples. Predictions from this work have much lower biases as well as scatter compared to predictions from LCFIT+Z. We also show that two common methods of condensing redshift PDFs into point estimates, max(PDF) and mean(PDF), resulting in similar errors for our model. Dotted lines are plotted at $\Delta z = \pm 0.02$ for reference.}
    \label{fig:sdss-delta-z}
\end{figure*}

\begin{figure*}
    \centering
    \includegraphics[scale=0.5]{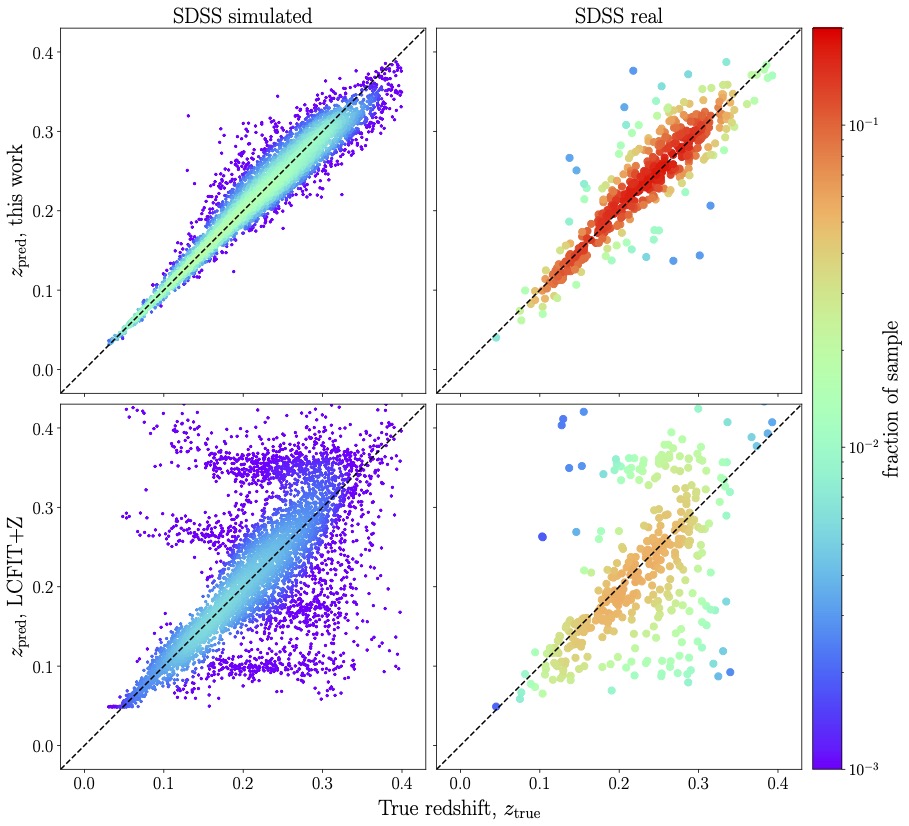}
    \caption{Predicted vs. true redshifts for the SDSS simulated and real SNe Ia samples colored by the fraction of each sample represented by each point. \textbf{(Top row)} Predictions from \name\ described in this work, \textbf{(bottom row}) predictions from LCFIT+Z. }
    \label{fig:sdss-scatter}
\end{figure*}

\begin{figure}
    \centering
    \includegraphics[scale=0.5]{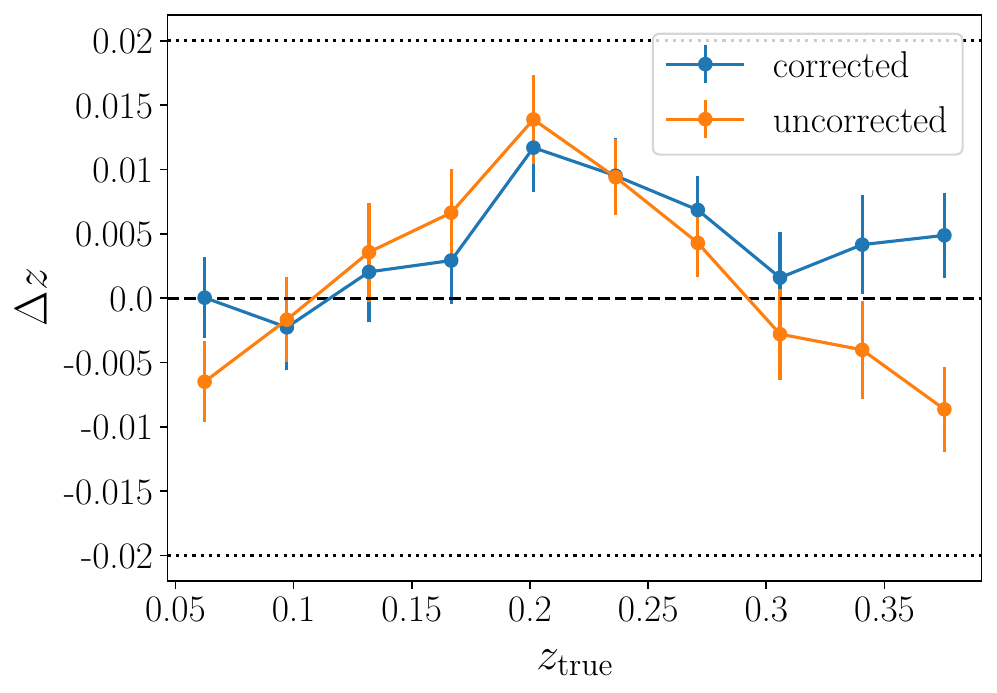}
    \caption{Predicted redshift error, $\Delta z$, as a function of true redshift, $z_{\mathrm{true}}$, for the SDSS real SNe Ia sample with and without bias correction computed from the simulated results. The point estimates used in this figure are computed using the mean(PDF) method. Dotted lines are plotted at $\Delta z = \pm 0.02$ for reference.}
    \label{fig:sdss-delta-z-corrected}
\end{figure}

We evaluate our model and LCFIT+Z on a test set of 5,274 simulated SDSS lightcurves and 489 real observed SDSS lightcurves with true redshift distributions shown in the right panel of Figure~\ref{fig:sdss-z-dist}. Table~\ref{tab:sdss-results} shows the evaluation metrics calculated with the mean and max point estimates on the simulated and observed SDSS datasets, as well as the LCFIT+Z point estimates. With the SDSS datasets, the similarity between the mean and max point estimates is more pronounced. We also note that though our model's performance degrades slightly between the simulated test set and the observed test set, all metrics still strongly favor our model as the better performer compared to LCFIT+Z. However, the $\sigma_{\mathrm{MAD}}$ values differ less between our model and LCFIT+Z compared to the PLAsTiCC test dataset, only offering a $2\times$ improvement as opposed to a $55-180\times$ improvement for the PLAsTiCC test dataset. This could be due to the narrower redshift range of SDSS, as the LCFIT+Z performance on the PLAsTiCC dataset degrades significantly after $z_{\mathrm{true}} \sim 0.4$. 

The mean binned residuals for our model and LCFIT+Z evaluated on the simulated and real SDSS datasets are shown in Figure~\ref{fig:sdss-delta-z}. The residuals from both models are much more constrained for SDSS than PLAsTiCC, as expected from the less significant differences in evaluation metrics. LCFIT+Z still exhibits larger mean $\Delta z$ values for both the simulated and real datasets as well as a larger spread in each bin. Both models generalize relatively well from simulations to real data, with our mean residuals staying within $|\Delta z| < 0.02$ and LCFIT within $|\Delta z| < 0.05$; however, an overall increase in $|\Delta z|$ values is noticeable between simulated and real data.

We show the comparison between predicted and true redshifts in Figure~\ref{fig:sdss-scatter} for the same datasets used in Figure~\ref{fig:sdss-delta-z}. Our model (top row) clearly produces more constrained predictions, as they lie much closer to the $z_{\mathrm{true}}=z_{\mathrm{pred}}$ line with a high density of points (shown in red) along the line. In contrast with the PLAsTiCC results, in which LCFIT+Z performed relatively well at lower redshifts, the performance over the full SDSS redshift range is poor. 

\subsubsection{Bias Correction}

Simulations not only allow deep learning methods such as \name\ to train on large datasets that would be infeasible with real data alone, but also give valuable estimates of the biases produced from those models that can be used to correct the results on real data. Here, we test this bias correction method on the results of \name\ on the SDSS simulated and real data. We compute the average $\Delta z$ values for the simulated SDSS sample in 10 bins of $z_{\mathrm{true}}$ values and bin the real SDSS sample in the same way. We then subtract the simulated $\Delta z$ value associated with the bin of each real $z_{\mathrm{pred}}$ estimate to produce the corrected curve in Figure~\ref{fig:sdss-delta-z-corrected}. Specifically, the corrected value of the $i^{\text{th}}$ SN in the real SDSS dataset belonging to bin $m$ is computed as 
\begin{equation}
z_{\text{corrected},i} = z_i - \langle \Delta z_{\text{sim},m} \rangle
\end{equation}
where $z_{\text{corrected},i}$ is the corrected prediction, $z_i$ is the uncorrected prediction, and $\langle \Delta z_{\text{sim},m} \rangle $ is the average $\Delta z$ from the simulated SNe in bin $m$.
The corrected $\Delta z$ values are computed as defined in \S\ref{subsec:metrics},
\begin{equation}
    \Delta z_{\text{corrected}} = \frac{z_{\text{corrected}}-z_{\text{true}}}{1+z_{\text{true}}}
\end{equation}
The corrected curve exhibits smaller biases than the uncorrected curve, showing that this bias correction method is valid in cases where the simulation is sufficiently representative of the real data.

\subsection{Further Experiments}
\subsubsection{Using \name\ photo-zs for Cosmology}
\begin{figure}
    \centering
    \includegraphics[scale=0.55]{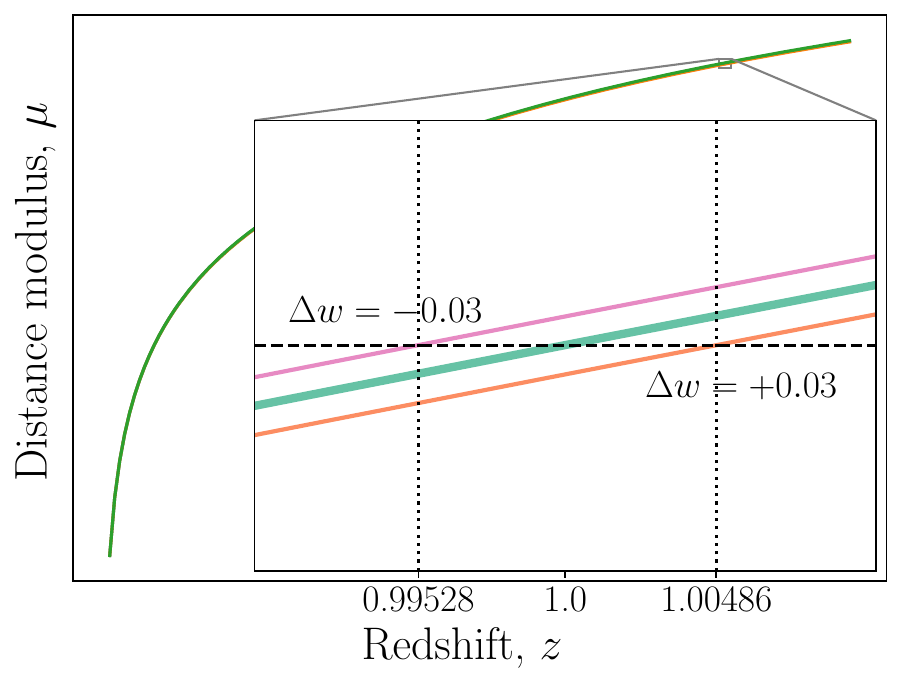}
    \includegraphics[scale=0.5]{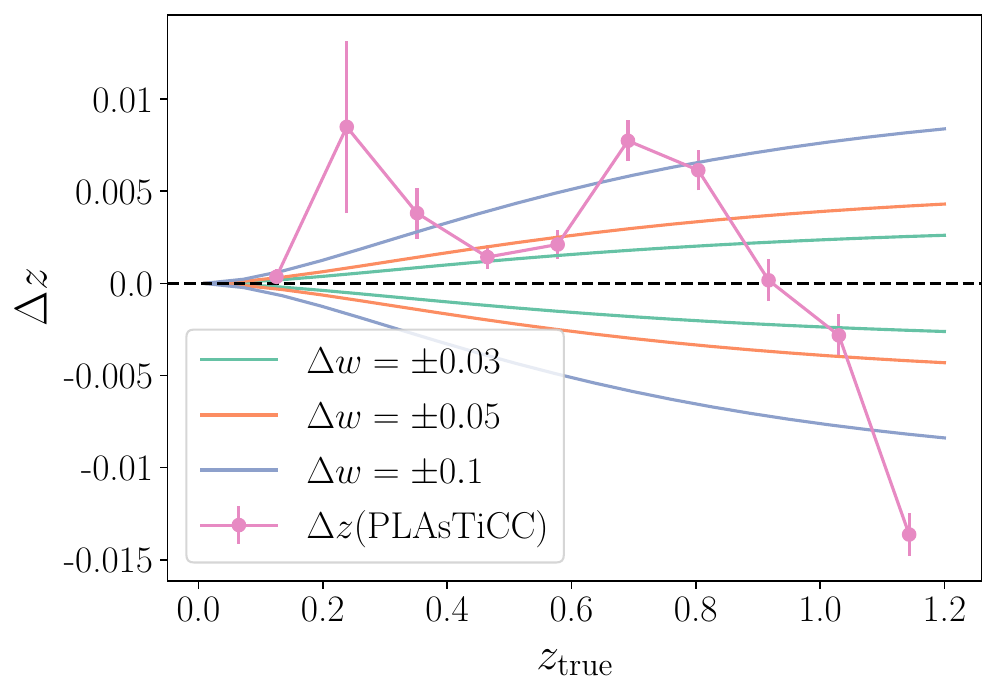}
    \caption{An illustration of our estimated cosmological biases arising from photo-$z$ errors. \textbf{(top)} A Hubble diagram zoomed in to the neighborhood of $z_{\text{model}}=1$ showing a fiducial cosmology (green: $w=-1, \Omega_m=0.3$) and two biased cosmologies (pink: $\Delta w = -0.03$, orange: $\Delta w = +0.03$). The dashed black line shows the distance modulus value at $z_{\text{model}}=1$ for the fiducial cosmology, $\mu_{\mathrm{model}}(z_{\text{model}}=1)$. The two dotted lines show the redshifts that correspond to the value of $\mu_{\mathrm{model}}(z_{\text{model}}=1)$ for the two biased cosmologies, i.e. $\mu_{\text{model}}(z_{\text{model}}=1) = \mu_{\Delta w = -0.03}(z=0.99528) = \mu_{\Delta w = +0.03}(z=1.00486)$. The values of these redshifts are labeled on the redshift axis. We approximate the redshift error required to create a bias of, e.g. $\Delta w = +0.03$, as the difference between the biased and fiducial redshift values, $d z_{\text{bias}} = 1-1.00486 = -0.00486$. \textbf{(bottom)} $\Delta z = dz / (1+z)$ values for various choices of $\Delta w$, compared to the mean binned residuals $\Delta z$ produced by the mean point estimates of \name\ PDFs (pink, reproduced from Figure~\ref{fig:plasticc-resids}).}
    \label{fig:hubble}
\end{figure}

Though producing a full cosmological analysis using photo-$z$s predicted by \name\ is outside the scope of this work, we provide some intuition for the quality of our estimates in the context of cosmology. The redshift error $dz_{\text{bias}}$ that results in a $w$ bias of $\Delta w$ at a particular redshift, $z_{\text{model}}$, can be approximated by the redshift difference 
\begin{equation}
    dz_{\text{bias}} = z_{\text{model}}-z_{\Delta w}
\end{equation} where $z_{\Delta w}$ is the redshift associated with the distance modulus $\mu_{\text{model}}(z_{\text{model}})$ for the biased cosmology, i.e.
\begin{equation}
    \mu_{\text{model}}(z_{\text{model}}) = \mu_{\Delta w}(z_{\Delta w})
\end{equation}
where $\mu_{\text{model}}, \mu_{\Delta w}$ are the distance moduli associated with the model and biased cosmologies, respectively.

A concrete example of this is illustrated in the top panel of Figure~\ref{fig:hubble}, which shows a zoomed-in portion of the Hubble diagram in the neighborhood of our chosen $z_{\text{model}}=1$ for our fiducial cosmology ($w=-1, \Omega_m=0.3$) as well as examples of biases on $w$ ($\Delta w \pm 0.03$) plotted on either side. Our choice of $z_{\text{model}}=1$ is motivated by the importance of high redshift SNe on the constraining power on $w$, so we compare with our PLAsTiCC results. We choose to focus on the $\Delta w = +0.03$ cosmology as our redshifts are slightly underestimated at $z=1$. In this example, $ z_{\Delta w = +0.03} = 1.00486$ at $z_{\text{model}} =1$ is shown as the dotted vertical line on the right.
We approximate $d z_{\text{bias}}$ for $\Delta w = +0.03$ at $z_{\text{model}} = 1$ to be \begin{equation}
    d z_{\text{bias}} = z_{\text{model}} - z_{\Delta w = +0.03} = 1 - 1.00486 = -0.00486
\end{equation} 
This corresponds to half the distance between the dotted vertical lines in the figure. The mean redshift error for our results on the PLAsTiCC dataset at $z_{\text{true}}=1$ are 
\begin{equation}
    \Delta z = \frac{z_{\text{pred}}-z_{\text{true}}}{1+z_{\text{true}}} = -0.002
\end{equation} translating to an expected redshift difference of
\begin{equation}
    z_{\text{pred}}-z_{\text{true}} = (1+z_{\text{true}}) \cdot -0.002 = -0.004
\end{equation}
This is below the expected $d z_{\text{bias}}$ associated with a $w$ shift of $\Delta w = +0.03$.

The bottom panel of Figure~\ref{fig:hubble} shows $\Delta z$ values for different choices of cosmological biases (here we choose to vary $w$) across the full LSST redshift range, along with the $\Delta z$ values produced by \name\ on the PLAsTiCC dataset. Note that the $y$ axis of this plot is $\Delta z = dz / (1+z)$, i.e. $dz_{\text{bias}}$ normalized by $(1+z_{\text{model}})$, for ease of comparison with \name\ mean residual $\Delta z$ values. This comparison over all redshifts shows that \name\ redshifts lack the precision to constrain cosmology to $\Delta w = \pm 0.03$, since the pink line representing \name\ redshift errors mostly does not lie in the area within the $\Delta w = \pm 0.03$ teal lines. However, this high-level analysis is an overestimate of the expected impact on cosmology, and a thorough cosmological analysis with \name\ photo-$z$'s will be the subject of a future work.

\subsubsection{Comparison with LCFIT+Z with Host Galaxy Redshift Prior}
\label{subsec:with-prior}

In \S\ref{subsec:plasticc} and \S\ref{subsec:sdss}, we showed results for \name\ and LCFIT+Z with no host galaxy information included. \name\ was formulated not to require a host redshift prior in order to prevent biases due to incorrect host redshifts. However, in a context in which host mismatches are rare and host redshifts are reliable, we want to use all available information to produce the most constrained SN photo-$z$ estimates.

We test both \name\ and LCFIT+Z with a prior on $z_{\text{pred}}$, the predicted redshift, given by the host galaxy photometric redshift estimate. We choose to use photometric redshifts as opposed to spectroscopic to more closely emulate a future scenario in which most host galaxies will not have spectroscopic information available.

We model this prior, $P(Z_{\text{host}})$, as a Gaussian centered on the photo-$z$ of the host galaxy, $z_{\text{host}}$, and use the estimated uncertainty, $\sigma_{z_{\text{host}}}$, as the standard deviation:
\begin{equation}
    Z_{\text{host}} \sim \mathcal{N}(z_{\text{host}}, \sigma_{z_{\text{host}}})
\end{equation}
We treat \name\ PDFs, $P(Z_{\text{Photo-$z$SN}})$, as Bayesian posteriors and apply the host prior using Bayes' theorem: 
\begin{equation}
\begin{aligned}
    P(Z_{\text{pred}}) = \frac{P(Z_{\text{Photo-$z$SN}} | Z_{\text{host}}) P(Z_{\text{host}})}{P(Z_{\text{Photo-$z$SN}})}\\
    = \frac{P(Z_{\text{Photo-$z$SN}})P(Z_{\text{host}})}{\sum_i P(Z_{\text{Photo-$z$SN},i})P(Z_{\text{host},i})}
\end{aligned}
\end{equation}
simplified by the fact that $P(Z_{\text{host}})$ and $P(Z_{\text{Photo-$z$SN}})$ are statistically independent.

LCFIT+Z uses a Markov chain Monte Carlo (MCMC) process to sample from the posterior distribution over a 5-dimensional parameter space: 4 SALT fit parameters color $c$, stretch-luminosity parameter $x_1$, time of peak brightness $t_0$, flux normalization parameter $x_0$; and redshift $z_{\text{phot}}$. The host prior constrains the search space, resulting in a higher likelihood of convergence to a global minimum.

\begin{figure}
    \centering
    \includegraphics[scale=0.47]{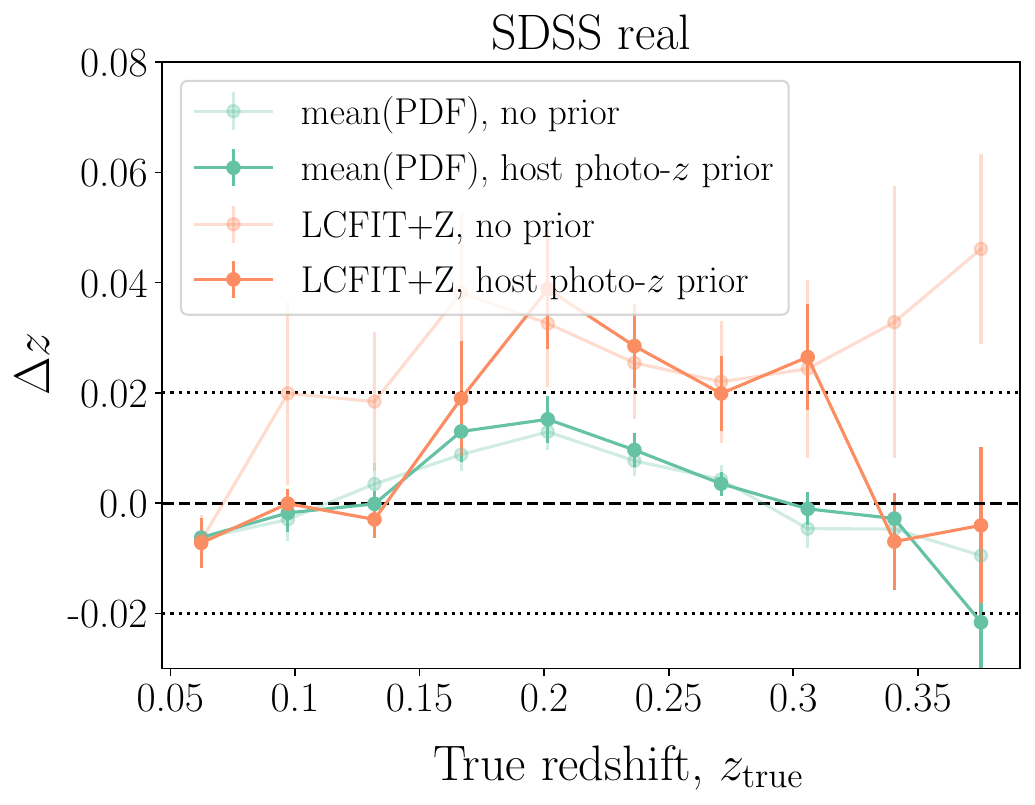}
    \caption{Mean binned residuals, $\Delta z \equiv \frac{z_{\mathrm{pred}} - z_{\mathrm{true}}}{1+z_{\mathrm{true}}}$, as a function of true redshift, $z_{\mathrm{true}}$, for the SDSS observed photometric SNe Ia sample. Errors from both \name\ and LCFIT+Z are shown with (darker) and without (lighter) a prior on $z_{\text{pred}}$ from the host galaxy photometric redshift. The lighter (no prior) curves are identical to those in Figure~\ref{fig:sdss-delta-z}, with the max(PDF) curves omitted for clarity.}
    \label{fig:sdss-delta-z-prior}
\end{figure}

In Figure~\ref{fig:sdss-delta-z-prior}, we show $\Delta z$ values for the SDSS photometric SNe Ia dataset for \name\ and LCFIT+Z with and without a host photo-$z$ prior. The $\Delta z$ results with no host prior are identical to those in the right panel of Figure~\ref{fig:sdss-delta-z}. The host prior noticeably improves LCFIT+Z results on $z_{\text{true}} < 0.2$ and $z_{\text{true}} > 0.3$, as well as the scatter across the full redshift range, shown in the smaller error bars. The \name\ results with and without host prior are very similar, however, as the host prior is often less constraining than the \name\ PDFs themselves. We show the bias, $\sigma_{\text{MAD}}$, and outlier rate for both models with and without the host prior in Table~\ref{tbl:sdss-prior}. These results show that \name\ outperforms LCFIT+Z even in the presence of a host galaxy redshift prior.

\begin{table}
    \centering
    \begin{tabular}{l c c c c}
        \toprule
        Metric & \multicolumn{2}{c}{this work} & \multicolumn{2}{c}{LCFIT+Z}\\
        \cmidrule(lr){2-3}  \cmidrule(lr){4-5}
        & no prior & with prior & no prior & with prior \\
        \midrule
         bias $\langle \Delta z \rangle$ & \textbf{0.0050} & 0.0061 & 0.027 & 0.021 \\
         $\sigma_{\mathrm{MAD}}$ & 0.018 & \textbf{0.017 }& 0.057 & 0.025\\
         outlier rate $\eta$ & 5.14\% & \textbf{4.93\% }& 26.1\% & 17.6\%\\
         \bottomrule
    \end{tabular}
    \caption{Evaluation metrics computed for the SDSS observed photometric SNe Ia sample both with and without a host galaxy photo-$z$ prior. The best results for each metric and dataset are shown in bold.}
    \label{tbl:sdss-prior}
\end{table}

%% file: chapters/photoz/conclusion.tex
In this work we presented \name, a convolutional neural network model that predicts full redshift PDFs from multi-band photometric SNe Ia lightcurves. We evaluated its performance on simulated SDSS and LSST lightcurves, as well as a photometrically confirmed SDSS SN Ia sample. We compared our results against LCFIT+Z, the most frequently used photometric redshift estimation method for SNe, and showed superior performance across all evaluation metrics. Our model also exhibits minimal redshift-dependent bias, which has plagued redshift estimators in the past, and generalizes well between simulated and observed data.

Though \name\ does not require host galaxy information to produce accurate SN photo-$z$ estimates, host galaxy redshifts can be incorporated as a prior to further improve our predictions. We tested \name\ and LCFIT+Z with a host galaxy photometric redshift prior to demonstrate the performance of these models in the LSST era when host galaxy spectroscopic redshifts will not be widely available. We show that while the host galaxy prior improves LCFIT+Z estimates, \name\ still produces more accurate photo-$z$ point estimates as well as constrained PDFs.

We envision \name\ to be useful for many tasks in supernova science, including precise volumetric rate calculations, discovery of incorrect host galaxy matches via redshift discrepancies, and photometric SN Ia cosmology. We briefly explored the cosmological constraints that can be expected from \name\ photo-$z$s and concluded that the bias on $w$ may be on the order of $\Delta w \sim 0.1$.

In future work, we intend to develop a framework for incorporating redshift PDFs and their associated uncertainties into the Hubble diagram and produce more accurately modeled cosmological constraints from \name\ redshift PDFs. We also plan to explore the use of \name\ redshift predictions as a method of identifying host confusion and potential mismatches for LSST. Though \name\ was developed for SNe Ia with cosmological applications in mind, the data processing and model architecture can be generalized for use with any astronomical time-domain events. 

Accurate photometric redshift estimation for SNe Ia will become vital in the imminent era of photometric SN Ia cosmology. We believe that the approach and model presented here will allow us to maximize the constraining power of these new datasets.

%% file: chapters/photoz/appendix.tex
\begin{figure}[h!]
    \centering
    \includegraphics[scale=0.47]{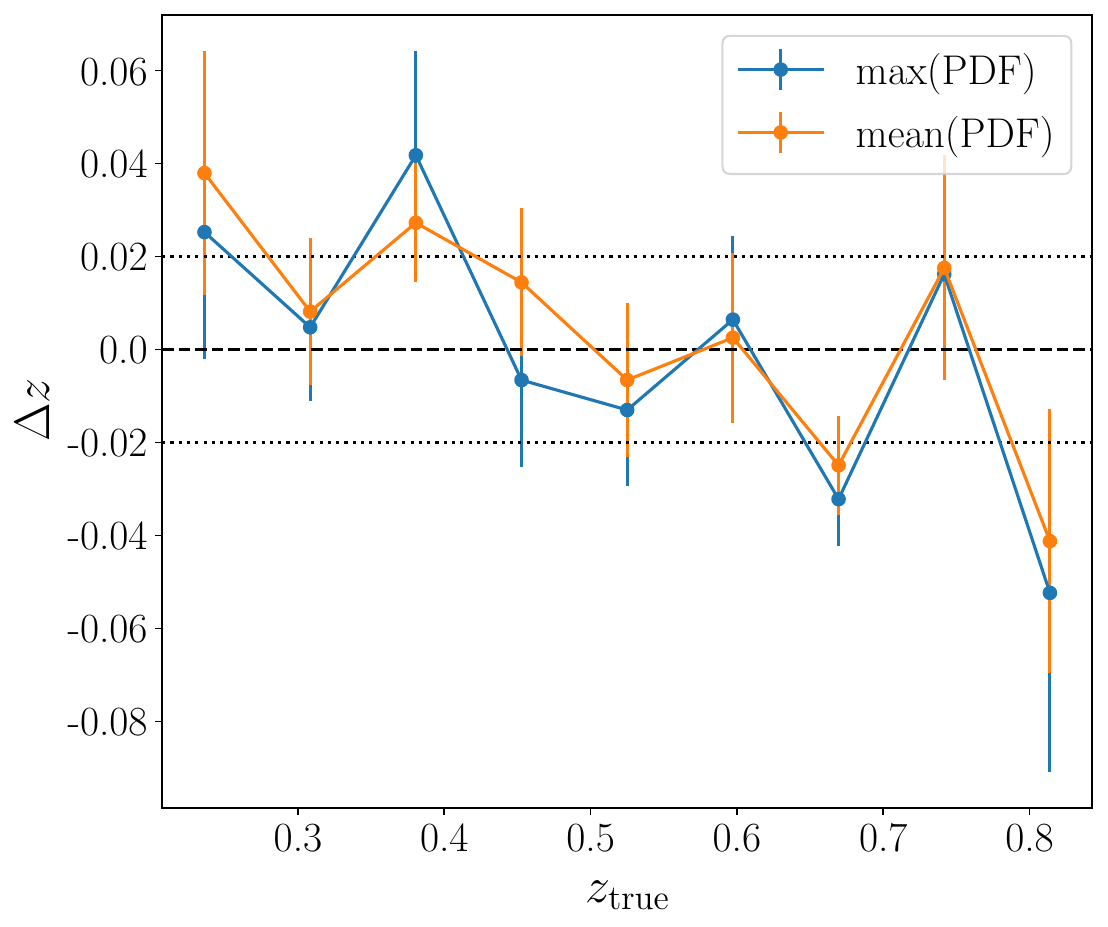}
    \caption{Mean binned residuals, $\Delta z \equiv \frac{z_{\mathrm{pred}} - z_{\mathrm{true}}}{1+z_{\mathrm{true}}}$, as a function of true redshift, $z_{\mathrm{true}}$, for the DES3YR SNe Ia sample produced by a model trained on the PLAsTiCC dataset. The max(PDF) and mean(PDF) methods of obtaining point estimates from \name\ PDFs are described in \S\ref{subsec:metrics}.}
    \label{fig:des-test}
\end{figure}

Due to the 2D Gaussian process regression in both wavelength and time dimensions that is performed in order to create the input images (details in \S\ref{subsec:lc-preprocess}), data from different surveys with different photometric bands can be processed into the same image format. This creates the potential for a survey-agnostic model that can be trained on data from a single survey and applied to data from others. To test this hypothesis, we evaluated our model trained on PLAsTiCC data, described in \S\ref{subsec:plasticc}, on 248 spectroscopically confirmed SNe Ia from the first 3 years of the Dark Energy Survey supernova program \citep[DES3YR,][]{Brout_2019}. We show these results in Figure~\ref{fig:des-test}.

The scatter and redshift-dependent bias are certainly more prominent in this result than the others presented in this work (Figures~\ref{fig:plasticc-resids} and~\ref{fig:sdss-delta-z}), but is nevertheless an impressive result considering that this 2D Gaussian process regression method uniquely makes cross-survey results possible. We also highlight that these observed SNe Ia cover a much larger redshift range than the SDSS observed sample, demonstrating the applicability of \name\ on high-redshift, deep sky surveys such as LSST. While training a fresh model with a full suite of Dark Energy Survey SNe Ia simulations would certainly produce improved results, we posit that improvements could be attained relatively inexpensively by "fine-tuning" an existing trained model using a small volume of SNe Ia simulated in the target survey. We leave these interesting extensions to future work.

\section{Redshift Prediction Outliers of the Real SDSS Dataset}
\begin{figure}
    \centering
    \includegraphics[scale=0.47]{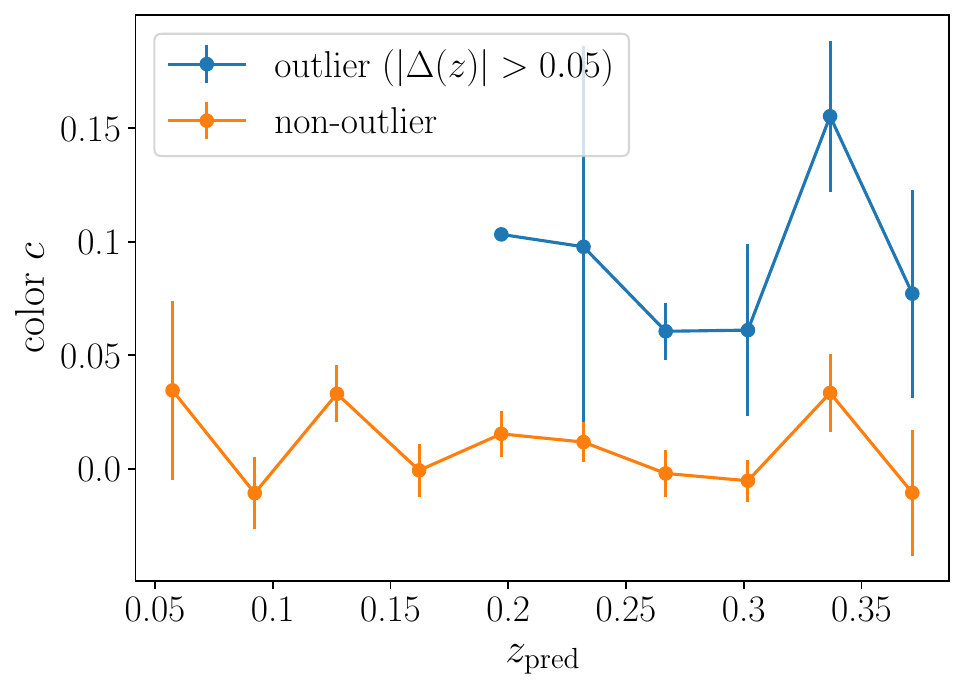}
    \caption{Mean binned SALT2 color, $c$, as a function of predicted redshift, $z_{\mathrm{pred}}$, for the \name\ prediction outliers and non-outliers in the SDSS observed photometric SN Ia sample.}
    \label{fig:outliers}
\end{figure}

We investigate the 29 \name\ prediction outliers, defined as objects with $\Delta z > 0.05$, in the observed SDSS dataset. We find that the population of outliers is noticeably redder than their non-outlier counterparts at the same predicted redshift (Figure~\ref{fig:outliers}), which could explain the prediction failure.

We also attempted to use the predicted PDF shapes to remove outliers, assuming that poor predictions may be correlated with wider PDFs, representing the model's lack of confidence in less accurate predictions. Although we found a slight correlation, it is not strong enough for this metric to be useful: 48\% of outliers have "wide" PDFs (defined as $\sigma > 0.025$ empirically) as well as 18\% of non-outliers.

%% file: chapters/connect_later.tex
\section*{Abstract}
Models trained on a labeled source domain (e.g., labeled images from wildlife camera traps) often generalize poorly when deployed on an out-of-distribution (OOD) target domain (e.g., images from new camera trap locations). In the domain adaptation setting where unlabeled target data is available, self-supervised pretraining (e.g., masked autoencoding or contrastive learning) is a promising method to mitigate this performance drop. Pretraining improves OOD error when the generic data augmentations used (e.g., masking or cropping) connect the source and target domains, which may be far apart in the input space. In this paper, we show on real-world tasks that standard fine-tuning after pretraining does not consistently improve OOD error over simply training from scratch on labeled source data. To better leverage pretraining for distribution shifts, we propose Connect Later: after pretraining with generic augmentations, fine-tune with \emph{targeted augmentations} designed with knowledge of the distribution shift. Pretraining learns good representations within the source and target domains, while targeted augmentations connect the domains better during fine-tuning. Connect Later improves average OOD error over standard fine-tuning and supervised learning with targeted augmentations on 4 real-world datasets: Connect Later achieves the state-of-the-art on astronomical time-series classification (\classification) by 2.5\%, wildlife species identification (\iwildcam) with ResNet-50 by 0.9\%, and tumor identification (\camelyon) with DenseNet121 by 1.1\%; as well as best performance on a new dataset for astronomical time-series redshift prediction (\redshifts) by 0.03 RMSE (11\% relative).\footnote{Code and datasets are available at \texttt{https://github.com/helenqu/connect-later}.}

\section{Introduction}

In many real-world scenarios, machine learning models are deployed on data that differ significantly from the training data \citep{quinonero2009dataset, koh2021wilds}. We focus on unsupervised domain adaptation \citep{shimodaira2000improving,blitzer2006domain,sugiyama2007covariate}, where we have labeled data from a source domain and unlabeled data from a target domain. We aim to learn a model that generalizes well to these out-of-distribution (OOD) target domain inputs.
A real-world example is wildlife identification, where the task is to identify animal species from static camera trap images.
However, human labels are only available for images from a small subset of these cameras.
Images from labeled cameras may not be representative of the habitats and characteristics of unlabeled camera images.

Pretraining on broad unlabeled data has shown promising results on improving OOD error in real-world problems\citep{caron2020swav, shen2022connect, radford2021clip, sagawa2022uwilds}.
In particular, contrastive pretraining has been shown to learn representations that transfer well across domains~\citep{shen2022connect,haochen2022beyond}.
In contrast to conventional domain adaptation methods that focus on learning domain-invariant features~\citep{ganin2016domain,kang2019contrastive,tzeng2017domain,saenko2010adapting,sun2016return,hoffman2018cycada}, contrastive pretraining learns representations that are not domain-invariant, but instead decompose the class and domain information, facilitating transfer across domains~\citep{shen2022connect}. A favorable decomposition depends on the generic data augmentations used during contrastive pretraining to connect the source and target domains in a structured way.
Intuitively, augmented (e.g. masked or cropped) source and target inputs are more likely to look similar if they are from the same class (e.g., cropping out the face of a lion in different habitats) than from different classes (e.g., no body parts of elephants and lions are alike).
However, these generic augmentations must connect the domains without knowledge of the distribution shift.

In this paper, we find on real-world benchmarks that standard fine-tuning after contrastive pretraining is not always effective for improving OOD error over purely supervised learning from scratch with labeled source data (Section~\ref{sec:pretraining}).
On the other hand, supervised learning with \emph{targeted augmentations} \citep{gao2023targeted} designed for the distribution shift improves OOD error over the supervised learning baseline on all datasets without access to any target unlabeled data.
Thus, pretraining does not always learn representations that transfer across domains with standard fine-tuning.

\begin{figure}
    \centering
    \includegraphics[scale=0.2]{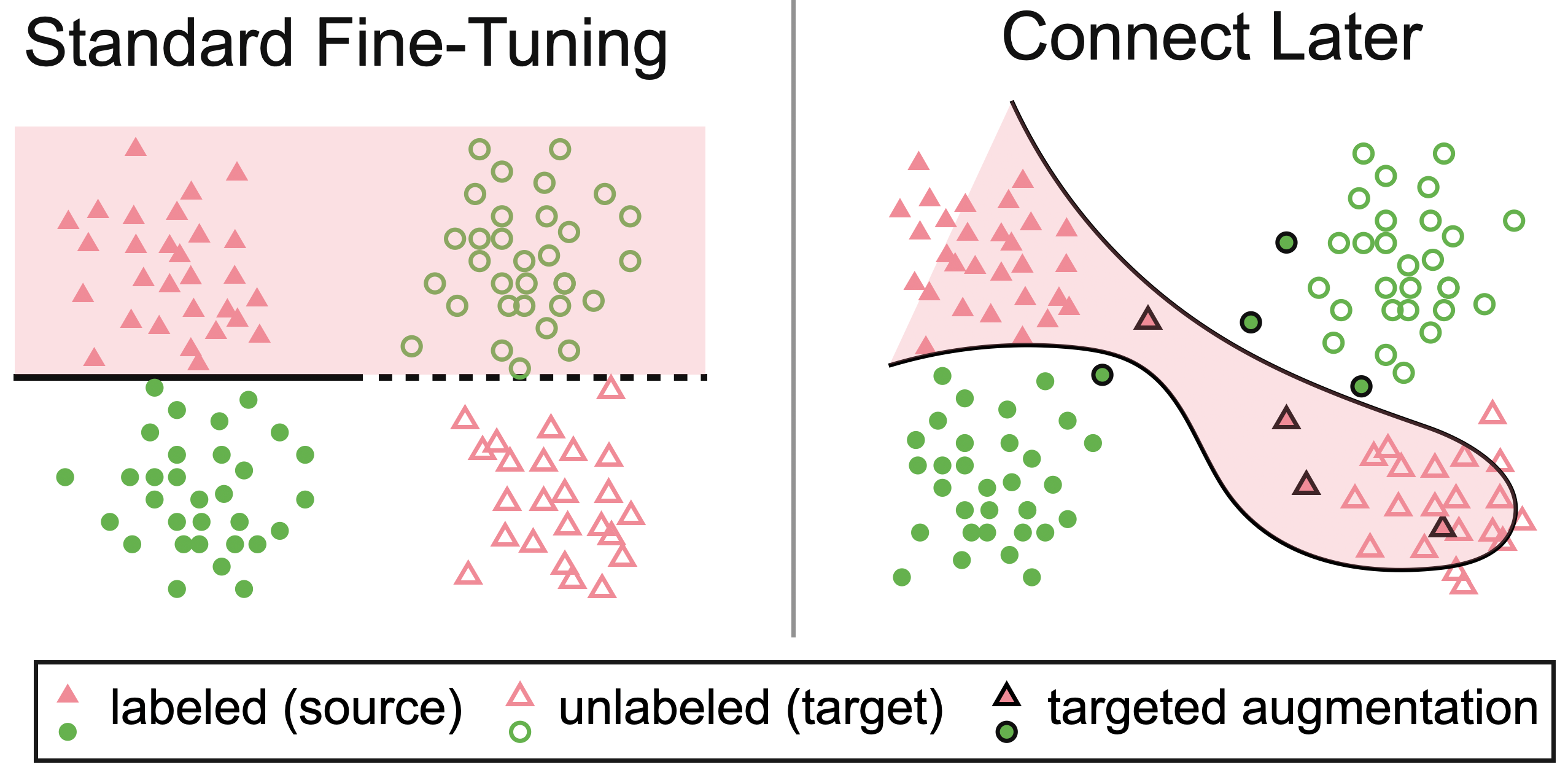}
    \caption{Overview of the Connect Later framework for domain adaptation applied to a toy binary classification problem with two domains (filled and unfilled points), showing the representations from contrastive pretraining in $\R^2$. \textbf{(Left)} After contrastive pretraining with generic augmentations, the classes within each domain are linearly separable in representation space. Since the domains are far apart in input space, generic augmentations may not connect the domains, resulting in misalignment in the pretrained representations. In this case, a classifier (with a linearly extrapolating decision boundary, dashed and solid line) learned on labeled source data will misclassify the target data. \textbf{(Right)} Connect Later employs targeted augmentations (filled points with black border), which are designed with knowledge of the distribution shift, to connect the domains better, resulting in a classifier that generalizes well to the target domain.}
    \label{fig:overview}
\end{figure}

To better leverage pretraining for domain adaptation, we propose the Connect Later framework (Figure~\ref{fig:overview}): after pretraining with generic augmentations, fine-tune with targeted augmentations (Section~\ref{sec:connect-later}).
Intuitively, pretraining learns good representations within each domain, while targeted augmentations better connect the domains.
On a simple theoretical example where contrastive pretraining fails, Connect Later generalizes well to the target domain.
We provide a general methodology for constructing these targeted augmentations, where the augmented inputs match the target distribution on a feature space where the domains differ.

We evaluate our framework on 4 real-world datasets: wildlife identification \citep[\iwildcam,][]{beery2020iwildcam, sagawa2022uwilds}, tumor detection \citep[\camelyon,][]{bandi2018detection, sagawa2022uwilds} and 2 astronomical time series tasks, \classification and \redshifts, which we curate from \citet{theplasticcteam2018photometric}.
In Section~\ref{sec:results}, we show that Connect Later improves ID and OOD performance over \sft or supervised learning with targeted augmentations across all datasets.
Although our understanding stems from contrastive learning, we empirically apply Connect Later to better leverage pretrained representations from both masked autoencoding and contrastive learning.
Connect Later achieves the state-of-the-art on three benchmarks, improving OOD accuracy on \classification by 3\% \citep{Boone_2019}, \iwildcam with ResNet-50 by 0.9\%, and \camelyon with DenseNet121 by 1.1\%. We also contribute the \redshifts dataset, on which Connect Later produces 11\% and 14\% relative improvement over \sft and supervised learning with targeted augmentations, respectively. 

\section{Setup}
We consider a prediction problem from an input space $\sX$ to a label space $\sY$, where $\sY = \{1,\dots,k\}$ for classification and $\sY \in \R$ for regression.

\paragraph{Domain adaptation.}
Let $P_S$ and $P_T$ be the source and target input distributions over $\sX$, respectively.
We consider unsupervised domain adaptation, where we have access to source inputs $x \sim P_S$, with corresponding labels $y\in\sY$ sampled from the label distribution $\pstar(\cdot \mid x)$, along with unlabeled target inputs sampled from the target distribution $P_T$. 
Let the unlabeled distribution $P_U = \beta P_S + (1-\beta) P_T$ be a mixture of the source and target, where $\beta \in [0,1]$. In practice, $P_U$ may also be a broader unlabeled distribution.
The goal is to learn a model $f: \sX \rightarrow \sY$ that minimizes error on the target domain $L_T(f) = \E_{x\sim P_T, y\sim \pstar(\cdot \mid x)}[\loss(f(x), y)]$. For example, $\loss:\sY\times \sY \rightarrow \R$ is the 0-1 loss in classification and squared loss in regression.

\paragraph{Augmentations.}
Augmented inputs $\inputxp \in \sX$ are drawn from an augmentation distribution $\sA(\cdot | x)$, given an input $x \in \sX$. 
Training with augmented inputs is often used to improve robustness \citep{hendrycks2019augmix, hendrycks2020many} and is a crucial part of the objective in contrastive pretraining \citep{caron2020swav,shen2022connect, devlin2019bert}.
In this work, we define two distinct augmentation distributions, $\aug$ and $\augft$, for the pretraining and fine-tuning steps, respectively. 
Typically, the pretraining augmentations $\aug$ are generic transformations, such as random cropping in vision or masking in NLP \citep{caron2020swav,chen2020simclr,he2020moco,radford2021clip,shen2022connect,he2022mae,devlin2019bert}. 
Fine-tuning augmentations $\augft$ have not been studied extensively and are typically also generic or simply the identity transformation \citep{sagawa2022uwilds, devlin2019bert}.

\paragraph{Contrastive pretraining for domain adaptation.}
Contrastive pretraining for domain adaptation consists of two steps: self-supervised pretraining on unlabeled data, then supervised fine-tuning on labeled source data \citep{shen2022connect}.
For simplicity below, we consider the population objectives.
Contrastive learning aims to learn an encoder which maps augmented views of the same input to similar features (``positive pairs'') and views of different inputs to dissimilar features (``negative pairs''), according to some distance metric. Formally, let $\pospairdist(\inputx, \posx)=\E_{\bar{\inputx}\sim \unlabeldist}[\aug(\inputx \mid \bar{\inputx})\aug(\posx \mid \bar{\inputx})]$ be the distribution over positive pairs, which are augmentations of a single input $\bar{\inputx}$. 
We pretrain an encoder $\encoder:\sX \rightarrow \R^\embeddim$ to minimize the distance $\dattract$ between positive pair embeddings and maximize the distance $\drepel$ between negative pair embeddings:
\begin{align}
\label{eqn:pretrain_objective}
\sL_{\text{pretrain}}(\encoder) = \E_{(\inputx,\posx)\sim \pospairdist}[\dattract(\encoder(\inputx), \encoder(\posx))] - \E_{\inputx,\inputxp\sim \unlabeldist}[\drepel(\encoder(\inputx), \encoder(\inputxp))].
\end{align}
The output of the pretraining step is a pretrained encoder $\empencoder=\argmin_\encoder \sL_{\text{pretrain}}(\encoder)$.


Fine-tuning then learns a prediction head $\head:\R^\embeddim\rightarrow \R^n$ 
(for regression, we let $n=1$) on top of the pretrained encoder using labeled source data with the objective
\begin{align}
\label{eqn:ft_objective}
    \sL_{\text{ft}}(\head) = \E_{\inputx \sim P_S, y\sim \pstar(\cdot \mid \inputx), \inputxp \sim \augft(\cdot | \inputx)} [\lossft(\head(\empencoder(\inputxp)),\;y;\; \theta)]
\end{align}
where $\lossft: \R^n \times \sY \rightarrow \R$ is a fine-tuning objective such as softmax cross entropy loss for classification or squared error for regression. The learned head is $\emphead=\argmin_{\head} \sL_{\text{ft}}(\head)$. In practice, we jointly fine-tune the head $\head$ and the encoder $\empencoder$.

\paragraph{\Sft.}
We refer to \textbf{\sft} as the pretraining+fine-tuning procedure where $\augft(x'\mid x)=1$ if $x'=x$ (no fine-tuning augmentations).
In our experiments, the standard fine-tuning procedure is linear probing then fine-tuning (LP-FT)~\citep{kumar2022finetuning}, which has been shown to improve ID and OOD performance over vanilla fine-tuning. In LP-FT, we first learn a linear predictor on top of frozen pretrained features before fine-tuning all the parameters jointly.

\paragraph{ERM with augmentations.}
As a baseline, we consider empirical risk minimization (ERM) with data augmentation, which optimizes the fine-tuning objective (Equation~\ref{eqn:ft_objective}) on labeled source data with randomly initialized parameters.
In this paper, we refer to \textbf{ERM} as the instantiation where $\augft(x'\mid x) = 1$ if $x'=x$ (no augmentations) and \textbf{ERM + targeted augmentations} as the instantiation with $\augft$ that is designed with knowledge of the distribution shift.


\section{Pretraining may not improve OOD performance}
\label{sec:pretraining} 

\begin{table}[]
    \centering
        \caption{Contrastive pretraining with standard fine-tuning substantially improves OOD performance for \camelyon but is not as effective for \iwildcam. Results are averaged over 15 trials for \iwildcam and 20 trials for \camelyon, and we report the 95\% confidence intervals on each mean estimate.}
        \begin{tabular}{lcccc}
            \toprule
            & \multicolumn{2}{c}{iWildCam (Macro F1, $\uparrow$)} & \multicolumn{2}{c}{Camelyon17 (Avg Acc, $\uparrow$)} \\
            & ID Test  & OOD Test & ID Val & OOD Test \\
            \midrule
            ERM & $46.4 \pm 0.5$ & $30.4 \pm 0.6$ & $89.3 \pm 0.9$ & $65.2 \pm 1.1$\\
            \Sft & $46.4 \pm 0.8$ & $31.2 \pm 0.6$ & $92.3 \pm 0.2$ & $91.4 \pm 0.9$ \\
            \bottomrule
        \end{tabular}
    \label{tbl:observations}
\end{table}

We compare ERM and \sft on two benchmark datasets, \iwildcam (wildlife species identification) and \camelyon (tumor detection).
In Table~\ref{tbl:observations}, we show that \sft on a model pretrained using SwAV contrastive learning~\citep{caron2020swav} makes minimal gains over ERM on \iwildcam ($46.4 \rightarrow 46.4$ ID, $30.4 \rightarrow 31.2$ OOD) compared to \camelyon ($89.3 \rightarrow 92.3$ ID, $65.2 \rightarrow 91.4$ OOD).
This result runs contrary to empirical results demonstrating that contrastive pretraining is an effective domain adaptation method \citep{caron2020swav, shen2022connect, radford2021clip, sagawa2022uwilds}.
We hypothesize that the generic pretraining augmentations connect the domains better for some tasks and distribution shifts than others.

\paragraph{Simple example with misaligned connectivity structure.}
To understand this phenomenon, we provide a simple binary classification example of when contrastive pretraining fails for domain adaptation, following a similar augmentation graph construction to~\citet{shen2022connect}, in Appendix~\ref{app:simple_example}.
When the connectivity structure misaligns the source and target domains, such that examples from the same class are less ``connected'' than examples from different classes across the domains, a linear classifier trained on these pretrained representations will not transfer from source to target.
This could happen, for example, when the source and target are far apart in input space and connectivity is low between examples from the same class across different domains.

\subsection{Robustness gains from pretraining depend on dataset connectivity}
To test this hypothesis, we empirically measure connectivity as defined in \citet{shen2022connect}.
We follow \citep{shen2022connect} and work in the augmentation graph setting, where nodes are inputs and edge weights are the positive-pair probabilities given by $\pospairdist$. 
We define connectivity between a class-domain pair $((y_1, d_1), (y_2, d_2))$ under four scenarios:
\begin{align}
    \begin{cases}
        \pairprobr & y_1 = y_2, d_1 = d_2~\text{~~(same class, same domain)}\\
        \pairproba & y_1 = y_2, d_1 \neq d_1\text{~~(same class, different domain)}\\
        \pairprobb & y_1 \neq y_2,d_1 = d_2\text{~~(different class, same domain)}\\
        \pairprobg & y_1 \neq y_2, d_1 \neq d_2\text{~~(different class and domain)}\\  
    \end{cases},
\end{align}
where each value is an average edge weight over the edges that satisfy each case.
\citet{shen2022connect} show in simple augmentation graphs that contrastive pretraining theoretically learns transferable representations when $\pairproba > \pairprobg$ and $\pairprobb > \pairprobg$, and that the ratios $\frac{\pairproba}{\pairprobg}$ and $\frac{\pairprobb}{\pairprobg}$ empirically correlate well with OOD accuracy.
Intuitively, the pretraining augmentations are less likely to change both the domain and class of an input than changing just domain or just class.

\begin{table}[]
    \centering
    \caption{Empirically estimated connectivity measures for \iwildcam and \camelyon. From~\citet{shen2022connect}, contrastive pretraining theoretically learns transferable representations for UDA when both across-domain ($\alpha$) and across-class ($\beta$) connectivity is greater than across-both ($\gamma$), using notation from ~\citep{shen2022connect}. In \iwildcam, across-both connectivity $>$ across-class, which violates the condition, while \camelyon satisfies the condition.}
    \begin{tabular}{lccc}
        \toprule
         & across-domain ($\alpha$) & across-class ($\beta$) & across-both ($\gamma$) \\ \midrule
        \iwildcam & 0.116 & 0.071 & 0.076 \\
        \camelyon & 0.16 & 0.198 & 0.152 \\ 
        \bottomrule
    \end{tabular}
    \label{tbl:connectivity}
\end{table}

\paragraph{Empirical evaluations of connectivity.}
To better understand why contrastive pretraining performs differently on \iwildcam and \camelyon, we empirically evaluate the connectivity measures for \iwildcam and \camelyon, following \citet{shen2022connect}. 
Using augmented inputs from 2 class-domain pairs, we train a binary classifier to predict the class-domain pair of each input, and interpret the test error of the classifier as an estimate for connectivity.
We average each connectivity value over 15 class-domain pairs (see Appendix~\ref{app:connectivity} for details).
Our results, summarized in Table~\ref{tbl:connectivity}, show that \iwildcam connectivity measures violate the condition for contrastive pretraining in the UDA setting, since across-both connectivity $>$ across-class ($\gamma > \beta$). 
This finding is consistent with our observation that contrastive pretraining fails for \iwildcam while producing gains for \camelyon, and further underscores the need for domain adaptation methods that correct the misaligned connectivity structure.

\section{Connect Later: Pretrain First, Targeted Augmentations Later}
\label{sec:connect-later}

Even when generic augmentations applied during pretraining misalign the connectivity structure, the pretrained representations are still useful since the classes are linearly separable \textit{within} each domain.
How do we leverage these pretrained representations when they may not transfer well across domains?
In this work, we propose the Connect Later framework (Figure~\ref{fig:overview}):
\begin{enumerate}
  \item Pretrain on unlabeled data with generic augmentations as in Equation~\ref{eqn:pretrain_objective}, producing a pretrained encoder $\empencoder$. 
  \item Design a targeted augmentation $\augft$ (discussed below) and use augmented source data to fine-tune the pretrained encoder $\empencoder$ jointly with a prediction head $h$ as in Equation~\ref{eqn:ft_objective}.
\end{enumerate}
  Intuitively, pretraining learns good representations of the source and target domains and allows us to reuse the pretrained model for multiple downstream tasks, while targeted augmentations better connect the domains for the particular task.
\paragraph{Simple example where Connect Later achieves 0 OOD error.} In our simple binary classification example in Appendix~\ref{app:simple_example}, we show that when the connectivity structure is misaligned, both \sft with contrastive pretraining and ERM + targeted augmentations have high OOD error, while Connect Later achieves 0 OOD error.
In this setting, ERM with targeted augmentations is unable to achieve 0 OOD error since some target inputs are ``unreachable'' via targeted augmentations of source inputs. The pretraining step in Connect Later uses unlabeled target data to learn representations where label information from source data can propagate to all target inputs.

\subsection{Real-world examples of targeted augmentations} 
\begin{figure}
    \centering
    \includegraphics[scale=0.3]{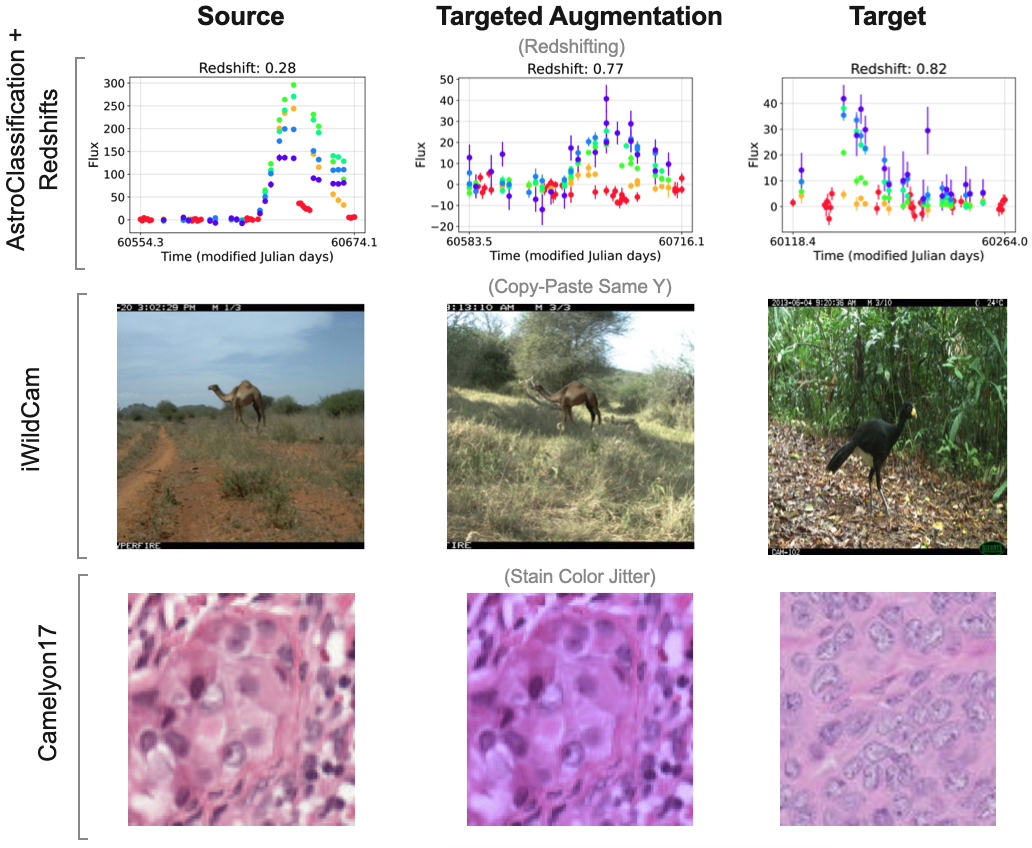}
    \caption{An example from the source dataset (left), an augmented version of the source example (middle), and an example from the target dataset (right) for our 3 tasks. (\textbf{Top row}) The target dataset in \classification and \redshifts is much higher redshift than the source dataset. We apply the redshifting augmentation to simulate placing source objects at a higher redshift to better match the target dataset. The flux errors and flux values of the augmented example (middle) show much better resemblance of the target example. (\textbf{Bottom row}) The \iwildcam target dataset is composed of images from cameras unseen during fine-tuning, which are in potentially new habitats, so we randomize the habitat background by applying the Copy-Paste Same Y augmentation. This algorithm places source dataset animals into empty backgrounds from other cameras that have observed the same species. \iwildcam image examples shown here are from \citet{gao2023targeted}.}
    \label{fig:datasets}
\end{figure}

We design targeted augmentations for 4 real-world tasks: species identification from wildlife camera trap images, tumor detection, and astronomical time-series classification and redshift prediction. We show examples from the source, augmented, and target datasets for these tasks in Figure~\ref{fig:datasets}. 

\paragraph{Wildlife species classification (\iwildcam).} For \iwildcam \citep{beery2020iwildcam,sagawa2022uwilds}, the task is to identify the wildlife species from static camera trap images. 
These cameras are placed in a wide variety of environments, which all have unique habitat conditions (e.g., African savannah vs. tropical rainforest) and camera characteristics (e.g., angles, resolutions). In this dataset, we use labeled data from 243 camera traps to learn a model that can generalize to data from 48 unseen camera traps. 
\begin{itemize}
    \item \textbf{Source:} 243 camera traps
    \item \textbf{Target:} 48 unseen camera traps
    \item \textbf{Targeted Augmentation:}
We augment the labeled dataset with the Copy-Paste Same Y algorithm, which uses image segmentation to copy-paste the animal onto different background images from cameras that have observed the same species \citep{gao2023targeted}. 
    \item \textbf{Task:} 182-class wildlife species classification
\end{itemize}

\paragraph{Tumor detection (\camelyon).} The task in \camelyon \citep{bandi2018detection} is to classify whether a patch of a histopathology slide contains a tumor. These slides are contributed from multiple hospitals, which use different stain colors and also vary in distributions of patient cancer stage.
\begin{itemize}
    \item \textbf{Source:} Hospitals 1-3.
    \item \textbf{Target:} Hospitals 4 and 5.
    \item \textbf{Targeted Augmentation:}
We augment the labeled dataset with the Stain Color Jitter algorithm, which jitters the color of the slide image in the hematoxylin and eosin staining color space \citep{tellez2018whole}. 
    \item \textbf{Task:} Binary classification of whether a slide contains a tumor.
\end{itemize}

\paragraph{Astronomical object classification (\classification).}
Astronomical object classification \citep{Boone_2019, allam2022paying} involves predicting the object type (e.g., type II supernova) from a time series of an object's brightness at multiple wavelengths (\textit{light curves}). We curate this dataset from the Photometric LSST Astronomical Time Series Classification Challenge \citep[PLAsTiCC,][]{theplasticcteam2018photometric} (details in Appendix~\ref{app:data}).

\begin{itemize}
    \item \textbf{Source:} Time-series of bright, nearby objects with expert labels
    \item \textbf{Target:} Time-series of all observed objects from the telescope, often faint and distant (higher redshift). Follow-up observation, which is required for expert labeling, is too expensive for these objects.
    \item \textbf{Targeted Augmentation:} 
We augment the labeled dataset by redshifting each object, i.e., simulating its observed properties as if it were further away (details in Appendix~\ref{app:targeted-augs}).
    \item \textbf{Task:} 14-class astronomical object classification
\end{itemize}

\paragraph{Redshift regression (\redshifts).}
Similar to object type, redshift information is also available only for bright, nearby objects. We predict the scalar redshift value of each object and minimize mean squared error. This task has been studied for individual object types, such as quasars \citep{nakoneczny2021photometric} and type Ia supernovae \citep{photozsnthesis}, but we consider a more realistic set of multiple object types. The labeled and unlabeled data are derived from the PLAsTiCC dataset. \redshifts is a new dataset that we contribute as part of this work.

\begin{itemize}
    \item \textbf{Source:} Time-series of bright, nearby labeled objects.
    \item \textbf{Target:} Time-series of all observed objects from the telescope, often faint and distant (higher redshift).
    \item \textbf{Targeted Augmentation:} Redshifting (same as \classification, Appendix~\ref{app:targeted-augs}).
    \item \textbf{Task:} Redshift regression
\end{itemize}

\subsection{Designing targeted augmentations}
How do we design these targeted augmentations? We provide a general methodology based on matching the target distribution on a feature space:
\begin{enumerate}
    \item Identify a feature space $\sZ$. We assume that we can label $z \in \sZ$ for each input and that the source and target domains largely differ on this feature space. One such example is the space of spurious, domain-dependent features (e.g., camera angle or resolution for \iwildcam), which is the approach followed by~\citet{gao2023targeted}. 
    \item Fit a transformed feature distribution $\hat{p_T}(\znew | z)$ to the target feature distribution. 
    \item Create a transformation distribution $T(x' | x, \znew)$ where $x'$ is the augmented version of $x$ with $z=\znew$. In this paper, we define $T$ with domain knowledge.
    \item Given an input $x$, generate augmentations by sampling a new feature $\znew$ from $\hat{p_T}(\znew \mid z)$, then sampling an augmentation from $T(x' | x, \znew)$. The resulting targeted augmentation probabilities are $\augft(x' \mid x) = \sum_{\znew} T(x' \mid x, \znew) \hat{p_T}(\znew \mid z)$.
\end{enumerate}

\paragraph{Targeted augmentation example.}
We follow the procedure outlined above to design a targeted augmentation for \classification and \redshifts, two astronomical time-series datasets (see Appendix~\ref{app:targeted-augs} for details). In these datasets, expert labels are only available for time-series of bright, nearby astronomical objects, while the unlabeled dataset contains mostly faint, distant objects.
We describe distances in terms of \textit{cosmological redshift}: nearby objects have lower redshift values than distant objects, causing the source and target redshift distributions to be mismatched (Appendix Figure~\ref{fig:z_dists}).
\begin{enumerate}
    \item The source and target domains primarily differ on their redshift distributions, so we identify this scalar feature as $z$.
    \item We roughly fit the target redshift distribution while constraining the transformed redshift value to not be too far from the original redshift $z$, such that $\hat{p_T}(\znew \mid z)$ is distributed as $\text{loguniform}(0.95z,\; \text{min}(1.5(1+z)-1, \;5z))$, following \citet{Boone_2019}.
    \item We define a transformation distribution $T(x' | x, \znew)$, where $x$ is a time-series of flux values at multiple wavelengths and $\znew$ is a new redshift value to transform to. We first fit a Gaussian process that models $x$ as a function of time and wavelength. Given $\znew$, we rescale the timestamps and wavelengths of the original input to account for the physical effects of the new redshift value. Then, we sample $\tilde{x'}$ from the Gaussian process at these new timestamps and wavelengths. Finally, we produce the transformed input $x'$ by scaling the flux values to account for $\znew$.
    \item We sample $\znew$ from $\hat{p_T}(\znew \mid z)$ and then sample augmentations $x'$ from $T(x' | x, \znew)$.
\end{enumerate}

\section{Experiments}
\label{sec:results}
We empirically test Connect Later with contrastive pretraining (\iwildcam, \camelyon) as well as pretraining with masked autoencoding (\classification, \redshifts) to demonstrate Connect Later as a general fine-tuning method.

\paragraph{Training procedure.} 

For \iwildcam, we use a ResNet-50 model pretrained on unlabeled ImageNet data with SwAV contrastive learning~\citep{caron2020swav}.
We use a DenseNet121 pretrained on the unlabeled data provided in \cite{sagawa2022uwilds} with SwAV for \camelyon.
To test Connect Later with other pretraining strategies, we use the masked autoencoding objective for \classification and \redshifts by masking 60\% of observations from each light curve (Appendix~\ref{app:experiments}). 
The same pretrained model is used for both tasks to demonstrate the reusability of pretrained features.
We note that masked autoencoding has been linked to contrastive learning \citep{zhang2022mask}. 
In particular, \citet{zhang2022mask} show that the masked autoencoding objective upper bounds the contrastive loss between positive pairs --- thus, masked autoencoding implicitly aligns the positive pairs induced by the masking augmentations.

We fine-tune the pretrained models with linear probing then fine-tuning (LP-FT)~\citep{kumar2022finetuning}, which has been shown to improve OOD performance.

\paragraph{Baselines.}
We evaluate our framework against three baselines: ERM, ERM+targeted augs, and \sft. We include Avocado \citep{Boone_2019}, the previous state-of-the-art model for \classification. We also include a self-training baseline for \classification and \redshifts, which has been shown to perform well on some real-world datasets \citep{sagawa2022uwilds}. For the self-training baseline, we pseudo-label the target dataset with a trained ERM+targeted augs model, then perform the same targeted augmentation on the pseudo-labeled target dataset.
We then train a model with the pseudo-labeled and augmented target dataset combined with the labeled source dataset. We include additional domain adaptation baselines for \iwildcam and \camelyon: domain-adversarial neural networks (DANN) \citep{ganin2016domain}, correlation alignment (CORAL) \citep{sun2016return}, Noisy Student \citep{xie2020selftraining}, and ICON\footnote{\label{icon}https://github.com/a-tea-guy/ICON}.

\subsection{Main results}
\label{sec:main-results}

\begin{table}[]
    \centering
        \caption{ID and OOD accuracy (\%) for \classification and RMSE for \redshifts of each method. Results are averaged over 5 trials and rows with means within 1 STD of the best mean are bolded.}
        \begin{tabular}{lcccc}
            \toprule
            & \multicolumn{2}{c}{AstroClassification} & \multicolumn{2}{c}{Redshift}\\
            & ID Test Acc ($\uparrow$) & OOD Acc ($\uparrow$) & ID Test RMSE ($\downarrow$) & OOD RMSE ($\downarrow$) \\
            \midrule
            
            ERM & $71.59 \pm 1.10$ & $61.26 \pm 1.10$ & $0.274 \pm 0.016$ & $0.320 \pm 0.009$ \\
            \Sft & $78.84 \pm 0.97$ & $67.84 \pm 0.70$ & $\mathbf{0.246 \pm 0.015}$ & $0.277 \pm 0.004$ \\
            ERM + targeted augs & $68.75 \pm 0.95$ & $67.54 \pm 0.32$ & $0.310 \pm 0.006$ & $0.286 \pm 0.007$ \\
            Self-Training & $77.72 \pm 0.59$ & $65.15 \pm 0.67$ & $0.304 \pm 0.010$ & $0.289 \pm 0.003$ \\
            Avocado \citep{Boone_2019} & - & $77.40$ & - & - \\
            Connect Later & $\mathbf{80.54 \pm 1.20}$ & $\mathbf{79.90 \pm 0.60}$ & $\mathbf{0.256 \pm 0.005}$ & $\mathbf{0.247 \pm 0.005}$ \\
            \bottomrule
        \end{tabular}

    \label{tbl:main-astro}
\end{table}

\begin{table}[]
    \centering
        \caption{ID and OOD performance for each method on \iwildcam and \camelyon. Results are averaged over 15 trials for \iwildcam and 20 trials for \camelyon, and we report 95\% confidence intervals on each mean estimate. Rows with means within 1 interval of the best mean are bolded.}
        \begin{tabular}{lcccc}
            \toprule
            & \multicolumn{2}{c}{iWildCam (Macro F1, $\uparrow$)} & \multicolumn{2}{c}{Camelyon17 (Avg Acc, $\uparrow$)} \\
            & ID Test  & OOD Test & ID Val & OOD Test \\
            \midrule
            ERM & $46.4 \pm 0.5$ & $30.4 \pm 0.6$ & $89.3 \pm 0.9$ & $65.2 \pm 1.1$ \\
            \Sft & $46.4 \pm 0.8$ & $31.2 \pm 0.6$ & $92.3 \pm 0.2$ & $91.4 \pm 0.9$\\
            ERM + targeted augs & $\mathbf{51.4 \pm 0.6}$ & $36.1 \pm 0.7$ & $96.7 \pm 0.0$ & $90.5 \pm 0.4$\\
             DANN ~\citep{sagawa2022uwilds} & $48.5 \pm 3.2$ & $31.9 \pm 1.6$ & $86.1 \pm 1.3$ & $64.5 \pm 1.2$ \\
            CORAL ~\citep{sagawa2022uwilds} & $40.5 \pm 1.6$ & $27.9 \pm 0.5$ & $92.3 \pm 0.7$ & $62.3 \pm 1.9$\\
            Noisy Student ~\citep{sagawa2022uwilds} & $47.5 \pm 1.0$ & $32.1 \pm 0.8$  & - & - \\
            ICON & $50.6 \pm 1.3$ & $34.5 \pm 1.4$ & $90.1 \pm 0.4$ & $93.8 \pm 0.3$ \\
            Connect Later & $\mathbf{51.7 \pm 0.8}$ & $\mathbf{36.9 \pm 0.7}$ & $\mathbf{98.5 \pm 0.0}$ & $\mathbf{94.9 \pm 0.4}$\\
            \bottomrule
        \end{tabular}

    \label{tbl:main-iwildcam}
\end{table}
Tables~\ref{tbl:main-astro} and~\ref{tbl:main-iwildcam} compare the results of Connect Later with baseline methods. On \iwildcam, Connect Later produces improvements in both ID and OOD performance despite the failure of \sft. Connect Later substantially outperforms all other variants on the OOD metric, including state-of-the-art performances on \classification by 3\% OOD, \iwildcam by 0.8\% OOD for ResNet-50, and \camelyon by 1.1\% OOD for DenseNet121.

\paragraph{iWildCam.} 
On \iwildcam, \sft does not improve over ERM in ID performance and minimally improves over ERM in OOD performance, while ERM+targeted augmentations improves by 6\% ID and OOD over both ERM and \sft. 
While 3 of the domain adaptation baselines made ID and OOD improvements over \sft (DANN: $46.4\% \rightarrow 48.5\%$ ID, $31.2\% \rightarrow 31.9\%$ OOD; Noisy Student: $46.4\% \rightarrow 47.5\%$ ID, $30.4\% \rightarrow 32.1\%$ OOD; ICON: $46.4\% \rightarrow 50.6\%$ ID, $30.4\% \rightarrow 34.5\%$), all fall short of ERM+targeted augs.
Connect Later improves over both \sft ($30.4\% \rightarrow 37.2\%$) and ERM+targeted augs ($36.3\% \rightarrow 37.2\%$) in OOD performance, achieving a new state-of-the-art performance for ResNet-50 on the \iwildcam benchmark.

\paragraph{Camelyon17.}
On \camelyon, \sft produces significant gains over ERM in both ID ($89.3\% \rightarrow 92.3\%$) and OOD ($65.2\% \rightarrow 91.4\%$) average accuracy. 
ERM+targeted augmentations outperforms \sft in ID accuracy ($92.3\% \rightarrow 96.7\%$), but does not improve OOD. 
DANN underperforms ERM in both ID and OOD accuracy, while CORAL produces similar ID accuracy as \sft but poor OOD performance. 
Connect Later sets a new state-of-the-art on \camelyon with DenseNet121, improving on the best ID performance by 1.8\% (ERM+targeted augs, $96.7\% \rightarrow 98.5\%$) and OOD performance by 1.1\% (ICON, $93.8\% \rightarrow 94.9\%$). 

\paragraph{AstroClassification.} 
For \classification, \sft provides a significant performance boost over ERM: $71.6\% \rightarrow 78.9\%$ ID, $61.3\% \rightarrow 67.8\%$ OOD. 
ERM+targeted augs underperforms in ID accuracy compared to ERM alone ($71.6\% \rightarrow 68.8\%$) and \sft ($78.9\% \rightarrow 68.8\%$), likely due to relatively strong targeted augmentations toward the redshift distribution of distant, faint objects present in the target distribution. 
However, OOD accuracy of ERM+targeted augs is competitive with \sft, outperforming ERM. 
Self-training improves both ID and OOD performance compared to ERM but underperforms \sft in both domains.
Connect Later outperforms the best baseline, \sft, by 12\% OOD and 2\% ID. The ID accuracy improves over \sft despite the drop in ID accuracy from adding targeted augmentations to ERM, showing a complementary benefit between pretraining and targeted augmentations. Connect Later sets a new state-of-the-art OOD performance on \classification by 3\% over Avocado, a heavily tuned random forest model with expert-designed features \citep{Boone_2019}.

\paragraph{Redshifts.} 
Similarly to \classification, \sft significantly improves over ERM in both ID ($0.27 \rightarrow 0.25$, 7\% relative) and OOD ($0.32 \rightarrow 0.28$, 13\% relative) RMSE. Self-training performs similarly to ERM+targeted augs, while Connect Later outperforms the best baseline variant, \sft, by 0.03 RMSE (11\% relative) with comparable ID error. We use the same pretrained model for both \classification and \redshifts for \sft and Connect Later, demonstrating the reusability of pretrained representations.



\subsection{Ablations}


\begin{table}[b]
    \centering
    \caption{Scaling up model size of Connect Later produces improvements in both ID and OOD performance on the \classification task. }
    \begin{tabular}{@{}lcc@{}}
        \toprule
         Number of Parameters & ID Accuracy ($\uparrow$) & OOD  Accuracy ($\uparrow$)\\
         \midrule
        21M (default) & 78.47 & 79.49 \\
        69M & 80.38 & 80.55 \\
        \bottomrule
    \end{tabular}

    \label{tbl:ablations-size}
\end{table}

\begin{table}[]
    \centering
    \caption{Linear probing (LP) in addition to fine-tuning (FT) hurts performance for the ERM+targeted augs model but improves performance for Connect Later (tested on the \classification task).}
    \begin{tabular}{@{}lcccc@{}}
        \toprule
         & \multicolumn{2}{c}{Connect Later} & \multicolumn{2}{c}{ERM+targeted augs} \\
         & ID Accuracy ($\uparrow$) & OOD Accuracy ($\uparrow$) & ID Accuracy ($\uparrow$) & OOD Accuracy ($\uparrow$) \\
         \midrule
        FT only & 78.07 & 78.6 & 77.88 & 68.43 \\
        LP-FT & 78.47 & 79.49 &  65.68	& 67.07  \\
        \bottomrule
    \end{tabular}

    \label{tbl:ablations-LP}
\end{table}

We performed ablations on the model size, strength of pretraining augmentations (masking percentage for masked autoencoding), and LP-FT on \classification. 
We find that downstream performance is quite robust to masking percentage, while scaling up model size and LP-FT improve performance for pretrained models.

\begin{figure}
    \centering
    \includegraphics[scale=0.4]{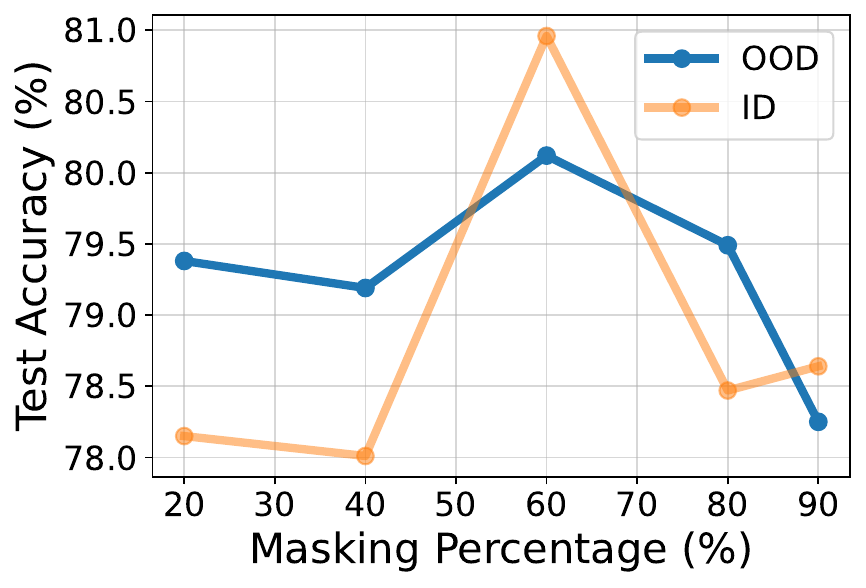}
    \caption{On the \classification task, Connect Later is relatively robust to pretraining masking percentage both ID and OOD, but 60\% masking performs best out of the percentages we tested.}
    \label{fig:ablations-masking}
\end{figure}


\paragraph{Model scale.} We tested Connect Later with a larger model ($\sim3\times$ the parameters of our model, $\text{21M} \rightarrow \text{69M}$), and find that scaling up model size improves both ID and OOD accuracy (Table~\ref{tbl:ablations-size}). This suggests that scaling up the model is a promising way to further improve performance with Connect Later.

\paragraph{Strength of pretraining augmentations (masking percentage).} We vary the strength of pretraining augmentations, which changes the connectivity between domains. This is most straightforward with the MAE objective, as augmentation strength is parameterized solely by masking percentage. We tested pretraining masking percentages \ \{20, 40, 60, 80, 90\}\% while keeping the masking strategy unchanged (replace 10\% of masked indices with random values from the lightcurve, another 10\% are kept unchanged, and 80\% are replaced with the mask token, which we choose to be 0). We show the ID and OOD test accuracy of each variant in Figure~\ref{fig:ablations-masking}. Both ID and OOD performance peak at 60\% masking, although we find that the performance of Connect Later is quite robust to the masking percentage, particularly for OOD performance. All of the masking percentages we tried improve on OOD performance over \sft or ERM with targeted augmentations. Particularly, even with the strongest pretraining augmentations (90\% masking), which should connect the domains more, the OOD performance did not improve over weaker augmentations. We hypothesize that increasing the strength of generic augmentations may indiscriminately increase the connectivity between all source and target examples, including examples from different classes that should not be strongly connected.

\paragraph{Linear probing then fine-tuning.} \citet{kumar2022finetuning} showed that linear probing (with fixed neural embeddings) and then fine-tuning (LP-FT) the entire model improves both ID and OOD performance. Intuitively, full fine-tuning with a randomly initialized linear probe can destroy the pretrained features, and training the linear probe first mitigates this. 
We test LP-FT against FT only (all model weights are fine-tuned) with the Connect Later model and the ERM+targeted augs baseline.
We find that LP-FT improves OOD accuracy by 0.9\% over FT only when applied to Connect Later on \classification (Table~\ref{tbl:ablations-LP}).
On the other hand, LP-FT decreased OOD accuracy by 1.4\% when applied to ERM+targeted augs, which uses random initialization (no pretraining).
As a result, we use LP-FT on pretrained models but not on ERM or ERM+targeted augs.

\section{Discussion and Related Work}

\paragraph{Augmentations for pretraining.}
Data augmentations such as cropping or masking have been vital to semi- and self-supervised learning objectives. Masking or noising the data and training a model to reconstruct the original inputs have been shown to produce useful pretrained representations across multiple modalities \citep{devlin2019bert, lewis2020bart, he2022mae, raffel2019exploring, chen2020simclr,he2020moco,caron2020swav}. 
In contrastive learning, models are trained to distinguish augmented ``views'' of the same input from views of a different input \citep{chen2020simclr, caron2020swav, he2020moco}. Our results demonstrating inconsistent OOD performance across datasets brings up the important future question of how to choose the best pretraining augmentation and algorithm for learning transferable representations.

\paragraph{Augmentations for robustness.}
Data augmentation has been used to improve model robustness and avoid catastrophic failures due to spurious, label-independent changes (e.g. translation or rotation in vision) \citep{hendrycks2019augmix, rebuffi2021data, ng2020ssmba}. The augmentation strategies used in prior work are generic perturbations that aim to increase the diversity of inputs \citep[e.g.,][]{simard2003best,krizhevsky2012imagenet,cubuk2019autoaugment,cubuk2020randaugment,devries2017improved,zhang2017mixup},
though a number of studies have shown that the type of data augmentations matters for performance \citep{chen2020simclr, xie2020unsupervised}.
Augmentations have also been leveraged in the self-training paradigm, which improves generalization to unseen data by training on the pseudo-labeled full dataset \citep{xie2020selftraining, sohn2020fixmatch, yang2021meanteacher}. 
We show that a self-training baseline with pseudo-labels from an ERM+targeted augs model does not outperform Connect Later, indicating that pretraining is an important component of the framework.
Connect Later exposes targeted augmentations as a design interface for improving robustness with knowledge of the distribution shift, while still leveraging pretrained representations.

\paragraph{Targeted augmentations.}
In problems with domain shift, \citet{gao2023targeted} show that targeted augmentations outperform generic augmentations on unseen data. They identify spurious domain-dependent, label-independent features in the source dataset and construct targeted augmentations by randomizing these features. \citet{gao2023targeted} consider the domain generalization setting, in which no data from the target dataset is available. We consider targeted augmentations in the domain adaptation setting, in which we can model the target distribution of these spurious features with the unlabeled target data. In general, designing targeted augmentations specific to each distribution shift may be difficult and require expert guidance. 
As part of the Connect Later framework, we provide a general methodology for the design of such augmentations. 
Certain aspects, such as the selection of feature space $z$ and transformation distribution $T$ could be learned from the unlabeled data itself, which we leave for future work. We show that targeted augmentations better leverage pretrained representations for complementary gains in OOD performance.



\section{Conclusion}

We show that pretraining with generic augmentations is not a panacea for all distribution shifts and tasks, and sometimes does not outperform supervised learning on labeled source data.
Pure supervised learning, however, does not use the unlabeled data or produce reusable representations. The Connect Later framework allows for better leverage of pretrained representations for OOD performance by applying targeted augmentations at fine-tuning time. Future work could focus on learning targeted augmentations from the source and target data distributions as well as further understanding of how the choice of pretraining augmentations affects downstream ID/OOD performance.


\section*{Appendix A. Additional Dataset Details}
\subsection*{A.1. AstroClassification, Redshifts Datasets}
\label{app:data}

The \classification and \redshifts datasets were adapted from the 2019 Photometric LSST Astronomical Time-Series Classification Challenge \citep{theplasticcteam2018photometric} \footnote{https://zenodo.org/record/2539456}. This diverse dataset contains 14 types of astronomical time-varying objects, simulated using the expected instrument characteristics and survey strategy of the upcoming Legacy Survey of Space and Time \citep[LSST][]{ivezic2019lsst} conducted at the Vera C. Rubin Observatory. It includes two overall categories of time-series objects: \textit{transients}, short-lived events such as supernovae, and \textit{variable} sources, those with fluctuating brightness such as pulsating stars. Specifically, the dataset includes the following transients: type Ia supernovae (SNIa), SNIax, SNIa-91bg, SNIbc, SNII, superluminous supernovae (SLSN), tidal disruption events (TDE), and single lens microlensing events ($\mu$Lens-Single); and the following variable objects: active galactic nuclei (AGN), Mira variables, eclipsing binary systems (EB), and RR Lyrae (RRL).  

Millions of potential new objects are discovered per observing night, and important metadata such as object type, redshift, or other physical parameters, require astronomers to take time-intensive \textit{spectra} of each object. Spectra are a granular brightness vs. wavelength measurement at a single point in time, and are typically only taken for bright, nearby objects which require less exposure time than faint, faraway objects. The vast majority of discovered objects, however, will not have spectra but instead a time series of imaging data taken in 6 broad wavelength ranges, or \textit{photometric bands}. The time-varying behavior of these objects in these coarse wavelength bands does offer important clues about these physical parameters, but expert interpretation of spectra are traditionally required for confident labeling. Thus, our labeled training data for both \classification and \redshifts come from the unrepresentative subset of objects with spectra.

In these tasks, we are specifically interested in predicting the object type (e.g. type II supernova) and the cosmological redshift of objects in the unlabeled dataset. \textit{Cosmological redshift} is a proxy for distance in the universe, and an important piece of metadata for understanding an object's physical processes as well as other applications, such as estimating the expansion rate of the universe with type Ia supernovae.

\paragraph{Problem Setting.} 
The task is to predict object type for \classification (redshift for \redshifts) from time-series of object brightness. The input $x$ consists of flux measurements and associated uncertainties at times $\vt$ and photometric band that each measurement was taken in $\vb$: $\{F(t_i,b_j)\}_{i=1,j=1}^{T,W}, \{F_{\text{err}}(t_i, b_j)\}_{i=1,j=1}^{T,W}$. For this work, we map each $b \in \vb$ to the central wavelength of the $b$ band, which we denote $\vw$. The domain $d$ is binary, corresponding to whether the object has a spectrum (and thus a label). The labels $y$ are available only for objects with spectra, and are one of 14 types of astronomical time-varying objects for \classification (redshift of the object for \redshifts). We seek to optimize performance on the unlabeled data, which are generally fainter and further away than the labeled subset. We evaluate on these examples as well as held-out examples from the labeled subset.

\paragraph{Data.}
The training set of 7,846 objects is designed to emulate a sample of objects with spectra and thus biased toward brighter, more nearby objects compared to the test set of 3,492,888 objects. A random subset of 10,000 test set objects was selected for evaluation.
\begin{enumerate}
    \item \textbf{Source:} 6,274 objects
    \item \textbf{ID Test}: 782 objects
    \item \textbf{OOD Test:} 10,000 objects
\end{enumerate}
All data were simulated with the SuperNova ANAlysis \citep[SNANA,][]{kessler2009snana} software library. Further details about the astrophysical models and LSST instrument characteristics used in the simulation can be found in \citet{Kessler_2019}. 

\section*{Appendix B. Data Augmentations}
\subsection*{B.1. Generic Augmentations for Pretraining}

\paragraph{AstroClassification and Redshifts.} For the \classification and \redshifts datasets, we randomly mask a subset of the input sequence using the masked language modeling paradigm introduced by \cite{devlin2019bert}. Given an unlabeled input sequence $x$, a training input $x'$ can be generated by randomly masking elements of $x$ while the associated label $y$ consists of the original, unmasked values. The model is trained to use contextual information (unmasked elements) to successfully reconstruct most of the sequence. From our ablation experiments, we find that a masking percentage of 60\% produces the best downstream results. We follow an existing implementation for astronomical time-series \citep{astromer} and set 80\% of the masked elements to 0, replace 10\% with a random element from the sequence, and keep the remaining 10\% unchanged.

\paragraph{iWildCam and Camelyon17.} 
For \iwildcam, we use a ResNet-50 model pretrained on ImageNet with SwAV, a contrastive learning algorithm~\cite{caron2020swav}. 
For \camelyon, we use a DenseNet121 pretrained with SwAV on the unlabeled \camelyon dataset from \citet{sagawa2022uwilds}.
SwAV uses random cropping augmentations of different resolutions.

\subsection*{B.2. Targeted Augmentations for Fine-Tuning}
\label{app:targeted-augs}

\paragraph{Redshifting for AstroClassification and Redshifts.} 

\begin{figure}
    \centering
    \includegraphics[scale=0.4]{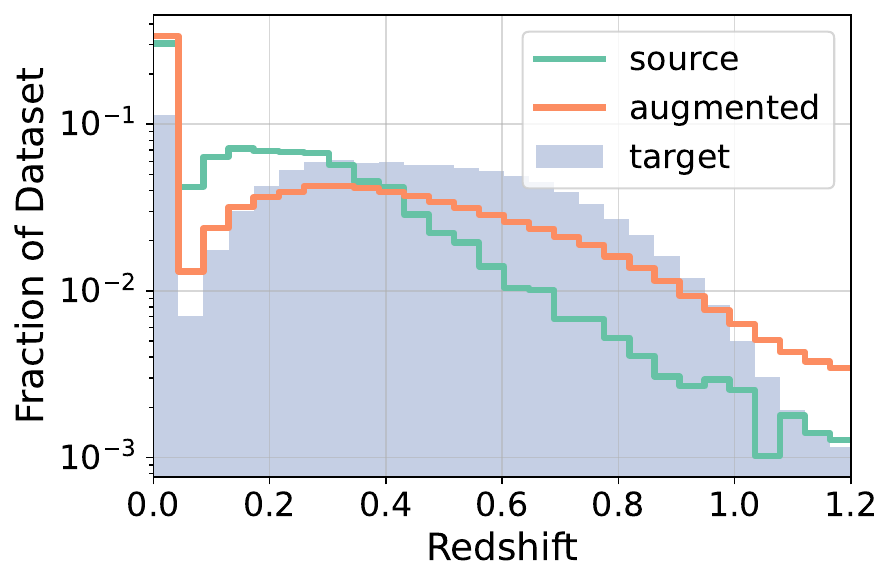}
    \caption{Redshift distributions of source, augmented, and target datasets for the \classification and \redshifts tasks.}
    \label{fig:z_dists}
\end{figure}

The OOD test set of the \classification and \redshifts datasets have many more high redshift objects than the source dataset, leading us to adopt an augmentation scheme to alleviate this shift. Figure~\ref{fig:z_dists} shows the redshift distributions of the source, augmented, and target datasets. Redshifting places each object at a new redshift and recomputes its light curve sampling, fluxes, and flux uncertainties accordingly. This augmentation algorithm was adapted from \citet{Boone_2019}. 

An input $\Xin \in \R^{T \times W}$ is a multivariate time series of flux values at specified times and observed wavelengths, $\{F(t_i, w_j)\}_{i=1,j=1}^{T,W}$.
We also have $\Xinerr \in \R^{T \times W}$, representing the flux errors corresponding to each element of $\mX$. We denote the elements of $\augXerr$ by $\{F_{\text{err}}(t_i, w_j)\}_{i=1,j=1}^{T,W}$.
Our goal is to model $F, F_{\text{err}}:\R\times\R \rightarrow \R$ at a new chosen redshift, $\znew$, to produce augmented inputs $\augX, \augXerr$.
\begin{itemize}
    \item We first construct a distribution from which to sample the new redshift, taking into account the current redshift of the object $\zorig$ as well as the target redshift distribution. We then sample a new redshift, $\znew \sim \text{loguniform}(0.95\zorig, \;\text{min}(1.5(1+\zorig)-1,\; 5\zorig))$.
    
    \item We fit a Gaussian process (GP) model for $F$ with training observations $\mX$ queried at the training input values $(\vt, \vw)$, and denote the predictive mean and variance of the GP as $F', F'_{\text{err}}$.

    \item Given the new redshift value $\znew$, we rescale the timestamps and wavelengths of the original observations to account for the physical effects of the new redshift value: $\vt_{\text{new}} = \frac{1+\znew}{1+\zorig}\vt$, $\vw_{\text{new}} = \frac{1+\znew}{1+\zorig}\vw$. We also randomly drop out 10\% as well as a large swath of $(\vt_{\text{new}}, \vw_{\text{new}})$ to simulate distinct observing seasons (telescope observing only occurs in the winter).

    \item We obtain GP predictions at test inputs $\{F'(t_{\text{new},i}, w_{\text{new},j})\}_{i=1,j=1}^{T,W}$, $\{F'_{\text{err}}(t_{\text{new},i}, w_{\text{new},i})\}_{i=1,j=1}^{T,W}$ and scale them by the log ratio of the new and original distances: 
    $$\tilde{\mX'}=10^{0.4(d(\znew)-d(\zorig))} \{F'(t_{\text{new},i}, w_{\text{new},j})\}_{i=1,j=1}^{T,W},$$ $$\tilde{\mX'}_{\text{err}}=10^{0.4(d(\znew)-d(\zorig))} \{F'_{\text{err}}(t_{\text{new},i}, w_{\text{new},j})\}_{i=1,j=1}^{T,W},$$ where $d(z)$ is the distance corresponding to redshift $z$.

    \item   We roughly model the observational noise of the telescope from the target data as a function of wavelength and sample $\epsilon \in \R^W$ from it. We define 
    $$\augX =\{\tilde{\augX}_{:,j} + \epsilon_j \}_{j=1}^{W}, \augXerr = \left\{\sqrt{\tilde{\mX'}_{\text{err},:,j}^2 + \epsilon_j^2} \right\}_{j=1}^{W}.$$

    \item We model the observational capabilities of the telescope to ensure that our augmented input $\augX, \augXerr$ does not fall below the threshold of detection. We ``accept" an augmented input $\augX, \augXerr$ if the signal-to-noise ratio (SNR) of at least two observations is over 5, i.e. $\text{SNR}(\augX_{i,j}, \mX'_{\text{err},i,j}) \geq 5$ for at least 2 of $i \in \{1,...,T\}, j \in \{1,...,W\}$. We define $\text{SNR}(x, x_{\text{err}})=\frac{|x|}{x_{\text{err}}}$.
\end{itemize}

\paragraph{Copy-Paste (Same Y) for iWildCam.}
This augmentation strategy randomizes the backgrounds of wildlife images to reduce the model's dependence on these spurious features for species classification. Specifically, a segmentation mask is applied to each image to separate the animal from the background, and the animal is ``copy-pasted" into a new background from a camera that has observed that animal species. This was the best performing augmentation strategy from \citet{gao2023targeted}.

\paragraph{Stain Color Jitter for Camelyon17.} 
This augmentation, originally from \citet{tellez2018whole}, alters the pixel values of the slide images to emulate different staining procedures used by different hospitals. The augmentation uses a pre-specified Optical Density (OD) matrix to project images from RGB space to
a three-channel hematoxylin, eosin, and DAB space before
applying a random linear combination. This was the best performing augmentation strategy from \citet{gao2023targeted}.

\section*{Appendix C. Experimental Details}
\label{app:experiments}

\paragraph{AstroClassification and Redshifts.}
For \classification and \redshifts, we pretrain with a masked autoencoding objective:

\begin{align}
\label{eqn:mae_objective}
\sL_{\text{MAE}}(\encoder) = \E_{\inputx\sim \unlabeldist,\inputxp\sim\aug(\cdot \mid \inputx)}[(\encoder(\inputxp)-\inputx)^2]
\end{align}
We use an encoder-only Informer model \citep{zhou2021informer} with 8 encoder layers of 12 attention heads each. The model hidden dimension was chosen to be 768 and the layer MLPs have hidden dimension 256. Due to the 2-dimensional position data (each element of the time-series has an associated time and photometric band/wavelength) and irregular sampling of our dataset, we train a positional encoding based on learnable Fourier features following~\citet{li2021learnable}. We also select a random window of length 300 from each example (and zero-pad examples with fewer than 300 observations) to produce inputs of uniform shape. We perform pretraining with a batch size of 256 and learning rate 1e-4 (selected from 1e-3 $\sim$ 1e-6) for 75,000 steps. We finetune the pretrained model with linear probing for 20,000 steps (for pretrained models only) and learning rate 1e-4, then fine-tuning for 10,000 steps at learning rate of 4e-5. We increase the learning rate for models without pretraining to 1e-4 for FT. The \redshifts task uses LP learning rate of 5e-4 and FT learning rate of 1e-4. We decrease the learning rate per step with a linear scheduler.

\paragraph{iWildCam.} 
For pretraining, we use ResNet-50 pretrained on ImageNet with SwAV~\citep{caron2020swav}. During fine-tuning, we train all models for 15 epochs with early stopping on OOD validation performance, following~\citet{gao2023targeted}. For pretrained models, we also do 10 epochs of linear probing before fine-tuning \citep[LP-FT,][]{kumar2022finetuning} for 15 epochs, where the linear probe is trained with Adam and the linear probe weights used to initialize the fine-tuning stage is chosen with OOD validation performance. To reduce the noise in OOD results, for all methods we select the epoch in the last 5 epochs with the best OOD validation performance and report OOD test results with that version of the model. Following~\citet{gao2023targeted}, we allow for 10 hyperparameter tuning runs, where we sample the following hyperparameters independently from the following distributions: the linear probe learning rate ($10^{\text{Uniform}[-3,-2]}$), fine-tuning learning rate ($10^{\text{Uniform}[-5,-2]}$), and probability of applying the augmentation ($\text{Uniform}[0.5, 0.9]$) and pick the hyperparameter configuration with the best OOD validation performance. For ERM and ERM+targeted augmentations, we use the tuned hyperparameters from~\citet{gao2023targeted}. To decrease the confidence interval due to an outlier seed, the reported performance of Connect Later is averaged over 15 seeds. All other results are averaged over 5 seeds.

\paragraph{Camelyon17.} For pretraining, we use DenseNet121 pretrained on the unlabeled \camelyon dataset presented in \citet{sagawa2022uwilds} with SwAV \citep{caron2020swav}. During fine-tuning, we train all models for 15 epochs with early stopping on OOD validation performance, following~\citet{gao2023targeted}. For pretrained models, we also do 10 epochs of linear probing before fine-tuning \citep[LP-FT,][]{kumar2022finetuning} for 15 epochs, where the linear probe is trained with Adam and the linear probe weights used to initialize the fine-tuning stage is chosen with OOD validation performance. To reduce the noise in OOD results, for all methods we select the epoch with the best OOD validation performance and report OOD test results with that version of the model. Following~\citet{gao2023targeted}, we allow for 10 hyperparameter tuning runs, where we sample the following hyperparameters independently from the following distributions: the linear probe learning rate ($10^{\text{Uniform}[-3,-2]}$), fine-tuning learning rate ($10^{\text{Uniform}[-5,-2]}$), probability of applying the augmentation ($\text{Uniform}[0.5, 0.9]$), and augmentation strength ($\text{Uniform}[0.05, 0.1]$), and pick the hyperparameter configuration with the best OOD validation performance. All results are averaged over 20 seeds.

\section*{Appendix D. Empirical Estimates of Connectivity}
\label{app:connectivity}

\begin{table}[]
    \centering
    \caption{Empirically estimated connectivity measures for \iwildcam, \classification, and \camelyon. \iwildcam and \classification results are averaged over 15 randomly selected class-domain pairs, while \camelyon results are averaged over all possible class-domain pairs.}
    \begin{tabular}{lccc}
        \toprule
         & across-domain & across-class & across-both \\ \midrule
        \iwildcam & 0.116 & 0.071 & 0.076 \\
        \classification & 0.287 & 0.159 & 0.097 \\
        \camelyon & 0.16 & 0.198 & 0.152 \\ 
        \bottomrule
    \end{tabular}
    \label{tbl:app-connectivity}
\end{table}

We empirically estimate connectivity measures for all of the datasets we tested on following the procedure outlined in Appendix D of \citet{shen2022connect}. 
Specifically, we train binary classifiers from scratch to predict the class-domain pair of a given input example. We randomly select 15 class-domain pairs for \iwildcam and \classification, while for \camelyon we use all class-domain pairs since \camelyon is a binary classification task. We label these class-domain examples following Appendix D of \citet{shen2022connect} and create a dataset with 80/10/10 train/validation/test split. We train using the same hyperparameters described in Appendix~\ref{app:experiments} for 3,000 steps with early stopping on the validation accuracy.
Our results are presented in Table~\ref{tbl:app-connectivity}.

\newcommand{\reachableset}{\sR}
\newcommand{\zpos}{z_{\posclass}}
\newcommand{\zneg}{z_{\negclass}}
\newcommand{\aaa}{a}
\newcommand{\bbb}{b}
\newcommand{\uone}{u_1}
\newcommand{\utwo}{u_2}

\section*{Appendix E. Simple construction where Connect Later improves over pretraining or targeted augmentations alone}
\label{app:simple_example}

We give a simple construction for constrastive pretraining based on the construction in Proposition 3 (Appendix A.2) of~\citet{shen2022connect}, where Connect Later improves over pretraining (\sft) or targeted augmentations alone.

\paragraph{Data distribution.}
We consider binary classification with 2 domains.
Let $\sS = \{ \inputx \in \inputspace: \domainx = 1 \}$ and $\sT = \{ \inputx \in \sT: \domainx = 2\}$, and assume that $\sourcedist$ and $\targetdist$ are uniform over $\sS$ and $\sT$.
The unlabeled distribution for pretraining is the uniform distribution over $\inputspace$.
The source domain $\sS=\{1,2\}$ contains 2 points and the target domain $\sT=\{3,4,5,6,7,8\}$ contains 6 points.
For simplicity, we let the labels $\labelx$ be a deterministic function of the input $\inputx$.
The label space is $\labelspace = \{\negclass, \posclass\}$. The label for $\inputx\in \{1,3,5,7\}$ is $\labelx=\posclass$ and the label for $\inputx\in \{2,4,6,8\}$ is $\labelx=\negclass$.
Only the source data is labeled.

\paragraph{ERM with targeted augmentations.} ERM with targeted augmentations learns a model on source labeled data. To specialize to this section, the ERM objective is
\begin{align}
    \label{eq:erm-object-setup}
    \Lerm(\clf) = \E_{\inputx \sim \sourcedist, \inputx' \sim \augft(\cdot \mid \inputx)}[ \ell(\clf(\inputx'), y_x) ].
\end{align}
ERM returns a classifier $\empclferm \in \argmin_{\clf} \Lerm(\clf)$.

\paragraph{Spectral contrastive learning.}
Following \citet{haochen2021spectral} and \citet{shen2022connect}, we analyze contrastive learning from an augmentation graph perspective, where inputs $\inputx$ are connected via augmentations with edge weights $\pospairdist(\inputx, \inputxp)$, which represent the probability of $\inputx, \inputxp$ being a  positive pair (augmentations of the same input $\inputx$).
For theoretical analysis, we analyze the spectral contrastive learning objective:
\begin{align}
    \label{eq:scl}
        \lpretrain(\encoder) = -2 \cdot &\E_{(\inputx, \posx) \sim \pospairdist}\left[\encoder(\inputx)^\top \encoder(\posx)\right] +\E_{\inputx, \inputx' \sim \unlabeldist}\left[\left(\encoder(\inputx)^\top \encoder(\inputx')\right)^2\right].
\end{align}
The result of pretraining to optimize the above objective is an encoder $\empencoder: \inputspace \to \R^\embeddim$.

\begin{figure}[t]
    \centering
	\includegraphics[width=0.6\linewidth]{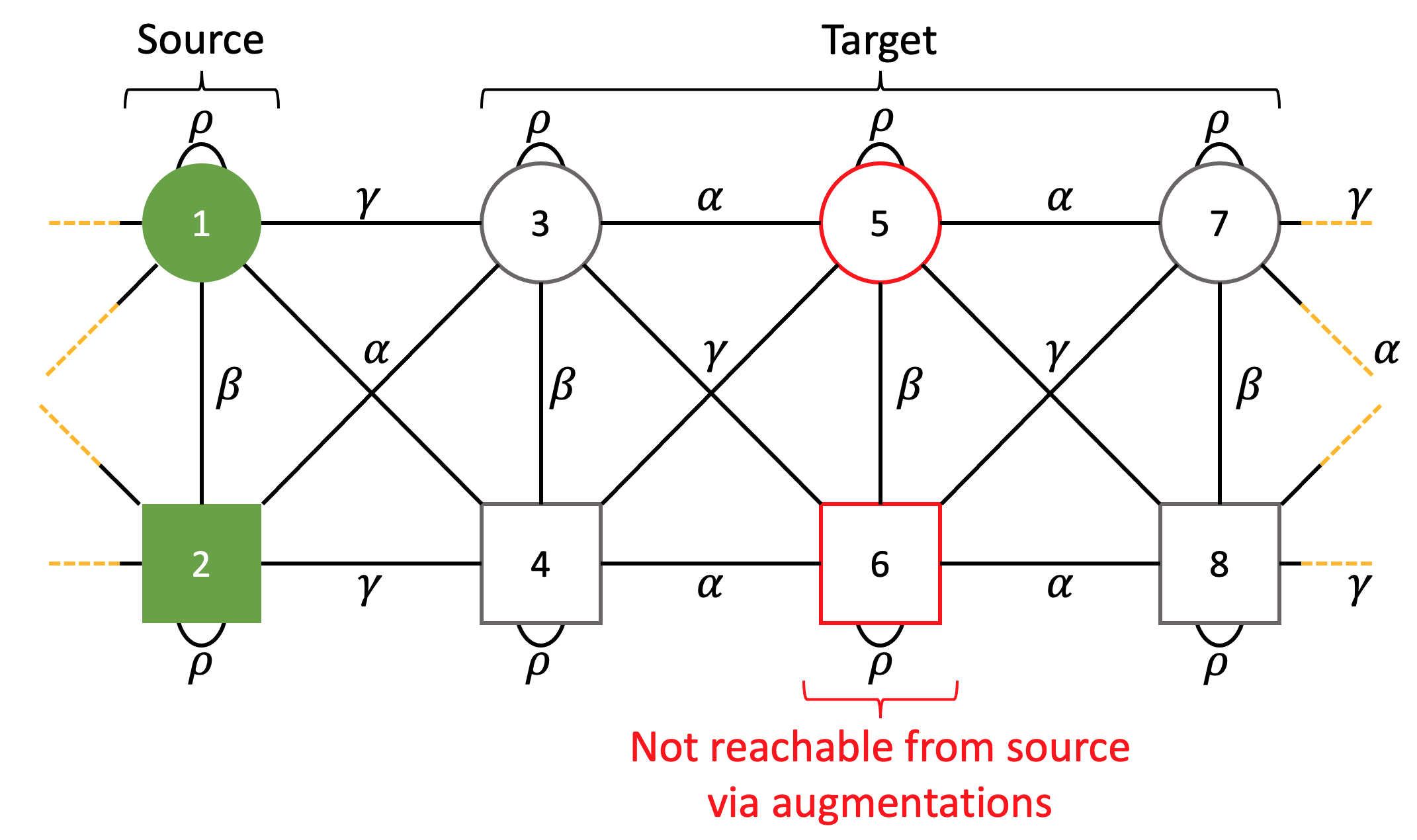}
	\caption{%
    Example distribution of data and augmentations for contrastive learning where Connect Later improves OOD performance over contrastive pretraining+standard fine-tuning and ERM+targeted augmentations. The augmentation graph is similar to~\citet{shen2022connect} except the edge weights connecting 1,2 and 3,4 are swapped. The shapes represent classes, while the labeled data is shaded in green. The generic augmentation probabilities are marked as edge weights, where we assume that $\alpha > \gamma + \beta$. Here, targeted augmentations which first swap inputs 1 and 2 before applying a generic augmentation help to align the source and target. However, some target inputs are not reachable via augmentations from source inputs. \Sft can generalize throughout the target domain, but only in conjunction with targeted augmentations that align the source and target. The orange dotted lines on the far ends connect to each other (the graph wraps around).}
    \label{fig:simple_example}
\end{figure}

\paragraph{Linear probing (fine-tuning step).}
Instead of analyzing fine-tuning, we follow~\citet{shen2022connect} and analyze linear probing on top of the pretrained representations from the encoder.
We train a linear model with parameters $\linmat \in \R^{\numcls \times \embeddim}$, where $\numcls$ is the number of classes.
We minimize the objective:
\begin{align}
    \label{eq:sq-loss}
    \lfinetune(\linmat) = \E_{\inputx \sim \sourcedist}\left[ \ell(\linmat \empencoder(\inputx), \labelx) \right] + \eta \| \linmat \|_F^2,
\end{align}
where $\ell$ is the squared loss and we take $\labelx\in \R^\embeddim$ to be a one-hot encoding of the class label.
The resulting classifier is $\empclf(\inputx) = \argmax_{i \in [\numcls]} (\emplinmat \empencoder(\inputx))_i$.

\paragraph{Pretraining augmentations (Figure~\ref{fig:simple_example})}
We define the pretraining augmentation distribution $\aug(\cdot \mid \inputx)$ to be
\begin{align}
        \aug(\inputxp \mid \inputx) =
    \begin{cases}
        \augprobr & \inputx = \inputxp\\
        \augproba & \{x',x\}\in \{\{1,4\}, \{3,5\}, \{5,7\}, \{2,5\}, \{4,6\}, \{6,8\}, \{1, 8\}, \{2, 7\}\\
        \augprobb & \{\inputxp, \inputx\} \in \{ \{1, 2\}, \{3, 4\}, \{5, 6\}, \{7, 8\} \} \\ 
        \augprobg & \{\inputxp, \inputx\} \in \{ \{1, 3\}, \{2, 4\}, \{3, 6\}, \{4, 5\}, \{5, 8\}, \{6, 7\}, \{1,7\}, \{2,8\} \} \\  
    \end{cases}. 
\end{align}
Notice that the weight between 1,3 is $\augprobg$ and the weight between 1,4 is $\augproba$, and the weights are similarly swapped for 2,4, and 2,5. 
We assume that $\augprobr, \augproba, \augprobb$, and $\augprobg$ are in $(0, 1)$ and are distinct. We also assume that the augmentation probabilities satisfy $\augprobr > \max\{\augproba, \augprobb\}$ and $\min\{\augproba, \augprobb\} > \augprobg$.
Following~\citet{shen2022connect}, we can convert these to positive pair probabilities $\pairprobr,\pairproba,\pairprobb,\pairprobg$ with similar properties by renormalizing.

Given the above setting, the following is a simplified form of Proposition 3 from~\citet{shen2022connect}, if we instead use the following augmentation distribution, which swaps the edge weight magnitudes that involve nodes 1 and 2:
\begin{align}
        \sA_{\text{prop}}(\inputxp \mid \inputx) =
    \begin{cases}
        \augprobr & \inputx = \inputxp\\
        \augproba & \{x',x\}\in \{\{1,3\}, \{3,5\}, \{5,7\}, \{2,4\}, \{4,6\}, \{6,8\}, \{1, 7\}, \{2, 8\}\\
        \augprobb & \{\inputxp, \inputx\} \in \{ \{1, 2\}, \{3, 4\}, \{5, 6\}, \{7, 8\} \} \\ 
        \augprobg & \{\inputxp, \inputx\} \in \{ \{1, 4\}, \{2, 3\}, \{3, 6\}, \{4, 5\}, \{5, 8\}, \{6, 7\}, \{1,8\}, \{2,7\} \} \\  
    \end{cases}. 
\end{align}
\begin{proposition}[\citet{shen2022connect}]
    \label{prop:separation}
     With the above construction for the input space $\inputspace$, unlabeled distribution $\unlabeldist$, and data augmentation $\sA_{\text{prop}}$, for some feature dimension $k \in \Z^+$ a linear probe trained on contrastive pre-trained features achieves 0 target error:
     $\Lzeroone(\empclf) = 0$.
    However, for all $\embeddim\in\Z^+$, there exists a minimizer $\empclf_\text{erm}$ of the ERM objective (with data augmentations according to $\sA_{\text{prop}}$) that has non-zero error: $\Lzeroone(\empclf_\text{erm}) = 1/3$.
\end{proposition}  

\paragraph{ERM with targeted augmentations can get high OOD error.}
In general, we proceed by defining the following targeted augmentation, which allows us to reduce to the setting of Proposition~\ref{prop:separation}:
\begin{align}
    \label{eqn:targeted_aug}
        \augft(\inputxp \mid \inputx) =
    \begin{cases}
        1 & \{\inputxp, \inputx\} \in \{1,4\}, \{2,3\} \\
        1 & \inputx=\inputxp \text{ and } \inputx \notin \{1,2\}\\
        0 & \text{otherwise}
    \end{cases}
\end{align}
which transforms input 1 to 4 and the input 2 to 3, while keeping all other inputs the same.
Since the ERM with augmentations objective will not contain a term involving inputs 5,6,7, or 8 and thus the prediction on these inputs do not affect the objective, there exists a minimizer of the ERM objective (Equation~\ref{eq:erm-object-setup}) that predicts the wrong label for inputs 5,6,7,8 and has target error 2/3.
This is because these nodes are unreachable via augmentations of the source inputs, and thus the ERM objective can be minimized with any arbitrary prediction on these inputs.

\paragraph{\Sft has high OOD error.}
By Proposition~\ref{prop:separation}, \sft after contrastive pretraining has zero target (OOD) error when the pretraining augmentations do not have swapped edges.
By symmetry, \sft (contrastive pretraining + linear probing) on our augmentation graph with pretraining augmentations $\aug$ outputs the opposite label for all target inputs, resulting in an OOD error of 1. This is because the source and target domains are misaligned in our augmentation graph.

\paragraph{Connect Later achieves zero OOD error.}
Connect Later applies targeted augmentations $\augft$ during the linear probing step (on top of contrastive pretrained representations). This choice of targeted augmentations reduces to the setting of Proposition~\ref{prop:separation} where the labeled source domain consists of the inputs 3,4 instead. By the symmetry of the graph and applying Proposition~\ref{prop:separation}, Connect Later achieves 0 OOD error.

%% file: chapters/conclusion.tex
Time domain astronomy and SN Ia cosmology are entering a new era of big data, with next-generation surveys on the horizon that will discover more time domain objects and SNe Ia than ever before. This unprecedented data volume will allow us to perform SN Ia cosmological analyses with remarkable precision, discover new and rare types of time domain objects, and deepen our understanding of the progenitor population and physical processes underlying these objects. However, it also presents immediate challenges to our traditional analysis framework, such as our heavy reliance on spectroscopic follow-up observation to determine SN properties. This thesis investigates and proposes solutions to these issues to ensure optimal science returns for these datasets.

We study photometric supernova classification, which is an important step towards reducing our reliance on spectroscopy for SN Ia cosmology. We present SCONE, a convolutional neural network model capable of finding type Ia with $>99$\% accuracy and classifying 6 SN types with $>95$\% accuracy. SCONE is currently an integral part of DES, LSST, and Roman analysis pipelines. We also demonstrate that SCONE can classify 6 SN types with up to 75\% accuracy on the night of trigger, which will be essential for optimal allocation of spectroscopic resources. We then turn our attention to measuring redshift for SNe Ia, typically a spectroscopic redshift measurement of the host galaxy. We investigate the directional light radius host galaxy matching technique for DES-like simulations and quantify the cosmological biases associated with incorrect host matches. We also present an alternative method for redshift estimation, Photo-zSNthesis, which predicts a redshift PDF from photometric SNe Ia data. Finally, we propose a new method for improving the robustness of machine learning algorithms trained on an unrepresentative training set, Connect Later, and demonstrate significant improvements on the Photometric LSST Astronomical Classification Challenge dataset as well as tumor detection and wildlife identification benchmark datasets.

The future is an exciting new frontier for time domain astrophysics and SN Ia cosmology. Large real-time data streams from surveys like LSST will necessitate the development of automated photometric anomaly detection methods in addition to models that classify into known types. Anomaly detection methods will be able to surface new and rare time domain events that we could observe for the first time thanks to the power of the Vera C. Rubin Observatory and Roman Space Telescope. In addition to photometric time-series data, we often have other modes of data available for a particular time domain object (e.g., spectra, images, auxiliary information about the host galaxy environment). A powerful perspective on multi-modal data is as a natural data augmentation, which disentangles spurious features (e.g., wavelength or type of observation) from intrinsic physical properties (e.g., type of object).
In astronomy, multi-modal models can combine different ``views" of the same object to construct a more complete understanding of the object's physical processes. Finally, the increasing reliance on machine learning models (e.g., for photometric classification or simulation-based inference) necessitates greater awareness of the robustness of these models and the dangers of unrepresentative training samples.

Looking forward for SN Ia cosmology, additional statistical infrastructure may allow us to maximize the constraining power of upcoming datasets. A statistical framework for incorporating uncertainties from photometric redshift estimates is needed for the LSST SN cosmology analysis, which will be the first to incorporate photometric SN Ia redshifts. In addition, current analysis pipelines implicitly assume a Gaussian likelihood for computing posteriors over cosmological parameters, while studies suggest that a more principled likelihood estimation can improve constraining power. Modern statistical techniques, such as simulation-based inference, is able to model arbitrary likelihood functions at a fraction of the computational cost of traditional methods and may offer a way forward.

This is an important moment in SN Ia and time domain science, as we are the ones catalyzing a multifaceted transition: from manual to automated, from spectroscopy- to photometry-oriented, from individual events to populations. The terabytes of data collected by LSST and the Roman Space Telescope will contain the answers to many pressing questions in cosmology and astrophysics, but we will need to build the tools to find them.